\documentclass[reviewcopy]{elsarticle}

\usepackage[reviewcopy]{adndt}
\usepackage{longtable}
\usepackage{inputenc}

\usepackage{amsmath}
\usepackage{multirow}
\usepackage{caption2}
\usepackage{amssymb}
\usepackage{dcolumn}
\usepackage[dvipdfm,bookmarks=true,colorlinks,%
            citecolor=blue,linkcolor=blue,hypertex, %
            breaklinks=true]{hyperref}
\biboptions{square,sort&compress}
\bibpunct[]{[}{]}{,}{n}{}{;}
\begin{document}
 \bibliographystyle{unsrt}

\begin{frontmatter}

\title{Systematics of capture and fusion dynamics in heavy-ion collisions}

\author[ZZU]{Bing Wang}

\author[ITP]{Kai Wen\fnref{x}}
\fntext[X]{Present address: Faculty of Pure and Applied Sciences,
           University of Tsukuba, Tsukuba 305-8571, Japan}

\author[ZZU]{Wei-Juan Zhao}

\author[ITP,NLHIA]{En-Guang Zhao}

\author[ITP,UCAS,NLHIA,CCNU]{Shan-Gui Zhou\corref{cor1}}
  \ead{sgzhou@itp.ac.cn}
  \cortext[cor1]{Corresponding author.}
%

\address[ZZU]{Department of Physics, Zhengzhou University, Zhengzhou
              450001, China}

\address[ITP]{CAS Key Laboratory of Frontiers in Theoretical Physics,\\
Institute of Theoretical Physics, Chinese Academy of Sciences, Beijing 100190,
China}

\address[NLHIA]{Center of Theoretical Nuclear Physics, National Laboratory
                of Heavy Ion Accelerator, Lanzhou 730000, China}

\address[UCAS]{School of Physics, University of Chinese Academy of Sciences,
Beijing 100049, China}

\address[CCNU]{Synergetic Innovation Center for Quantum Effects and Application,

              Hunan Normal University, Changsha, 410081, China}

\begin{abstract}
We perform a systematic study of capture excitation functions by using
an empirical coupled-channel model. In this model, a barrier distribution is
used to take effectively into account the effects of couplings between the
relative motion and intrinsic degrees of freedom. The shape of the barrier
distribution is of an asymmetric Gaussian form.
The effect of neutron transfer channels is also included in the
barrier distribution. Based on the interaction potential between the
projectile and the target, empirical formulas are proposed to determine the
parameters of the barrier distribution. Theoretical estimates for barrier
distributions and calculated capture cross sections together
with experimental cross sections of 220 reaction systems
with  $182 \leqslant Z_{\rm P}Z_{\rm T} \leqslant 1640$ are tabulated. The
results show that our empirical formulas work quite well in the energy region
around the Coulomb barrier. This model can provide prediction of capture
cross sections for the synthesis of superheavy nuclei as well as
valuable information on capture and fusion dynamics.
\end{abstract}

\end{frontmatter}

\newpage

\tableofcontents
\listofDfigures
\listofDtables
\clearpage

\section{Introduction}

In recent decades, the study of capture and fusion dynamics in heavy-ion
collisions has been a subject of intense experimental and theoretical
interests because the heavy-ion fusion not only is of central importance for nucleosynthesis
but also can reveal rich interplay between nuclear structure and reaction dynamics
\cite{Frobrich1984_PR116-337,Balantekin1998_RMP70-77,Dasgupta1998_ARNPS48-401,
Canto2006_PR424-1,Diaz-Torres2007_PLB652-255,
Zhang2011_SciChinaPMA54S1-6,Back2014_RMP86-317,Canto2015_PR596-1}.
Experimentally, the capture cross sections cannot be measured directly.
If the fast fission is neglected, the
capture cross sections are measured as a sum of the fusion and quasifission (QF)
cross sections,
\begin{equation}
 \sigma_\mathrm{capture} = \sigma_\mathrm{fusion} + \sigma_\mathrm{QF}.
\end{equation}
After two nuclei fuse, the newly formed excited compound nucleus cools down
through emitting particles or fission. Consequently, the fusion cross section is
the summation of the evaporation residual (EvR) and fusion-fission (FuF) cross
sections,
\begin{equation}
 \sigma_\mathrm{fusion} = \sigma_\mathrm{EvR} + \sigma_\mathrm{FuF}.
\end{equation}
For light and medium-heavy systems, each capture event leads to the formation
of a compound nucleus. Meanwhile, the fission is forbidden owing
to the high fission barrier of the compound nucleus. In this case,
$\sigma_\mathrm{capture} \simeq \sigma_\mathrm{fusion}\simeq
\sigma_\mathrm{EvR}$, thus the measured fusion
excitation functions can be used to study the capture and fusion processes
simultaneously. For heavy systems, especially for more
symmetric ones or those leading to superheavy compound nuclei, the
QF process comes in and competes with the fusion
process~\cite{Antonenko1993_PLB319-425,Adamian1997_NPA618-176,
Adamian1998_NPA633-409,
Adamian1999_PLB451-289,Diaz-Torres2000_PLB481-228,Zagrebaev2001_PRC64-034606,
Diaz-Torres2001_PRC64-024604,Aritomo2004_NPA744-3,Diaz-Torres2006_PRC74-064601,
Knyazheva2007_PRC75-064602,Huang2010_PRC82-044614,
Huang2011_PRC84-064619,Wang2012_PRC85-041601R,
Zhang2013_NPA909-36,Wang2014_JPCS515-012022,Banerjee2015_PRC91-034619}.
In some reactions, only a small number of the capture events go to fusion, hence
the fusion cross section is only part of the capture cross section.
In this case, it is necessary to clearly distinguish the capture and fusion processes
\cite{Siwek-Wilczynska2002_APPB33-451,Wang2008_PRC77-014603,Zhu2014_PRC89-024615}.
The whole process of the synthesis of superheavy nuclei can be
divided into three stages: i) the capture process in which
the projectile is captured by the target and
then a composite system is formed; ii) the process of the formation of a
compound nucleus, which competes against the quasifission;
iii) the survival process in which the excited compound nucleus cools down
through emitting neutrons or fission. Therefore,  it's very important to
examine carefully
these three steps in the study of the synthesis mechanism of
superheavy nuclei. In this work, we will only focus on the capture process.

Theoretically, the capture process is often treated
as a barrier penetration problem, i.e., the system overcomes or penetrates
through the potential barrier between the two interacting nuclei, then the
projectile is captured by the target, forming a composite system.
The single barrier penetration models (SBPM) have been applied to describe
successfully the capture (fusion) excitation functions for light reaction
systems. However, for medium-heavy and heavy systems,
a significant enhancement of sub-barrier fusion cross sections was
observed in comparison to predictions of SBPM \cite{Stokstad1980_PRC21-2427}.
The enhancement can be explained by the strong coupling between the relative
motion and intrinsic degrees of freedom (the vibration of nuclear surfaces, the
rotation of deformed nuclei, etc.) and the coupling to the nucleon transfer
channels
\cite{Dasso1983_NPA405-381,Dasso1983_NPA407-221,Broglia1983_PRC27-2433R}.
To account for these couplings, a more quantal approach, the coupled-channel
(CC) model, has been developed
\cite{Thompson1988_CPR7-167,Hagino1999_CPC123-143,Hagino2012_PTP128-1061}.
The CC model is very successful in describing fusion
excitation functions near the Coulomb barrier
\cite{Beckerman1988_RPP51-1047}.
However, for heavy systems, it is necessary to
take into account a large number of channels which is not easy to realize in the CC model. Moreover, the
structure information of the interacting nuclei are needed as inputs.
Therefore, full CC calculations become intractable in many cases
including many fusion reactions leading to superheavy nuclei.
In the eigenchannel framework,  the
couplings to other channels split the original single barrier
into a set of discrete barriers
\cite{Dasgupta1998_ARNPS48-401,Hagino2012_PTP128-1061}.
These barriers distribute around the original single barrier, each of them
has a certain weight representing the probability of encountering the
corresponding barrier. Therefore, the probability of fusion substantially
changes owing to the barrier distribution as compared with that from the SBPM.
Those barriers lower than the original single barrier are responsible for the
enhancement of the sub-barrier fusion cross section in comparison to the
predictions of SBPM. Based on the concept of the barrier distribution, several
empirical CC approaches have been developed
\cite{Siwek-Wilczynska2002_APPB33-451,Zagrebaev2001_PRC65-014607,
Zagrebaev2004_PAN67-1462,Liu2006_NPA768-80,Zhu2014_PRC90-014612,
Zhu2014_PRC89-024615}. Different shapes of the barrier
distribution were assumed in these empirical CC approaches
\cite{Siwek-Wilczynska2002_APPB33-451,Zagrebaev2001_PRC65-014607,
Zagrebaev2004_PAN67-1462,Liu2006_NPA768-80} and different methods
were proposed to determine the parameters of the barrier distributions
\cite{Reisdorf1985_NPA438-212,Siwek-Wilczynska2002_APPB33-451,
Zagrebaev2001_PRC65-014607,Zagrebaev2004_PAN67-1462,Liu2006_NPA768-80,
Atta2014_PRC90-064622}. In Refs. \cite{Zagrebaev2001_PRC65-014607,
Zagrebaev2004_PAN67-1462}, an asymmetric Gaussian shape was assumed for the
barrier distribution and dynamical deformations of the projectile and target
were introduced to explain the enhancement of fusion cross
sections at sub-barrier energies. In Ref. \cite{Liu2006_NPA768-80}, the authors
proposed that the weighting function of the barrier is the superposition of two
Gaussian functions. Actually, these empirical CC approaches provide an
alternative to the full CC calculations, especially in the cases where the full CC
calculations become intractable. The main shortcoming of empirical CC approaches
is the choice of the distribution function of barriers which are responsible
for couplings of relative motion
to intrinsic degrees of freedom. In most of the empirical CC approaches, the
barrier distribution has only one
maximum, while the experimental barrier distributions extracted from the capture
excitation functions, i.e.,
$d^2(E\sigma_{\rm
capture})/dE^2$~\cite{Rowley1991_PLB254-25,Rowley1992_NPA538-205}, usually
exhibit more complicated structure.
Note that besides the SBPM and the full and empirical CC
approaches,
several microscopic dynamics models, such as the time-dependent
Hartree-Fock (TDHF) theory
\cite{Umar2006_PRC74-021601R,Umar2012_PRC85-017602} and the quantum molecular
dynamics (QMD) model
\cite{Aichelin1991_PR202-233,Wang2002_PRC65-064608,Wang2004_PRC69-034608,
Zhang2015_SciChinaPMA58-112002}, have
been also used to explore the fusion dynamics
\cite{Feng2005_NPA750-232,Feng2008_NPA802-91,Guo2007_PRC76-014601,
Guo2008_PRC77-041301R,
Bian2008_PLB665-314,Bian2009_NPA829-1,Wen2013_PRL111-012501,Zhu2013_NPA915-90,
Wen2014_PRC90-054613,Dai2014_SciChinaPMA57-1618,Dai2014_PRC90-044609,
Wang2014_PRC90-054610,Jiang2014_PRC90-064618,Oberacker2013_PRC87-034611,
Wang2015_SciChinaPMA58-112001,Bourgin2016_PRC93-034604}.
In recent years, Sargsyan {\it et
al.} developed a quantum diffusion approach
\cite{Sargsyan2009_PRC80-034606,Sargsyan2010_EPJA45-125,
Sargsyan2011_PRC84-064614} for describing the capture process, which is based
on the quantum master equation for the reduced density matrix. In the
quantum diffusion approach, fluctuation and dissipation effects in collisions of
heavy ions are taken into account to model the couplings to various channels.
For addressing both dissipation
and decoherence effects on the capture (fusion) process, a coupled-channel
density-matrix approach based on the
Lindblad equation has been developed \cite{Diaz-Torres2008_PRC78-064604,
Diaz-Torres2010_PRC82-054617}. It was shown that the influence of
decoherence on capture (fusion) and inelastic scattering, which is caused by
the irreversible dissipation of flux or probability to the environment, is
strong. Moreover, energy-shifting formulas for reaction and
capture probabilities have been proposed for simplifying (or avoiding
impracticable) coupled-channel calculations~\cite{Diaz-Torres2014_PLB739-348}.
Based on reaction theory, a useful method were suggested for extracting capture cross sections
from observed quasi-elastic backscattering excitation functions~\cite{Sargsyan2013_PRC88-044606,Sargsyan2014_PRC90-064601,
Sargsyan2015_PRC92-054620}.

In the last forty years, a large number of fusion excitation functions, mostly
for light and medium-heavy systems, have been measured. This provides us a
possibility to make a systematic study on the capture and fusion dynamics.
In the present work, we collect and compile the measured capture (fusion)
excitation functions. We perform a systematic study of these capture
excitation functions by using an empirical coupled-channel (ECC) model in which
the coupled-channel effects are treated effectively by introducing a barrier
distribution. The shape of the barrier distribution is chosen to be an
asymmetric Gaussian form~\cite{Zagrebaev2001_PRC65-014607,
Zagrebaev2004_PAN67-1462}. Therefore, the barrier distribution is determined by
three parameters, i.e., the left width, the right width, and the central value.
The coupling effects of the neutron transfer channels are also considered in the
barrier distribution. The purpose of this systematic study is to find an
empirical way to determine the parameters of barrier distribution in the
ECC model. We will show that this ECC model can give a reasonably adequate
and systematic description
of the capture cross sections. Note that this model has been used to study the
CC effects in fusion reactions $^{32}$S + $^{94,96}$Zr and $^{40}$Ca +
$^{94,96}$Zr \cite{Wang2016_SciChinaPMA59-642002} and extended to describe the
complete fusion cross sections for the reactions involving weakly bound
nuclei at above-barrier energies~\cite{Wang2016_PRC93-014615}.

The paper is organized as follows. In Section~\ref{sec:ecc}, the
ECC model, the interaction potential between the two interacting
nuclei, and the empirical formulas determining the parameters of the barrier
distribution are introduced. The results and discussions of studying the
capture excitation functions are shown in Section~\ref{sec:res}
where the effect of the neutron transfer are also investigated.
A summary is given in Section~\ref{sec:sum}. In the appendix, the
calculated results are shown in Graphs 1--19. The parameters used in the
calculations are tabulated in Table~\ref{tab:total}. The collected data are
compiled in Table~\ref{tab:data} where the calculated capture cross sections are
also included.

\section{The empirical coupled-channel model}\label{sec:ecc}

First we give a sketch of our empirical coupled-channel model. In this model, as
what is done in several other similar approaches, the capture cross sections are
calculated in the framework of quantum tunneling through the potential barrier
between the two interacting nuclei. The nuclear potential is taken to be the
deformed Woods-Saxon form and the axially quadrupole shapes of the two nuclei
are considered. The barrier distribution of an asymmetric Gaussian form is used
to take effectively into account the effects of couplings between the relative
motion and intrinsic degrees of freedom. Empirical formulas
are proposed for the parameters of the barrier distribution according to the
static and dynamical deformations of the two colliding nuclei.
In our model, the coupling effects of the positive $Q$-value neutron transfer
(PQNT) channels are also considered by modifying the barrier distribution.
We give the details of the
empirical coupled-channel model in the following Subsections.

\subsection{The capture cross section}

The capture cross section at a given center-of-mass energy $E_{\rm c.m.}$
can be written as the sum of the cross section for
each partial wave $J$,
\begin{equation}\label{eq:sig_cap}
\sigma_{\rm capture}({E_{\rm c.m.}})=\frac{\pi\hbar^2}{2\mu E_{\rm c.m.}}
                \sum_{J}^{J_{\rm max}}(2J+1)T(E_{\rm c.m.},J),
\end{equation}
where~$\mu$~denotes the reduced mass of the reaction system and $T(E_{\rm
c.m.},J)$ denotes the
penetration probability. A ``pocket'' in the interaction
potential may appear due to the competition between the long-range
repulsive Coulomb interaction and the short-range attractive nuclear force. The
``pocket'' becomes shallower with the increase of angular momentum $J$ because
the contribution of the centrifugal potential becomes stronger. $J_{\rm max}$
is the critical angular momentum: For the partial wave with angular
momentum larger than $J_{\rm max}$, the ``pocket'' of the interaction
potential disappears \cite{Newton2004_PRC70-024605,Newton2004_PLB586-219}. The
interaction potential around the Coulomb barrier can be approximated by an
``inverted'' parabola. The analytical expression for the penetration
probability is given by the well-known Hill-Wheeler formula
\cite{Hill1953_PR089-1102}
\begin{equation}\label{eq:HW}
 T_{\rm HW}(E_{\rm c.m.},J) = \left\{1+\exp\left[
                    \frac{2\pi}{\hbar\omega(J)}
                    \left(\frac{\hbar^2J(J+1)}{2\mu R_{\rm B}^{2}(J)}
                         + B - E_{\rm c.m.}\right)
                         \right]\right\}^{-1},
\end{equation}
where $R_{\rm
B}(J)$~and~$\hbar\omega(J)$~are the position of the barrier and the curvature
for the $J$th partial wave, respectively. $\hbar\omega(J)$~is calculated as
\begin{equation}\label{eq:omega}
 \hbar\omega(J)=\left.\sqrt{-\frac{\hbar^2}{\mu}\frac{\partial^2}{\partial
 R^2}V(R,J)}\right|_{R=R_{\rm B}(J)}.
\end{equation}
Note that for deep sub-barrier penetration, Eq.~(\ref{eq:HW}) is not
valid because of the long tail of the Coulomb potential.
In Ref.~\cite{Li2010_IJMPE19-359} a new barrier penetration formula was
proposed for potential barriers containing a long-range Coulomb interaction,
this formula is especially appropriate for the barrier penetration
with penetration energy much lower than the Coulomb barrier.
An enhancement in the
fusion cross section by several orders of magnitude was observed at
sub-barrier energies compared with the predictions of SBPM. This enhancement
can be attributed to the coupling between the relative motion of the two nuclei
and other degrees of freedom and couplings to PQNT channels.
These couplings lead to a distribution of barriers rather than a single
barrier. Therefore, a barrier distribution $f(B)$ is introduced to take
into account the coupled-channel effects in an empirical way. Then, the
penetration probability is calculated as
\begin{equation}\label{eq:Tran}
  T(E_{\rm c.m.},J) =\int f(B)T_{\rm HW}(E_{\rm c.m.},J,B){\rm d}B.
\end{equation}
In the present work, the barrier distribution is taken to be
an asymmetric
Gaussian function
\begin{equation}\label{eq:distri}
f(B)=\left\{
      \begin{array}{cc}
       \dfrac1N\exp\left[-\left(\frac{B-B_{\rm m}}{\varDelta_1}\right)^2\right],
                                                  \quad & B < B_{\rm m}, \\[1em]
       \dfrac1N\exp\left[-\left(\frac{B-B_{\rm m}}{\varDelta_2}\right)^2\right],
                                                          \quad & B > B_{\rm m}.
      \end{array}
     \right.
\end{equation}
$f(B)$ satisfies the normalization condition $\int
f(B)dB=1$. $N =\sqrt{\pi}(\varDelta_1+\varDelta_2)/2 $~is a normalization
coefficient. The left width $\varDelta_1$, the right width $\varDelta_2$, and the
central value $B_{\rm m}$ of the barrier distribution will be discussed in
Subsections \ref{sec:para_dis} and \ref{sec:para_nutr}.

Note that up to now, there is not a proof or mathematical derivation 
of Eq.~(\ref{eq:Tran}) and Eq.~(\ref{eq:distri}) based on the coupled Schr\"odinger equations. In the present work, we will use these simple expressions and
investigate its validity and usefulness by comparing theoretical results with
available data.

\subsection{The interaction potential}

\begin{figure}[htbp]
\centering{\includegraphics[width=.4\linewidth,height=1.5in]{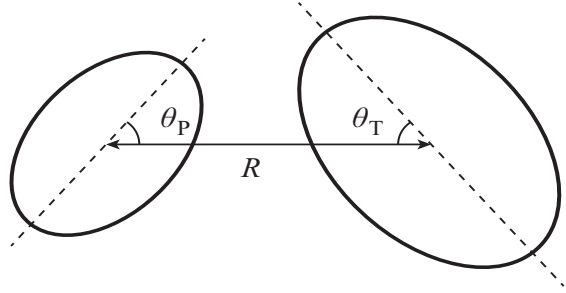}}
\caption{Relative position of deformed projectile and
target nuclei. $R$ is the relative distance between the two nuclei.
$\theta_{\rm P}$ ($\theta_{\rm T}$) denotes the angle between the
symmetry axis of the deformed projectile (target) and the collision
axis.}\label{fig:position}
\end{figure}

The interaction potential is one of the key quantities in the calculation of the
penetration probability. In the present work the effective interaction
potential of the two nuclei consists of the long-range Coulomb
repulsive potential, the attractive short-range nuclear potential and the
centrifugal potential,
 \begin{equation}\label{eq:Totpot}
V(R,J,\beta_{\rm P}^0,\beta_{\rm T}^0,\theta_{\rm P},\theta_{\rm T})
 =  V_{\rm C}(R,\beta_{\rm P}^0,\beta_{\rm T}^0,\theta_{\rm P},\theta_{\rm T})
 +  V_{\rm N}(R,\beta_{\rm P}^0,\beta_{\rm T}^0,\theta_{\rm P},\theta_{\rm T})
 +  \frac{\hbar^2J(J+1)}{2\mu R^2}.
 \end{equation}
Here~$R$~is the distance between the centroids of the two interaction
nuclei and $\beta_{i}^0$'s~($i={\rm P,T}$) are the ground state quadrupole
deformations of the projectile and target nuclei.  $\theta_i$'s~($i={\rm P,T}$)
denote the
angles measured between the symmetry axes of deformed nuclei and the collision
axis, as shown in Fig.~\ref{fig:position}.

For the nuclear part of the potential between two deformed nuclei, we use the
three-parameter Woods-Saxon potential \cite{Wong1973_PRL31-766} which is
widely used in the description of the heavy-ion collisions,
\begin{eqnarray}\label{eq:Npot}
V_{\rm N}(R,\beta_{\rm P}^0,\beta_{\rm T}^0,\theta_{\rm P},\theta_{\rm T})
       =\frac{-V_0}{1+\exp\left[(R-R_{\rm P}-R_{\rm T})
                                 /a\right]},
\end{eqnarray}
with
 \begin{eqnarray}
R_{\rm P} & = & R_{\rm 0P}[1+(5/4\pi)^{1/2}\beta_{\rm P}^0P_2(
                  {\rm cos}\theta_{\rm P})] \nonumber,\\
R_{\rm T} & = & R_{\rm 0T}[1+(5/4\pi)^{1/2}\beta_{\rm T}^0P_2(
                  {\rm cos}\theta_{\rm T})]  \nonumber, \\
   R_{0i} & = & r_0A_i^{1/3}, \quad i={\rm P, T},
\end{eqnarray}
where $V_0$ and $a$ denote the
depth and the diffuseness parameters of the potential, respectively. $r_0$ is
the radius parameter. The Coulomb interaction
between two deformed nuclei is
given as \cite{Wong1973_PRL31-766}
 \begin{equation}\label{eq:Cpot}
V_{\rm C}(R,\beta_{\rm P}^0,\beta_{\rm T}^0,\theta_{\rm P},\theta_{\rm T})
   = \mbox{} \frac{Z_1Z_2e^2}{R}
   +   \left(\frac{9}{20\pi}\right)^{\tfrac{1}{2}}
       \left(\frac{Z_1Z_2e^2}{R^3}\right) \sum_{i={\rm P,T}}
         R_{0i}^2\beta_i^0P_2({\rm {\rm cos}}\theta_i)
   +   \left(\frac{3}{7\pi}\right)\left(\frac{Z_1Z_2e^2}{R^3}\right)
       \sum_{i={\rm P,T}}R_{0i}^2[\beta_i^0P_2({\rm cos}\theta_i)]^2,
 \end{equation}
where~$P_2({\rm cos}\theta_i)$~ is the Legendre polynomial.

The depth~$V_0$, the radius parameter $r_0$, and the diffuseness
parameter~$a$~are chosen as,
   \begin{eqnarray}\label{eq:pot_para}
     V_0 & = & 80~{\rm MeV}, \nonumber\\
     r_0 & = & 1.16~{\rm fm},\nonumber\\
      a  & = &\left\{1.17\left[1+0.53\left(A_{\rm P}
              ^{-1/3}+A_{\rm T} ^{-1/3}\right)\right]
\right\}^{-1}~{\rm fm}.
\end{eqnarray}

\subsection{The parameters of the barrier distribution}
\label{sec:para_dis}

Based on Eqs.~(\ref{eq:HW}-\ref{eq:Totpot}), we fit the parameters of the
barrier distribution to the experimental values and get the optimal parameter
sets in Subsection \ref{subs:fitting}. Furthermore, we also propose empirical
formulas to calculate the parameters of the barrier distribution in Subsection
\ref{subs:cal}.

\subsubsection{Barrier distribution parameters from fitting}
\label{subs:fitting}

For the reactions with measured capture
excitation functions, we can get the three parameters of the barrier
distributions by fitting the data. From these fitted parameters,
one may
learn some systematics of the parameters of the barrier distributions.
The three parameters of the barrier distribution
are fitted by using the Levenberg-Marquardt method
\cite{Press1996_Nume_Recipes_2}. In the present work, we use a $\chi^2$ merit
function to determine the best-fit parameters by searching the minimum of
$\chi^2$, the $\chi^2$ merit function is
\begin{eqnarray}\label{eq:chi}
\chi^2 = \sum_{i=1}^{N}
         \left[\frac{\sigma_{\rm th}(E_i) -\sigma_{\rm exp}(E_i)}
               {\delta\sigma_{\rm exp}(E_i)} \right]^2.
\end{eqnarray}
For all the reaction systems, the error of the
cross section $\delta\sigma_{\rm exp}$ is assumed to be 3$\%$ of the
corresponding measured cross section $\sigma_{\rm exp}$. The fitted results
for the three parameters are listed in Table \ref{tab:total}.

\subsubsection{Barrier distribution parameters from empirical formulas}
\label{subs:cal}

From fitting, one can only obtain the parameters of the barrier distributions
for those reactions with measured capture excitation functions available.
To also make predictions, we propose empirical
formulas to calculate the parameters of the barrier distribution. In the
ECC model, the barrier distribution is related to couplings to
rotational states and low-lying collective vibrational states.
Nuclear rotational states are related to static deformations,
the vibrational modes are connected to the change of nuclear
shape. Furthermore, when the two nuclei come close enough to each other,
both nuclei are distorted owing to the attractive nuclear force and the
repulsive Coulomb force, thus dynamical deformations develop
\cite{Wang2012_PRC85-041601R}.
Therefore, the barrier distribution parameters for a reaction system can be
determined by the shapes of the projectile and target
and the development of dynamical deformations. In the
present work, we focus on a two-dimensional PES and examine the interaction
potential of the two nuclei. The interaction potential of a reaction system is
given as
  \begin{equation}\label{eq:pot_dy}
 V(R,\beta_{\rm P},\beta_{\rm T},\theta_{\rm P},\theta_{\rm T})
 = V_{\rm C}(R,\beta_{\rm P},\beta_{\rm T},\theta_{\rm P},\theta_{\rm T})
 + V_{\rm N}(R,\beta_{\rm P},\beta_{\rm T},\theta_{\rm P},\theta_{\rm T})
 + \frac{1}{2}\sum_{i={\rm P,T} }C_i\Bigl(\beta_i-\beta_i^0\Bigr)^2,
  \end{equation}
where~$\beta_{i}$'s~($i={\rm P,T}$) denote quadrupole deformations of the
projectile and the target.~$\beta_{i}^0$'s~are the ground
state quadrupole deformations. The total deformation
of the reaction system is defined as $\beta= \beta_{\rm P}+ \beta_{\rm T}$
and the dynamical deformations are defined as
$\delta \beta_{i}= \beta_{i}-\beta_{i}^0$.
In the present work, we assume a tip-tip orientation between the two deformed
nuclei,
i.e., $\theta_i = 0^\circ~(90^\circ)$ if the nucleus is prolate (oblate).
In order to reduce the number of parameters, we assume that the dynamical
deformation energies of the two nuclei are proportional to their mass number:
$C_{\rm P}\delta\beta_{\rm P}^2/C_{\rm T}\delta\beta_{\rm T}^2=A_{\rm
P}/A_{\rm T}$ \cite{Wang2012_PRC85-041601R}. Thus
only one parameter $\delta
\beta=\delta\beta_{\rm P}+\delta\beta_{\rm T}$~is
introduced. $C_i$'s ($i={\rm P,T}$) are the stiffness parameters of the nuclear
surface, which can be calculated with the liquid drop model
\cite{Bohr1998_Nucl_Structure_2}
\begin{eqnarray}\label{eq:sur_para}
C_i=(\lambda-1)\left[(\lambda+2)R_{0i}^2\sigma-\frac{3}{2\pi}\frac{Z^2_i e^2}
                     {R_{0i}(2\lambda+1)} \right],
\end{eqnarray}
where $R_{0i}$~is the radius of the nucleus. Here, we only take into account the
quadrupole deformation ($\lambda=2$). $\sigma$ is the coefficient of surface
tension which satisfies~$4\pi R^2_{0i}\sigma=a_sA^{2/3}_i$ and $a_s=18.32$ MeV is the
surface energy parameter \cite{Bohr1998_Nucl_Structure_2}.

\begin{figure}[ht!]
\centering{\includegraphics[width=.4\linewidth,height=2in]{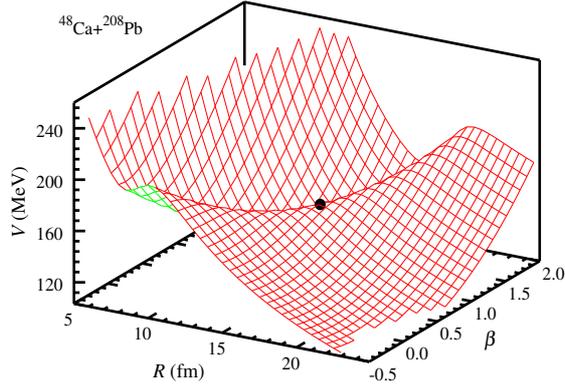}}
\caption{Potential energy surface calculated with Eq. (\ref{eq:pot_dy}) for the
reaction system $^{48}{\rm Ca}+{}^{208}{\rm Pb}$. The black dot denotes the
saddle point.}
\label{fig:PES}
\end{figure}

\begin{figure}[ht!]
\centering{\includegraphics[width=.35\linewidth,height=2in]{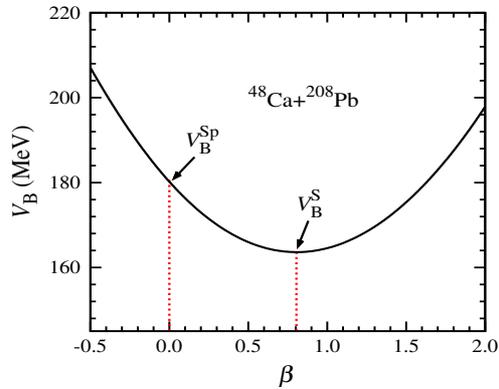}}
\caption{Coulomb barrier for the reaction system $^{48}{\rm Ca}+{}^{208}{\rm
Pb}$ as a function of the total deformation, i.e., the potential energy at the
ridge of the potential energy surface (see Fig. \ref{fig:PES}). $V_{\rm B}^{\rm
Sp}$ is the Coulomb barrier of the two spherical nuclei ($\beta_{\rm P}=0$ and
$\beta_{\rm T}=0$). $V_{\rm B}^{\rm S}$ is the
Coulomb barrier at the saddle point.}
\label{fig:R_PES}
\end{figure}

Figure \ref{fig:PES} shows the PES of the reaction system
$^{48}{\rm Ca}+{}^{208}{\rm Pb}$. One can find that the Coulomb barrier
changes with the total deformation.
The ridge of the PES, i.e., the Coulomb barrier as a function of
the total deformation of the system, is shown in Fig. \ref{fig:R_PES}.
Note that both of the two nuclei are doubly magic and spherical in their
ground states, but the total dynamical deformation of the system is 0.81 at the
saddle point. This large dynamical deformation develops due to the combined
effect of the attractive nuclear force and the repulsive Coulomb force. As one
can see, with $\beta$ increasing for the prolate configuration, the Coulomb
barrier first decreases and then increases. A minimum appears in the curve which
is the saddle point of the PES. Two characteristic configurations $V_{\rm
B}^{\rm Sp}$ and $V_{\rm B}^{\rm S}$ in the PES are also shown in Fig.
\ref{fig:R_PES}. $V_{\rm B}^{\rm Sp}$ denotes the Coulomb barrier of the
configuration with two spherical nuclei. $V_{\rm B}^{\rm S}$ denotes the Coulomb
barrier at the saddle point.

For $^{48}{\rm Ca}+{}^{208}{\rm Pb}$, $V_{\rm B}^{\rm Sp}$ is
180.2 MeV and $V_{\rm B}^{\rm S}$ is 163.6 MeV. The difference of the Coulomb
barriers at these two configurations is 16.6 MeV. It is quite large, similar as
the deformation energy at the saddle point which is 12.44 MeV.
The barriers can distribute in a wide range due to the development of the
dynamical deformation, leading to large widths
of the barrier distribution. In the present work, we assume that the central
value $B_{\rm m}$ of the barrier distribution is between $V_{\rm B}^{\rm
Sp}$ and $V_{\rm B}^{\rm S}$. These two characteristic configurations are used
to calculate $B_{\rm m}$. The widths of the barrier distributions are related
to the deformation energy at the saddle point. Finally the parameters of the
barrier distribution are chosen as follows
\begin{eqnarray}\label{eq:disp}
     B_{\rm m}& = &a V_{\rm B}^{\rm Sp}+ (1-a)V_{\rm B}^{\rm S},  \\
   \varDelta_1& = & bE_{\rm D} ,  \\
   \varDelta_2& = & cE_{\rm D},
\end{eqnarray}
with
\begin{eqnarray}\label{eq:dy_en}
E_{\rm D} = \frac{1}{2}\sum_{i={\rm P,T}}C_i\left(\beta_i^{\rm S}\right)^2.
\end{eqnarray}
Here $E_{\rm D}$ is the deformation energy of the reaction system at the saddle
point. $\beta_i^{\rm S}$'s are the deformation parameters of the projectile and
the target at the saddle-point configuration. For systems in
which both of the two nuclei are spherical, we found that the ECC model
with the following set of the parameters $a, b, {\rm and}~c$ can give a good
description of the data,
\begin{eqnarray}\label{eq:disp_s}
   a & = & \left\{
           \begin{array}{cc}
             0.26, \quad  &  Z_{\rm P}Z_{\rm T}<1150,  \\[1em]
             0.5, \quad  &  Z_{\rm P}Z_{\rm T}\ge1150,
           \end{array}
           \right.   \nonumber \\
   b & = & 0.32, \nonumber \\
   c & = & 0.93.
\end{eqnarray}
For deformed systems, the ECC model with a more asymmetric shape of
the barrier distribution works better,
\begin{eqnarray}\label{eq:disp_d}
   a & = & \left\{
           \begin{array}{cc}
             0.23, \quad  &  Z_{\rm P}Z_{\rm T}<1150,  \\[1em]
             0.37, \quad  &  Z_{\rm P}Z_{\rm T}\ge1150,
           \end{array}
           \right.    \nonumber \\
   b & = & 0.12, \nonumber \\
   c & = & 1.12.
\end{eqnarray}

\subsection{The positive $Q$-value neutron transfer channels}
\label{sec:para_nutr}

A large enhancement of sub-barrier fusion cross sections have been
experimentally observed in many reaction systems
with positive $Q$-value neutron transfer
channels~\cite{Beckerman1980_PRL45-1472,Timmers1997_PLB399-35,
Stefanini2007_PRC76-014610,
Zhang2010_PRC82-054609,Kolata2012_PRC85-054603,Kohley2013_PRC87-064612,
Jia2014_PRC89-064605}. Many efforts have been made to study the effect of the
neutron transfer \cite{Beckerman1980_PRL45-1472,Zagrebaev2003_PRC67-061601R,
Zagrebaev2004_PTPSupp154-122,Rachkov2014_PRC90-014614,
Stelson1990_PRC41-1584,Jia2012_PRC86-044621,Jiang2014_PRC89-051603R,
Scamps2015_PRC91-024601,Zhang2014_PRC89-054602,Sargsyan2012_PRC86-014602,
Sargsyan2013_PRC88-064601,Sargsyan2015_PRC91-014613}. Until now the role of the
PQNT effect is still not very clear. Different viewpoints are proposed to
understand
the role of the PQNT effect.
In Refs.~\cite{Zagrebaev2003_PRC67-061601R,Zagrebaev2004_PTPSupp154-122,
Rachkov2014_PRC90-014614}, the authors suggested that the PQNT channels can
provide a gain in the kinetic energy, and thus fusion is favored.
In Ref.~\cite{Stelson1990_PRC41-1584}, it was proposed that
a neutron flow between the projectile and the target nuclei before fusion could
promote neck formation which provides a force strong enough to overcome the
Coulomb force. Sargsyan {\it et al.} suggested that the deformations of
the interacting nuclei change owing to the PQNT
\cite{Sargsyan2012_PRC86-014602,Sargsyan2013_PRC88-064601,
Sargsyan2015_PRC91-014613}. Thus, the influence
of the PQNT channels on fusion is accompanied by the change of the nuclear
deformations. In the present work, the effect of the coupling to the PQNT
channels is simulated by broadening the barrier distribution.
Furthermore, only
one neutron pair transfer channel is considered in the present model. When the
$Q$ value for one neutron pair transfer is positive, the widths of the barrier
distribution are calculated as
\begin{eqnarray}\label{eq:NeuT}
  \varDelta_1&\rightarrow& gQ(2n)+\varDelta_1,\nonumber  \\
  \varDelta_2&\rightarrow& gQ(2n)+\varDelta_2,
  \end{eqnarray}
where $g$ is a dimensionless constant, $Q(2n)$ is the $Q$ value for one
neutron pair transfer, and $\varDelta_1$ and $\varDelta_2$ are calculated with
Eqs.~(\ref{eq:disp}-\ref{eq:disp_d}). For all reactions with
positive $Q$ value for one neutron pair transfer channel, $g$ is taken
as $0.32$. For one neutron pair transfer from target to projectile,
$Q(2n)$ is given as
\begin{equation}
Q(2n) = M(Z_{\rm P},A_{\rm P})  + M(Z_{\rm T},A_{\rm T})
       -M(Z_{\rm P},A_{\rm P}+2)- M(Z_{\rm T},A_{\rm T}-2),\nonumber
\end{equation}
and for one neutron pair transfer from projectile to target,  $Q(2n)$ is
given as
\begin{equation}\label{Q2n}
Q(2n) = M(Z_{\rm P},A_{\rm P})  + M(Z_{\rm T},A_{\rm T})
       -M(Z_{\rm P},A_{\rm P}-2)- M(Z_{\rm T},A_{\rm T}+2).
\end{equation}
Nuclear masses are taken from Refs.~\cite{Audi2012_ChinPhysC36-1287,Wang2012_ChinPhysC36-1603}.

\section{Results and Discussions}\label{sec:res}

We have collected 220 sets of capture excitation functions for reactions with  $182
\leqslant Z_{\rm P}Z_{\rm T} \leqslant 1640$. Among these 220
reaction systems, there are 89 with positive $Q$ value for
one neutron pair transfer channel. In the present work, the nuclear
deformation parameters are taken from Ref.~\cite{Moller1995_ADNDT59-185}. We
assume that the involved nuclei are prolate except for
$^{12}$C ($\beta^0=-0.32$ from the multi-dimensionally constrained relativistic
mean field
model~\cite{Lu2012_PRC85-011301R,Lu2014_PRC89-014323,Zhou2016_PS91-063008}),
$^{27}$Al, and
$^{28}$Si. In order to achieve a better agreement with the experiment, the
ground state quadrupole deformation parameters
of $^{194,198}$Pt  are taken from Ref.~\cite{Goriely2002_PRC66-024326} and for
$^{35}$Cl, $\beta^0=0.1$ is used. For each reaction system, the quadrupole
deformation parameters of the projectile and the target nuclei,
the position and curvature of the barrier
between the two nuclei, the fitted (c.f. Subsection \ref{subs:fitting})  and
calculated (c.f. Subsection \ref{subs:cal}) results for the three
parameters of the barrier distribution, and the $Q$ value for one neutron
pair transfer are tabulated in Table~\ref{tab:total}. In this Section, we first
introduce the collection of the data. Then we define an average deviation to
represent the agreement between the data and fitted (or calculated) values of
the cross sections. In the last two Subsections, we show the comparison between
the calculated capture cross sections and the experimental values for these 220
reactions and discuss in detail several typical examples. For convenience, we
call the cross sections from our ECC model with fitted parameters
``fitted'' cross sections and those with parameters determined by
Eqs.~(\ref{eq:disp}-\ref{eq:disp_d}) ``calculated'' cross sections.

\begin{table}[!htb]
\centering{
\renewcommand{\arraystretch}{0.6}
\footnotesize
\caption{\footnotesize  Two reactions $^{18}$O+$^{74}$Ge and
$^{17}$O+$^{144}$Sm used to test our procedure for extracting the data from
the authors' graphs in Refs. \cite{Leigh1995_PRC52-3151,Jia2012_PRC86-044621}.
The two reactions and the corresponding references are tabulated in the first
column. The experimental values of the capture (fusion) cross sections
taken from the authors' tables are listed in columns 2--5 (the incident
energies $E^{\rm Exp.}$ in the center-of-mass frame for the former and in the
laboratory frame for the latter, the fusion cross section $\sigma_{\rm F}^{\rm Exp.}$
and the errors of the cross section $\delta\sigma_{\rm F}^{\rm Exp.}$). The
experimental values of the capture (fusion) cross sections extracted from the
authors' graphs are listed in columns 6 and 7 (the incident energies $ E^{\rm
Ext.}$ and the fusion cross section $\sigma_{\rm F}^{\rm Ext.}$). The last
column shows relative deviations between the tabulated
values of the cross sections and those extracted from the authors'
graphs ($\Delta\sigma=\sigma_{\rm F}^{\rm Exp.}-\sigma_{\rm F}^{\rm Ext.}$).}
\label{test}
\begin{tabular}
{@{\extracolsep\fill}l
D{.}{.}{-1}D{.}{.}{-1}D{.}{.}{-1}D{.}{.}{-1}D{.}{.}{-1}D{.}{.}{-1}D{.}{.}{-1}}
\hline\noalign{\vskip2pt}
 \multirow{2}{*}{Reaction}                                &
 \multicolumn{1}{c}{ $ E_{\rm c.m.}^{\rm Exp.} $  }       &
 \multicolumn{1}{c}{ $ E_{\rm lab}^{\rm Exp.} $  }        &
 \multicolumn{1}{c}{$\sigma_{\rm F}^{\rm Exp.}$}          &
 \multicolumn{1}{c}{$\delta\sigma_{\rm F}^{\rm Exp.}$}    &
 \multicolumn{1}{c}{$ E^{\rm Ext.} $ }                    &
 \multicolumn{1}{c}{$\sigma_{\rm F}^{\rm Ext.}$}          &
 \multicolumn{1}{c}{$\Delta\sigma/\sigma_{\rm F}^{\rm Exp.}$}        \\

                                               &
  \multicolumn{1}{c}{   (MeV) }                &
  \multicolumn{1}{c}{   (MeV) }                &
  \multicolumn{1}{c}{  (mb) }                  &
  \multicolumn{1}{c}{  (mb) }                  &
  \multicolumn{1}{c}{ (MeV) }                  &
  \multicolumn{1}{c}{ (mb) }           &
  \multicolumn{1}{c}{ (\%) } \\ \noalign{\vskip2pt} \hline\noalign{\vskip2pt}
$^{18}$O+$^{74}$Ge \cite{Jia2012_PRC86-044621}
&30.2&  & 0.14  & 0.01   & 30.27  & 0.14 &  0.00\\
&30.5&  & 0.25  & 0.01   & 30.50  & 0.25 &  0.00\\
&31.0&  & 0.89  & 0.03   & 31.01  & 0.90 & -1.12\\
&31.5&  & 1.84  & 0.05   & 31.52  & 1.85 & -0.54\\
&32.0&  & 4.56  & 0.08   & 32.03  & 4.55 &  0.22\\
&32.5&  & 10.0  & 0.1    & 32.50  & 10.02 & -0.20\\
&33.0&  & 16.7  & 0.2    & 33.01  & 16.71 & -0.06\\
&33.5&  & 31.5  & 0.3    & 33.49  & 31.54 & -0.13\\
&34.0&  & 46.9  & 0.4    & 34.00  & 47.17 &  -0.56\\
&34.5&  & 63.1  & 0.6    & 34.51  & 63.32 & -0.35\\
&35.0&  & 82.6  & 0.8    & 35.02  & 83.67 & -1.30\\
&35.5&  & 99.4  & 0.9    & 35.53  & 100.76 & -1.37\\
&36.0&  & 136.5  & 1.2   & 36.00  & 135.23 &  0.93\\
&36.5&  & 156.7  & 1.3   & 36.54  & 157.88 & -0.75\\
&37.0&  & 190.1  & 1.6   & 37.02  & 190.12 & -0.01\\
&37.5&  & 211.9  & 2.0   & 37.56  & 211.89 & 0.005\\
&38.0&  & 254.8  & 1.8   & 38.04  & 255.17 & -0.15\\
&38.9&  & 292.5  & 2.7   & 38.99  & 293.33 & -0.28\\
&39.9&  & 371.3  & 2.3   & 39.97  & 370.04 &  0.34\\
&40.9&  & 409.9  & 3.3   & 40.96  & 412.41 & -0.61\\
&41.9&  & 497.9  & 3.7   & 41.97  & 496.63 & 0.26\\
&43.0&  & 545.6  & 4.4   & 42.99  & 553.50 & -1.45\\
&44.0&  & 593.4  & 4.8   & 43.98  & 598.06 & -0.79\\
&46.0&  & 686.5  & 7.4   & 45.99  & 687.51 & -0.15\\
&46.9&  & 731.6  & 7.6   & 46.96  & 731.44 & 0.02\\
&49.0&  & 847.9  & 7.9   & 49.00  & 853.96 & -0.71\\
$^{17}$O+$^{144}$Sm \cite{Leigh1995_PRC52-3151}
&  & 60.88  & 0.07     & 0.05    &  60.87 & 0.07 & 0.00\\
&  & 61.38   & 0.11     & 0.07    &  61.31 &  0.11 & 0.00 \\
&  & 61.88   & 0.19     & 0.09    &  61.78 &  0.19 & 0.00\\
&  & 62.38   & 0.40     & 0.09    &  62.36 &  0.40 &  0.00\\
&  & 62.88   & 0.81     & 0.23    &  62.84 &  0.82 & -1.23\\
&  & 63.38   & 1.2      & 0.2     &  63.31 &  1.18 & 1.67\\
&  & 63.88   & 2.4      & 0.2     &  63.79 &  2.45 & -2.08   \\
&  & 64.88   & 5.7      & 0.3     &  64.79 &  5.75 & -0.88\\
&  & 65.88   & 13.9     & 0.5     &  65.79 &  14.08 & -1.29\\
&  & 66.88   & 30.8     & 0.6     &  66.79 &  31.56 & -2.47\\
&  & 67.88   & 57.2     & 0.7     &  67.79 &  58.46 & -2.20\\
&  & 68.88   & 90.5     & 0.9     &  68.79 &  93.52 & -3.34\\
&  & 70.01   & 137      & 1       &  69.90 &  141.06 & -2.96   \\
&  & 70.88   & 171      & 1       &  70.74 &  173.24 & -1.31 \\
&  & 71.88   & 209      & 1       &  71.75 &  215.91 & -3.31\\
&  & 72.88   & 253      & 2       &  72.75 &  261.31 & -3.28\\
&  & 73.88   & 296      & 2       &  73.80 &  302.63 & -2.24\\
&  & 74.88   & 345      & 2       &  74.75 &  350.48 & -1.59\\
&  & 75.88   & 383      & 2       &  75.75 &  394.15 & -2.91\\
&  & 76.88   & 428      & 2       &  76.81 &  443.27 & -3.57\\
&  & 77.88   & 463      & 3       &  77.75 &  477.03 & -3.03\\
&  & 78.88   & 508      & 3       &  78.76 &  520.95 & -2.55 \\
&  & 79.88   & 548      & 3       &  79.81 &  560.63 & -2.30 \\
&  & 89.88   & 852      & 4       &  89.82 &  870.82 & -2.21 \\
&  & 99.88   & 1072     & 7       &  99.84 &  1101.37 & -2.74 \\
\noalign{\vskip2pt} \hline
\end{tabular}
}
\end{table}

\subsection{Collection of the capture excitation functions}
\label{subs:coll}

220 sets of capture excitation functions are collected in the present work.
There are more reaction systems for which the capture cross sections have been
measured and some can be found in the NRV website \cite{NRVmisc}.
Since we focus on the capture and fusion dynamics
around the Coulomb barrier, only the reaction systems for which capture
excitation functions are measured in this energy region are considered in this
work. Note that reaction systems with weakly bound nuclei involved are
also excluded, because the coupling to the breakup channel is very
complicated, especially in the sub-barrier region
\cite{Diaz-Torres2002_PRC65-024606,Diaz-Torres2007_PRL98-152701,
Gomes2009_PRC79-027606,Gomes2012_JPG39-115103,Gomes2010_NPA834-151c,
Wang2014_PRC90-034612,Zhang2011_NPA864-128,Fang2013_PRC87-024604,
Fang2015_PRC91-014608,Guo2015_PRC92-014615,Hu2015_PRC91-044619,
Hu2016_PRC93-014621,Fang2016_PRC93-034615,Cortes2015_JPCS630-012017}. Actually, in Ref.~\cite{Wang2016_PRC93-014615}, 
we extended our ECC model by including the breakup effect which is described by
a prompt-breakup probability function. For the reactions induced by $^{9}$Be,
the extended ECC model together with the measured prompt-breakup
probabilities can give a good description of the complete fusion cross
sections at above-barrier energies~\cite{Wang2016_PRC93-014615}.

Among these 220 reactions, there are 71 reactions for which the values of the
cross sections are not given
in the corresponding references, but the measured
cross sections are shown in graphs. So we extract the data from the
authors' graphs for these reaction systems. For part of these 71 reactions, the
graphs are obtained from scanned PDF files. To test the reliability of our
procedure for extracting the data from the authors' graphs, we take  two
reactions, $^{18}$O+$^{74}$Ge and $^{17}$O+$^{144}$Sm for which the data are
given both in authors' tables and in authors' graphs
\cite{Leigh1995_PRC52-3151,Jia2012_PRC86-044621}, as examples. Note that, the
PDF file for Ref. \cite{Leigh1995_PRC52-3151} was a scanned one and that
for Ref. \cite{Jia2012_PRC86-044621} has a better quality. We compare the
extracted and tabulated
values of fusion cross sections in Table~\ref{test}. The tabulated
experimental values are listed in columns 2--5. Columns 6 and 7 show the fusion
cross sections extracted from the authors' graphs. The last column shows the
relative deviations $\Delta\sigma/\sigma_{\rm F}^{\rm Exp.}$ between them. The
deviation $\Delta\sigma$ is defined as
$\Delta\sigma=\sigma_{\rm F}^{\rm Exp.}-\sigma_{\rm F}^{\rm Ext.}$. For
$^{18}$O+$^{74}$Ge, the relative deviations are mostly within $1\%$. For
$^{17}$O+$^{144}$Sm, the relative deviations are around $2\%$ and the largest
one is $3.57\%$. This means that our procedure of
extracting the data from the authors' graphs is reliable.

\subsection{Average deviation of the fitted and calculated
cross sections from experimental values}

To show quantitatively the agreement between the fitted (or calculated) from
measured capture excitation functions, we introduce
an average deviation $\mathcal{D}$,
\begin{equation}\label{eq:dev}
\mathcal{D}=\frac{1}{n}\sum_{i=1}^n\left\{ \log\left[
            \frac{\sigma_{\rm th}(E_i)}{\sigma_{\rm exp}(E_i)}\right]\right\}^2.
\end{equation}
Here $n$ denotes the number of energy points for each set of experimental
capture excitation function. $\sigma_{\rm th}(E_i)$ and $\sigma_{\rm exp}(E_i)$
are the fitted (or calculated) capture cross section and the experimental value,
respectively.

\begin{figure}[htbp!]
\centering{\includegraphics[width=.4\linewidth,height=2in]{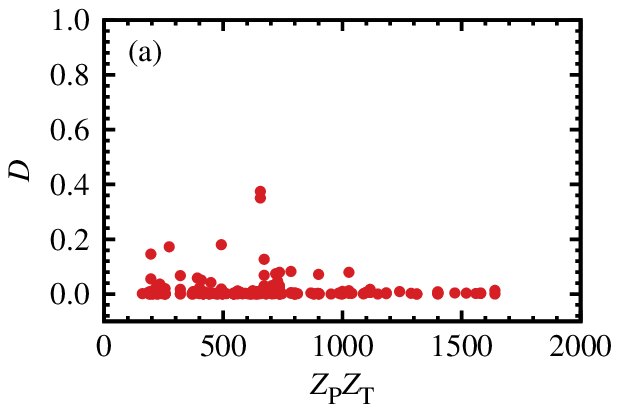}
\includegraphics[width=.4\linewidth,height=2in]{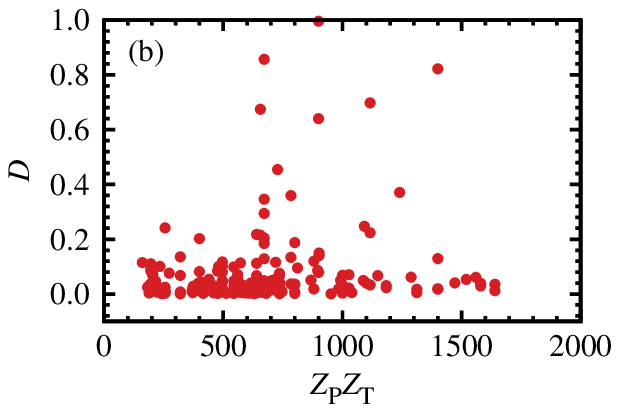}}
\caption{The average deviation $\cal D$ of the fitted (a) and calculated (b)
cross sections from the data for a total of 220
reaction systems as a function of $Z_{\rm P}Z_{\rm T}$.}
\label{fig:devi}
\end{figure}

Figure \ref{fig:devi} shows the average deviations of the fitted
[Fig.~\ref{fig:devi}\textcolor{blue}{(a)}] and calculated
[Fig.~\ref{fig:devi}\textcolor{blue}{(b)}] cross sections from
experimental values for these 220 reaction systems.
From Fig.~\ref{fig:devi}\textcolor{blue}{(a)}, one can see that most of the capture
excitation functions can be well reproduced by the ECC model with
the fitted barrier distribution parameters. For several
reaction systems, the data in the deep sub-barrier region are also used in the
fitting procedure. In some of these cases, we find that the fitted results can not
reproduce the data in the deep sub-barrier region, e.g.,
$^{16}$O+$^{204,208}$Pb. However, it works quite well in the energy region around the
Coulomb barrier. Figure~\ref{fig:devi}\textcolor{blue}{(b)} shows that our
ECC model with the barrier distribution calculated by
Eqs.~(\ref{eq:disp}-\ref{eq:NeuT}) can also describe well most
of the capture excitation functions; the average deviations for most reaction
systems are quite small. There
are some reaction systems for which the deviations of the calculated results
from the experiments are noticeable; further discussions about typical examples
of these systems will be given in the following two Subsections.

\subsection{Reaction systems with negative $Q$ value for one neutron pair
transfer}

Graphs 1--11 show the calculated capture cross sections of 131 reaction
systems for which the $Q$ values for one neutron pair transfer are negative.  In
these reactions, the neutron pair transfer channel is closed. The $Q$ values for
one neutron pair transfer, i.e., $Q(2n)$s are listed in Table \ref{tab:total}.
The data are represented by solid squares or circles and the corresponding
references are listed in Table~\ref{tab:data}. The arrow indicates the central
value of the barrier distribution, i.e., $B_\mathrm{m}$ in Eq. (\ref{eq:disp}).
The solid line denotes the calculated cross sections. In these
reactions, the vibrational and rotational couplings are responsible for the
enhancement of cross sections in the sub-barrier region as compared with
predictions of SBPM. From Graphs 1--11, it can be seen that the calculated
capture excitation functions of most of these 131 reaction
systems are in good agreement with the measured capture excitation functions.
As mentioned in Subsection \ref{subs:coll}, we focus on the capture and fusion
dynamics around the Coulomb barrier and collect capture excitation
functions measured in this energy region. Generally, for
reactions with negative $Q$ value for one neutron pair transfer channel,
the present model can describe the capture excitation functions very well at
energies around the Coulomb barrier. For some reactions, the deviation of the
calculated results from the experiment is noticeable. Next we discuss some
typical examples among these 131 reactions in detail.

\begin{figure}[htb]
\centering{\includegraphics[width=.6\linewidth]{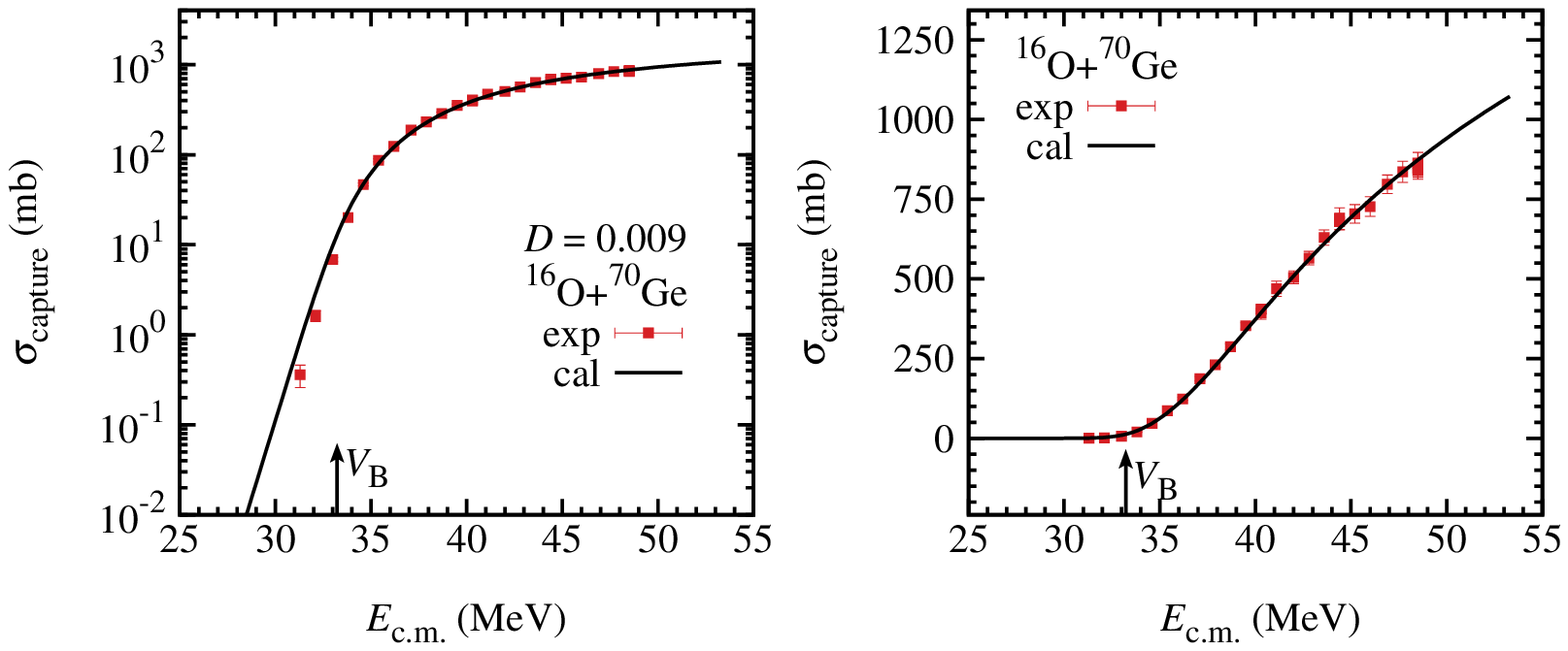}\\
\includegraphics[width=.6\linewidth]{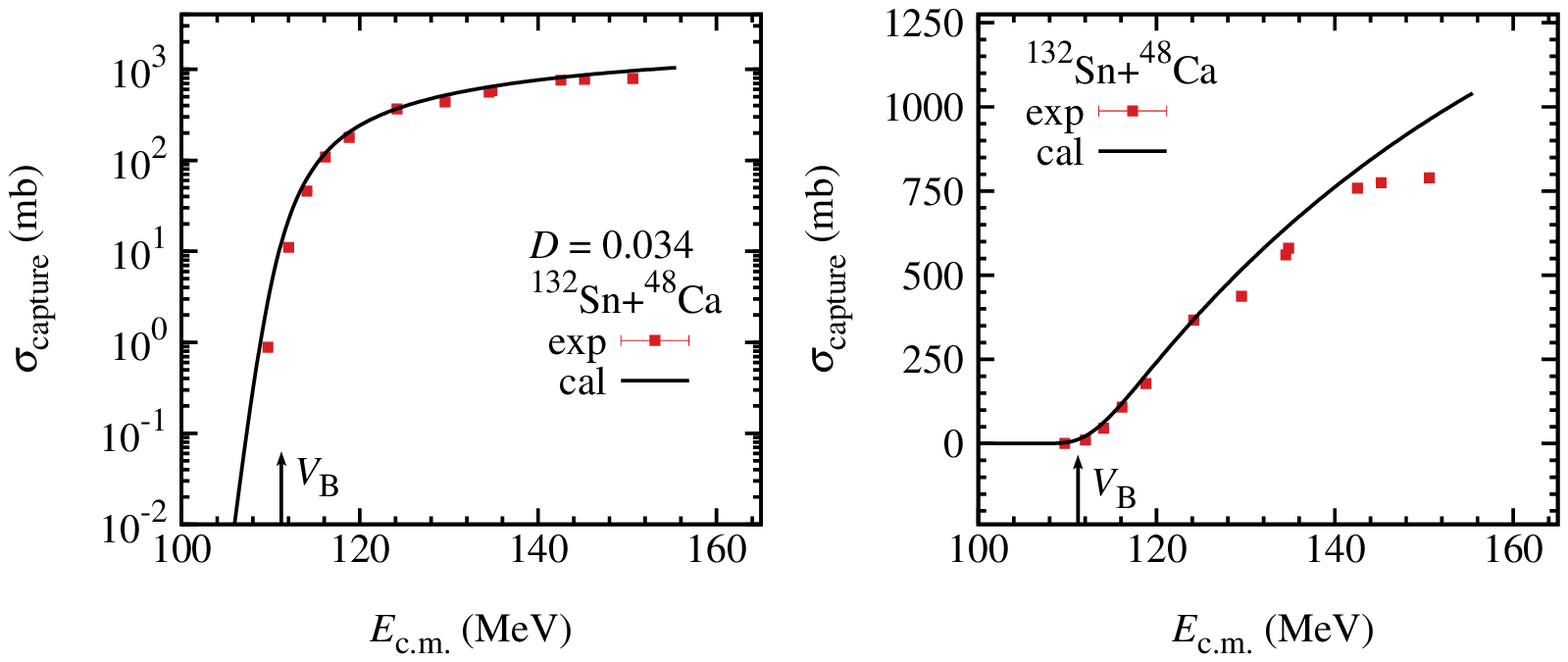}}
\caption{The experimental and calculated capture excitation functions
for $^{16}$O+$^{70}$Ge and $^{132}$Sn+$^{48}$Ca. The results are
shown in logarithmic scale (the left panel) and linear scale (the right panel).
The solid line denotes the calculated cross sections. The arrow indicates the
central value of the barrier distribution $B_\mathrm{m}$ given in Eq.
(\ref{eq:disp}). The solid squares show the experimental values which are given
in Table \ref{tab:data}. $\mathcal{D}$ denotes the average deviation of the
calculated cross sections from the experimental values.}
\label{fig:oge}
\end{figure}

\subsubsection{Typical examples}

For the reactions for which the calculated capture excitation functions are in
good agreement with the measured ones,
we select a light system $^{16}$O + $^{70}$Ge with $ Z_{\rm P}Z_{\rm T} = 256$
and a heavy system $^{132}$Sn + $^{48}$Ca with $Z_{\rm P}Z_{\rm T} = 1000$ as
examples. In the reaction $^{16}$O + $^{70}$Ge the target
$^{70}$Ge is deformed with the quadrupole deformation parameter $\beta=0.241$
(see Table~\ref{tab:total}). For the reaction $^{132}$Sn + $^{48}$Ca, both the
projectile and the target are doubly magic nuclei and their quadrupole
deformation parameters are zero (see Table~\ref{tab:total}). Therefore, $^{16}$O
+ $^{70}$Ge is a deformed reaction system and $^{132}$Sn + $^{48}$Ca is a
spherical reaction system. In Fig.~\ref{fig:oge}, the comparison of the
calculated capture cross sections with the experimental values for $^{16}$O +
$^{70}$Ge and $^{132}$Sn + $^{48}$Ca reactions is shown. The solid line denotes
the calculated cross sections. The solid squares are the experimental values
which are listed in Table \ref{tab:data}. The arrow indicates the central value
of the barrier distribution $B_\mathrm{m}$ given in Eq. (\ref{eq:disp}). For
both reactions, we can see that the calculated capture cross sections are in
good agreement with the data. For $^{132}$Sn + $^{48}$Ca, in
the well above-barrier region, the calculated cross sections overestimate the
experimental values. This may be attributed to the fact that only the EvR cross
sections are measured and the FuF cross sections is not taken into account at
high energies (See Table \ref{tab:data}). The average deviations
$\cal{D}$ defined in Eq. (\ref{eq:dev}) for $^{16}$O + $^{70}$Ge and $^{132}$Sn
+ $^{48}$Ca are 0.009 and 0.034, respectively.

\begin{figure}[htb]
\centering{\includegraphics[width=.6\linewidth]{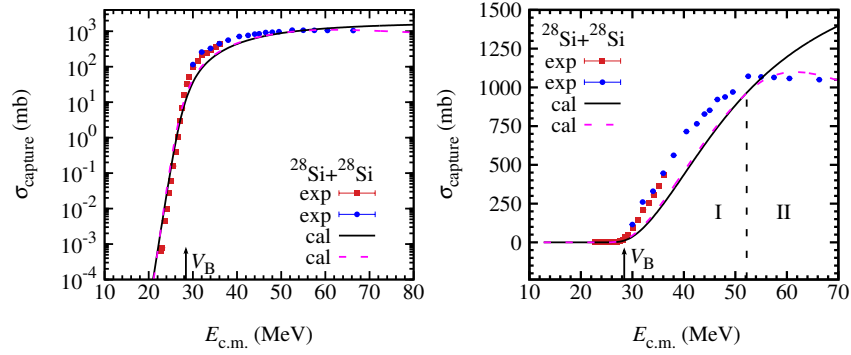}}
\caption{Same as Fig. \ref{fig:oge} but for $^{28}$Si+$^{28}$Si and the dash
line denotes the calculated results obtained by adjusting the parameters $V_0$
and $r_0$ of the nuclear potential.}
\label{fig:sis}
\end{figure}
\begin{figure}[htb]
\centering{\includegraphics[width=.6\linewidth]{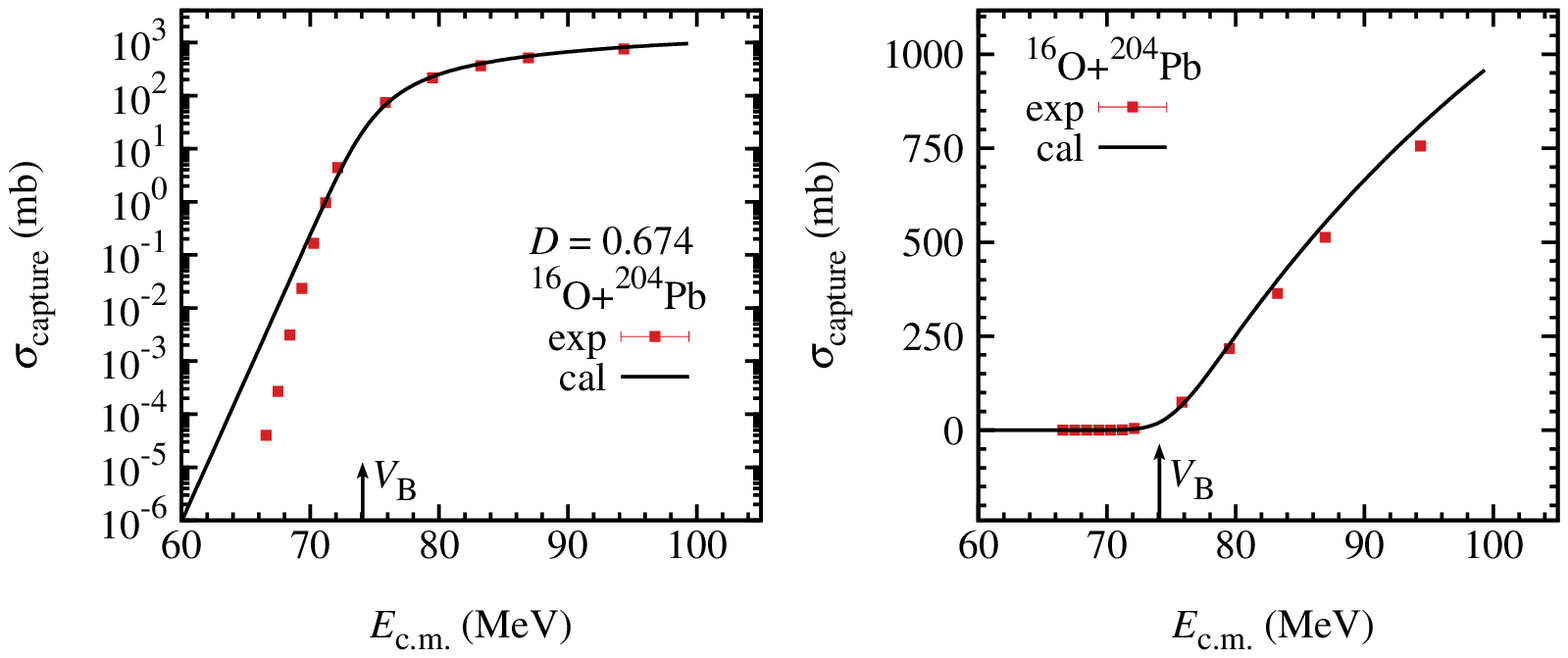}\\
\includegraphics[width=.6\linewidth]{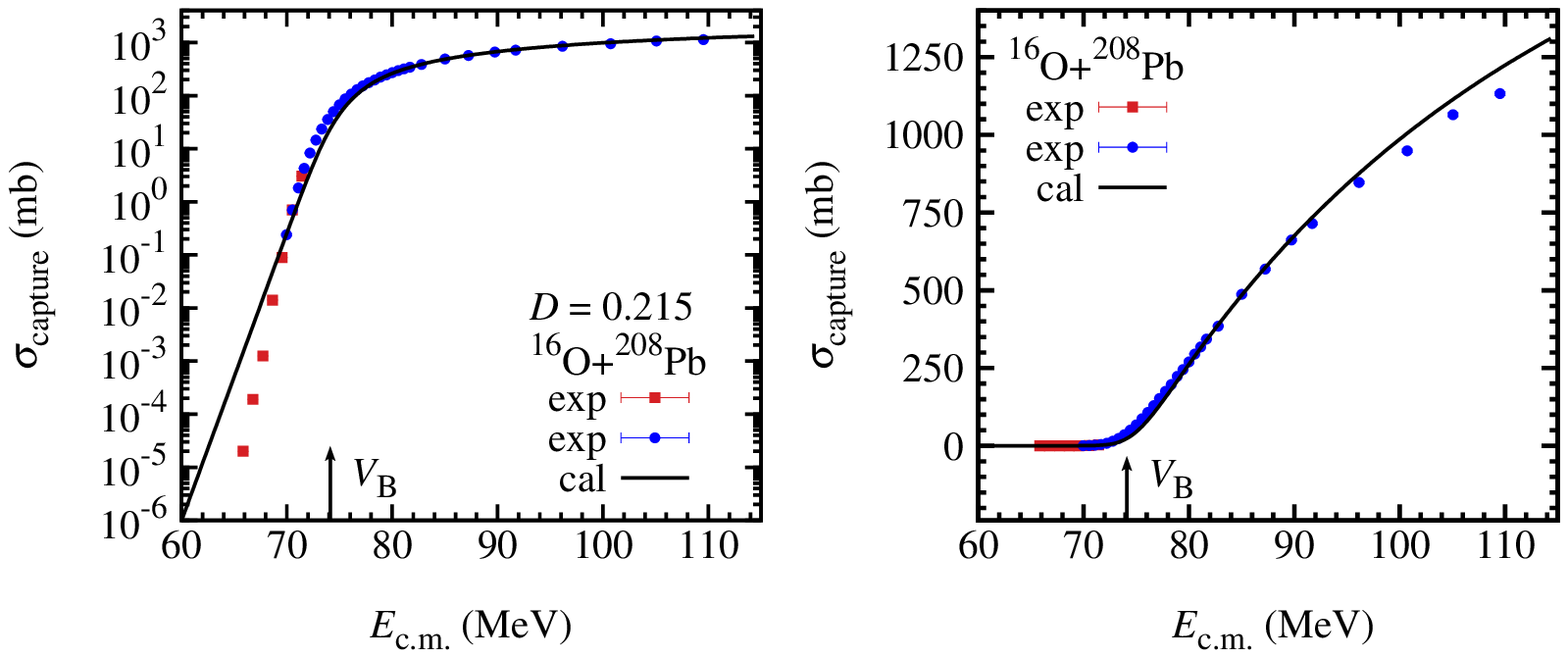}}
\caption{Same as Fig. \ref{fig:oge} but for $^{16}$O+$^{204,208}$Pb.}
\label{fig:opb}
\end{figure}

\subsubsection{Entrance channel effect and fusion hindrance in the deep
sub-barrier region}
A fusion excitation function can be divided into three distinct regimes which
are
referred to as regions I, II and III
\cite{Nagashima1986_PRC33-176,Eudes2014_PRC90-034609}. In low energy region
(region I), the fusion cross section increases with the energy increasing.
In region II, the fusion cross section saturates, whereas in region
III the fusion cross section regularly diminishes. The behavior of fusion
excitation function in region II can be explained by the entrance
channel effect because the fusion cross section is limited by the
disappearance of the ``pocket'' in the interaction potential. In the present
work, this entrance channel effect is taken into account by making a cut-off
at a critical angular momentum $J_{\rm max}$ as seen in Eq.~(\ref{eq:sig_cap}).
$J_{\rm max}$ is dependent on
the behavior of the potential inside the barrier radius which is sensitive
to the parameters of the nuclear potential. The ``pocket'' becomes shallower
with the depth of the nuclear potential decreasing
\cite{Newton2004_PRC70-024605}. To describe the behavior of the capture and
fusion excitation function in region II, it is needed to adjust the parameters
of the nuclear potential. In the present work,
the parameters of the nuclear potential are fixed as Eq.~(\ref{eq:pot_para}).
Hence, for some reactions, the predictions in region II do not agree with the
data well. For example, the results for $^{28}$Si+$^{28}$Si
reaction are shown in Fig.~\ref{fig:sis}. The data show that the cross section
saturates at energies above 1.7 times of the Coulomb barrier, while the result
of our model denoted by the solid line shows that the cross section still
increases in this region; actually the calculated cross section reaches
saturation at a large energy which is beyond the scale of Fig.~\ref{fig:sis}.
By adjusting the parameters $V_0$ and $r_0$ of
the nuclear potential to be $25$ MeV and $1.3$ fm, the results of our
model represented by the dash line can reproduce the behavior of the capture
(fusion) excitation function in region II. This implies that the depth of the
nuclear potential might be smaller than that given in Eq.~(\ref{eq:pot_para})
for light reaction systems.

\begin{figure}[htb!]
\centering{\includegraphics[width=.6\linewidth]{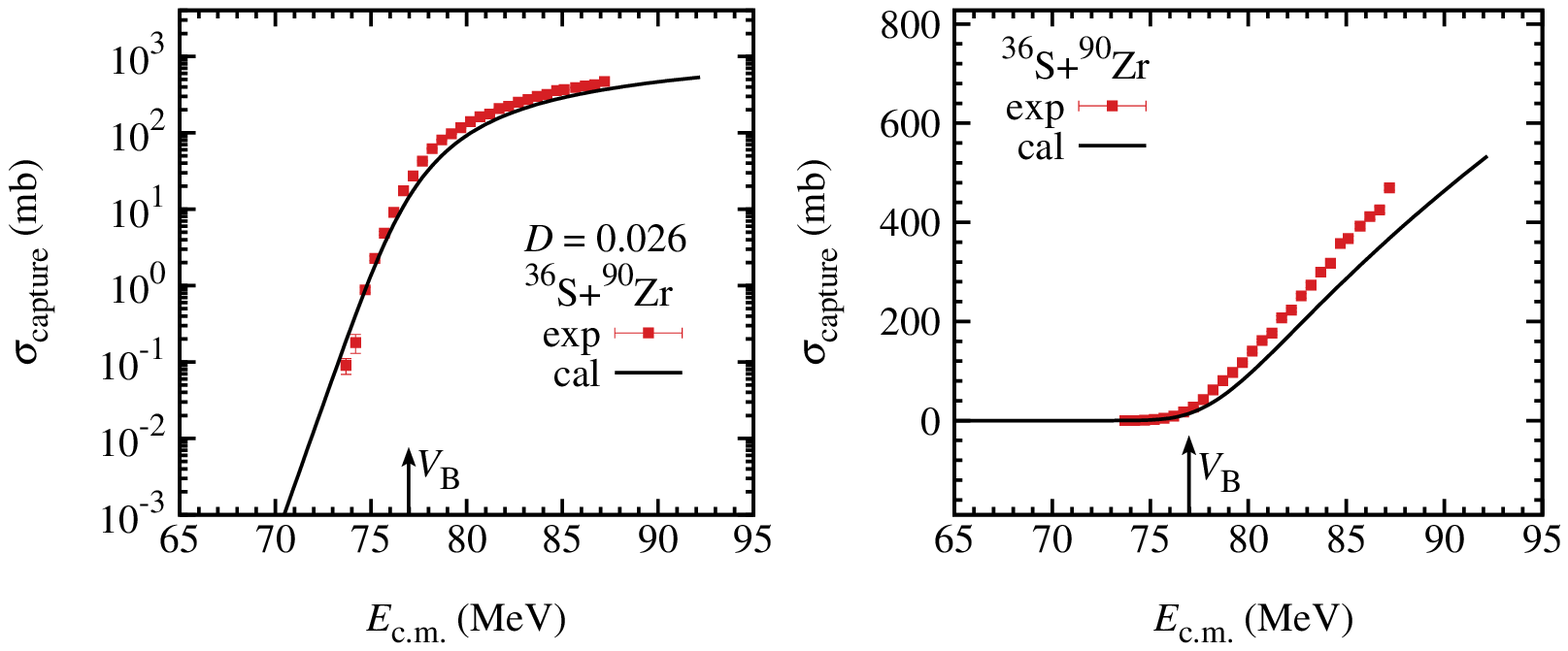}\\
\includegraphics[width=.6\linewidth]{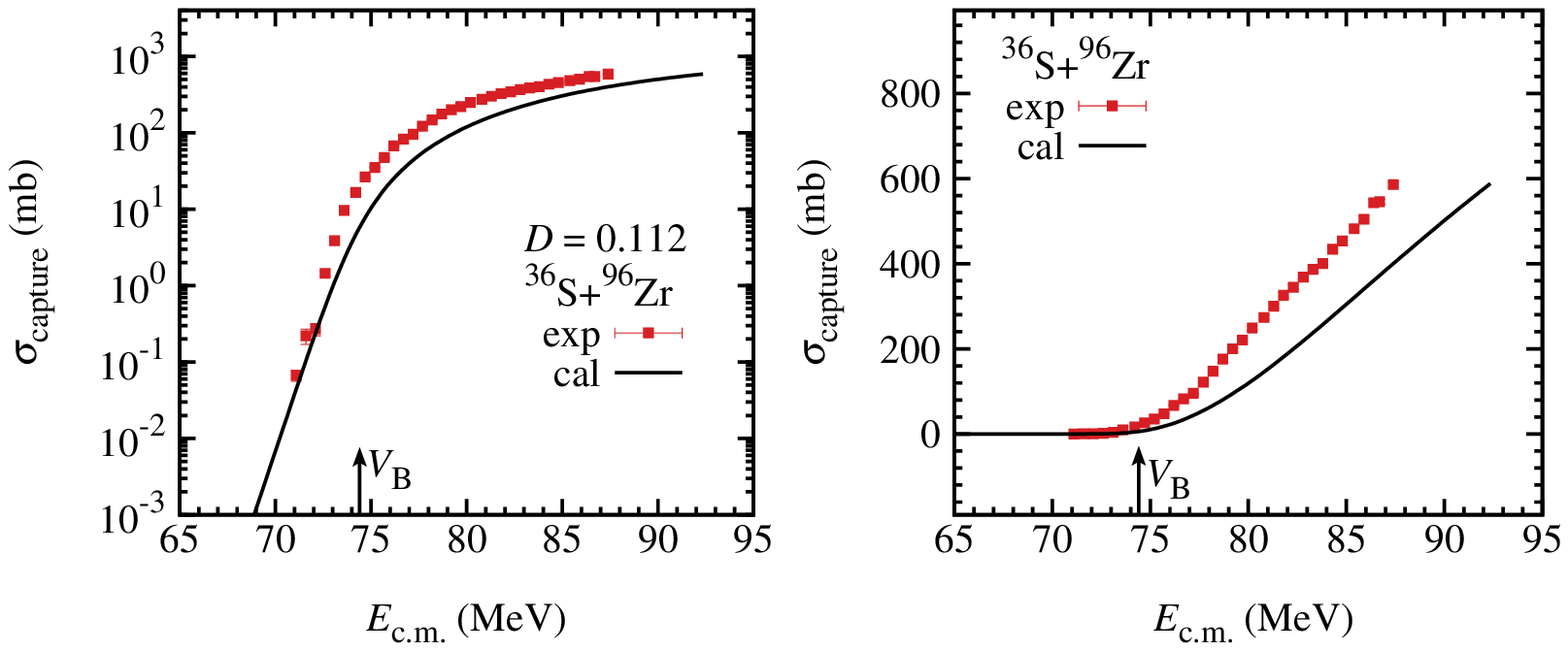}}
\caption{ Same as Fig. \ref{fig:oge} but for  $^{36}$S+$^{90,96}$Zr.}
\label{fig:szr}
\end{figure}

The comparison between the calculated capture cross sections and the
experimental values for $^{16}$O+$^{204,208}$Pb reaction systems is shown in
Fig. \ref{fig:opb}. One can find that the theoretical
predictions overestimate the capture cross sections at sub-barrier energies
below a certain energy. As mentioned in Section~\ref{sec:ecc},
Eq.~(\ref{eq:HW}) has a shortcoming and is not valid for deep sub-barrier
penetration because of the long tail of the Coulomb potential. We notice that
the new barrier penetration formula proposed in
Ref.~\cite{Li2010_IJMPE19-359} is more proper for deep sub-barrier penetration.
The implementation of this barrier penetration formula in the ECC
model is in progress. In addition, this shortcoming can also be avoided in the
quantal CC model. However, the standard CC calculations
overestimate the cross sections in the very deep sub-barrier region and are
unable to explain the steep falloff in fusion cross sections.
Different mechanisms are proposed to explain this hindrance phenomenon
\cite{Misicu2006_PRL96-112701,Dasgupta2007_PRL99-192701,
Ichikawa2007_PRC75-057603,
Diaz-Torres2008_PRC78-064604,
Diaz-Torres2010_PRC82-054617,Ichikawa2009_PRL103-202701,
Denisov2014_PRC89-044604}. In the very deep
sub-barrier region, the fusion mechanism is different from that at energies
around the Coulomb barrier. In the present work,
we focus on the capture and fusion dynamics around the Coulomb barrier and
the hindrance phenomenon is beyond the scope of this work.

\subsubsection{Further discussions}

For the reactions with noticeable deviations of the calculated results from the
experimental values, we select the  $^{36}$S+$^{90,96}$Zr as
examples. The comparison of the calculated capture
cross sections with the experimental values for $^{36}$S+$^{90,96}$Zr reactions
is shown in Fig. \ref{fig:szr}.
The deviations of the calculated cross sections from experimental values are
quite large, especially at energies above the Coulomb barrier, the predictions
underestimate the cross sections. The average deviation
$\cal{D}$ defined in Eq. (\ref{eq:dev}) for $^{36}$S+$^{90}$Zr is 0.026, which
is much smaller than that of $^{36}$S+$^{96}$Zr. The reason is that the agreement
between the calculated results and the data in the sub-barrier
region for $^{36}$S+$^{90}$Zr is better than that for $^{36}$S+$^{96}$Zr. In the
present work, we use the logarithm of the ratio of the calculated cross section
to the corresponding experimental value in the definition of the average
deviation $\cal{D}$
[see Eq. (\ref{eq:dev})]. The capture cross section grows exponentially
with energy in the sub-barrier region while linearly with energy at
energies above the barrier. Therefore, the average deviation $\cal{D}$ is more
sensitive to the agreement between the calculated capture cross sections and the
data in the sub-barrier region than that at energies above the Coulomb barrier.

\begin{figure}[ht]
\centering{
\includegraphics[width=.6\linewidth]{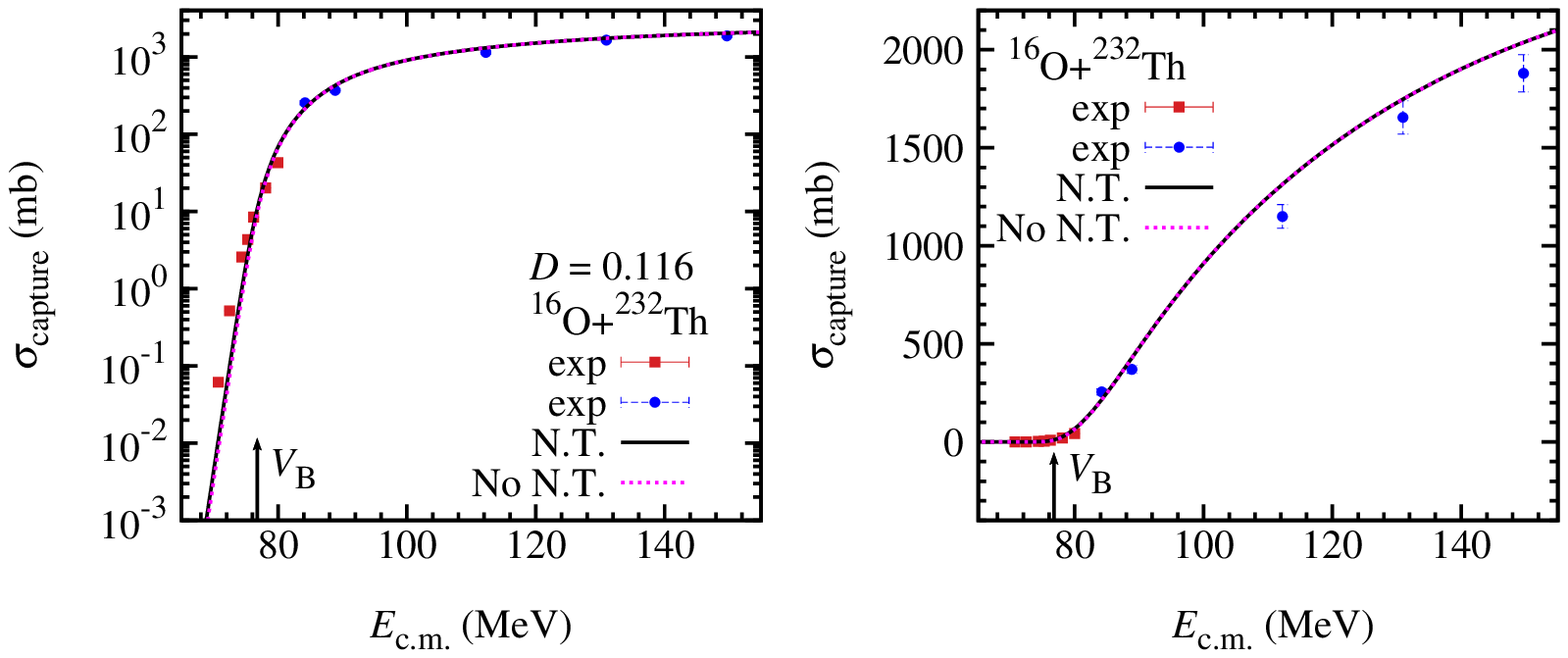}
\includegraphics[width=.6\linewidth]{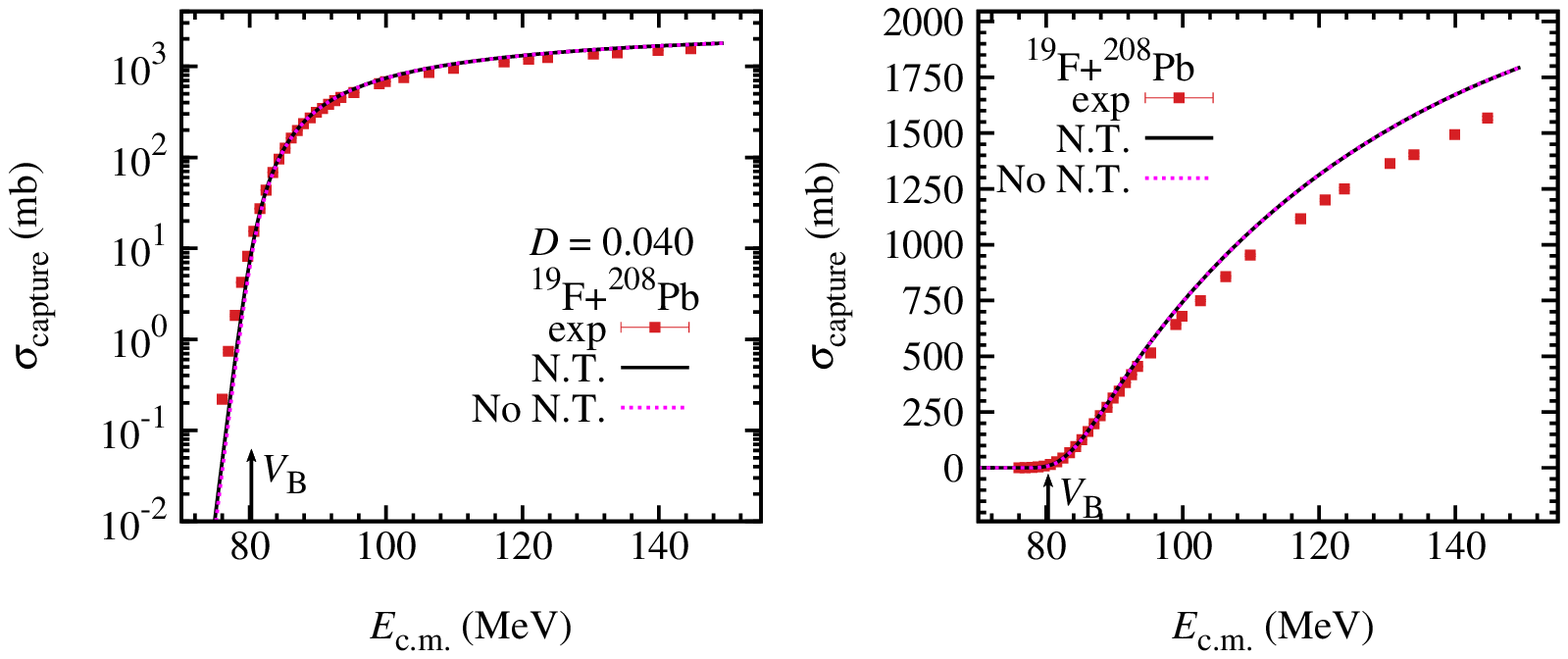}}
\caption{The experimental and calculated capture excitation functions
for $^{16}$O+$^{232}$Th and  $^{19}$F+$^{208}$Pb. The results are shown in
logarithmic scale (the left panel)
and linear scale (the right panel). The solid line denotes the
calculated cross section with the effect of one neutron pair
transfer taken into account. The dotted line denotes the calculated cross
section with the effect of neutron transfer neglected.
The arrow indicates the central value of the barrier distribution $B_\mathrm{m}$
given in Eq. (\ref{eq:disp}).
The solid squares and circles show the
experimental values  which are given in Table \ref{tab:data}.
$\mathcal{D}$ denotes the average deviation of the calculated cross sections
with the effect of one neutron pair transfer taken into account from the
experimental values.}
\label{fig:NTig}
\end{figure}

\subsection{Reaction systems with positive $Q$ value for one neutron pair
transfer}

Among the 220 reaction systems we have collected, there are 89 reaction systems
with positive $Q$ value for one neutron pair transfer channel. The comparison
between the calculated and measured capture excitation functions for these 89
reaction systems is illustrated in Graphs 12--19.
The solid line denotes the results with the effect of the one
neutron pair transfer taken into account. The dotted line denotes the results
with the effect of the neutron transfer neglected. The
arrow indicates the central value of the barrier distribution $B_\mathrm{m}$
given in Eq.~(\ref{eq:disp}). The data are represented by solid squares or
circles and the corresponding references are listed in Table \ref{tab:data}. The
$Q$ values for one neutron pair transfer, i.e., $Q(2n)$s are also given in
Table \ref{tab:total}. From Graphs 12--19, it can be seen that the calculated
capture cross sections of most of the 89 reaction systems are in good
agreement with the experimental values. Generally, for the reactions
with positive $Q$ value for one neutron pair transfer, the present model
can describe the capture excitation functions quite well. Next, we first give
some general discussions on the influence of the neutron transfer on capture
cross sections. Then we discuss the reactions for which the deviation of the
calculated results from the experiment is noticeable in detail by taking some
typical examples.

\subsubsection{Reaction systems with negligible effects of PQNT couplings}

In the present model, the coupling to the PQNT channels is simulated by
broadening the barrier distribution given in Eq.~(\ref{eq:NeuT}). Therefore,
the effects of PQNT couplings is related to the changes of the widths of the
barrier distribution, especially the change of the left width $\varDelta_1$.
If $gQ(2n)$ is large, but the ratio of $gQ(2n)$ to
$\varDelta_1$ is small, i.e., the change of the left width $\varDelta_1$ is
quite small, then the influence of one neutron
pair transfer on capture and fusion cross sections can be ignored.
In addition, for the reactions with small positive $Q(2n)$s (smaller than 1 MeV,
see Table~\ref{tab:total}), the changes of the widths of the barrier
distributions are also small, the
influence of one neutron pair transfer on capture and
fusion cross sections can also be ignored. For example, the calculated capture
cross
section and the experimental values of the $^{16}$O+$^{232}$Th and
$^{19}$F+$^{208}$Pb reactions are shown in Fig. \ref{fig:NTig}. The
$Q(2n)$s for the $^{16}$O+$^{232}$Th and $^{19}$F+$^{208}$Pb reactions are
0.63 MeV and 0.60 MeV, respectively. The solid line denotes the results with
the effect of one neutron pair transfer taken into account, the dotted line
denotes the results with this effect neglected.
One can find that the solid line is very close to the dotted one, both
the results with and without taking into account the effect of one neutron
pair transfer show good agreement with the data. In other reaction systems with
small positive $Q(2n)$s, the influence of one neutron pair transfer on capture
and fusion cross sections can also be ignored.

\begin{figure}[htb]
\centering{
\includegraphics[width=.6\linewidth]{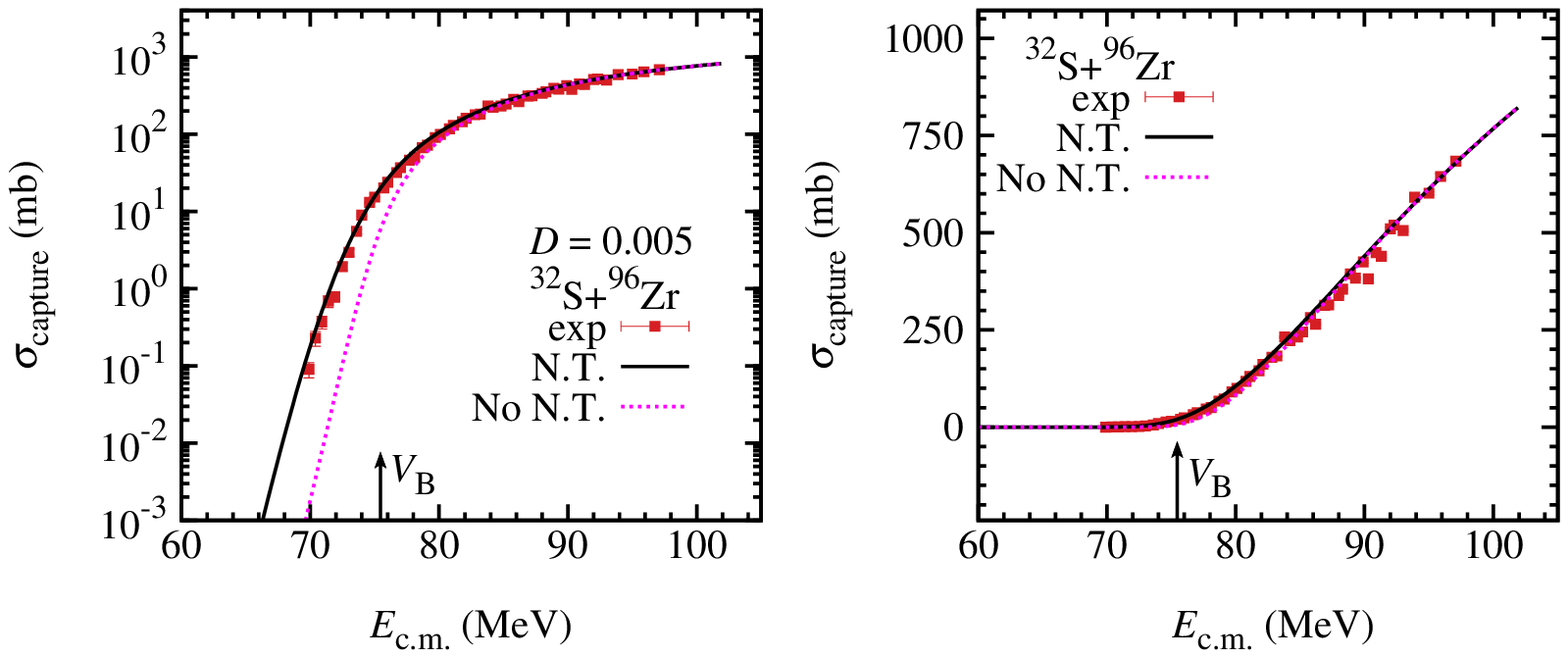}
\includegraphics[width=.6\linewidth]{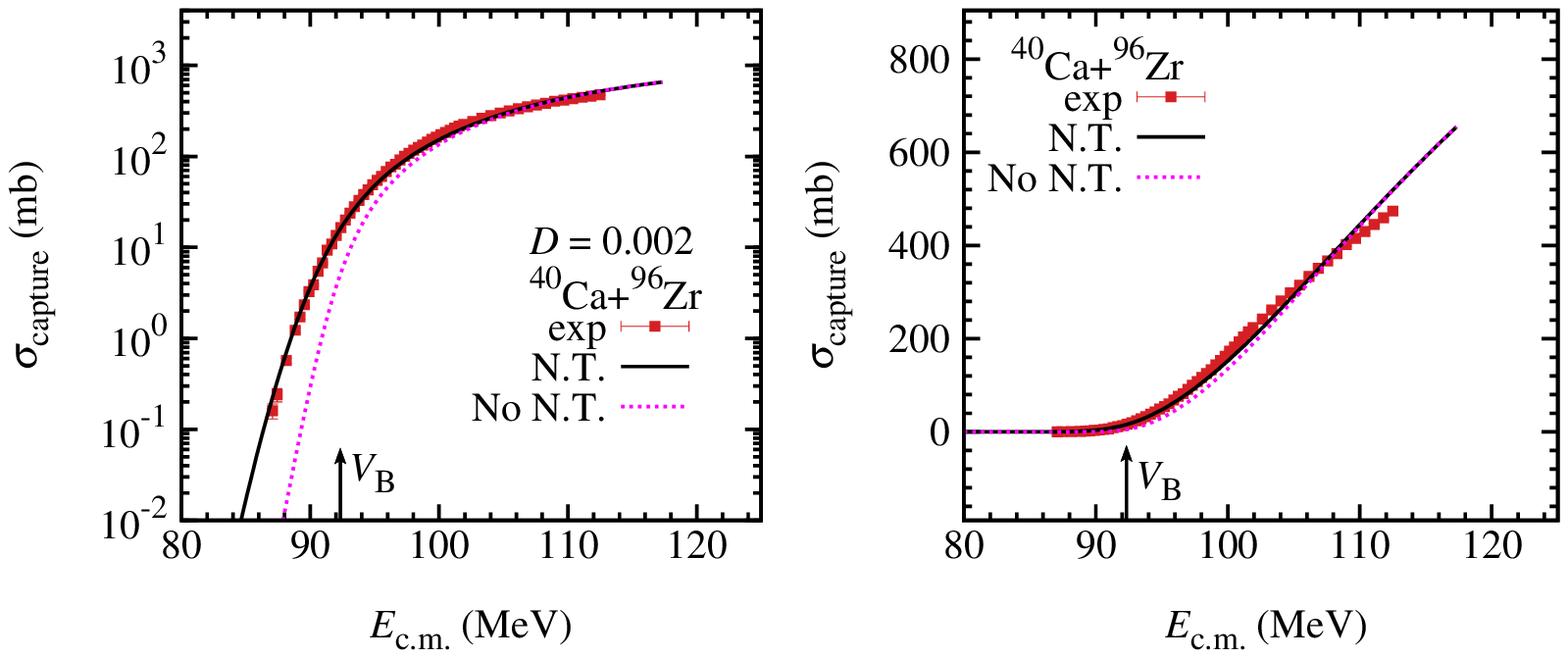}
}
\caption{Same as Fig. \ref{fig:NTig} but for $^{32}$S+$^{96}$Zr and
$^{40}$Ca+$^{96}$Zr.}
\label{fig:NTlar}
\end{figure}

\subsubsection{Reaction systems with large positive $Q(2n)$: Effects of PQNT
couplings}

In reaction systems with large positive $Q(2n)$s,
the distributions with the neutron transfer couplings taken into account become
much broader as compared with those with the effect of neutron transfer
neglected. Therefore, the calculated cross sections with this effect
considered are much larger than those with this effect neglected
in the sub-barrier region. Meanwhile, the change of the slope of the
capture excitation function in the sub-barrier energy region owing to the
coupling to one neutron pair transfer channel with a large positive
$Q$ value is significant. We take the reaction systems of
$^{32}$S+$^{96}$Zr and $^{40}$Ca+$^{96}$Zr as examples. The $Q(2n)$s for
$^{32}$S+$^{96}$Zr and $^{40}$Ca+$^{96}$Zr reactions are 5.74 MeV and 5.53 MeV,
respectively. After taking into account the coupling to PQNT channels, the
changes of the widths of the barrier distributions given in Eq.~(\ref{eq:NeuT})
are about 1.8 MeV for these two reactions. The calculated capture
cross sections and the experimental values of these two reactions are shown in
Fig.~\ref{fig:NTlar}. The solid line denotes the results with the effect of
one neutron pair transfer taken into account. The dotted line denotes the
results with the effect of the neutron transfer neglected. One
can find that the measured capture cross sections show a large enhancement as
compared with the results without considering the coupling effects of PQNT
channels (the dotted line) in the sub-barrier region.
While after taking this effect into account, the calculated values are in
good agreement with the data. In the region above
the Coulomb barrier, the difference between the solid line and the dotted line
becomes smaller and smaller as the energy increase. Therefore the couplings
to PQNT
channels mainly affect the capture cross sections in the sub-barrier region.

\subsubsection{Further discussions}

\begin{figure}[htb!]
\centering{\includegraphics[width=.6\linewidth]{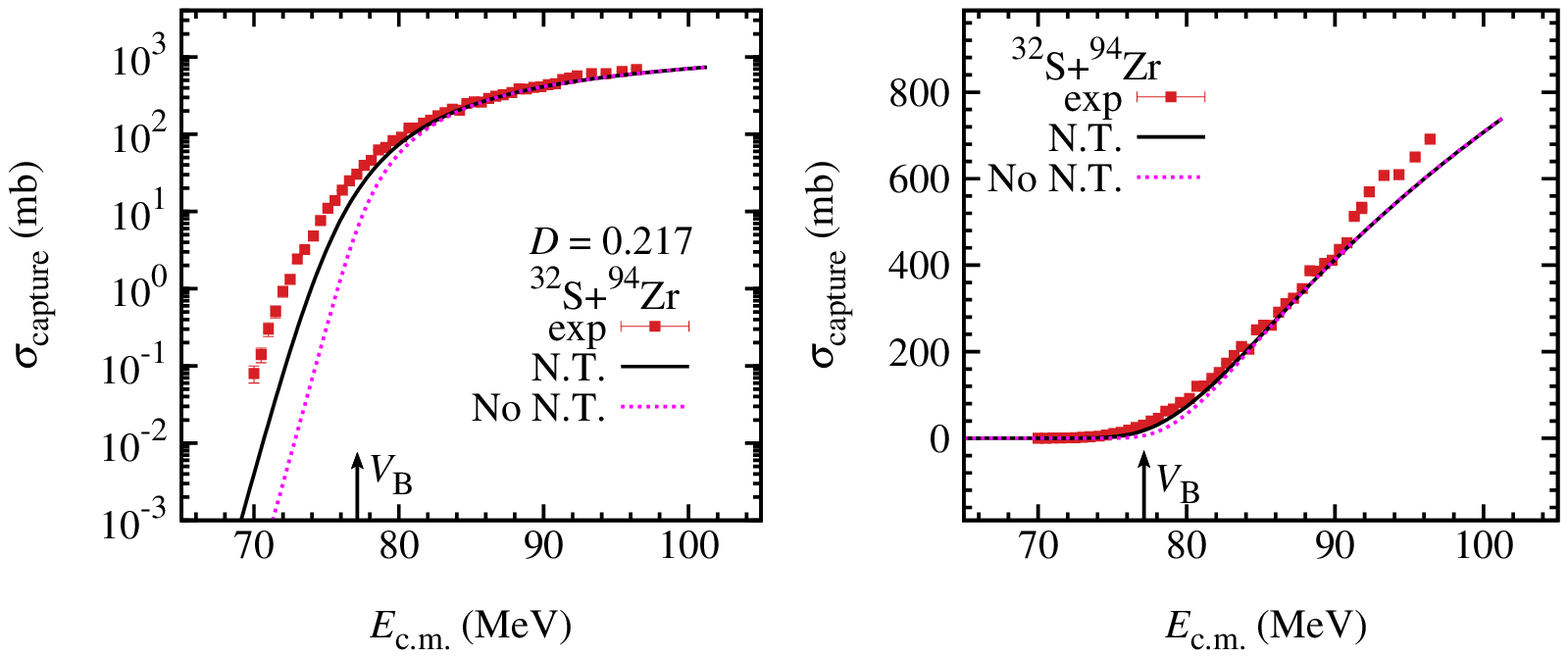}\\
\includegraphics[width=.6\linewidth]{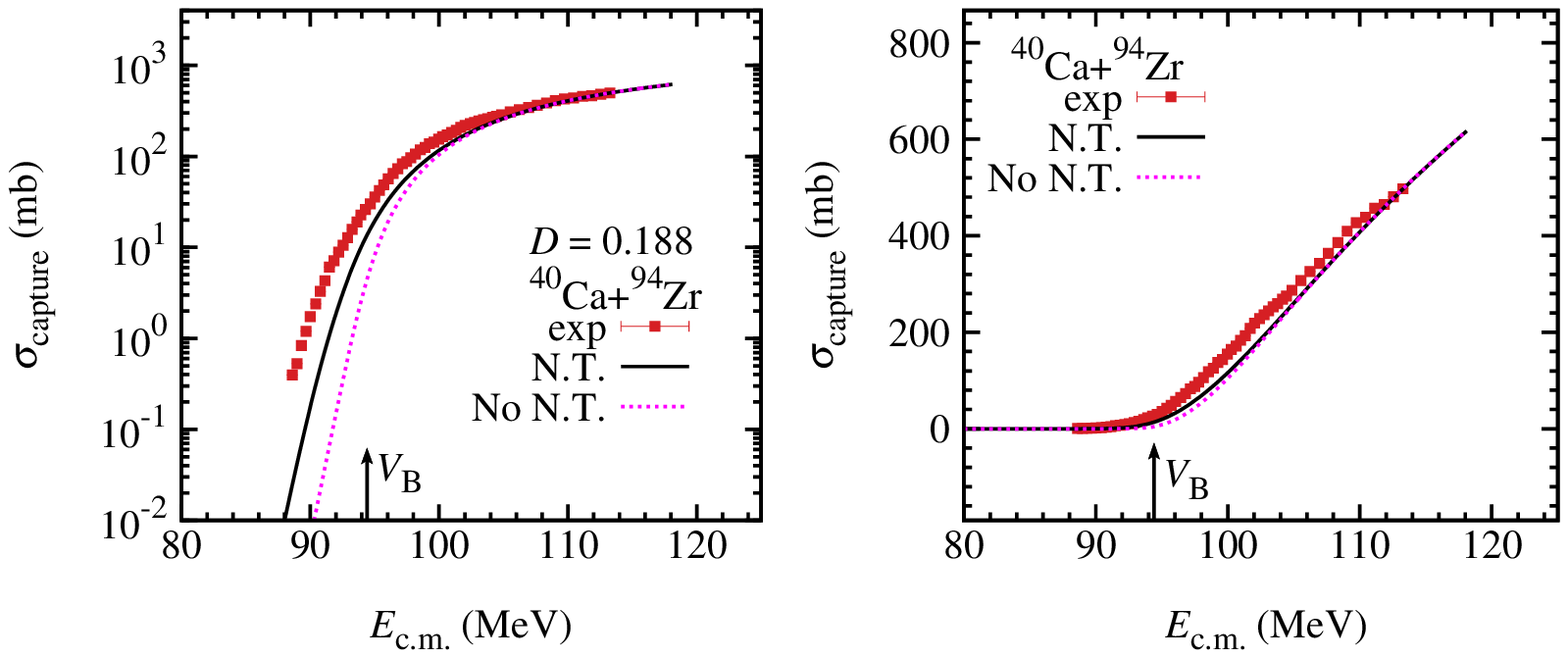}}
\caption{Same as Fig. \ref{fig:NTig} but for $^{32}$S+$^{94}$Zr and
$^{40}$Ca+$^{94}$Zr.}
\label{fig:94zr}
\end{figure}
\begin{figure}[htb!]
\centering{\includegraphics[width=.35\linewidth]{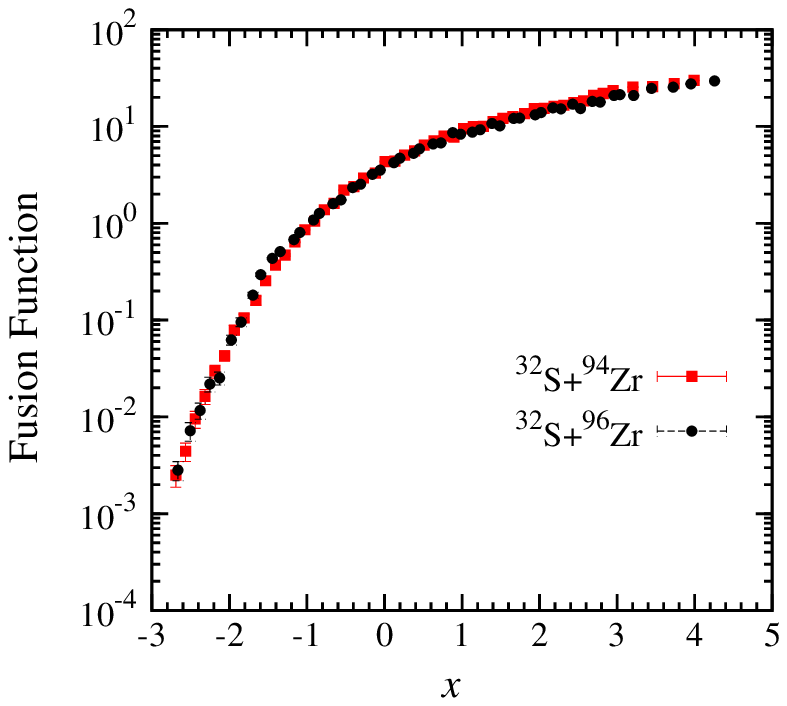}
\includegraphics[width=.35\linewidth]{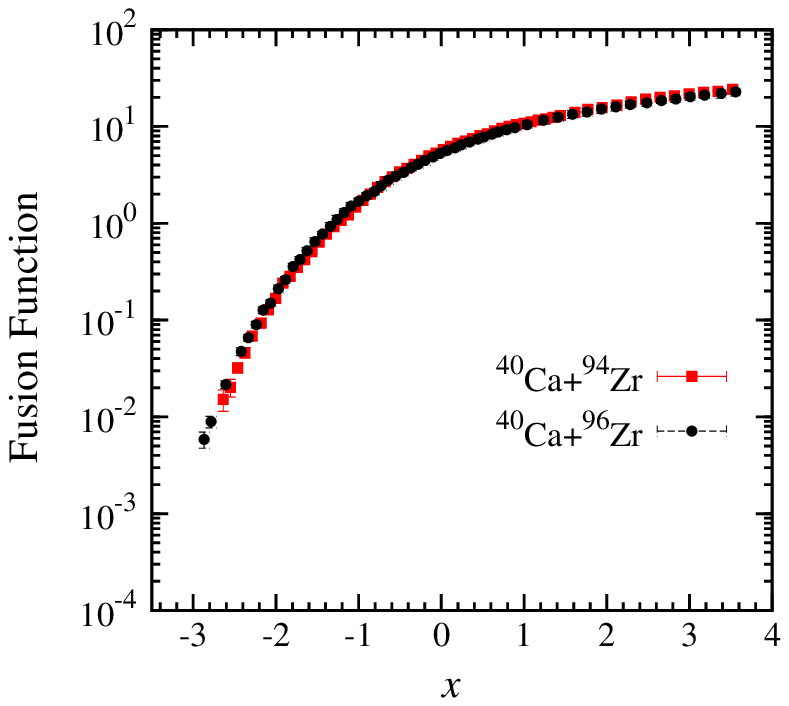}}
\caption{The reduced capture excitation
functions of the reactions $^{32}$S+$^{94,96}$Zr and
$^{40}$Ca+$^{94,96}$Zr.}
\label{fig:red}
\end{figure}

\begin{figure}[htb!]
\centering{\includegraphics[width=.6\linewidth]{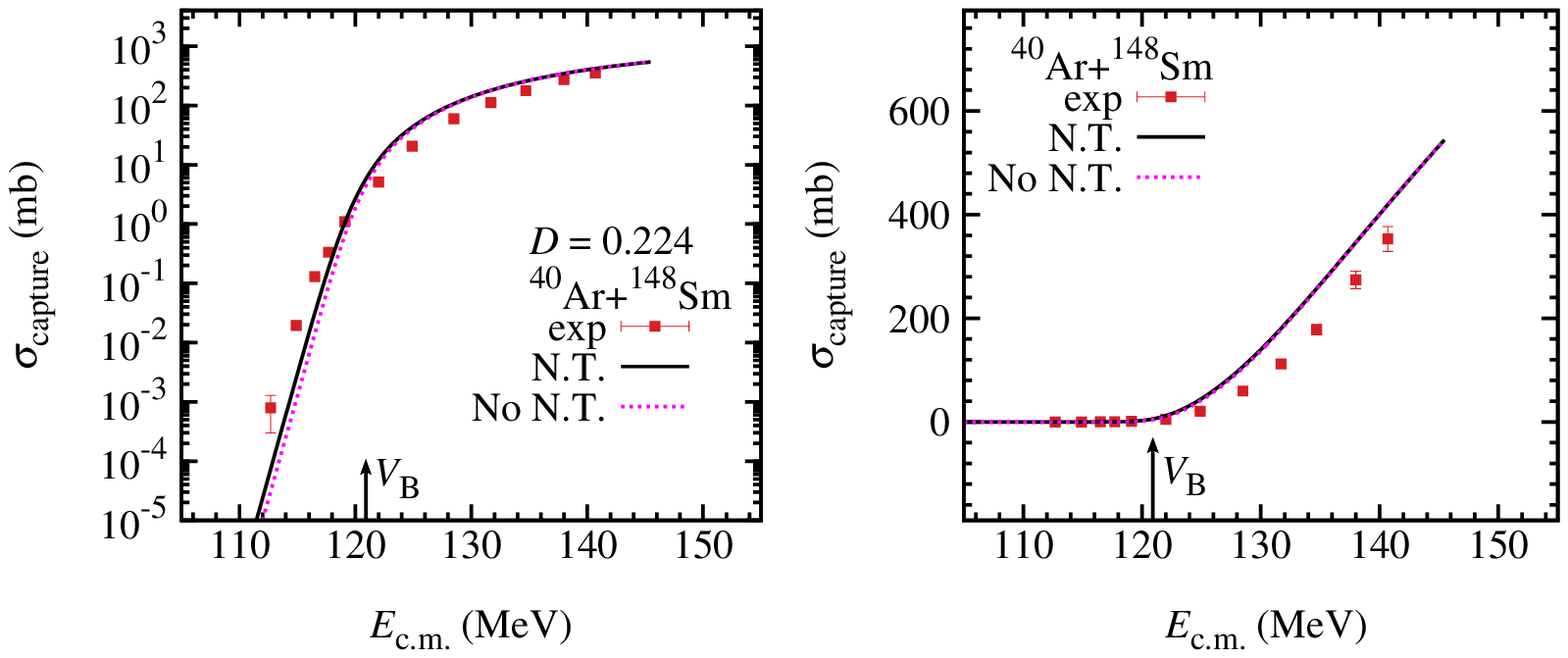}\\
\includegraphics[width=.6\linewidth]{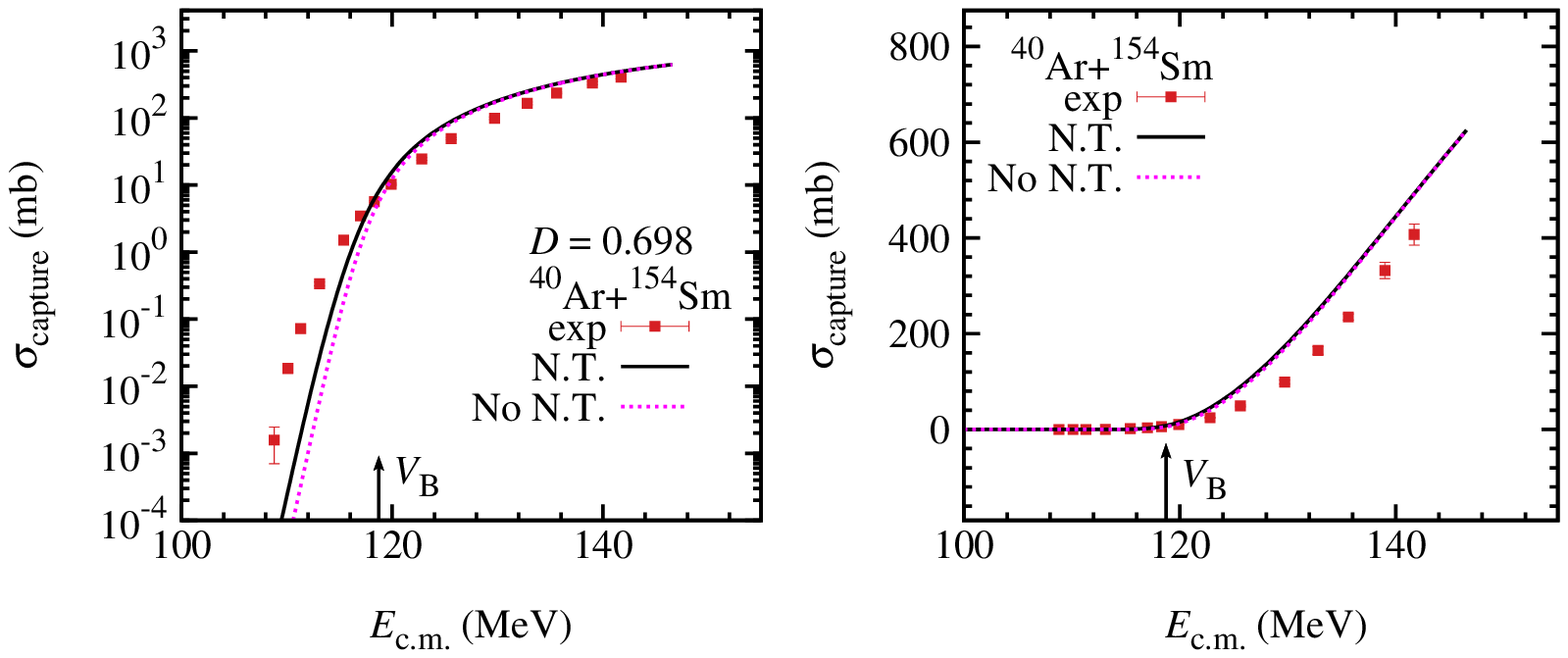}
}
\caption{Same as Fig.~\ref{fig:NTig} but for $^{40}$Ar+$^{148,154}$Sm.}
\label{fig:arsm}
\end{figure}

Figure \ref{fig:94zr} shows the results for $^{32}$S+$^{94}$Zr and
$^{40}$Ca+$^{94}$Zr. It can be seen that with the effect
of one neutron pair transfer taken into account, the results from our
calculations underestimate the data in
the sub-barrier region. For the four reactions $^{32}$S+$^{94,96}$Zr and
$^{40}$Ca+$^{94,96}$Zr, the $Q(2n)$s are 5.10 MeV, 5.74 MeV, 4.89 MeV, and 5.53
MeV, respectively. On one hand, in the framework of the present model, one may
conclude that the influences of one neutron pair transfer on capture are similar
for these four reactions because of similar $Q(2n)$s.
On the other hand, only the data for $^{32}$S+$^{96}$Zr
and $^{40}$Ca+$^{96}$Zr are well reproduced by taking into account the effect
of neutron transfer (see Fig. \ref{fig:NTlar}).
These findings drive us to investigate in detail the influence of various
coupling channels on
the capture cross section. To this end, we first reduce the data to
eliminate the geometrical factors and static effects of the potential.
Several reduction methods
\cite{Beckerman1982_PRC25-837,DiGregorio1989_PRC39-516,Gomes2005_PRC71-017601,
Canto2009_JPG36-015109,Canto2009_NPA821-51,Wolski2013_PRC88-041603R,
Qu2014_PRC90-064603} can be used. In the present work,
the fusion function
method~\cite{Canto2009_JPG36-015109,Canto2009_NPA821-51}
is adopted to reduce the capture excitation
functions~\cite{Canto2015_PRC92-014626,Gomes2016_FBS57-205,
Gomes2015_FBS57-165}.
The barrier parameters $R_{\rm B}$, $V_{\rm B}$, and $\hbar\omega$ are obtained
from the double folding and parameter-free S\~ao Paulo potential (SPP)
\cite{CandidoRibeiro1997_PRL78-3270,Chamon1997_PRL79-5218,
Chamon2002_PRC66-014610}. The reduced capture excitation functions of the
reactions $^{32}$S+$^{94,96}$Zr and $^{40}$Ca+$^{94,96}$Zr are shown
in Fig.~\ref{fig:red}. One can find that the behaviors of the reduced capture
excitation functions of $^{32}$S+$^{94}$Zr and $^{32}$S+$^{96}$Zr are very
similar. The situation is the same for the reactions $^{40}$Ca+$^{94,96}$Zr.
This means that
the effects of the couplings (to the inelastic excitations and the neutron
transfer channel) in $^{32}$S+$^{94}$Zr are similar to that in
$^{32}$S+$^{96}$Zr. The same conclusion can be drawn for the reactions
$^{40}$Ca+$^{94,96}$Zr.
In the present work, the static quadrupole deformation parameters of
$\beta^0=0.062$ and $\beta^0=0.217$ are used for $^{94}$Zr and $^{96}$Zr,
respectively \cite{Moller1995_ADNDT59-185}. The underestimate of the data for
the reactions $^{32}$S+$^{94}$Zr and $^{40}$Ca+$^{94}$Zr might come from the
smaller deformation parameter for $^{94}$Zr used in our calculations. Further
study has been done with our ECC model to elucidate which of these two
effects (the coupling to the
PQNT channel or the structure of $^{94}$Zr) is responsible for the underestimate
of the capture cross section of the reactions $^{32}$S+$^{94}$Zr and
$^{40}$Ca+$^{94}$Zr \cite{Wang2016_SciChinaPMA59-642002}. In addition, in
Ref.~\cite{Karpov2015_PRC92-064603}, the
QCC model with a semiclassical consideration of neutron rearrangement
has been used to study these reactions and it was pointed out that the
sub-barrier fusion enhancements owing to neutron rearrangement and excitation of
the collective vibrational and/or rotational states are not additive.
Esbensen {\it et al.} suggested that the couplings to
positive $Q$-value proton transfer channels are also important to explain the
fusion data for the reaction
$^{40}$Ca+$^{96}$Zr~\cite{Esbensen2016_PRC93-034609}.

In Fig.~\ref{fig:arsm}, the comparison of the calculated capture cross sections
to the experimental values for $^{40}$Ar+$^{148,154}$Sm reactions is
illustrated. The $Q(2n)$s for these two reactions are 1.04 MeV and 1.69 MeV,
respectively. One can find that the results with the effect
of neutron transfer taken into account overestimate the data at energies above
the Coulomb barrier and underestimate the data in the sub-barrier region. This
indicates that the widths of the barrier distributions are not appropriate.
From the fitted results of the parameters of the barrier distribution listed in
Table \ref{tab:total}, one can find that much broader barrier distributions than
those given in Subsection~\ref{subs:cal} are
needed to describe the capture excitation functions for these two reactions.

\section{Summary}\label{sec:sum}

In this work, we use an empirical coupled-channel (ECC) model to systematically
study the
capture and fusion excitation functions.
In this model, a barrier distribution of asymmetric Gaussian form
is used to take into account the effects of couplings between
the relative motion and intrinsic degrees of freedom, as well as the effect of neutron transfer channels.
Based on the interaction potential between the
projectile and the target, empirical formulas are proposed to calculate
the parameters for the barrier distribution.
We collect and compile 220 sets of capture excitation
functions for reaction systems with $182 \leqslant Z_{\rm P}Z_{\rm T} \leqslant
1640$ and for 89 of them, the $Q$ values for one neutron pair
transfer channel are positive.
We tabulate the parameters of barrier
distributions, the calculated capture cross sections, and the experimental cross sections of these 220 reaction systems.
The results show that for most of the reactions,
this empirical coupled-channel model can describe the capture (fusion)
cross sections very well
in the energy region around the Coulomb barrier.
We expect that this model will provide prediction of capture
cross sections for the synthesis of superheavy nuclei as well as
useful information on capture and fusion dynamics.

\ack
Helpful discussions with G. G. Adamian, P. R. S. Gomes, Alexander Karpov,
Cheng-Jian Lin, Ning Wang,
Huan-Qiao Zhang, Zhen-Hua Zhang, and Jie Zhao are gratefully acknowledged.
We also thank P. R. S. Gomes and Huan-Qiao Zhang for a careful reading of this
manuscript.
This work has been partly supported by
the National Key Basic Research Program of China (Grant No. 2013CB834400),
the National Natural Science Foundation of China (Grants
No. 11121403,
No. 11175252,
No. 11120101005,
No. 11275248,
No. 11475115, and
No. 11525524),
and
the Knowledge Innovation Project of the Chinese Academy of Sciences (Grant No.
KJCX2-EW-N01).
The computational results presented in this work have been obtained on
the High-performance Computing Cluster of SKLTP/ITP-CAS and
the ScGrid of the Supercomputing Center, Computer Network Information Center of
the Chinese Academy of Sciences.
We also point out the usefulness of the NRV website \cite{NRVmisc}
which is particularly helpful in the initial stage of this work.


\newpage
\GraphExplanation
\label{GraphExp}

\begin{center}
 \renewcommand{\Dfiguresname}{}
\renewcommand{\theDfigures}{Graphs 1--11}
\begin{Dfigures}[!ht]
\begin{minipage}[t]{0.2\textwidth}
\vspace{-6em}
 \caption[\hspace{1.2cm} Excitation functions for 131 reaction systems with
negative $Q$ value for one neutron pair transfer]{}
\end{minipage}
\begin{minipage}[t]{0.8\textwidth}
 {\begin{tabular}{ll}
  &  \multirow{1}{13cm}{\footnotesize {\bf Excitation
functions for 131 reaction systems with negative $Q$ value for one neutron
pair transfer} \\For each reaction system, the
results are shown in logarithmic scale (the left panel) and linear scale (the
right panel). The solid line denotes the calculated cross section.
The arrow indicates the central value of the barrier distribution
$B_\mathrm{m}$ given in Eq.~(\ref{eq:disp}). The solid squares and circles are
the experimental values which are given in Table~\ref{tab:data}. $\mathcal{D}$,
defined in Eq.~(\ref{eq:dev}), denotes the average deviation of the calculated
cross sections from the experimental values. }\\
 & \\
 & \\
 & \\
 & \\
 & \\
 \end{tabular}
 }\end{minipage}
\end{Dfigures}

 \renewcommand{\Dfiguresname}{}
\renewcommand{\theDfigures}{Graphs 12--19}
 \begin{Dfigures*}[!ht]
 \begin{minipage}[t]{0.2\textwidth}
\vspace{-7em}
\caption[\hspace{1.2cm}  Excitation
functions for 89 reaction systems with positive $Q$ value for one neutron
pair transfer]{}
\end{minipage}
\begin{minipage}[t]{0.8\textwidth}
  \begin{tabular}{ll}
   &   \multirow{9}{13cm}{\footnotesize {\bf Excitation
functions for 89 reaction systems with positive $Q$ value for one neutron
pair transfer} \\For each reaction system, the
results are shown in logarithmic scale (the left panel) and linear scale (the
right panel). The solid line denotes the calculated cross section with the
effect of one neutron pair transfer taken into account. The dotted line
denotes the calculated cross section with the effect of neutron transfer
neglected. The arrow indicates the central value of the barrier
distribution $B_\mathrm{m}$ given in Eq.~(\ref{eq:disp}). The solid squares and
circles are the experimental values  which are given in Table~\ref{tab:data}.
$\mathcal{D}$, defined in Eq.~(\ref{eq:dev}),
denotes the average deviation of the calculated cross sections
with the effect of one neutron pair transfer taken into account from the
experimental values.}\\
 & \\
 & \\
 & \\
 & \\
 & \\
 & \\
 & \\
 & \\
\end{tabular}
\end{minipage}
\end{Dfigures*}

\end{center}
\begin{Dfigures}[!ht]
 \centerline{\includegraphics[width=0.47\textwidth]{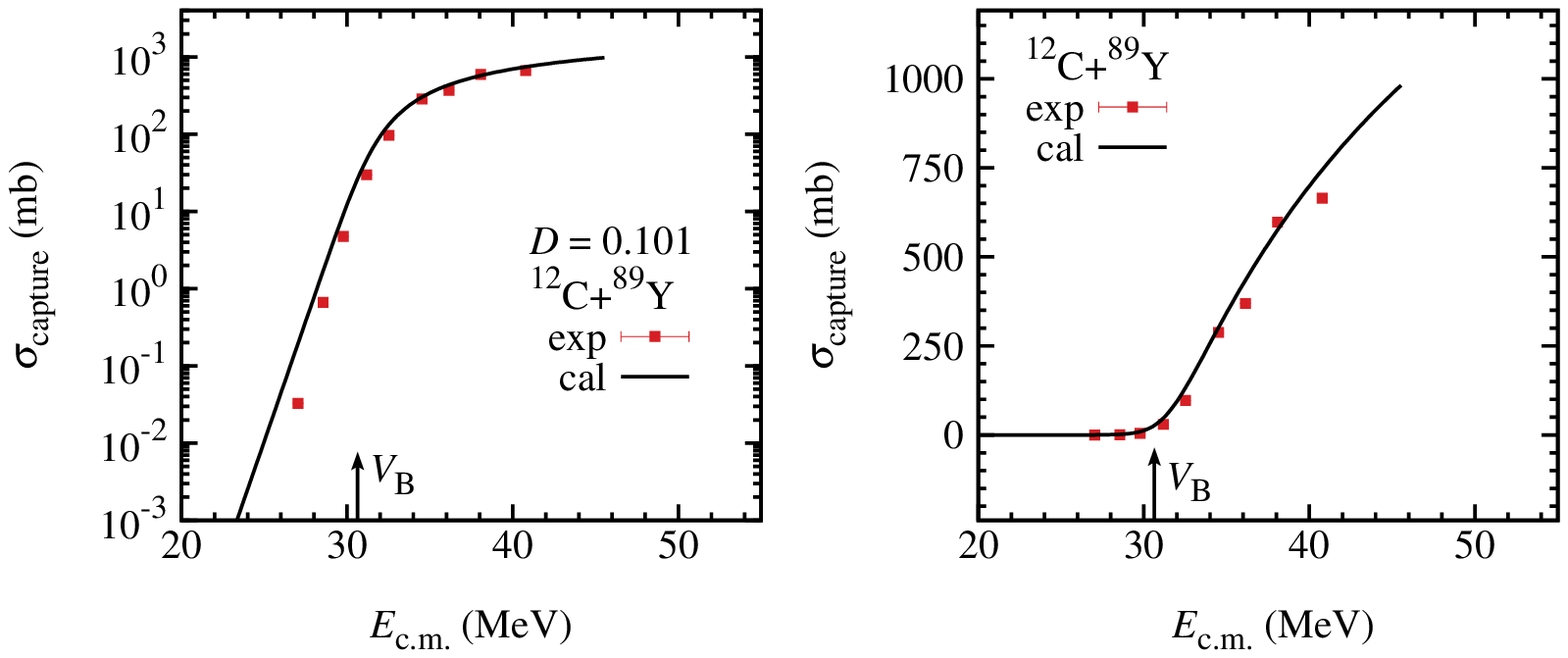}
  \includegraphics[width=0.47\textwidth]{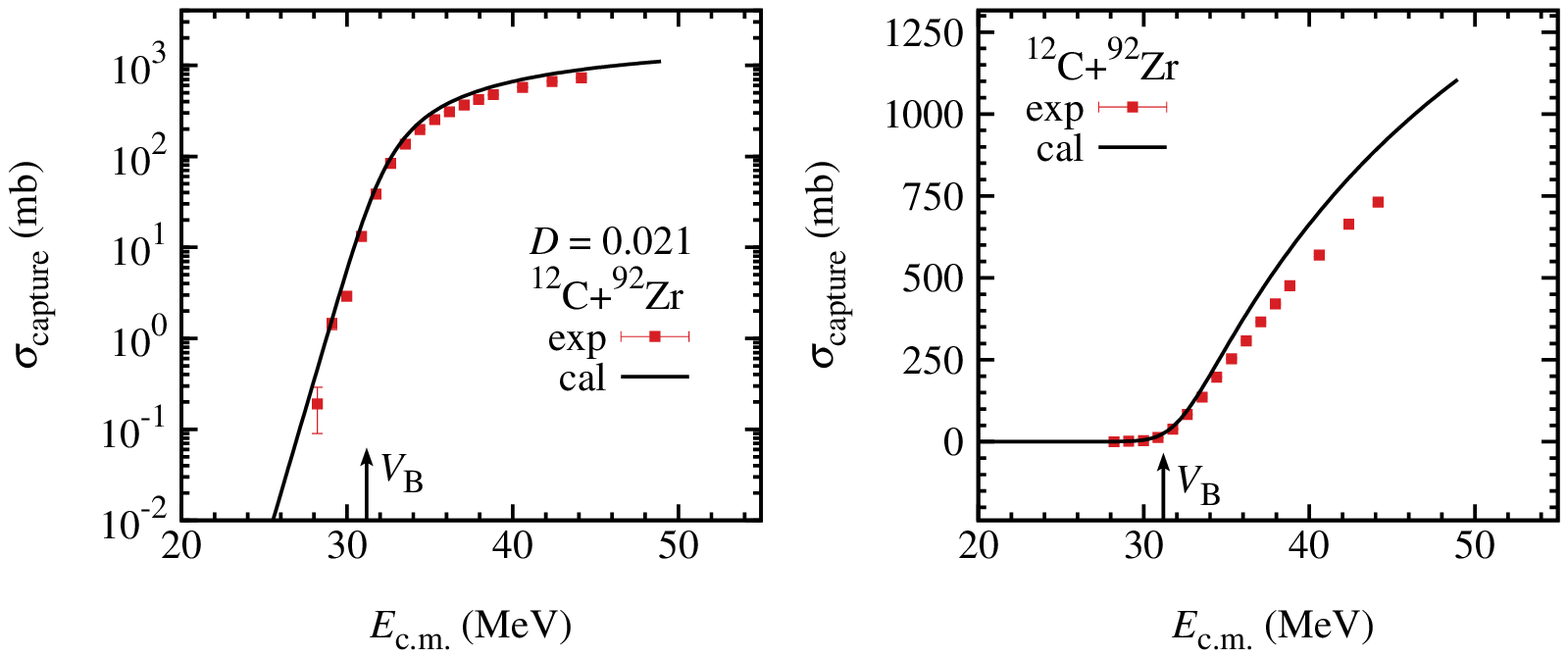}}
 \centerline{\includegraphics[width=0.47\textwidth]{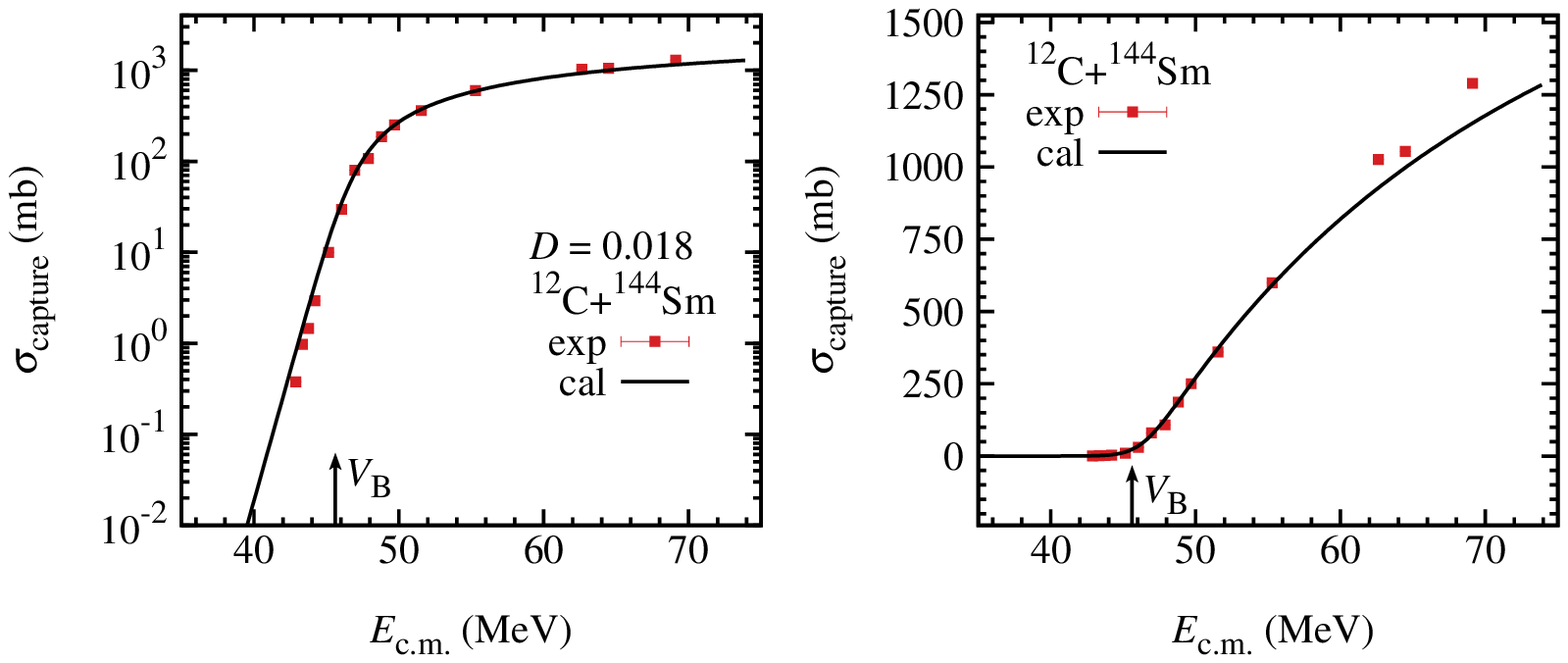}
  \includegraphics[width=0.47\textwidth]{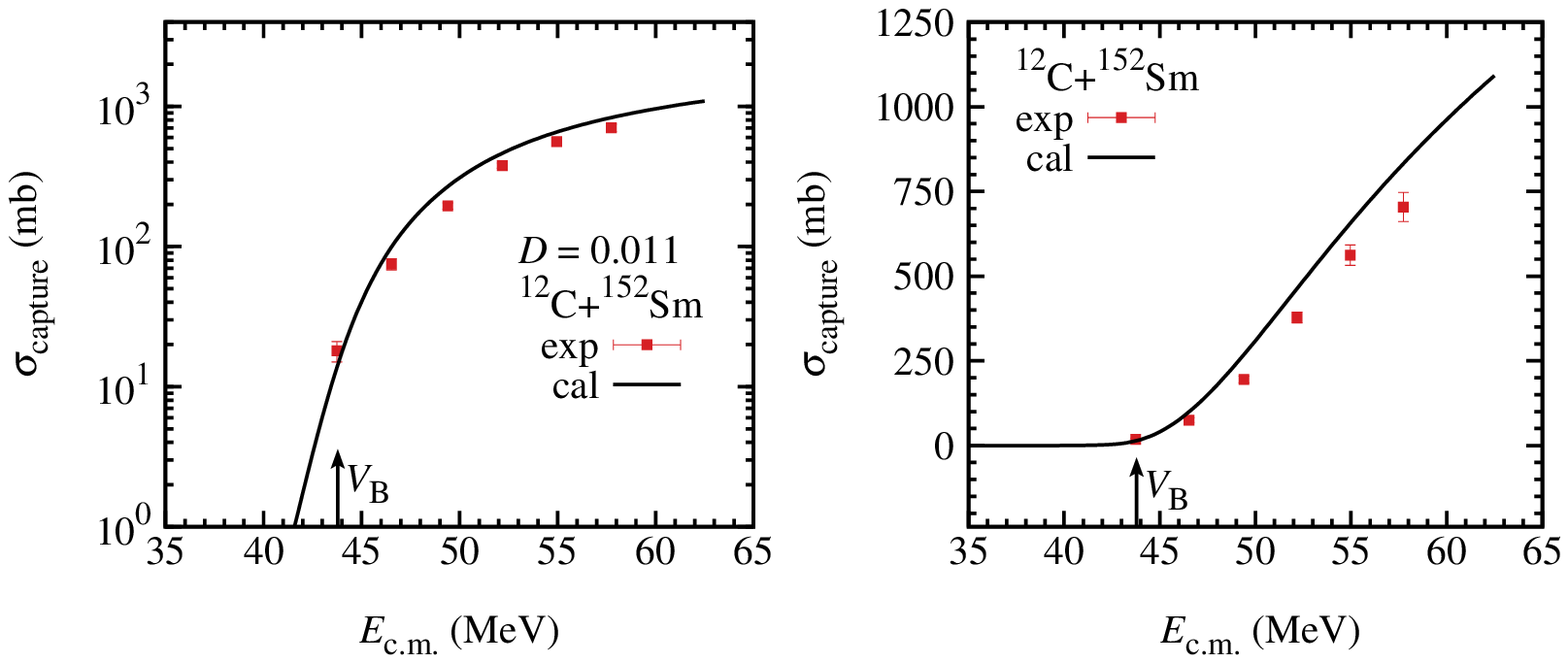}}
 \centerline{\includegraphics[width=0.47\textwidth]{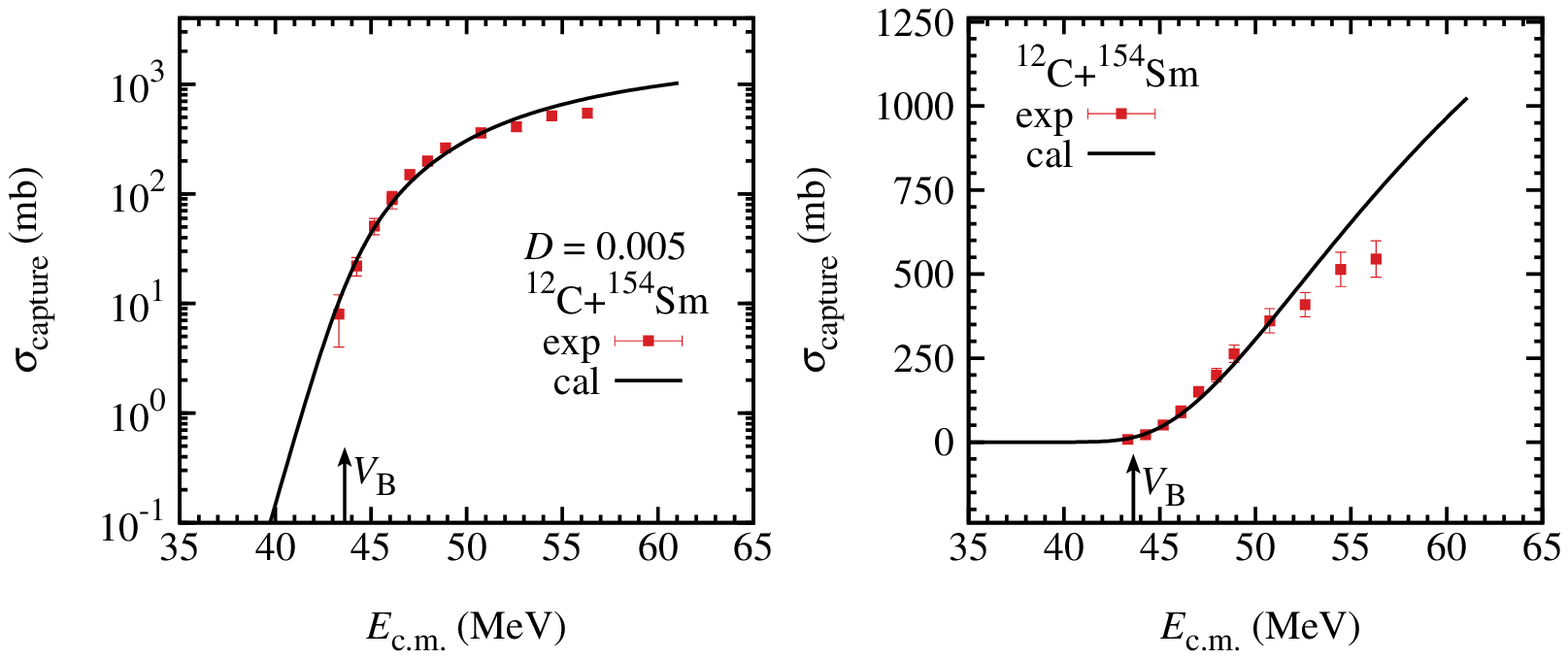}
  \includegraphics[width=0.47\textwidth]{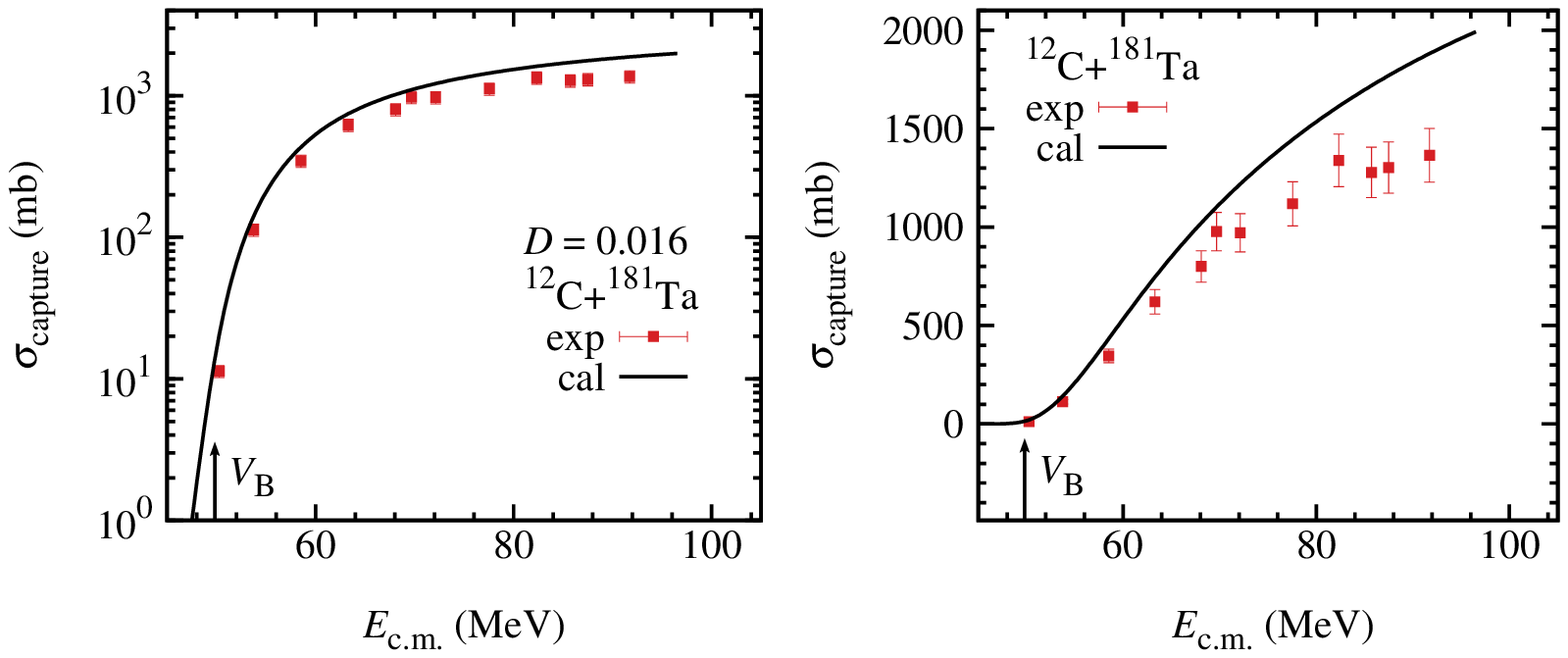}}
 \centerline{\includegraphics[width=0.47\textwidth]{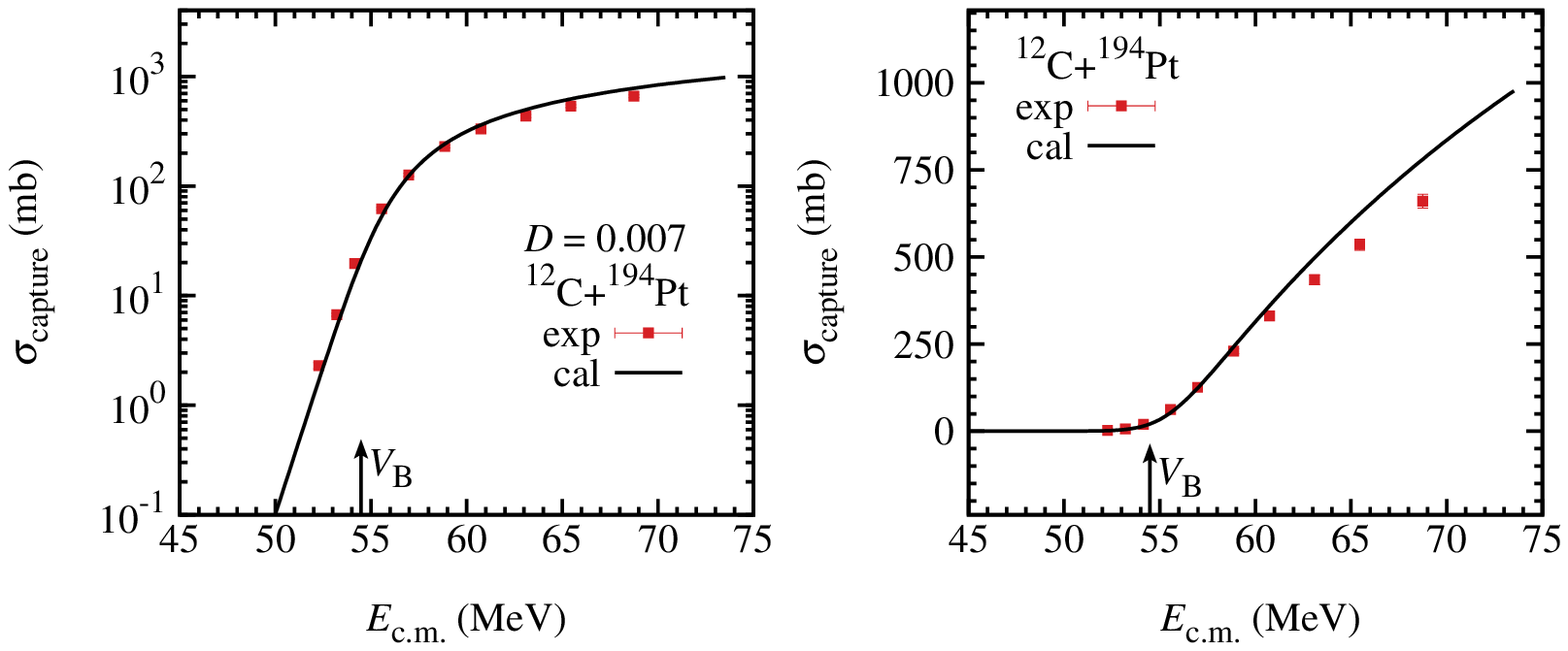}
  \includegraphics[width=0.47\textwidth]{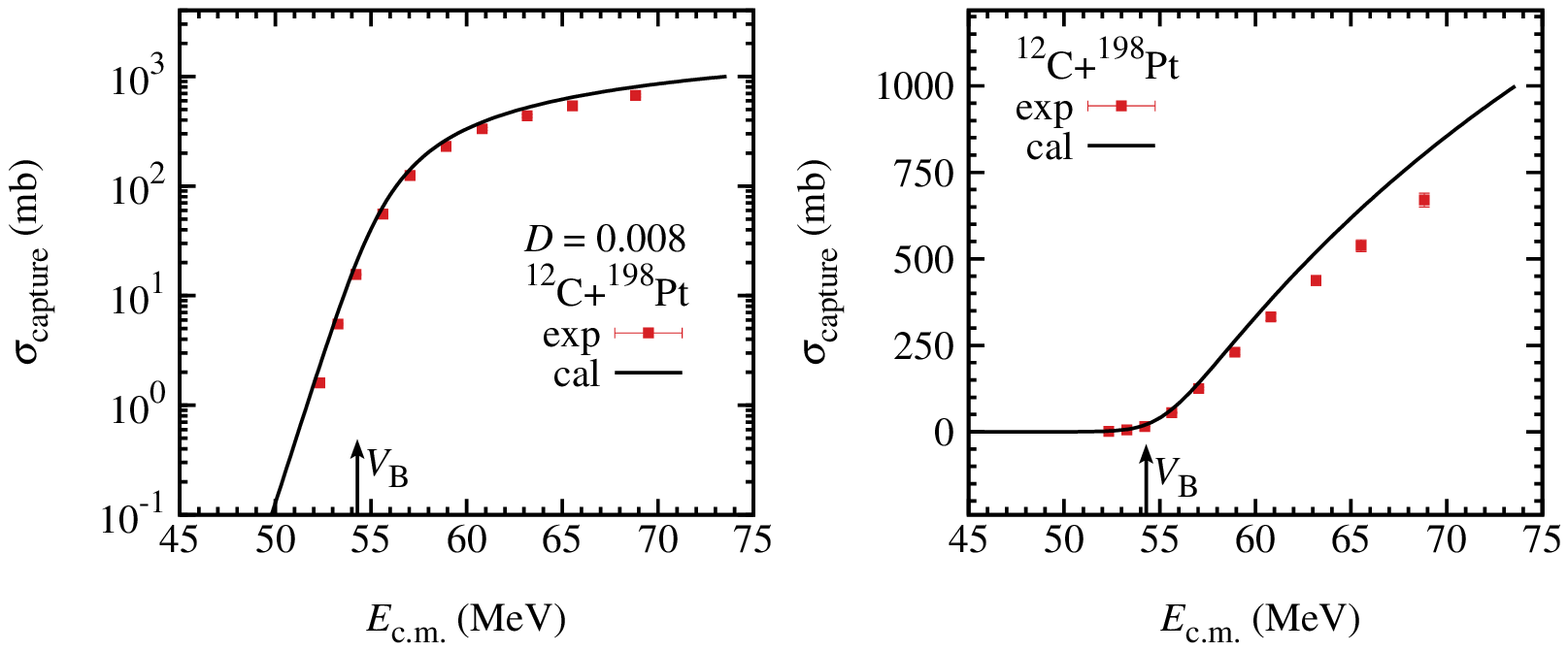}}
 \centerline{\includegraphics[width=0.47\textwidth]{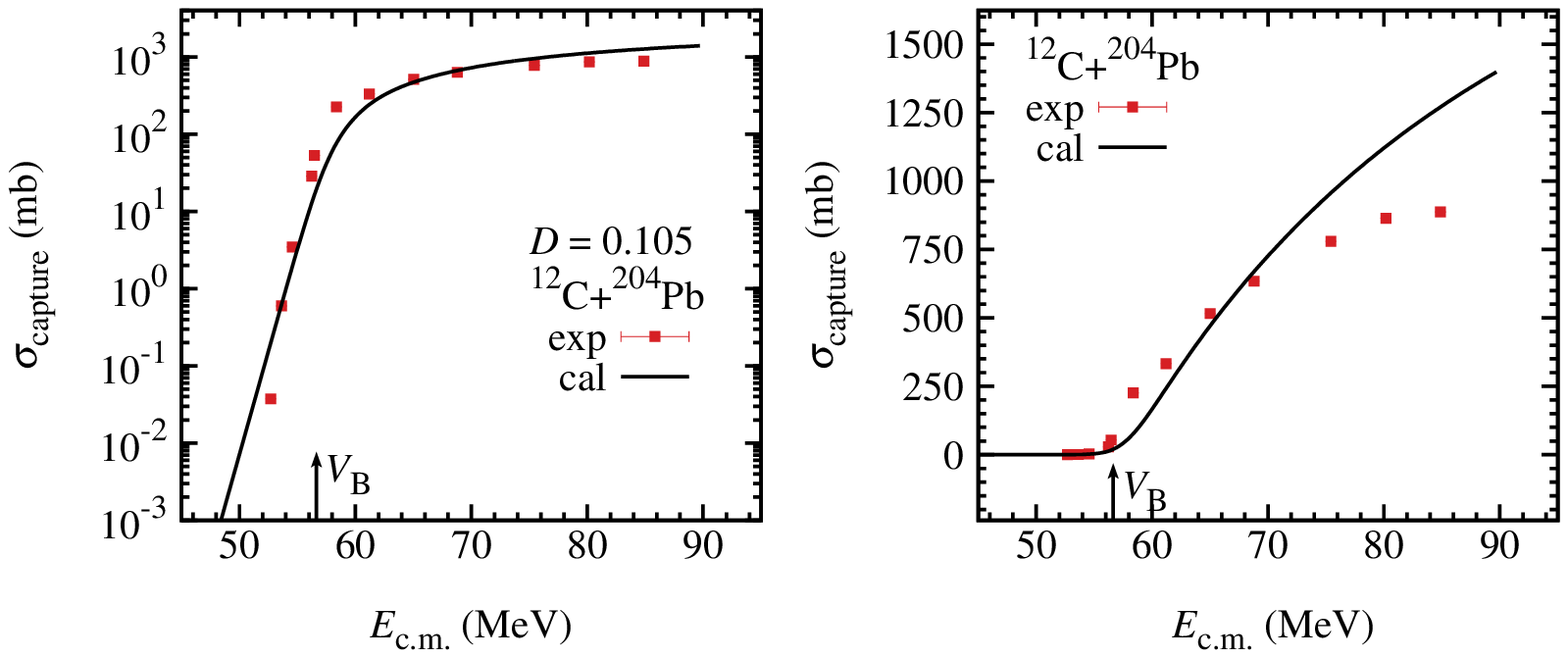}
  \includegraphics[width=0.47\textwidth]{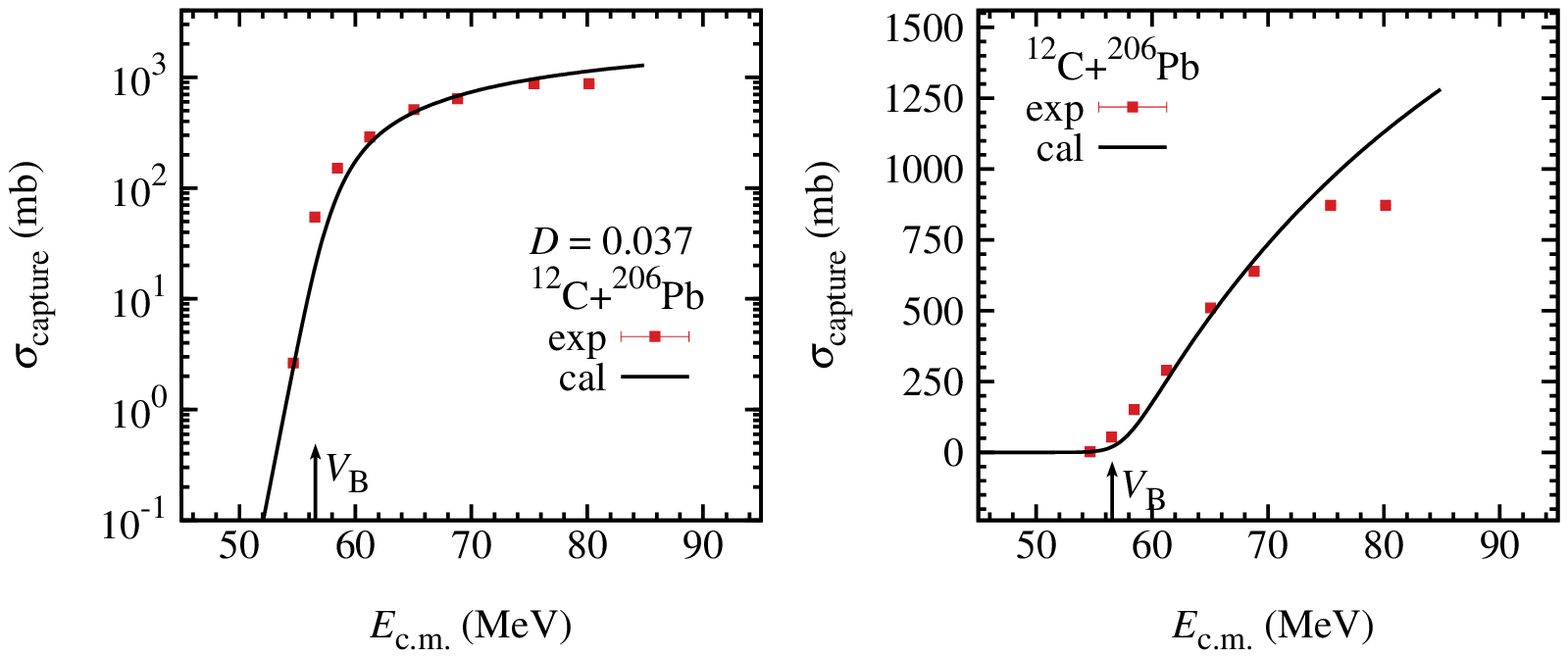}}
 \centerline{\includegraphics[width=0.47\textwidth]{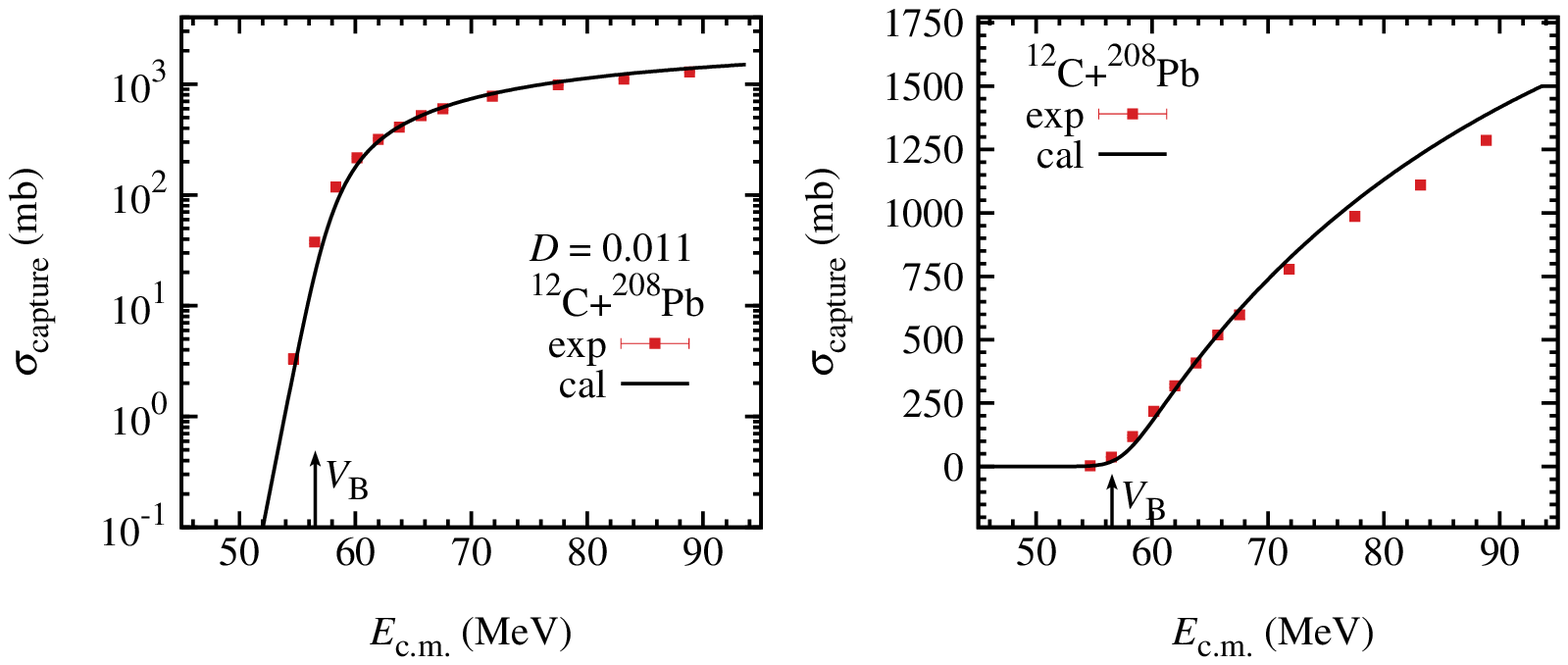}
  \includegraphics[width=0.47\textwidth]{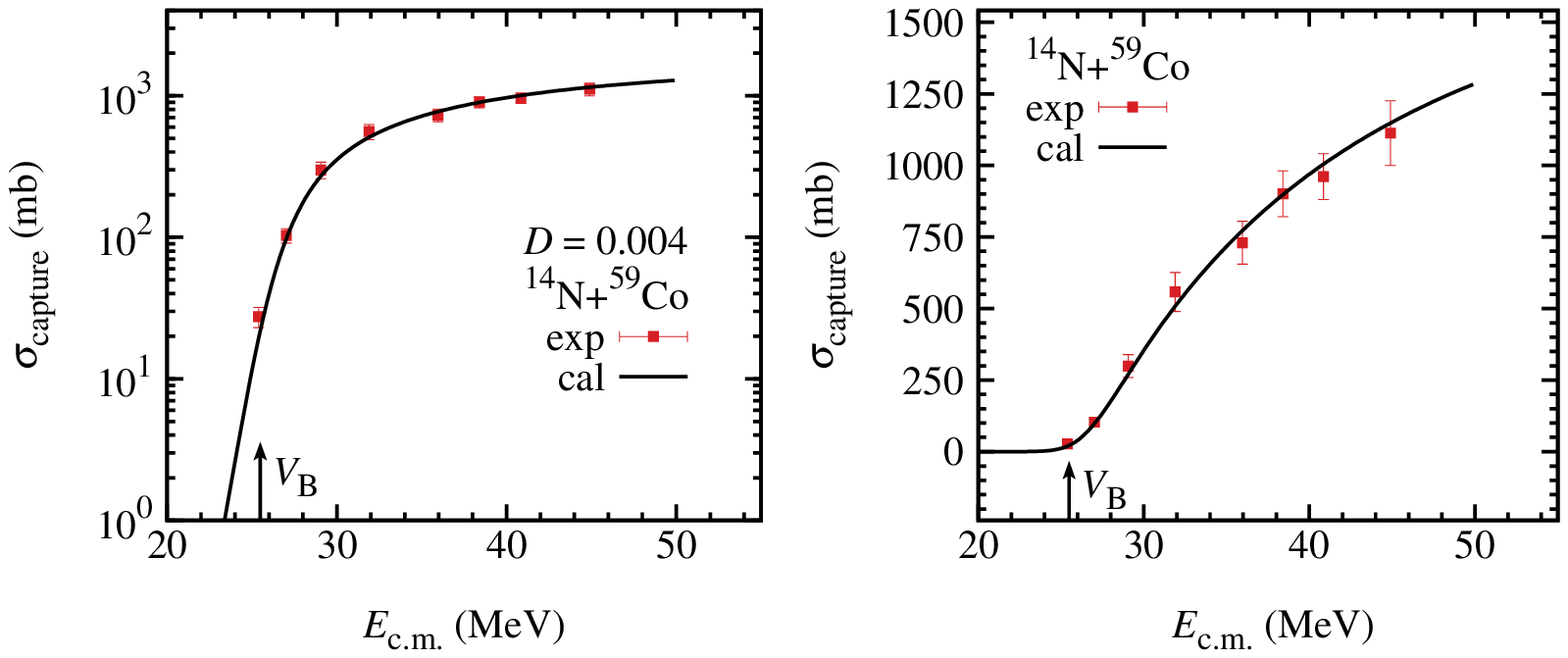}}
  \centerline {Graph 1}
 \end{Dfigures}
 \begin{Dfigures}[!ht]
 \centerline{\includegraphics[width=0.47\textwidth]{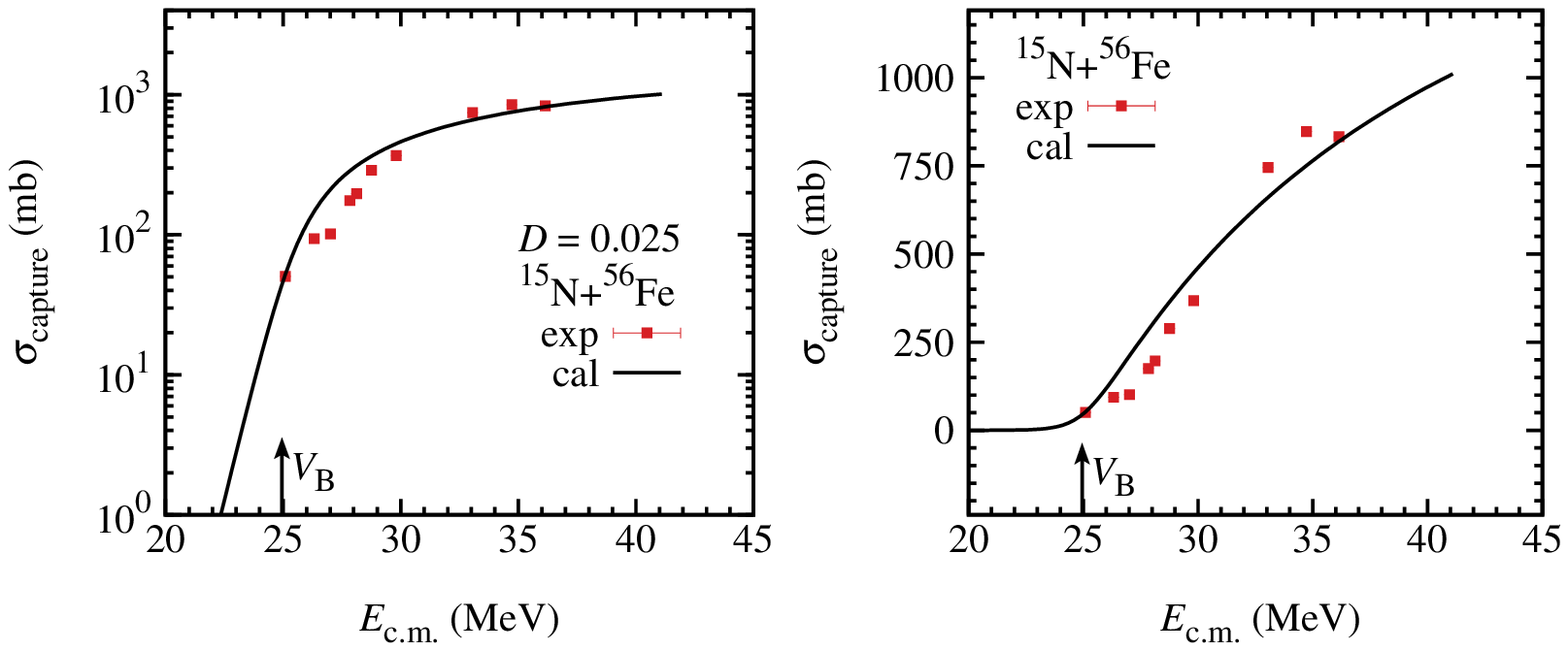}
  \includegraphics[width=0.47\textwidth]{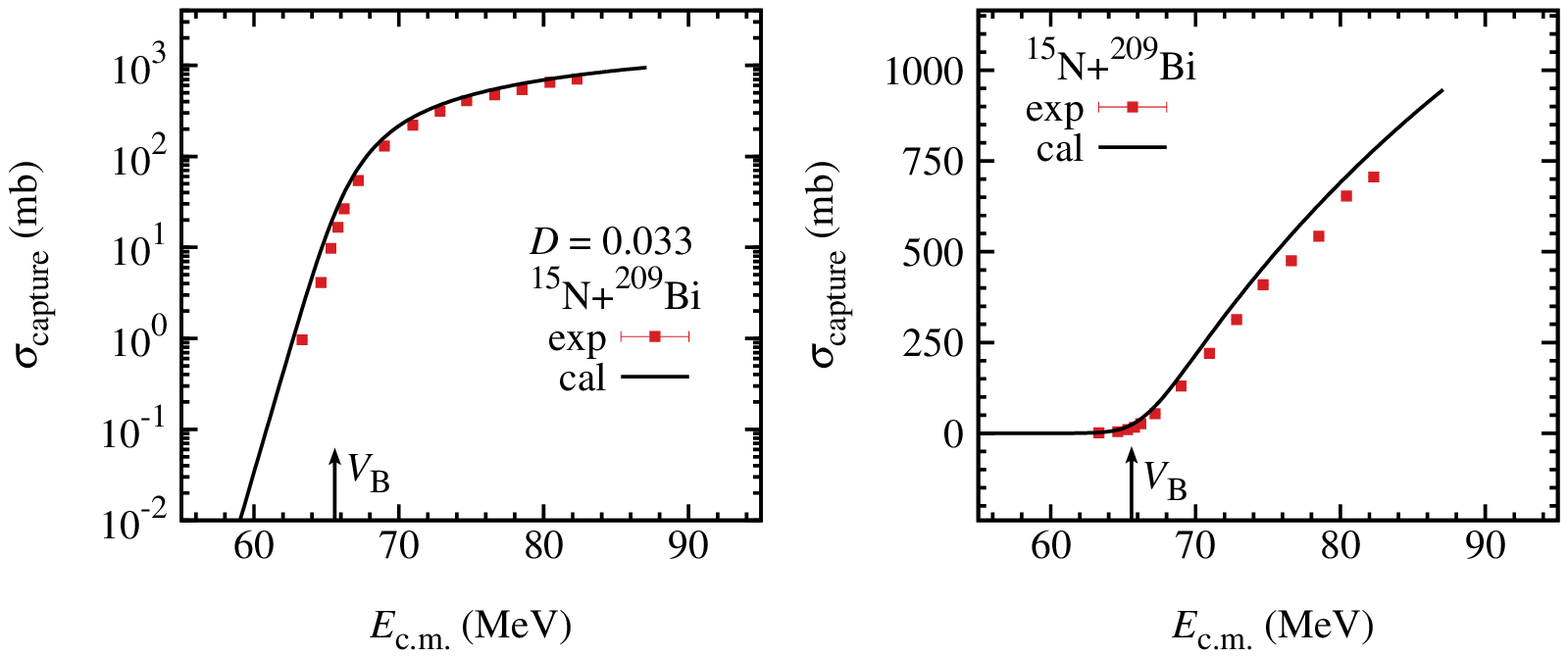}}
 \centerline{\includegraphics[width=0.47\textwidth]{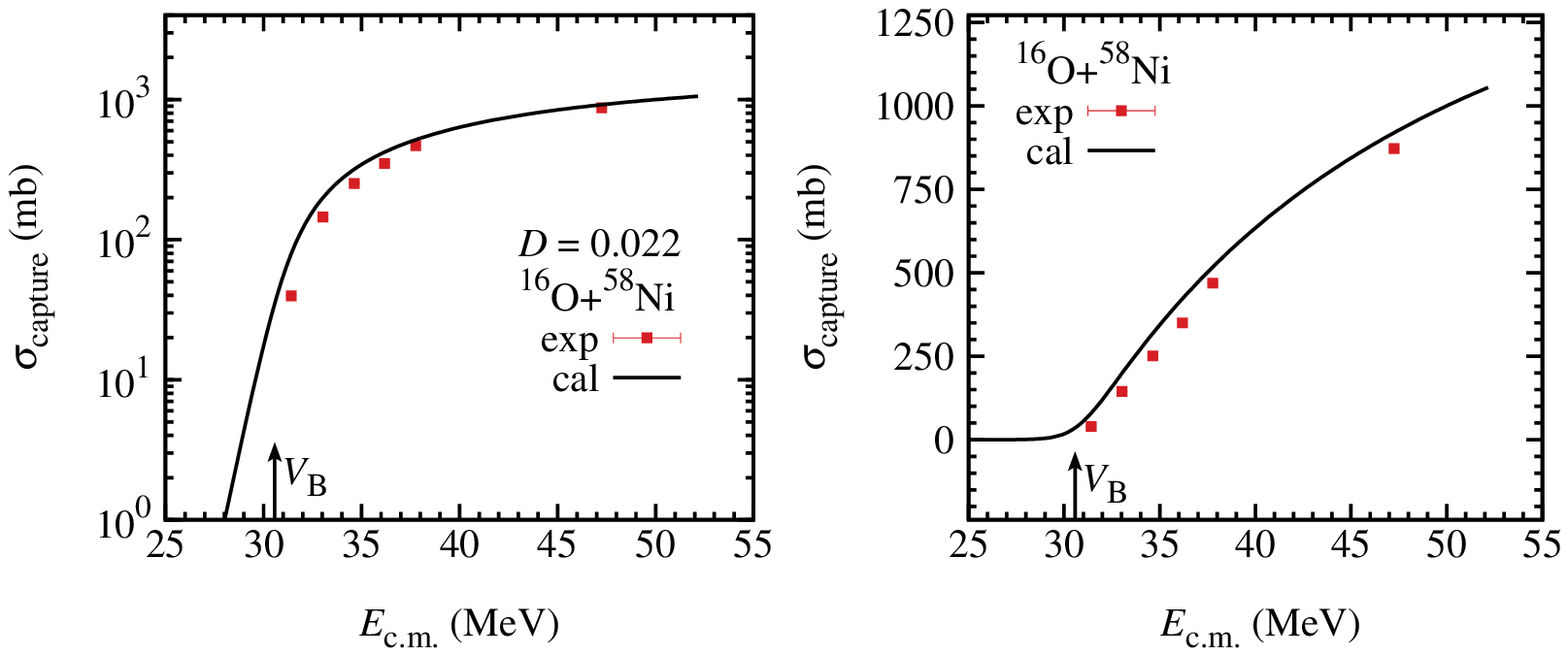}
  \includegraphics[width=0.47\textwidth]{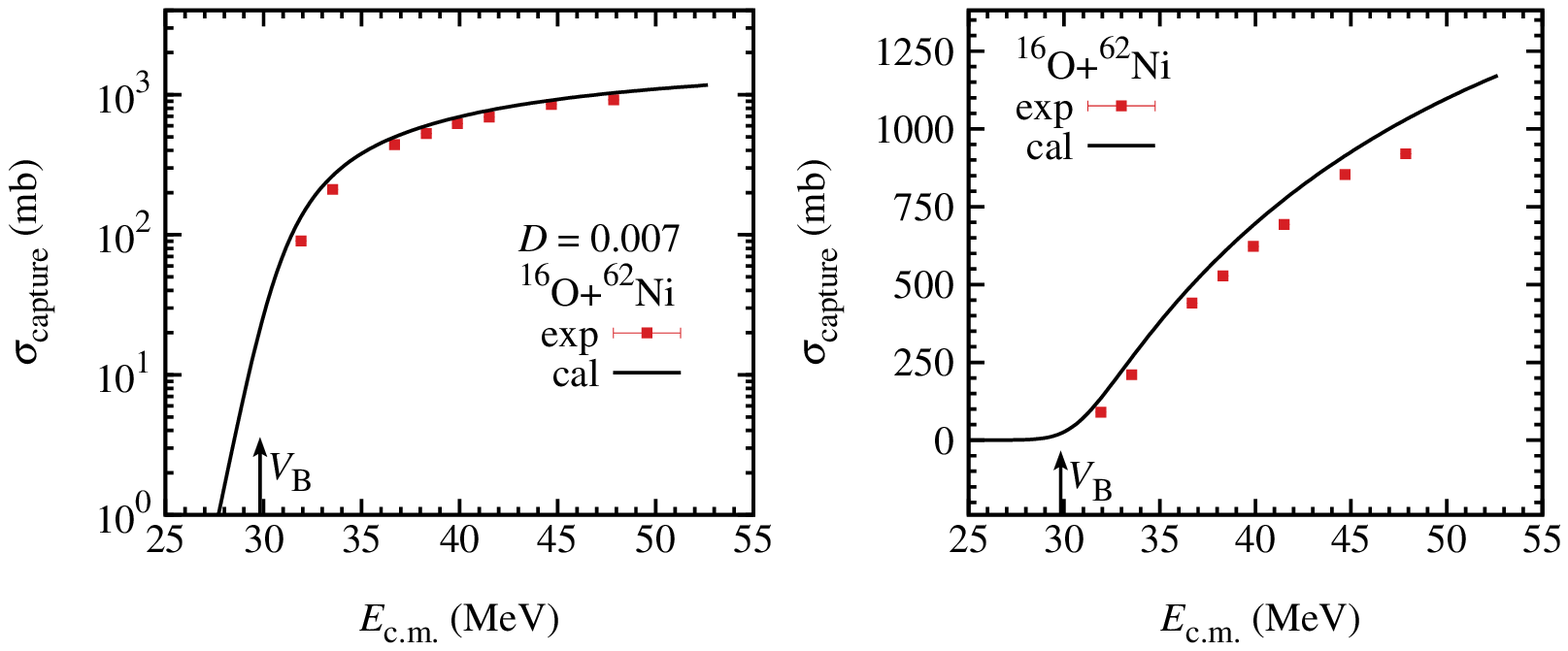}}
 \centerline{\includegraphics[width=0.47\textwidth]{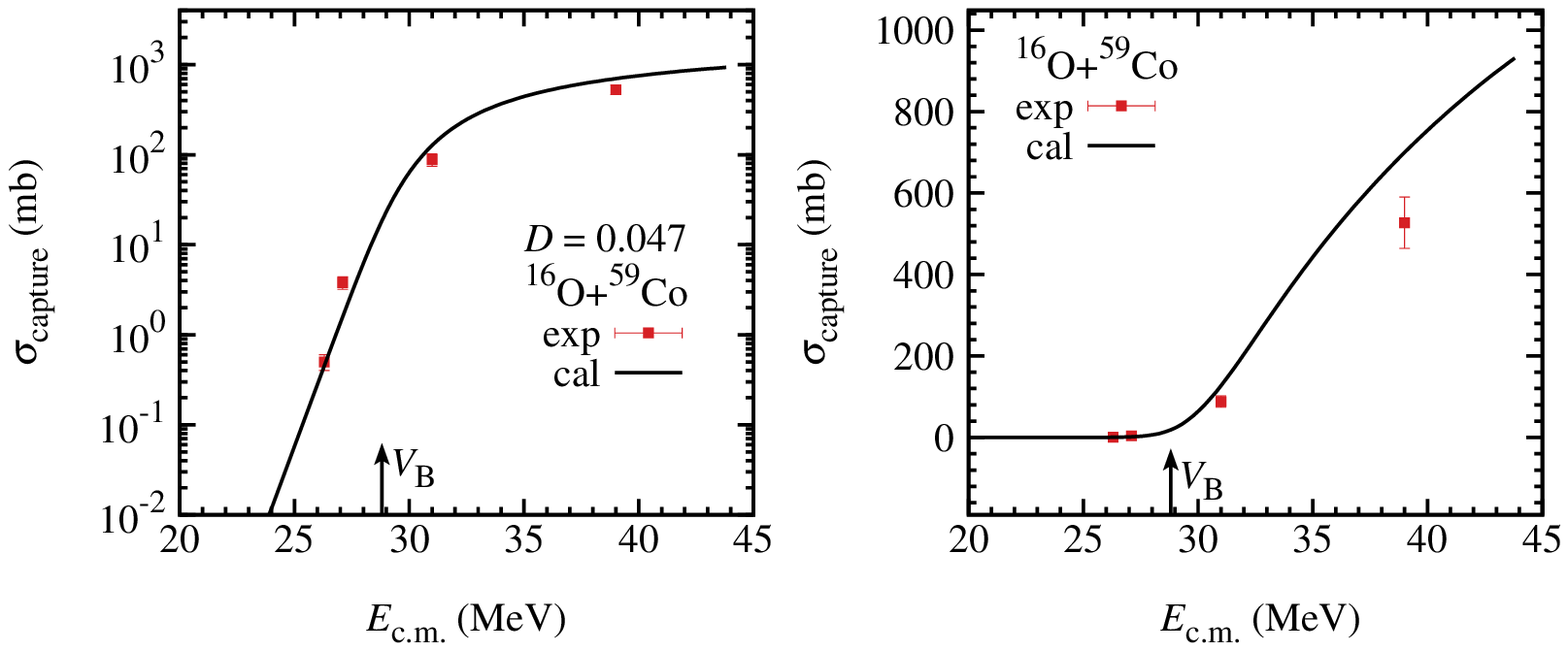}
  \includegraphics[width=0.47\textwidth]{16O70Ge.eps}}
 \centerline{\includegraphics[width=0.47\textwidth]{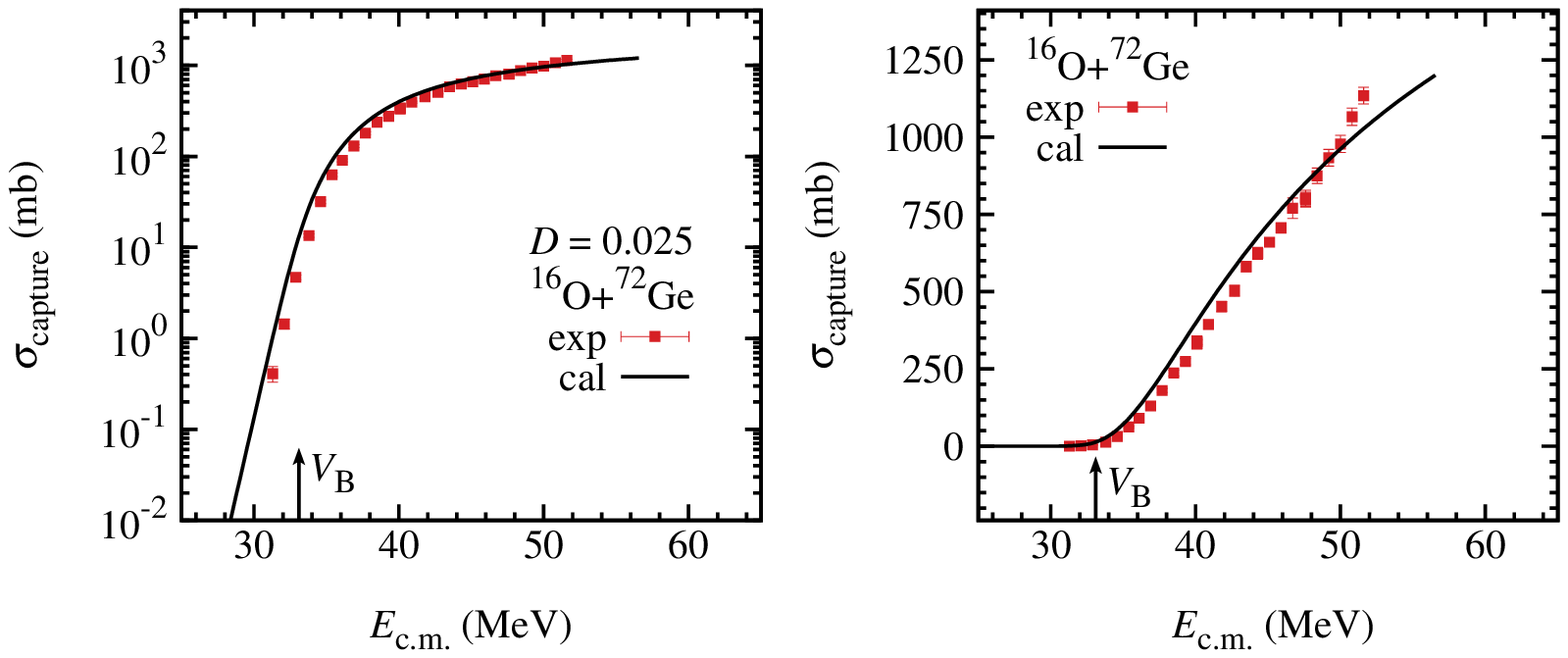}
  \includegraphics[width=0.47\textwidth]{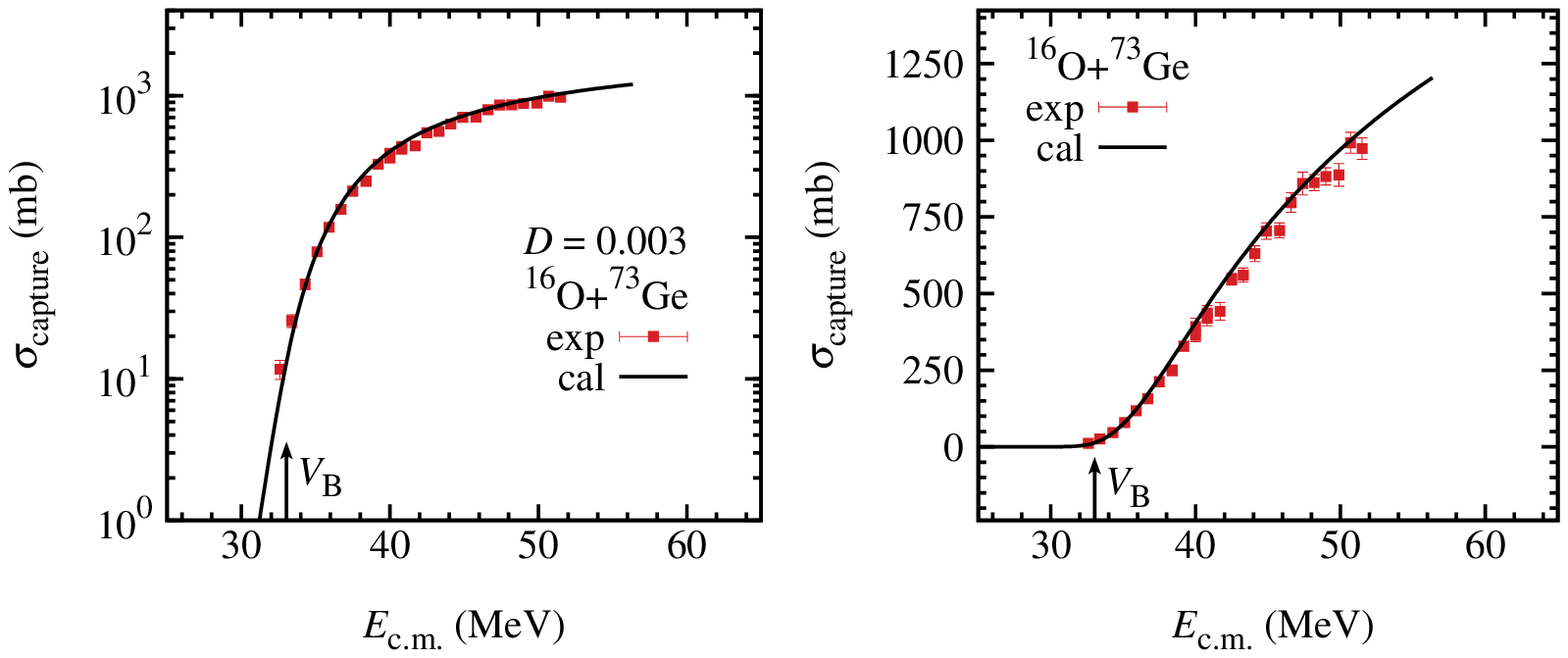}}
 \centerline{\includegraphics[width=0.47\textwidth]{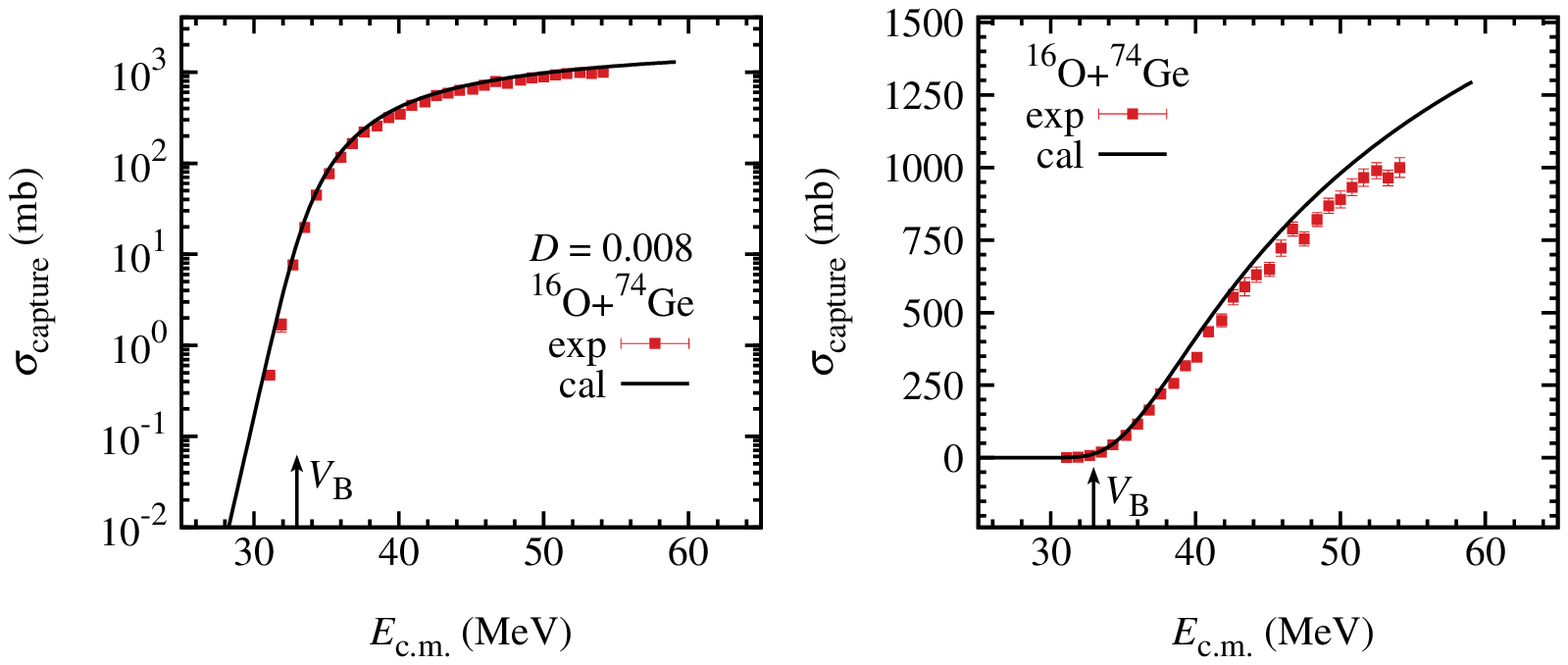}
  \includegraphics[width=0.47\textwidth]{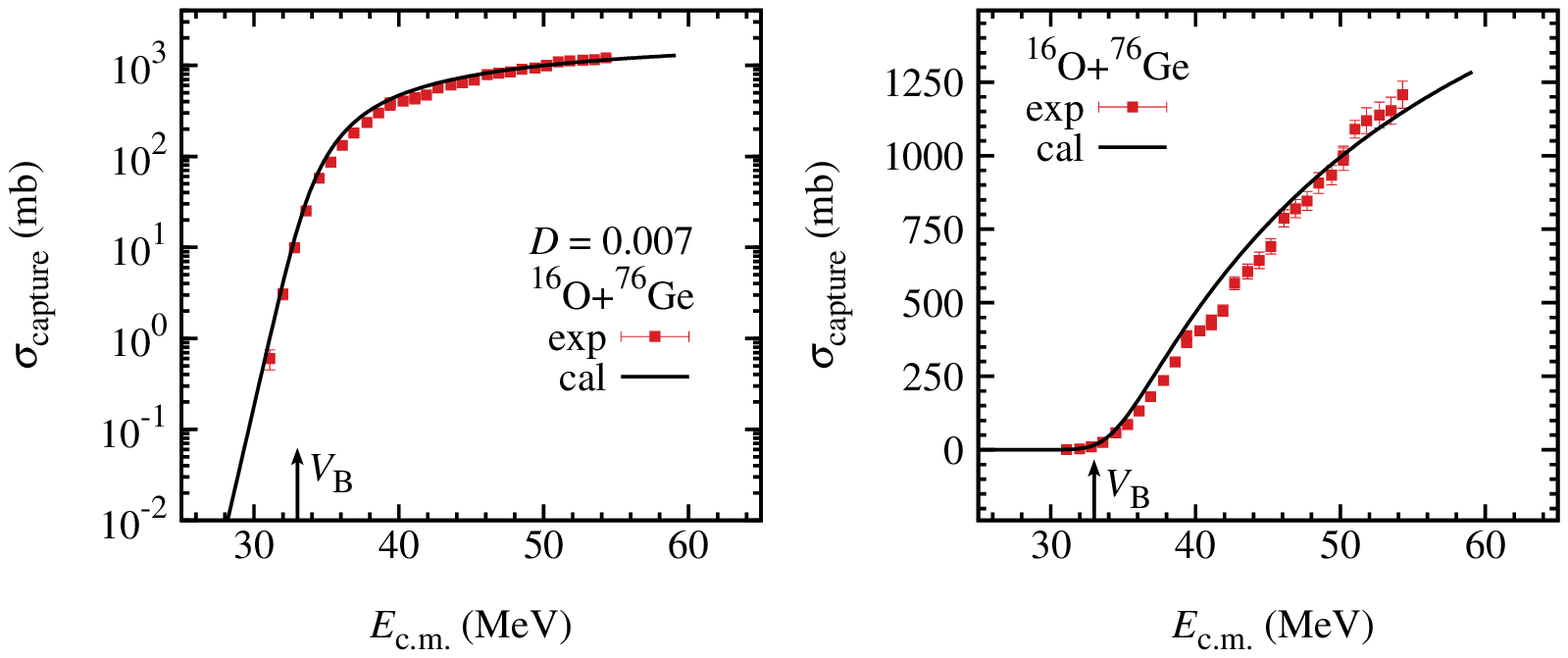}}
 \centerline{\includegraphics[width=0.47\textwidth]{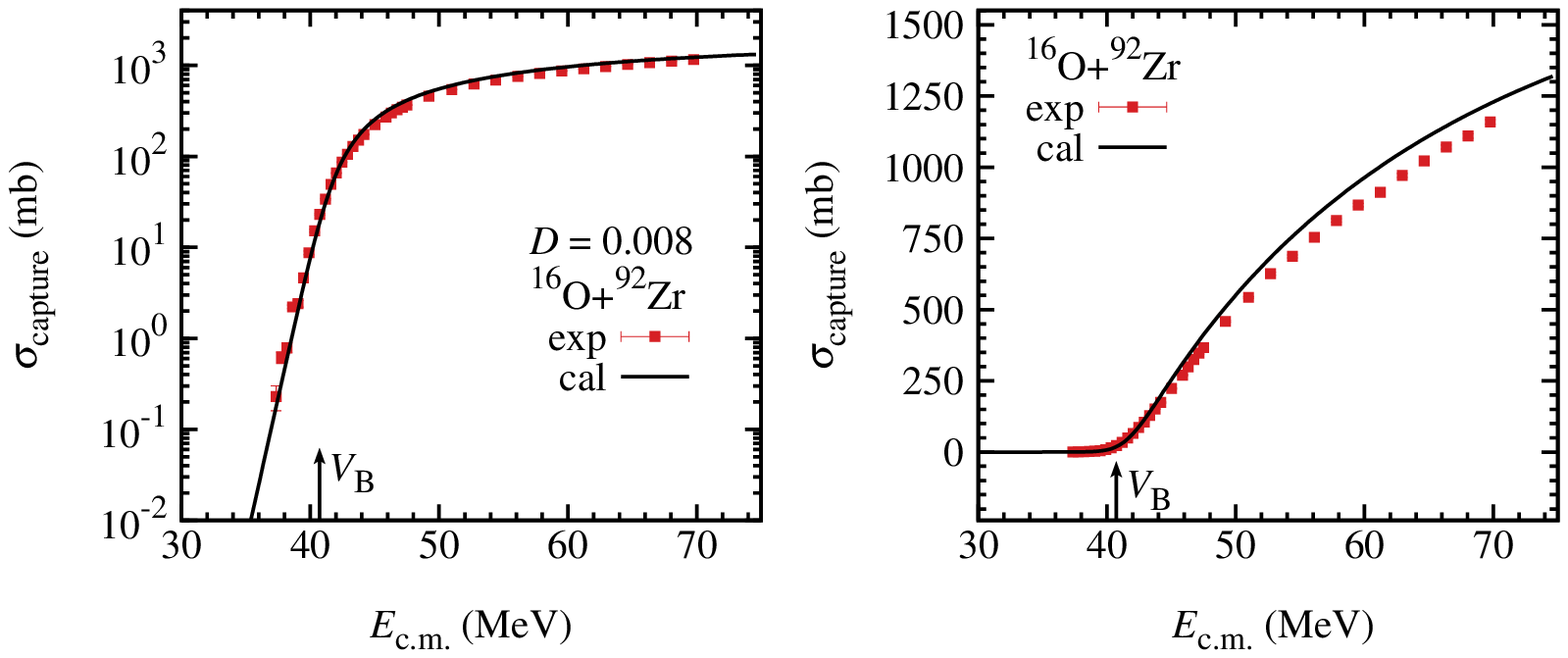}
  \includegraphics[width=0.47\textwidth]{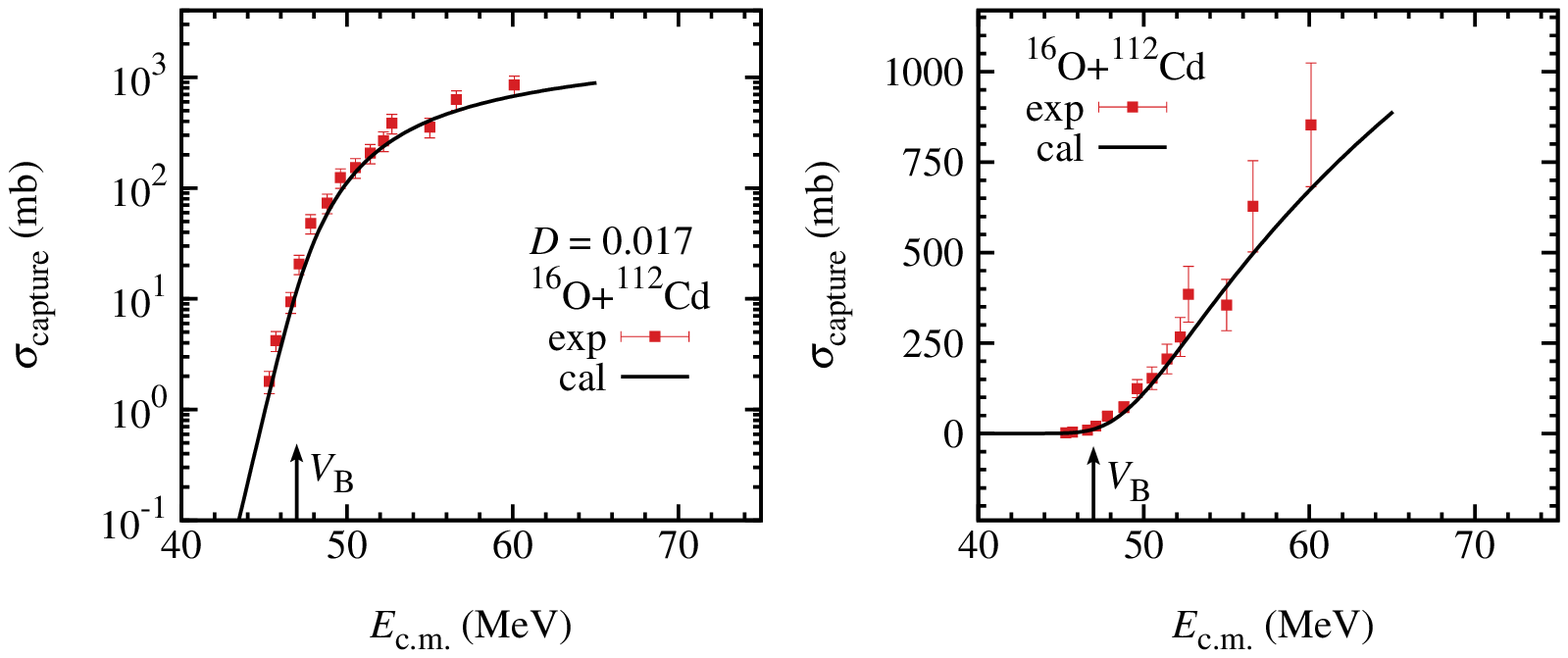}}
  \centerline {Graph 2}
 \end{Dfigures}
 \begin{Dfigures}[!ht]
 \centerline{\includegraphics[width=0.47\textwidth]{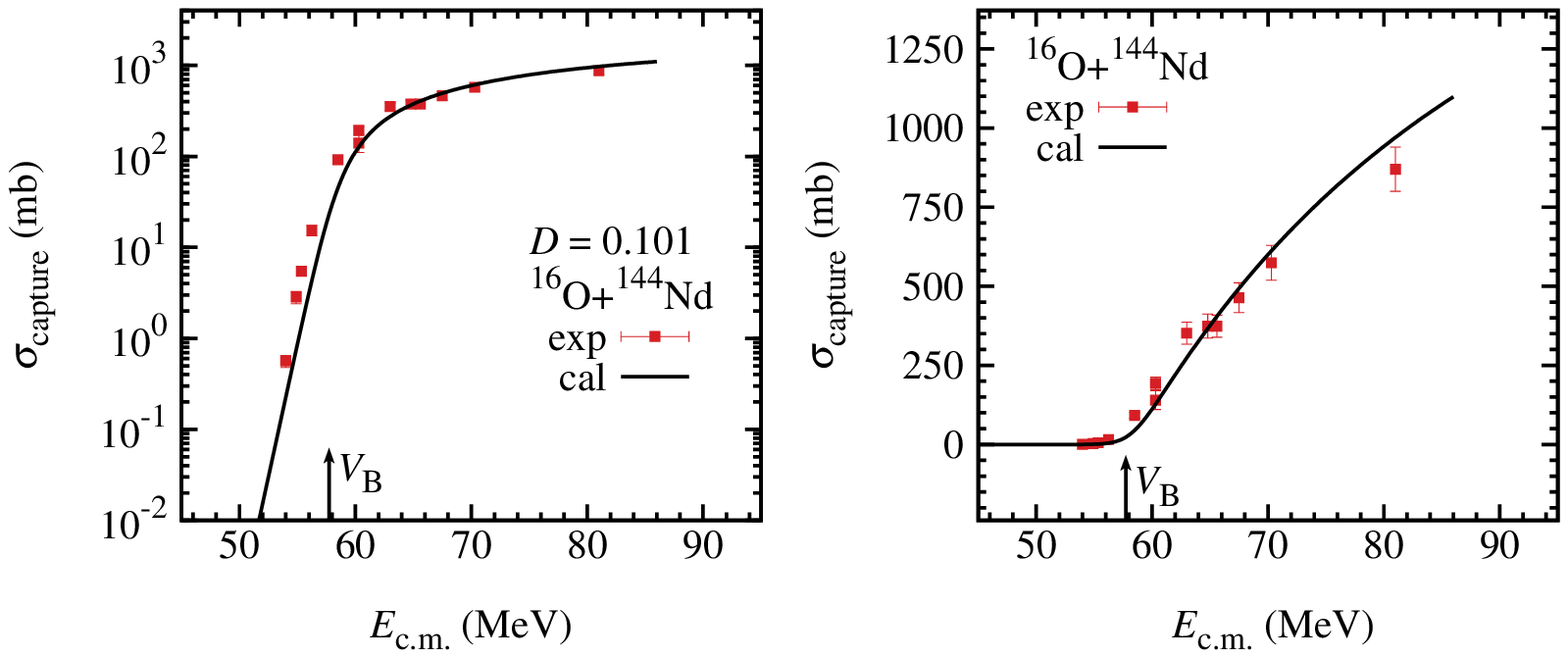}
  \includegraphics[width=0.47\textwidth]{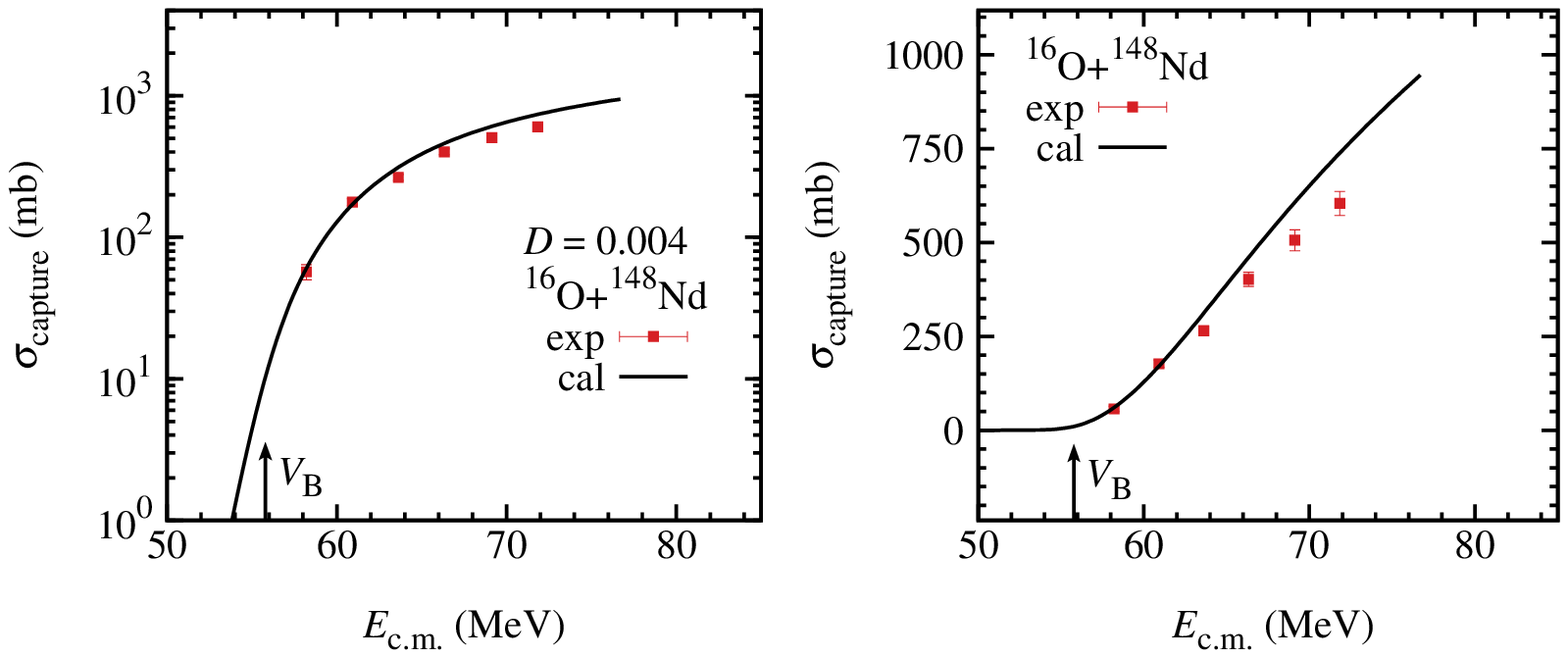}}
 \centerline{\includegraphics[width=0.47\textwidth]{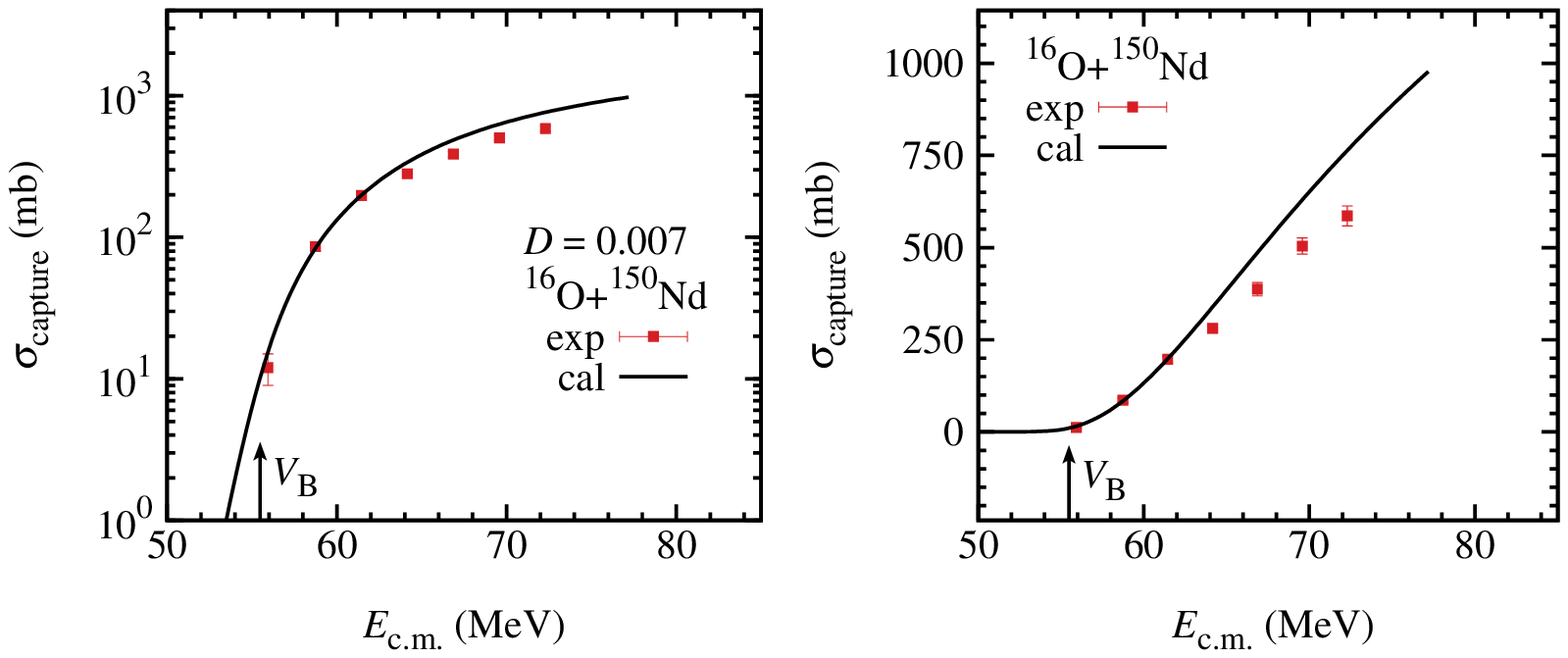}
  \includegraphics[width=0.47\textwidth]{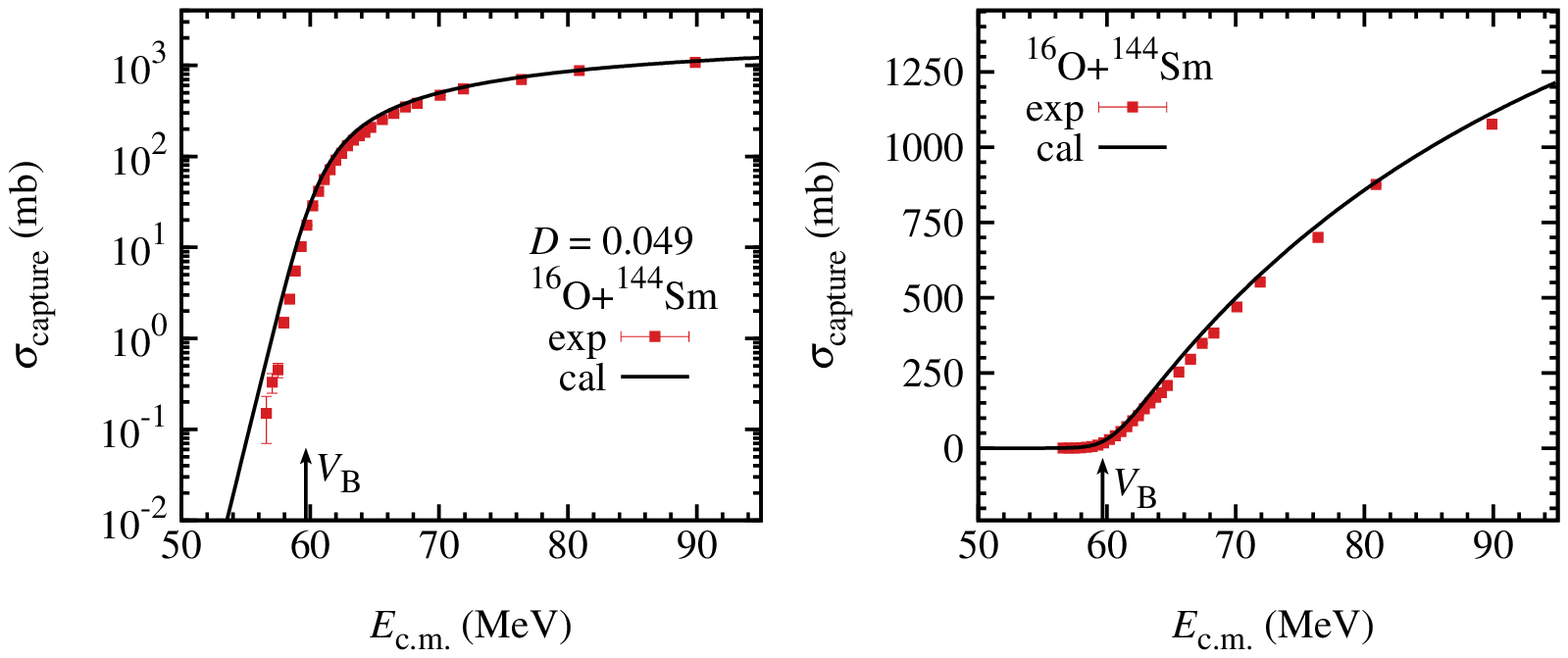}}
 \centerline{\includegraphics[width=0.47\textwidth]{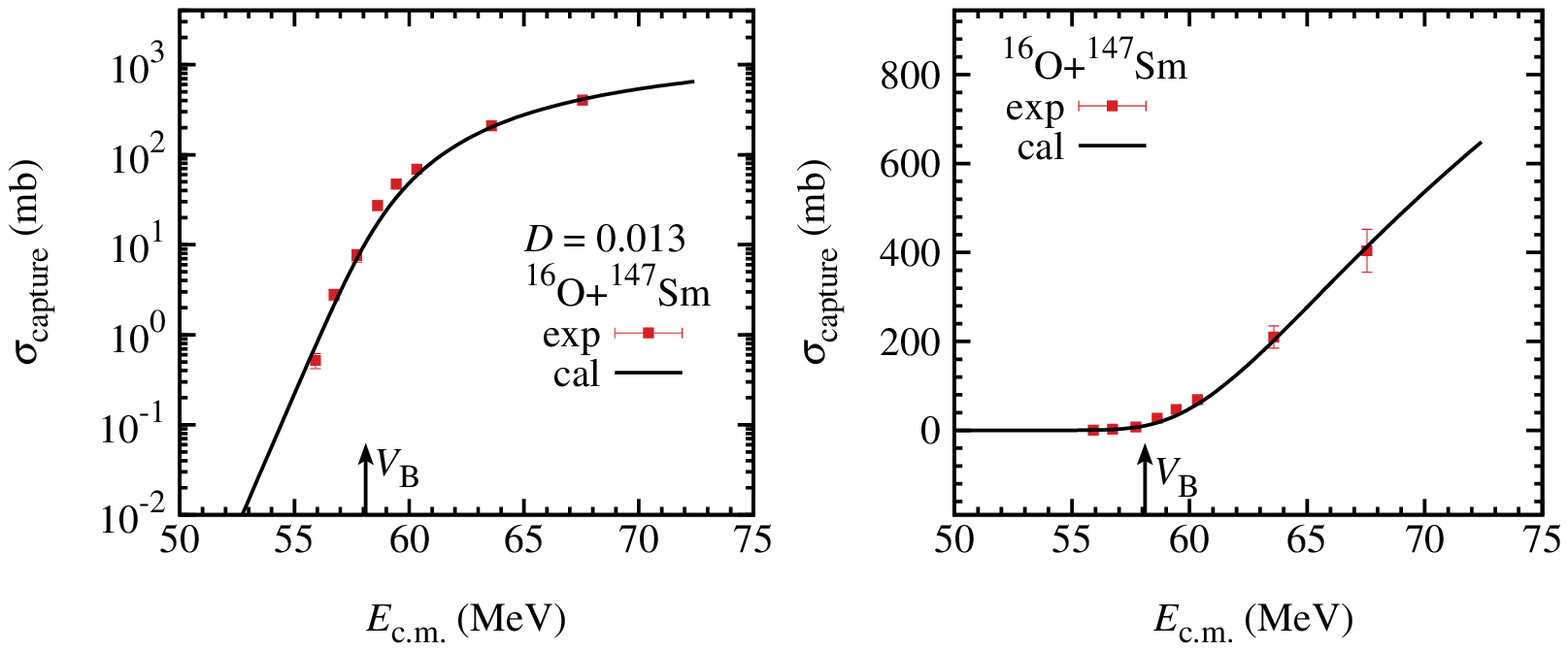}
  \includegraphics[width=0.47\textwidth]{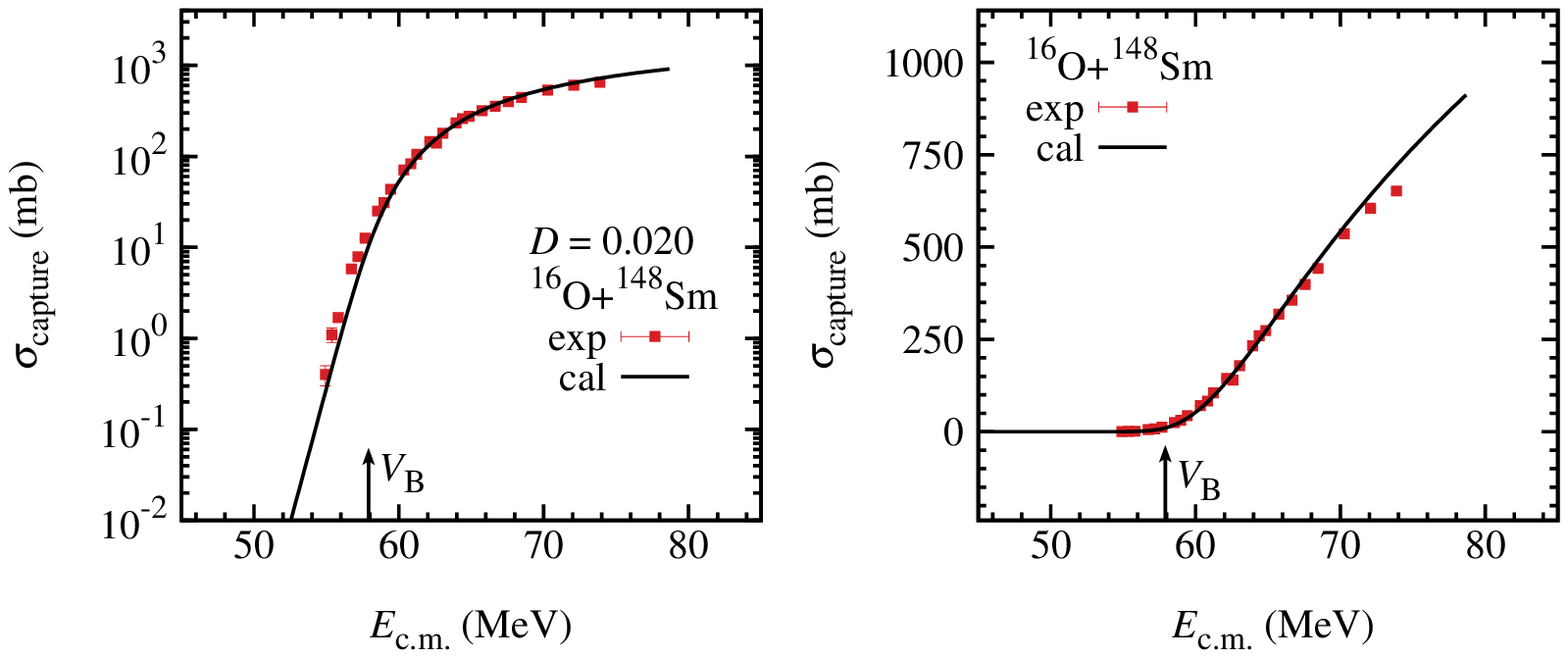}}
 \centerline{\includegraphics[width=0.47\textwidth]{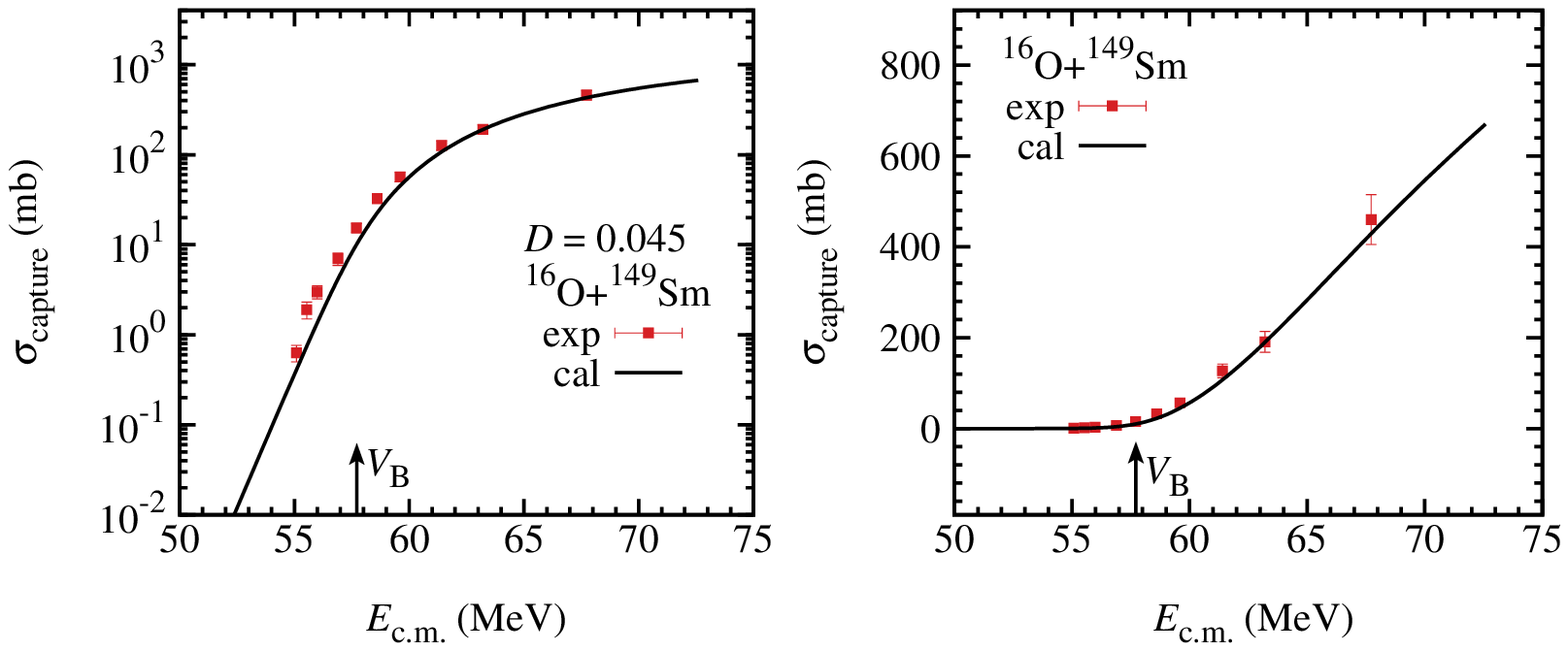}
  \includegraphics[width=0.47\textwidth]{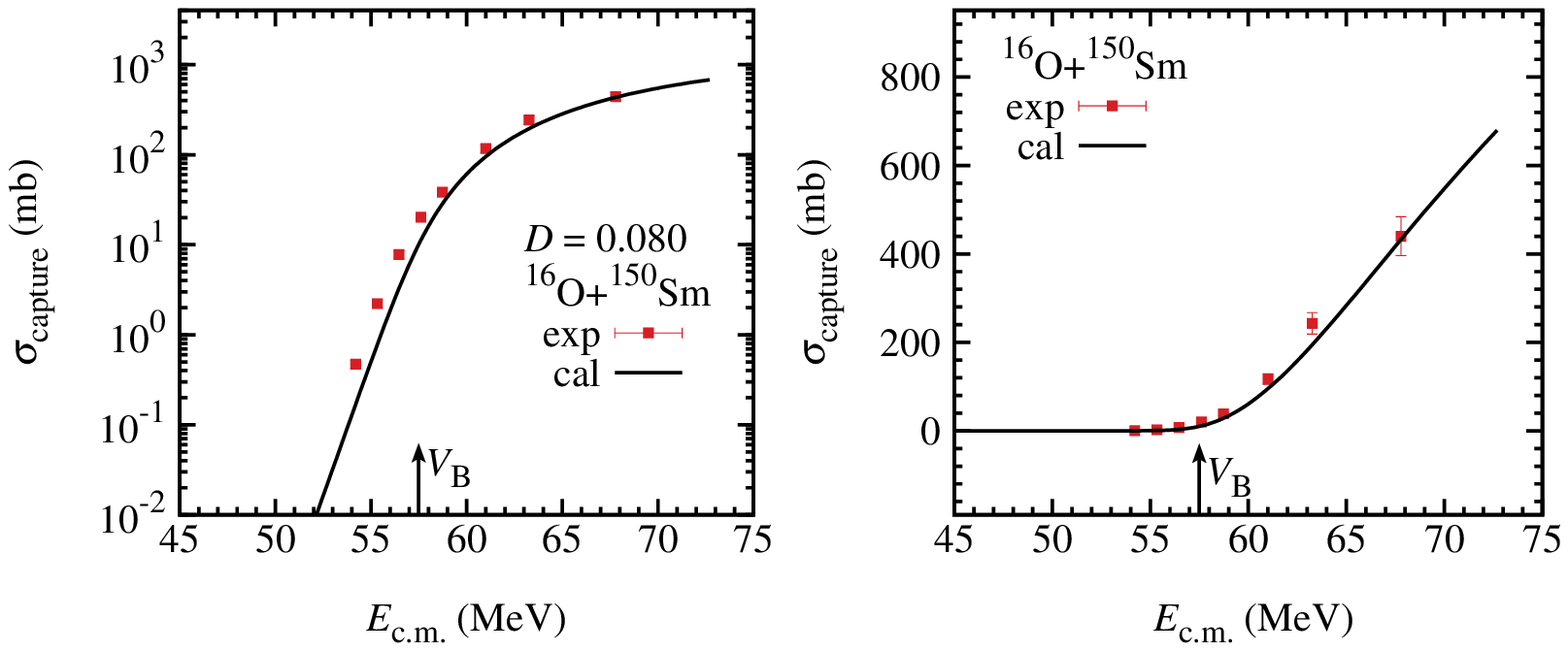}}
 \centerline{\includegraphics[width=0.47\textwidth]{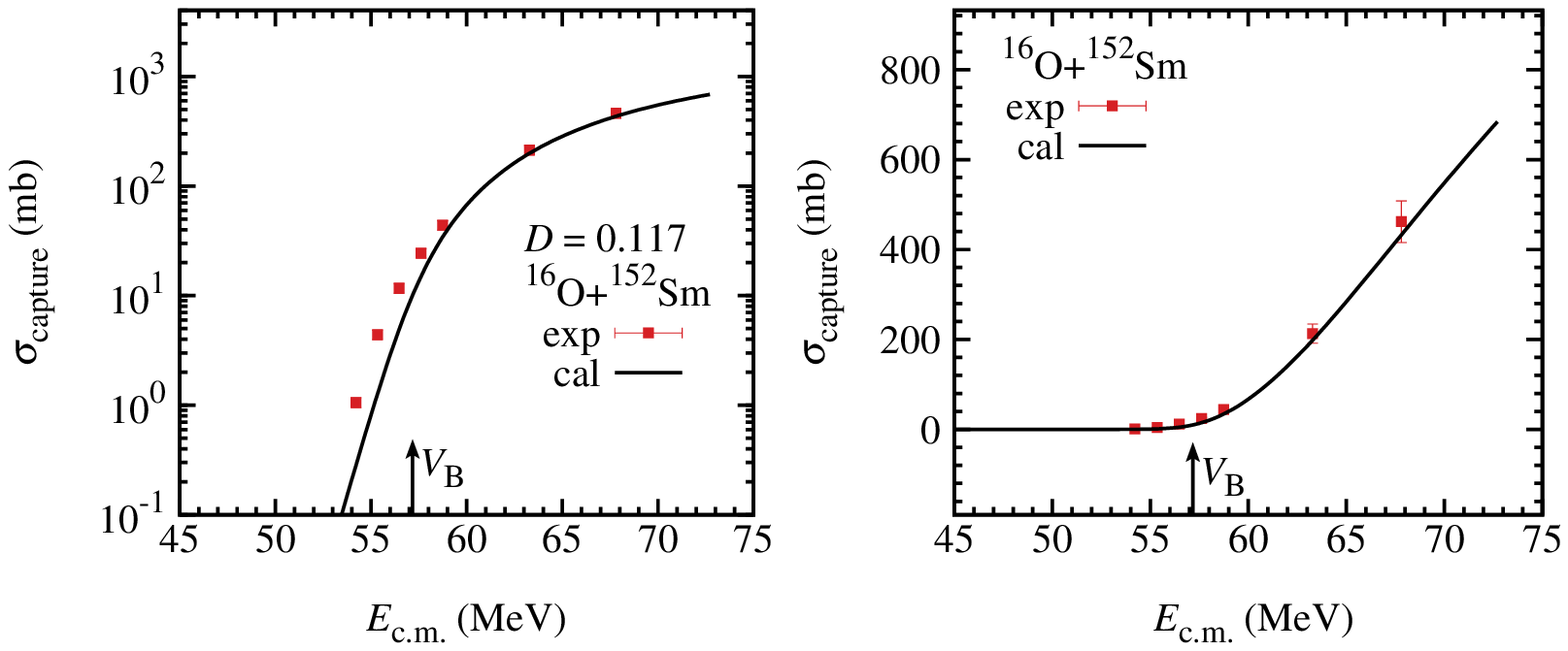}
  \includegraphics[width=0.47\textwidth]{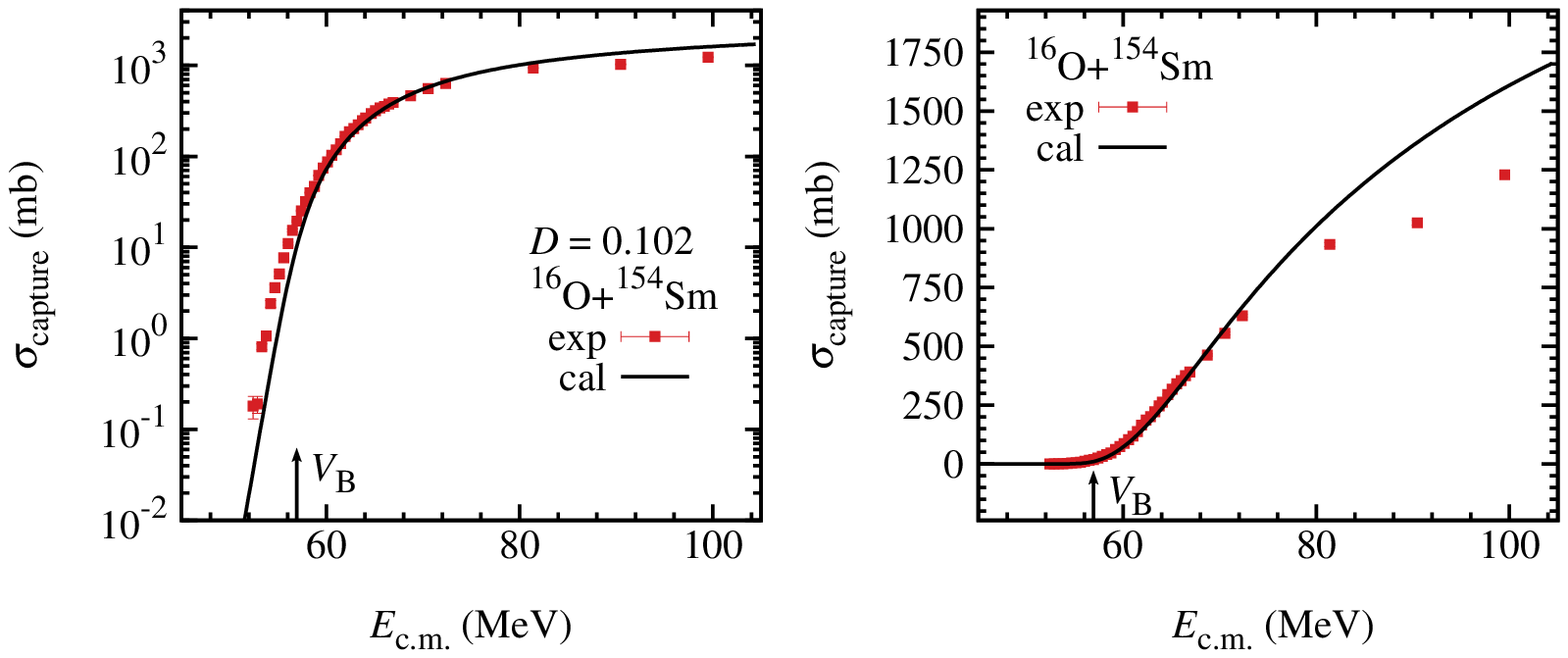}}
 \centerline{\includegraphics[width=0.47\textwidth]{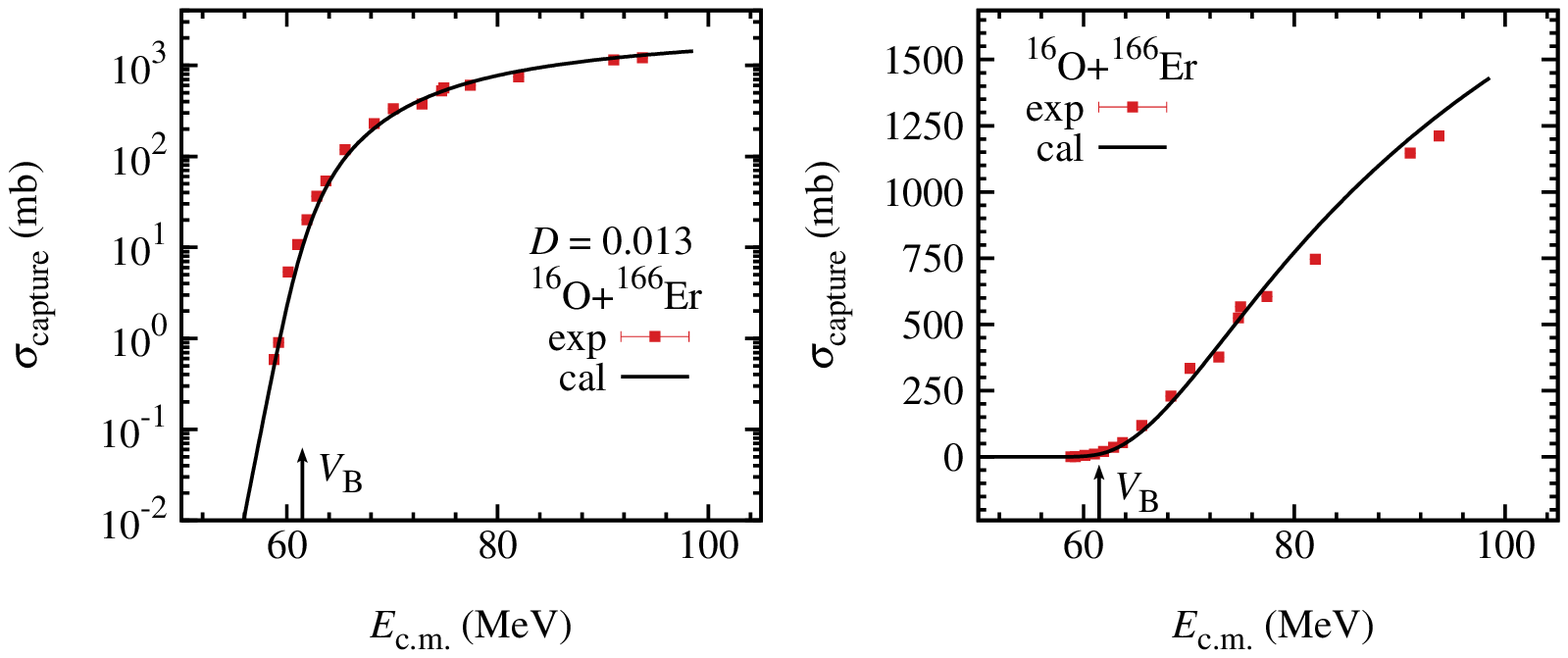}
  \includegraphics[width=0.47\textwidth]{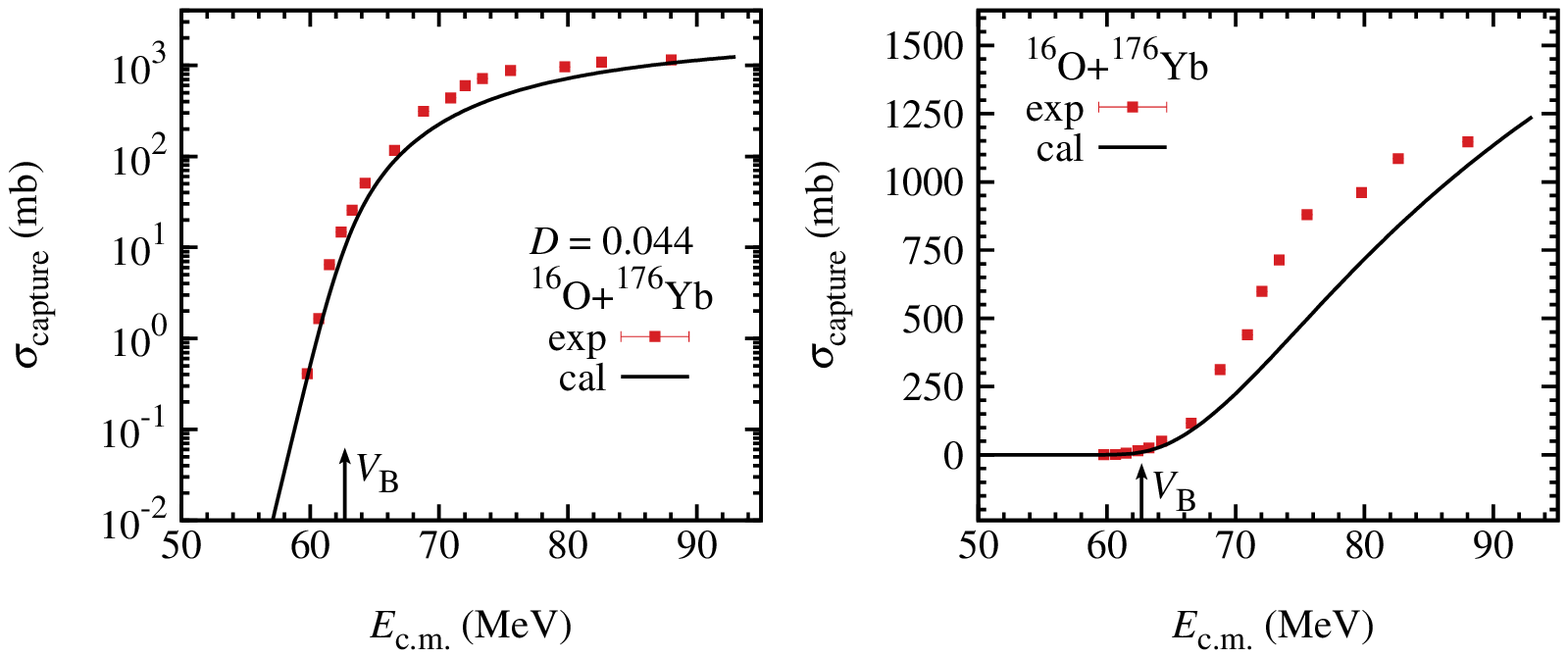}}
  \centerline {Graph 3}
 \end{Dfigures}
 \begin{Dfigures}[!ht]
 \centerline{\includegraphics[width=0.47\textwidth]{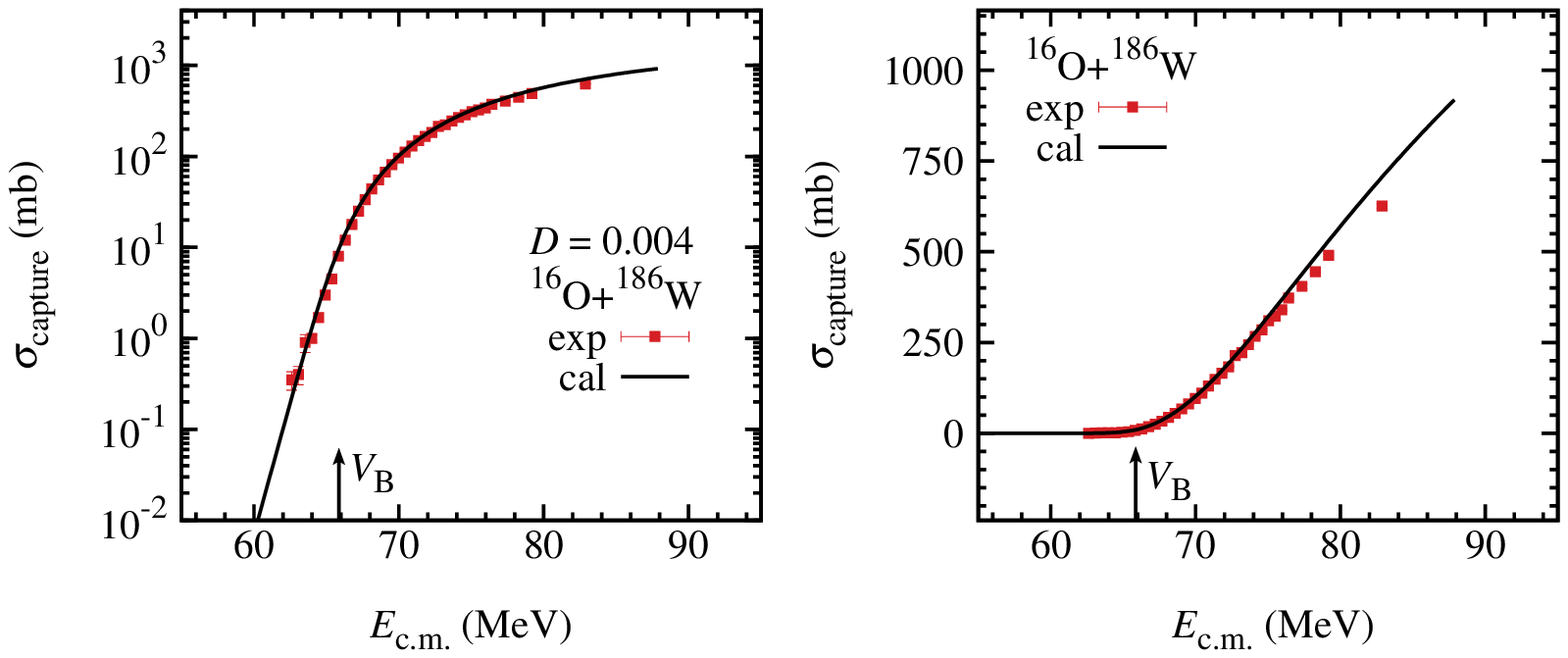}
  \includegraphics[width=0.47\textwidth]{16O204Pb.eps}}
 \centerline{\includegraphics[width=0.47\textwidth]{16O208Pb.eps}
  \includegraphics[width=0.47\textwidth]{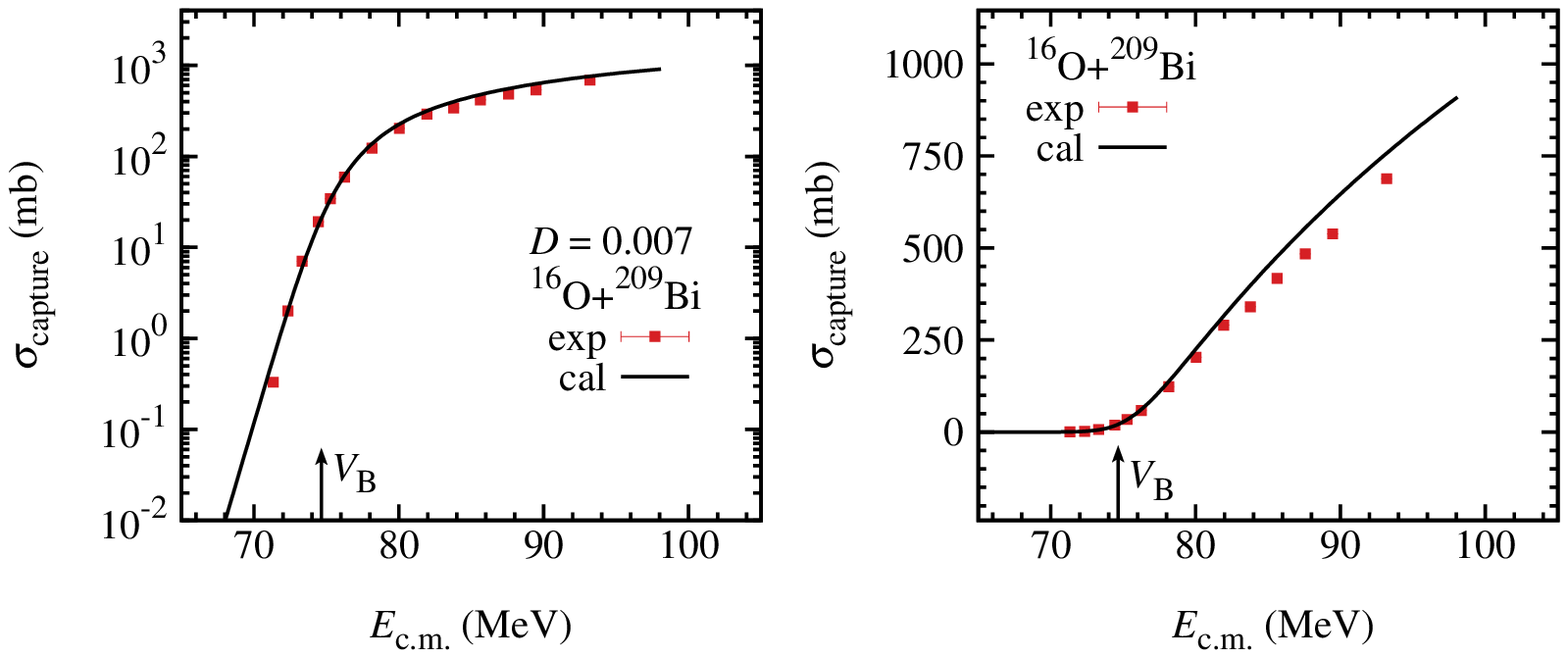}}
 \centerline{\includegraphics[width=0.47\textwidth]{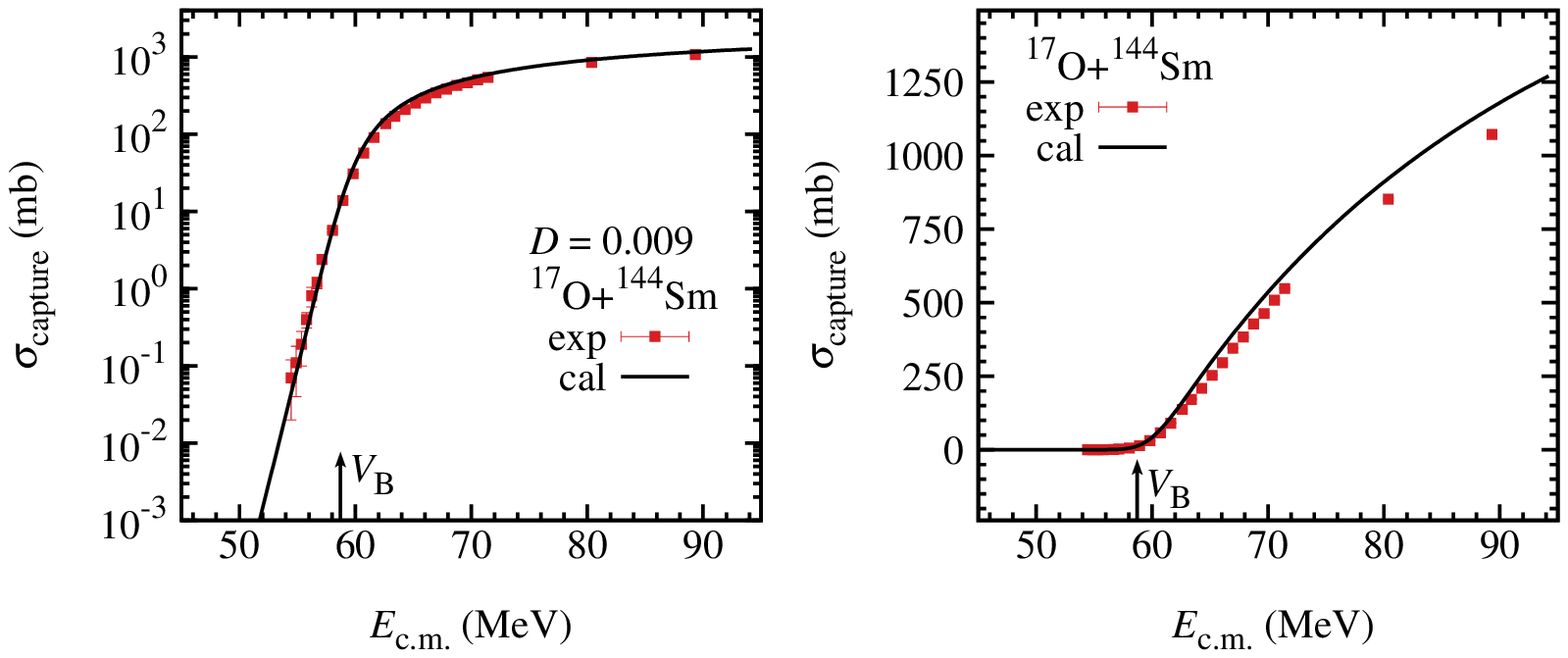}
  \includegraphics[width=0.47\textwidth]{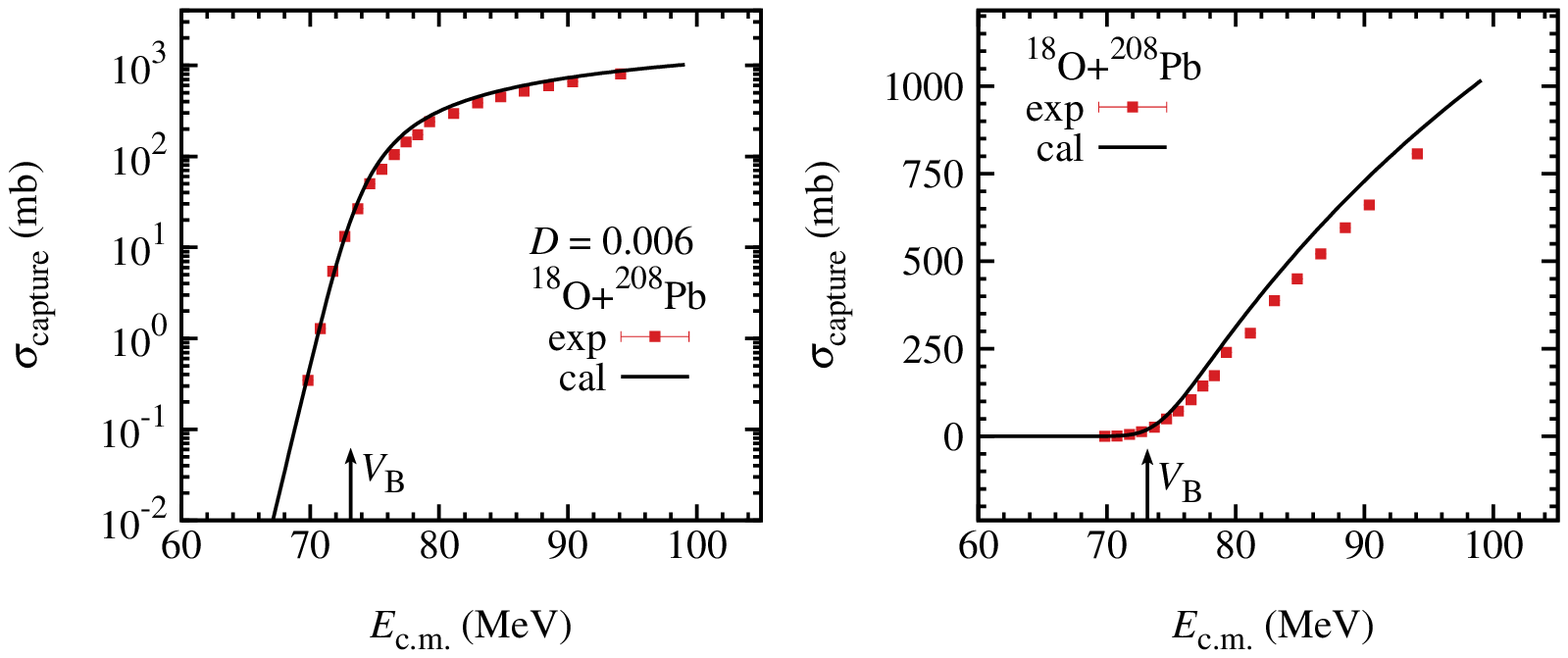}}
 \centerline{\includegraphics[width=0.47\textwidth]{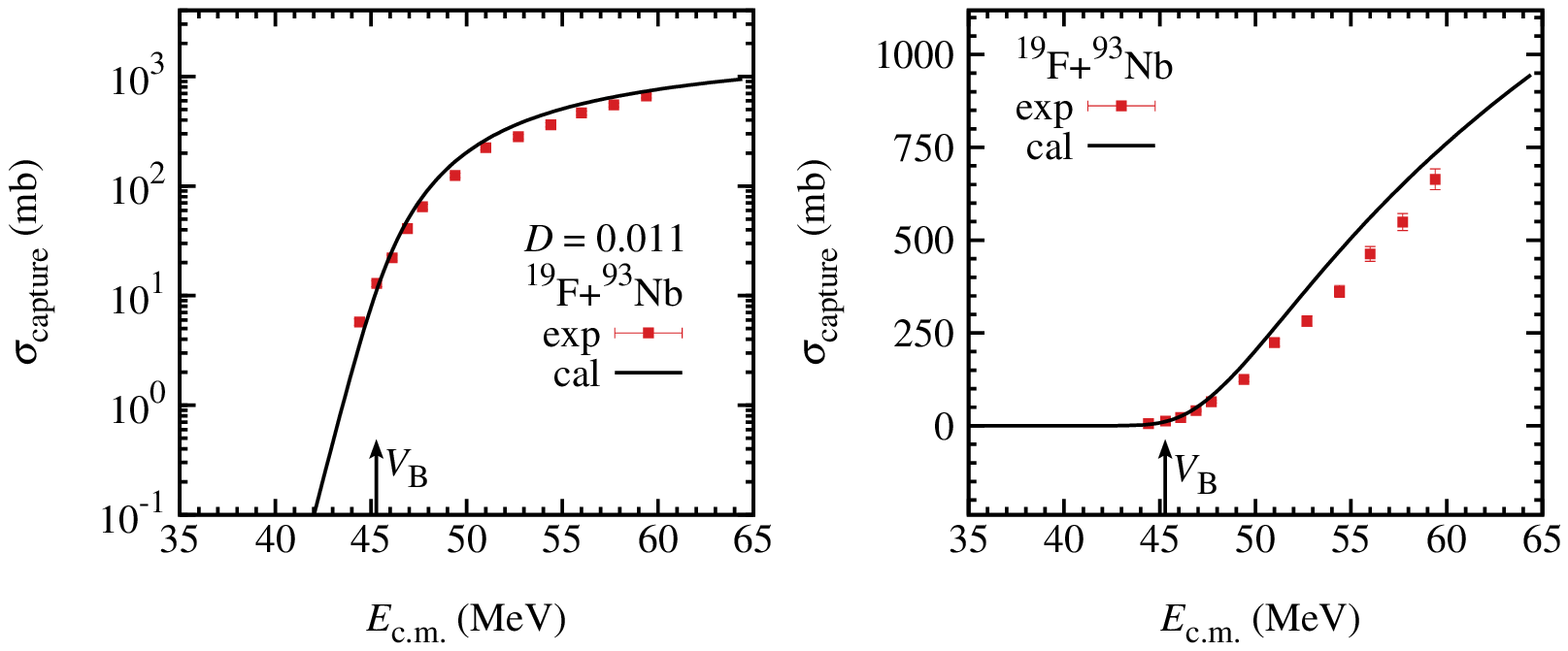}
  \includegraphics[width=0.47\textwidth]{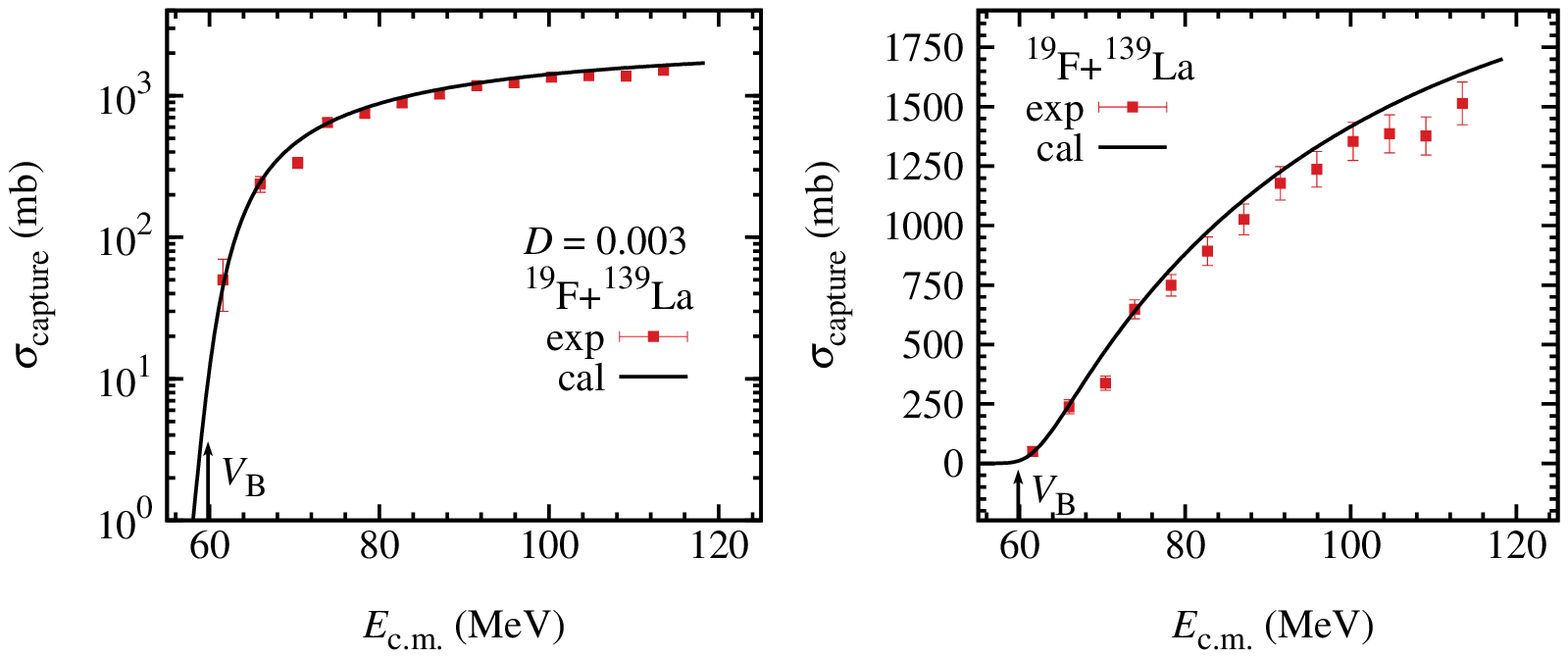}}
 \centerline{\includegraphics[width=0.47\textwidth]{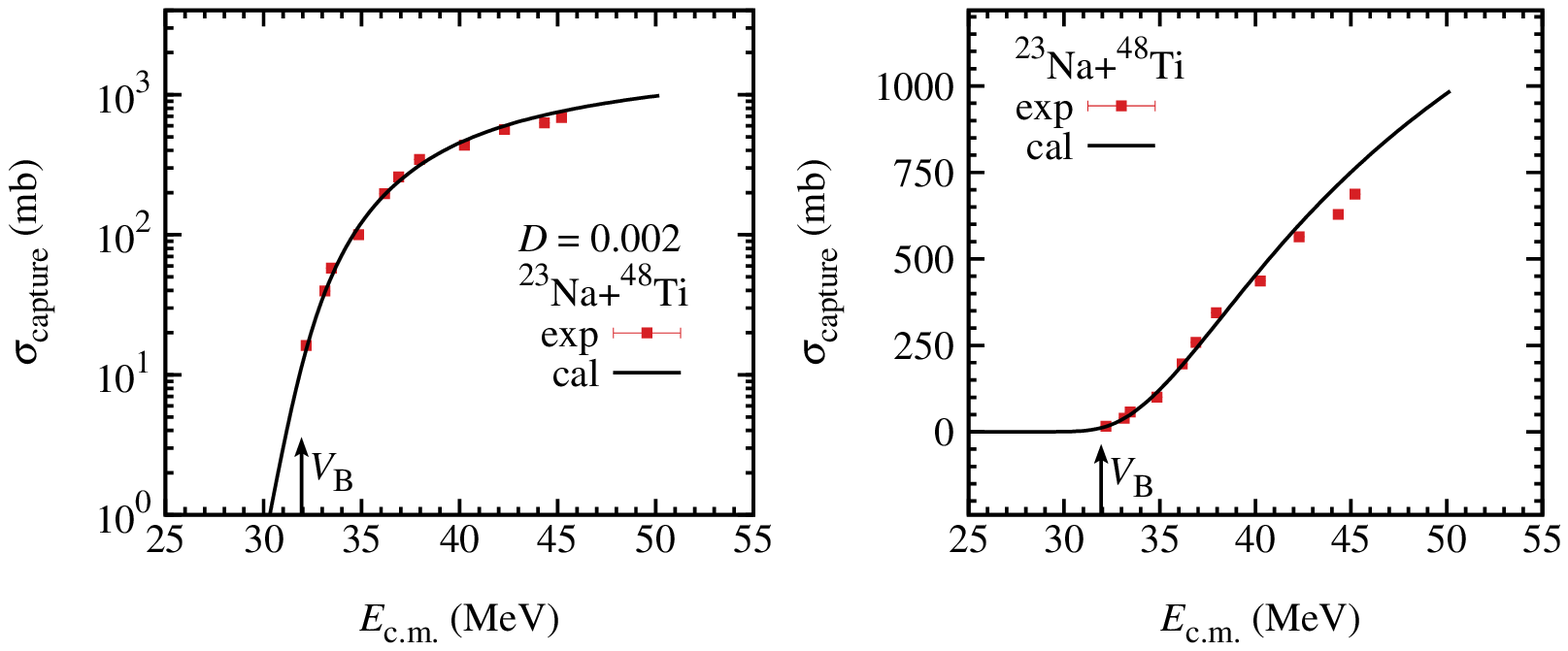}
  \includegraphics[width=0.47\textwidth]{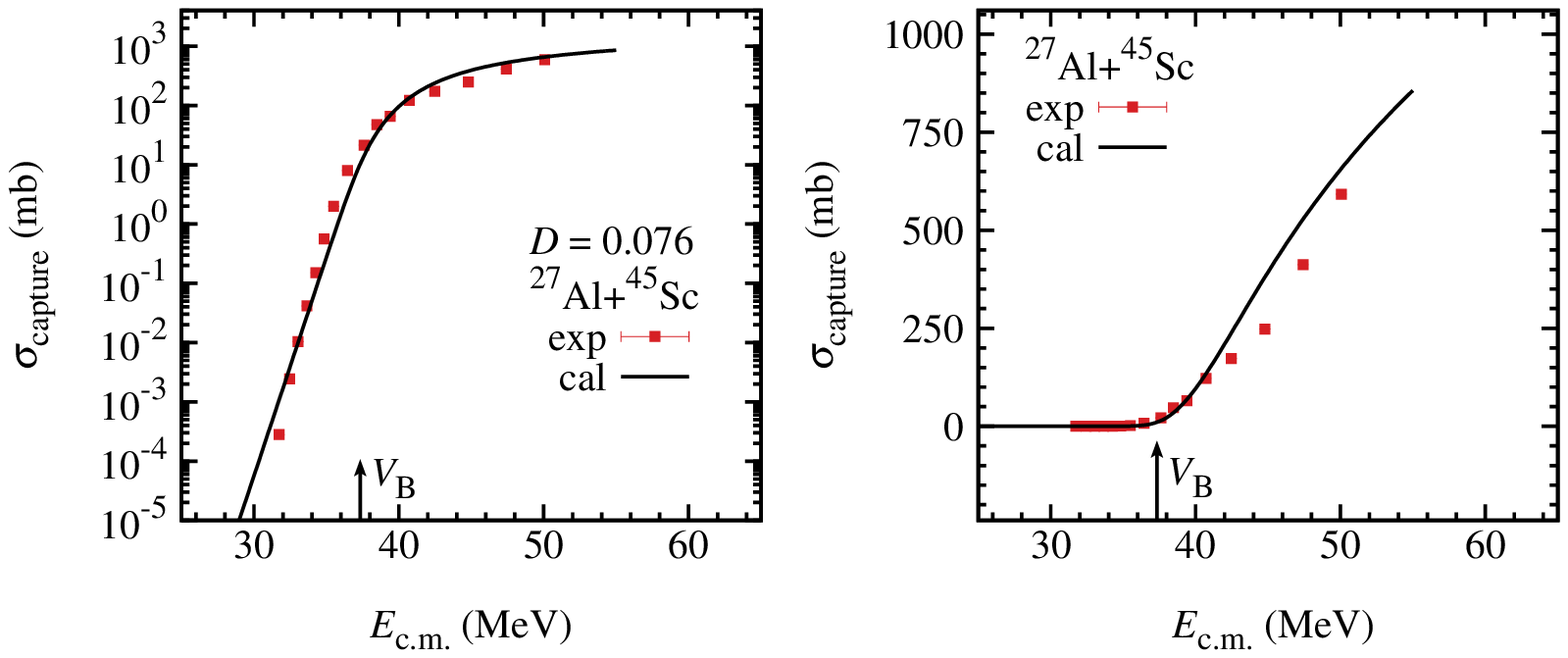}}
 \centerline{\includegraphics[width=0.47\textwidth]{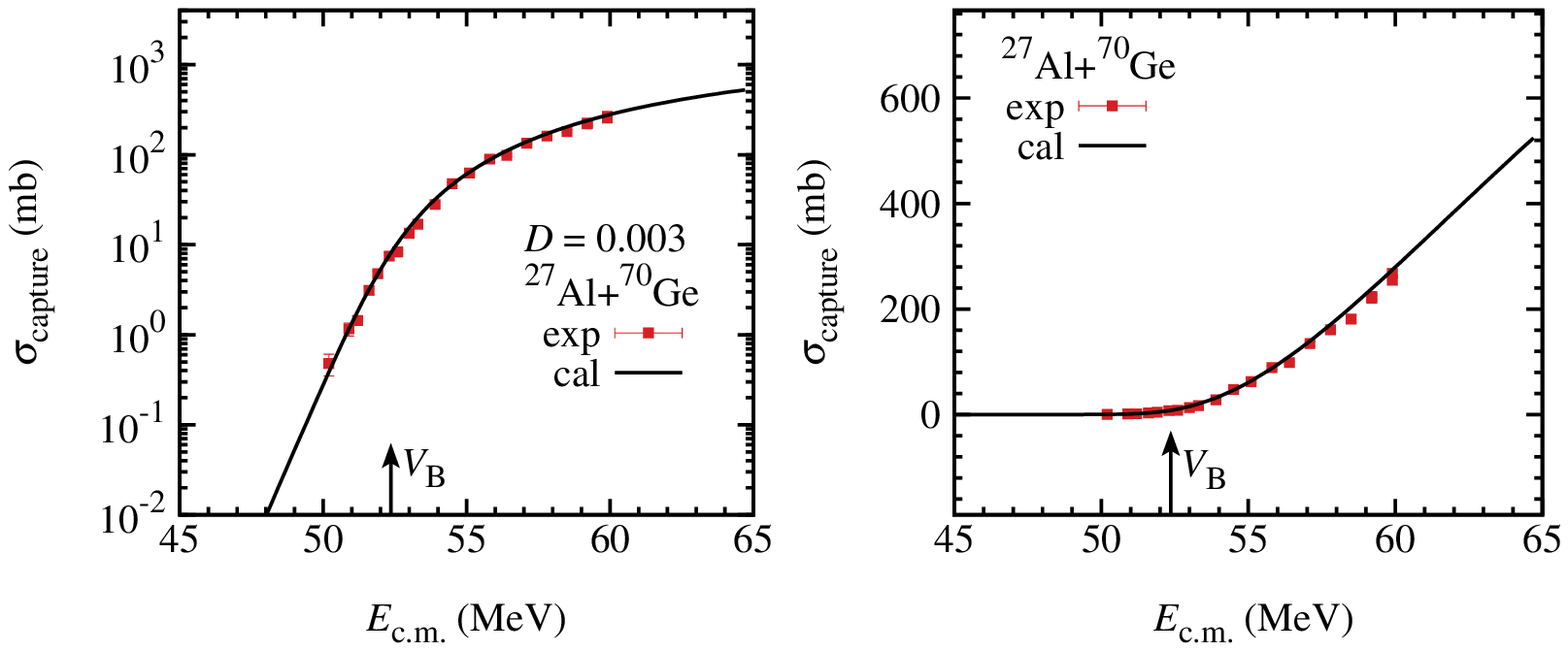}
  \includegraphics[width=0.47\textwidth]{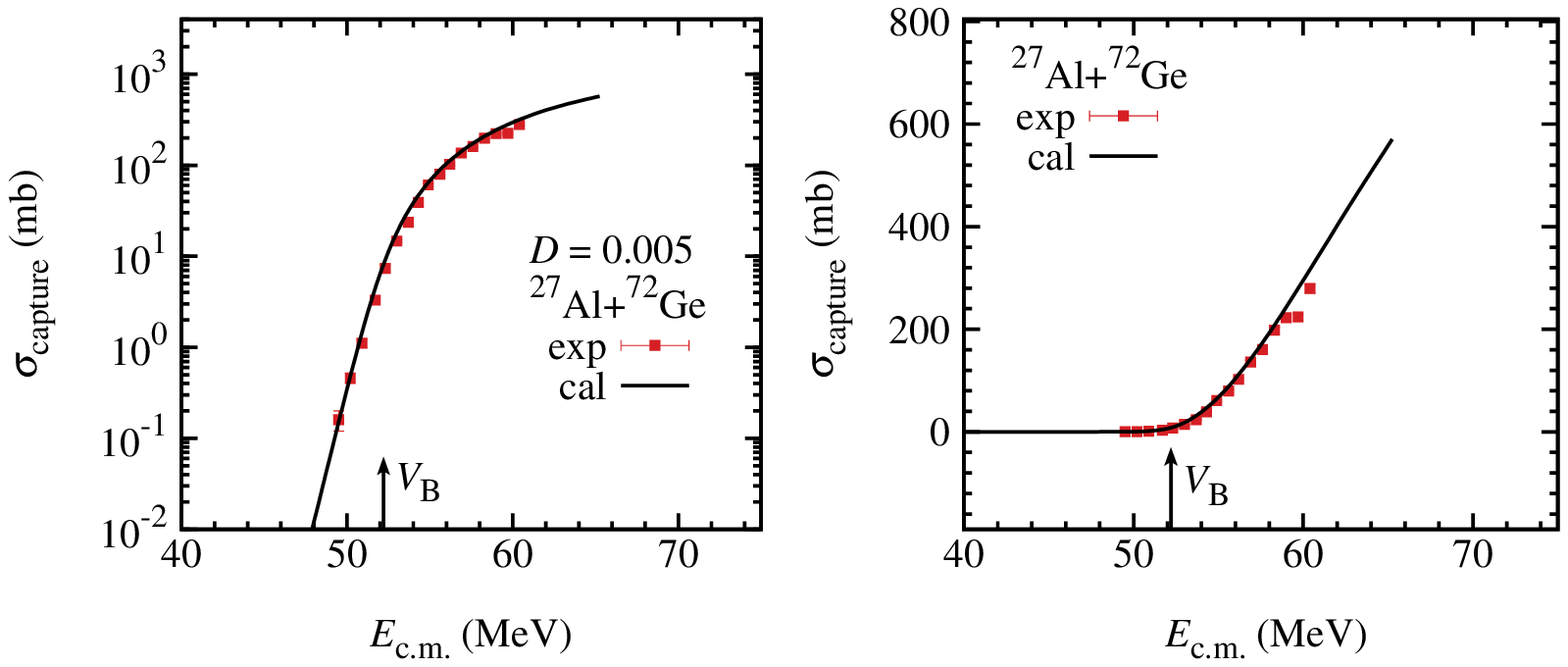}}
  \centerline {Graph 4}
 \end{Dfigures}
 \begin{Dfigures}[!ht]
 \centerline{\includegraphics[width=0.47\textwidth]{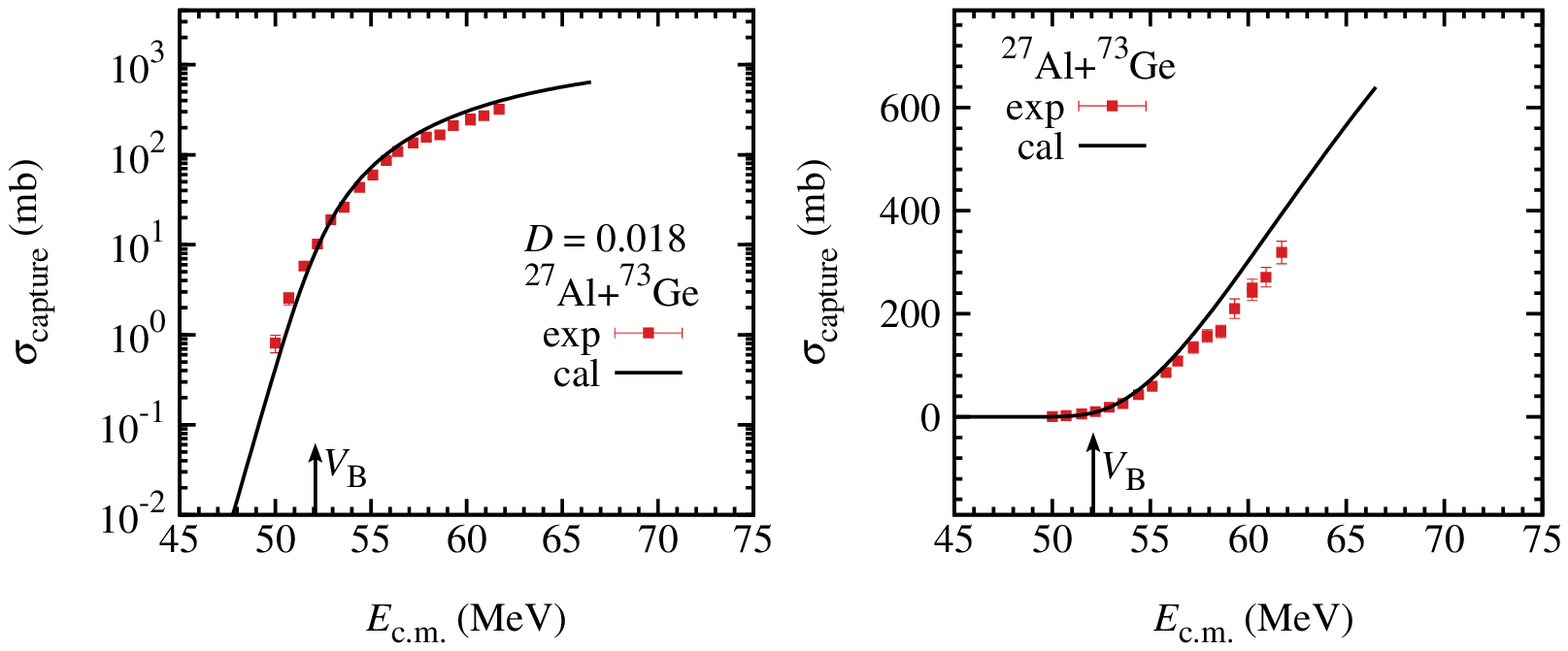}
  \includegraphics[width=0.47\textwidth]{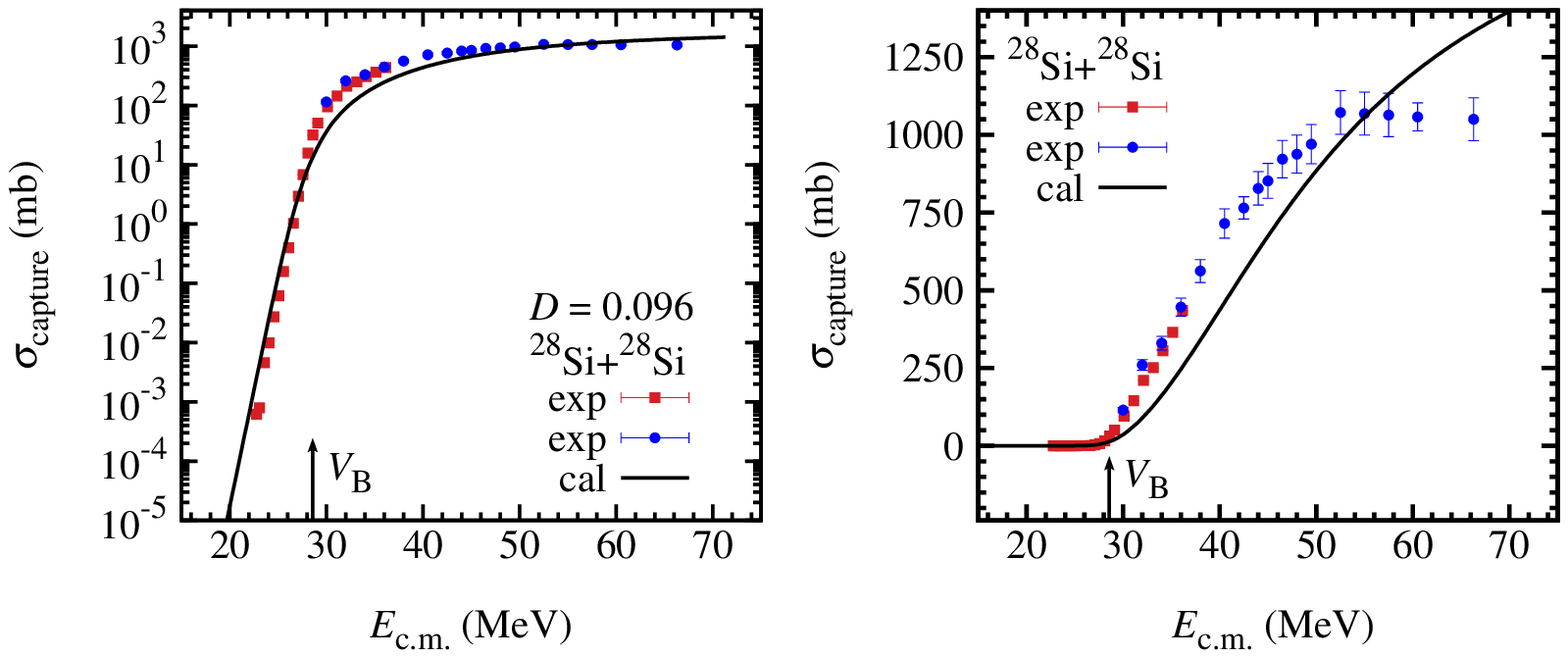}}
 \centerline{\includegraphics[width=0.47\textwidth]{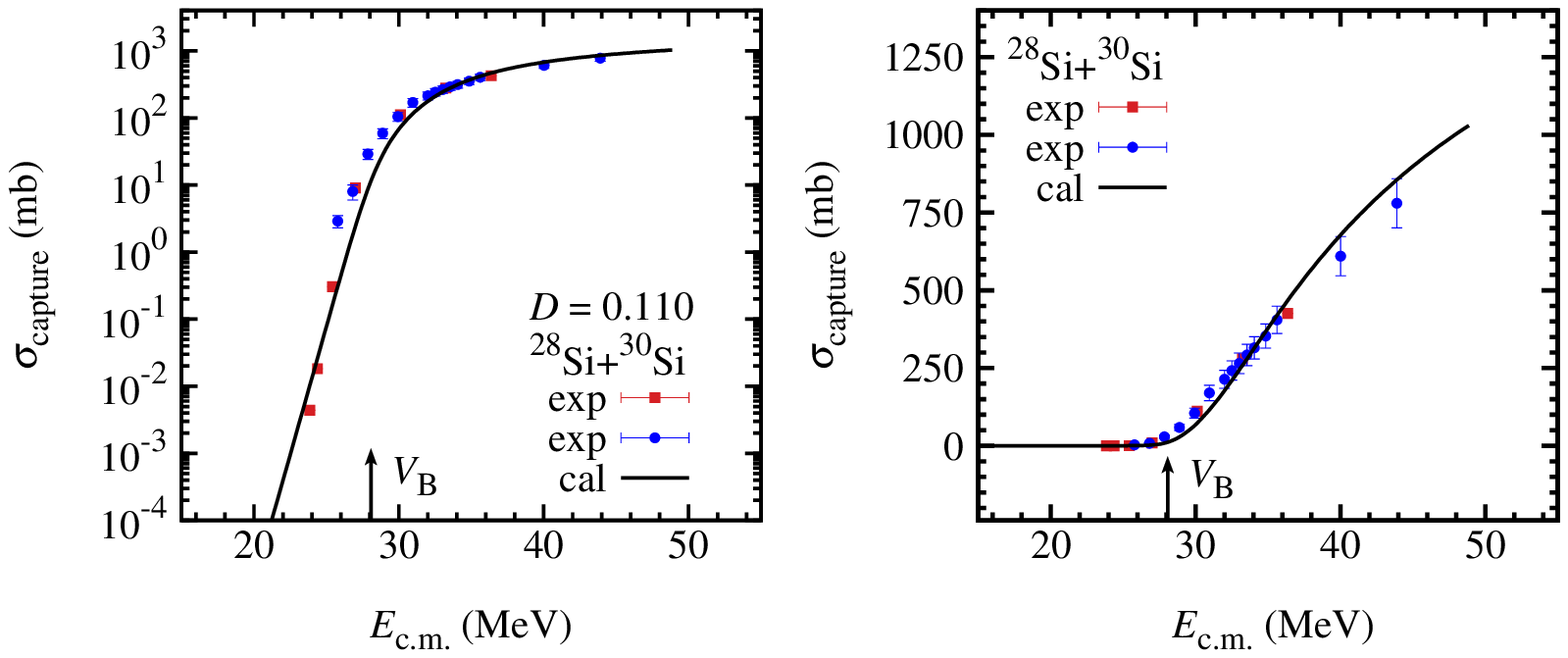}
  \includegraphics[width=0.47\textwidth]{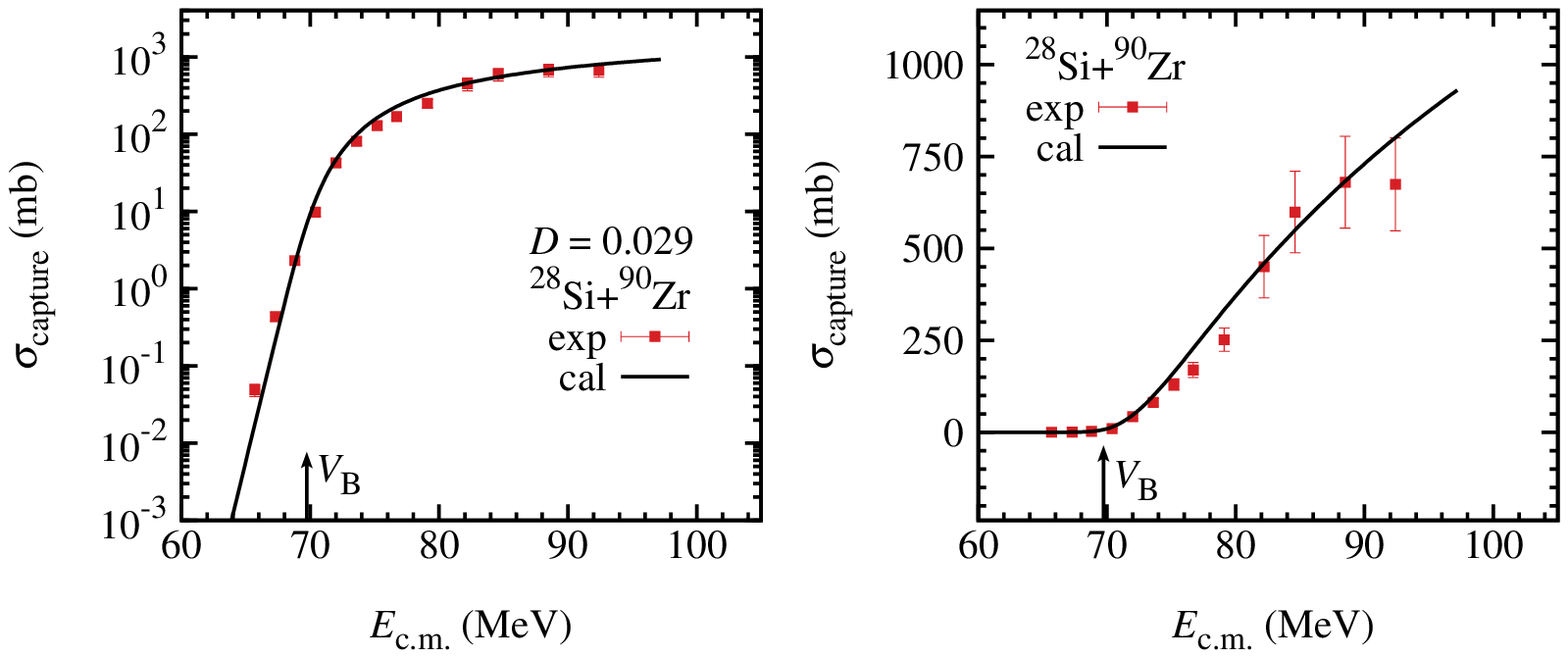}}
 \centerline{\includegraphics[width=0.47\textwidth]{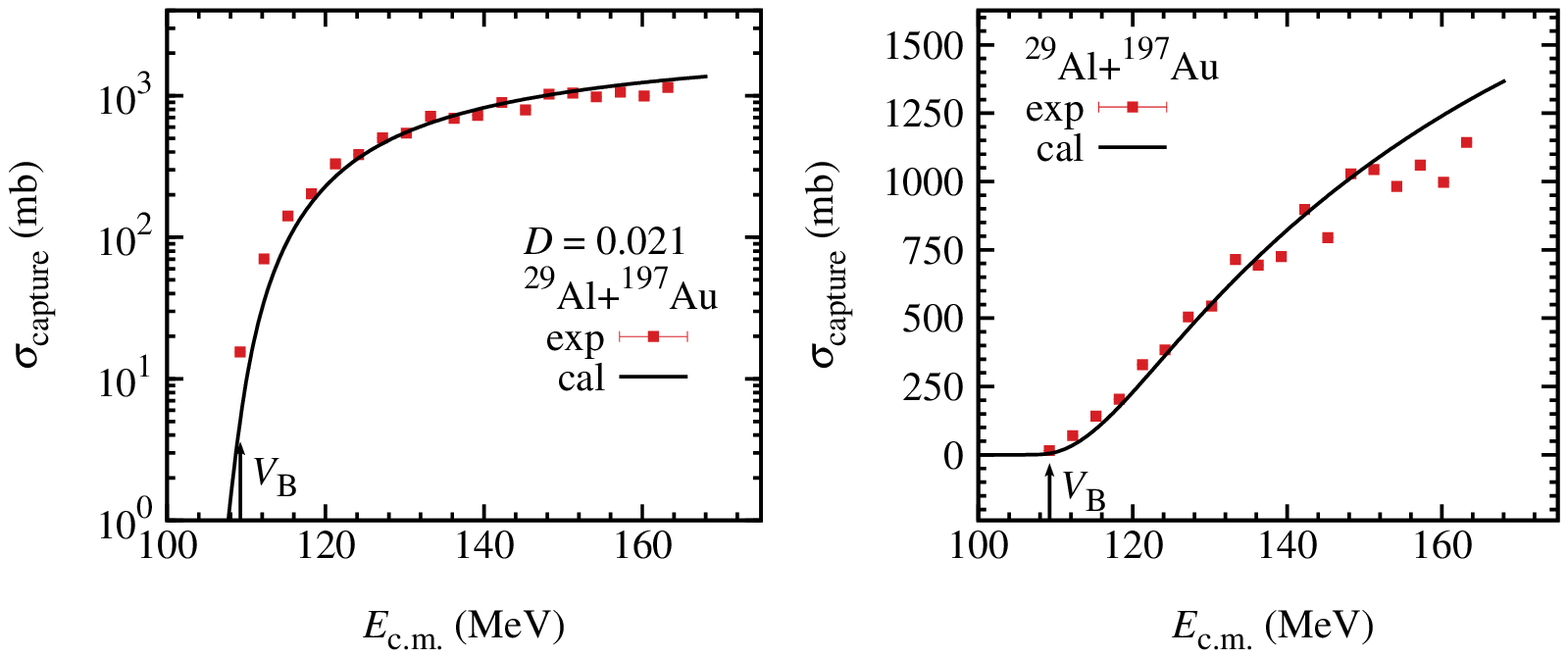}
  \includegraphics[width=0.47\textwidth]{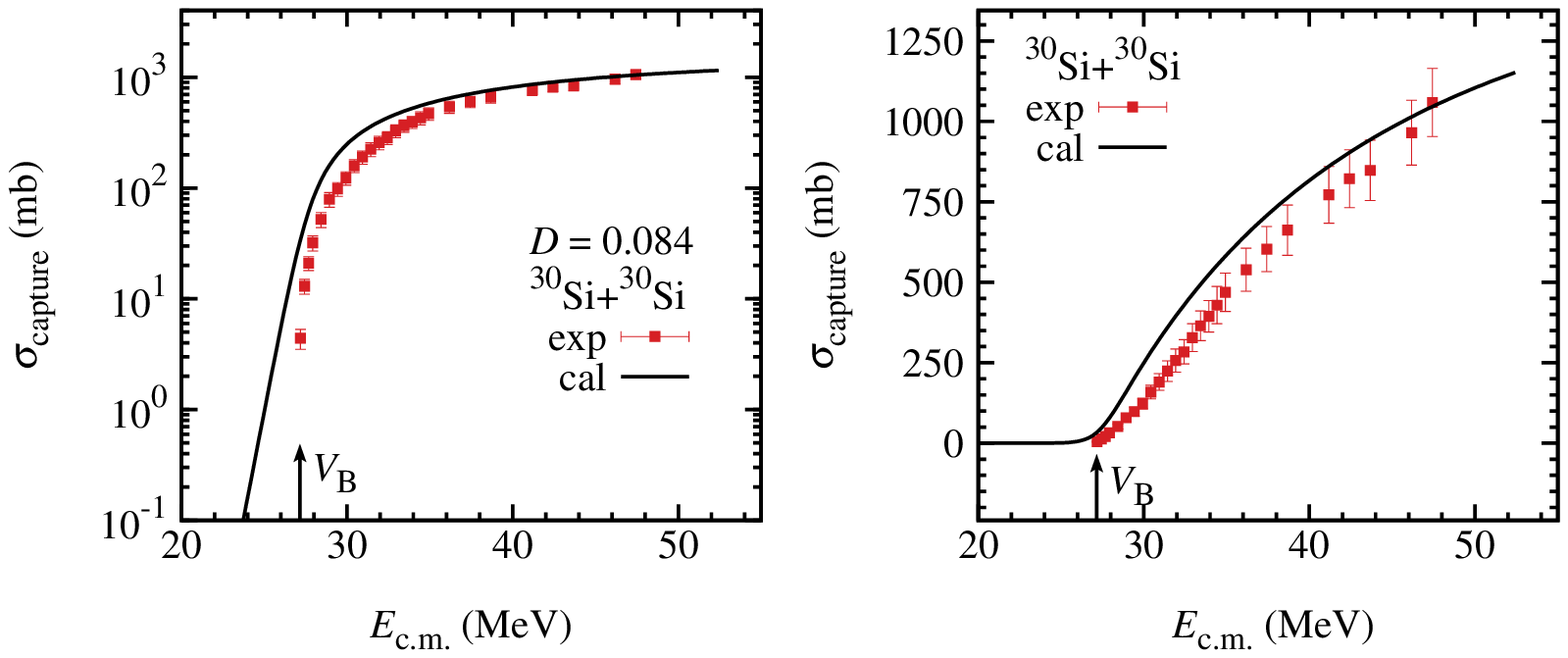}}
 \centerline{\includegraphics[width=0.47\textwidth]{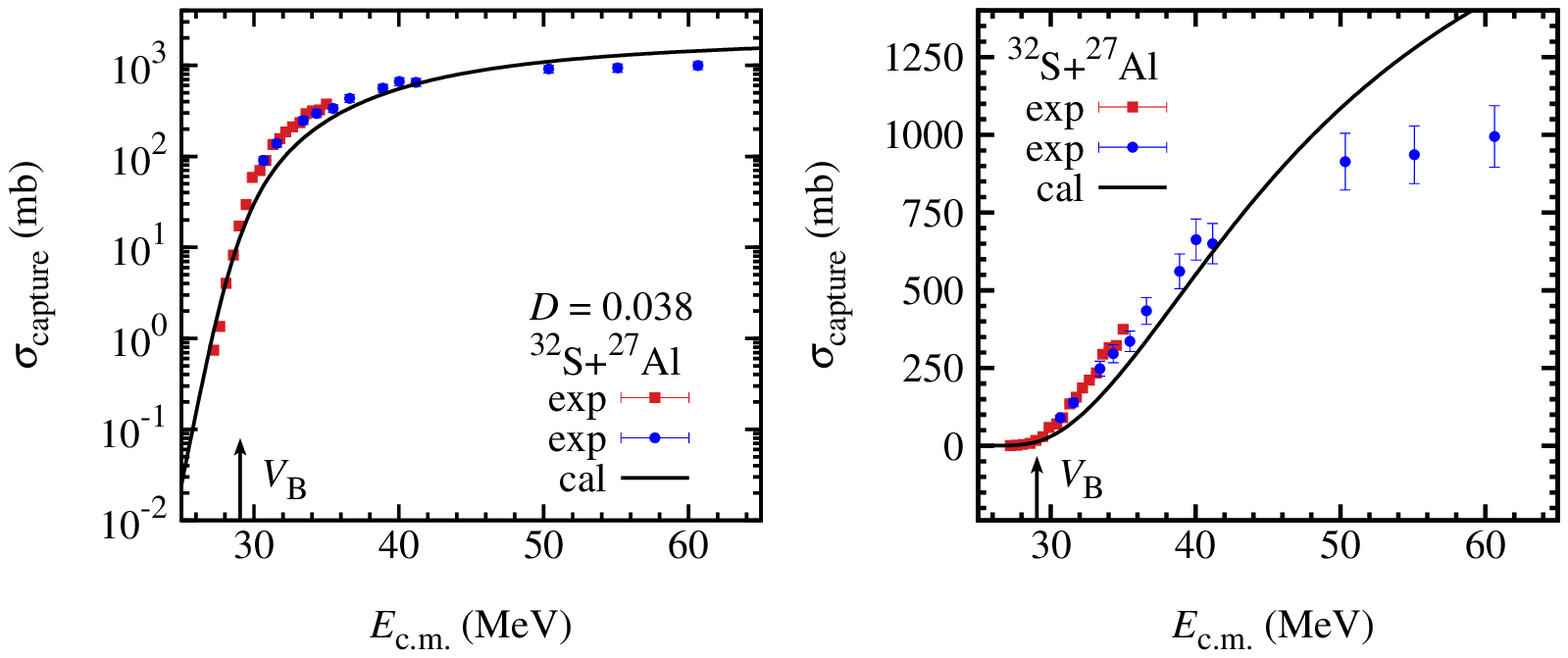}
  \includegraphics[width=0.47\textwidth]{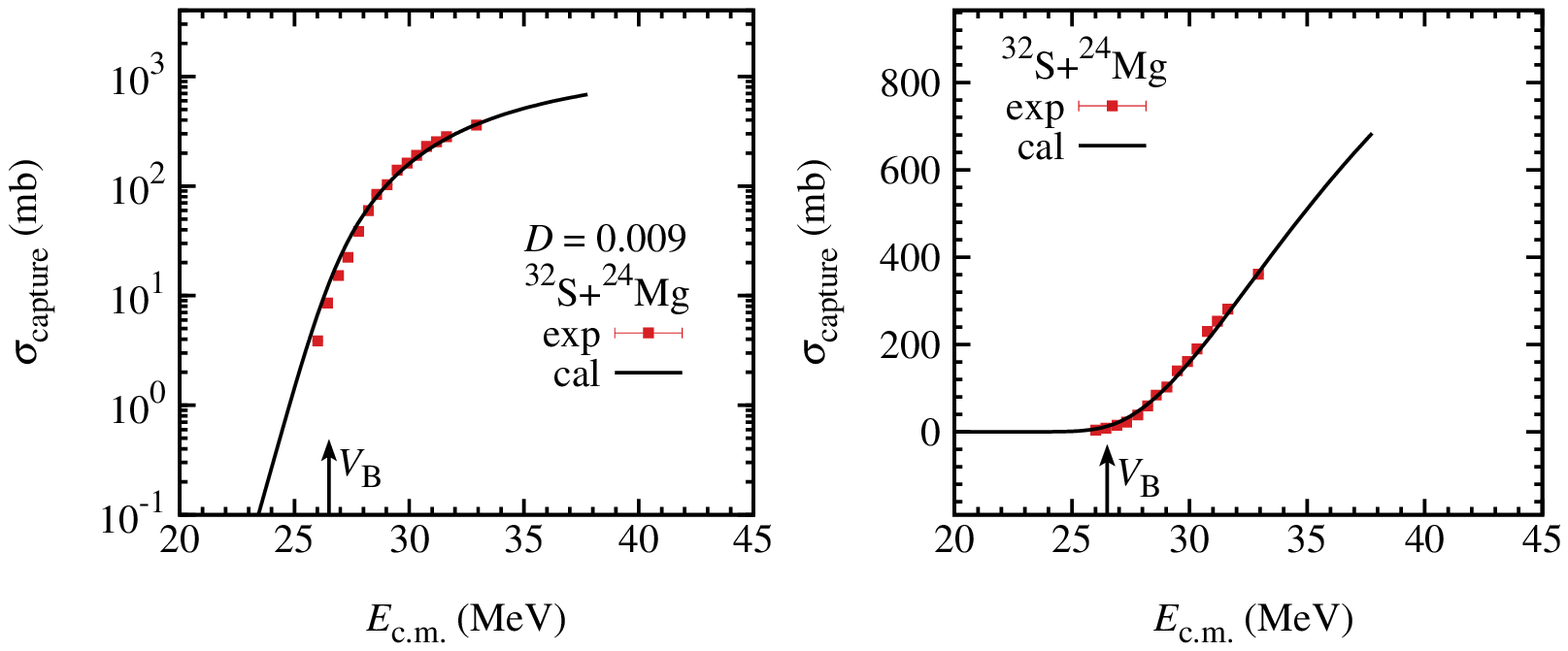}}
 \centerline{\includegraphics[width=0.47\textwidth]{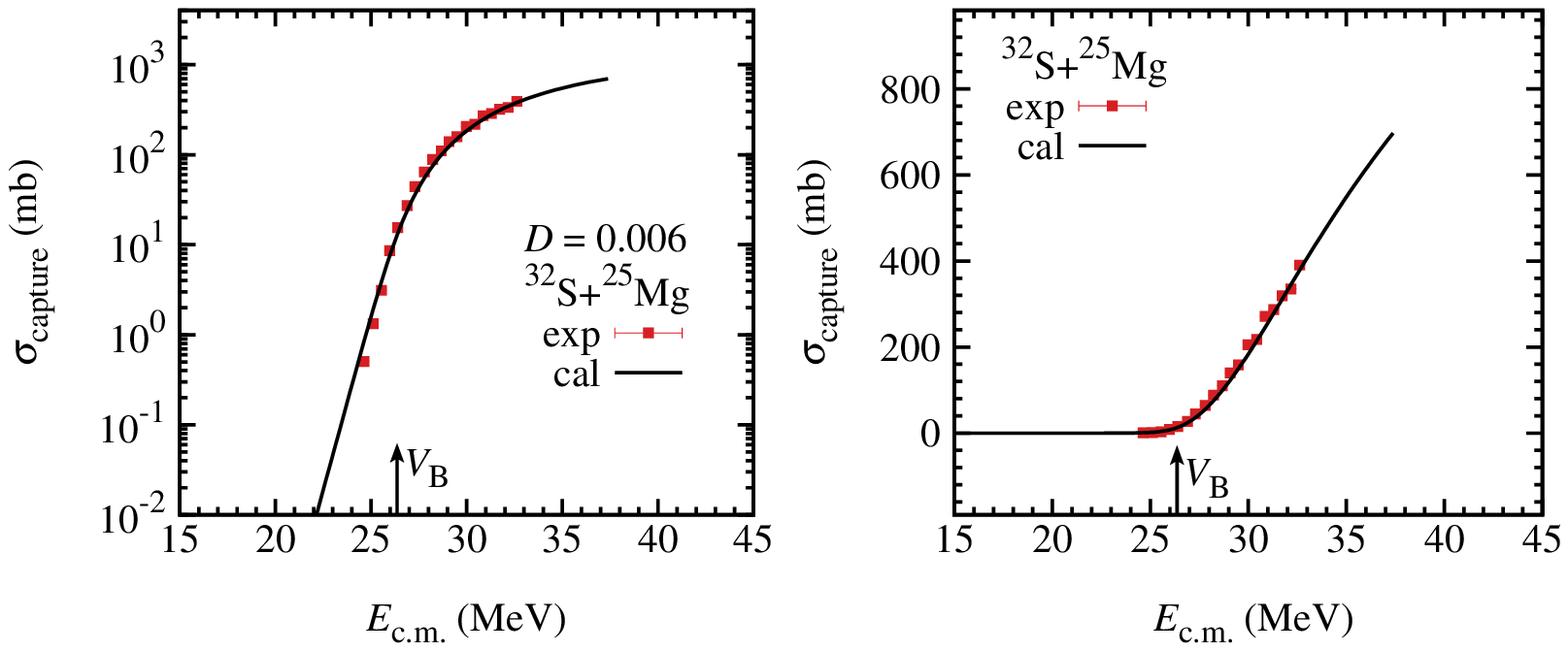}
  \includegraphics[width=0.47\textwidth]{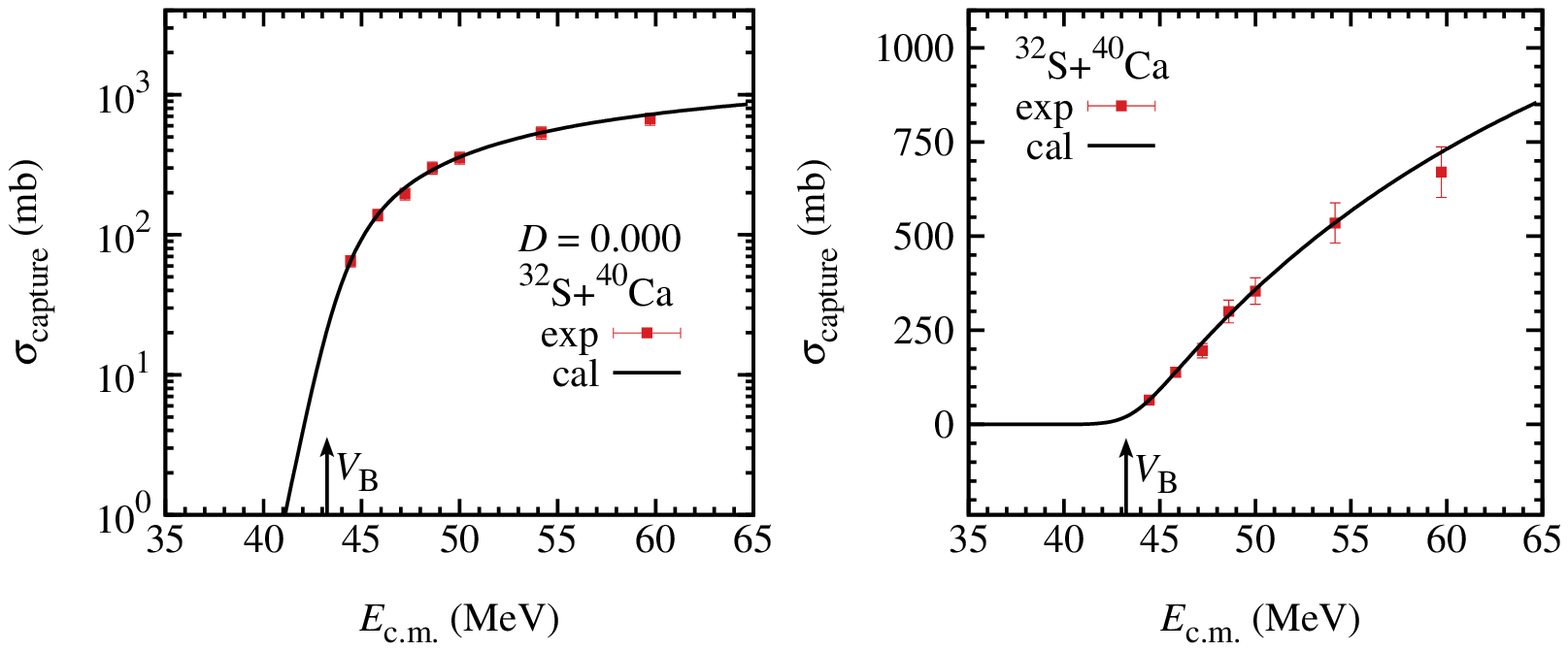}}
 \centerline{\includegraphics[width=0.47\textwidth]{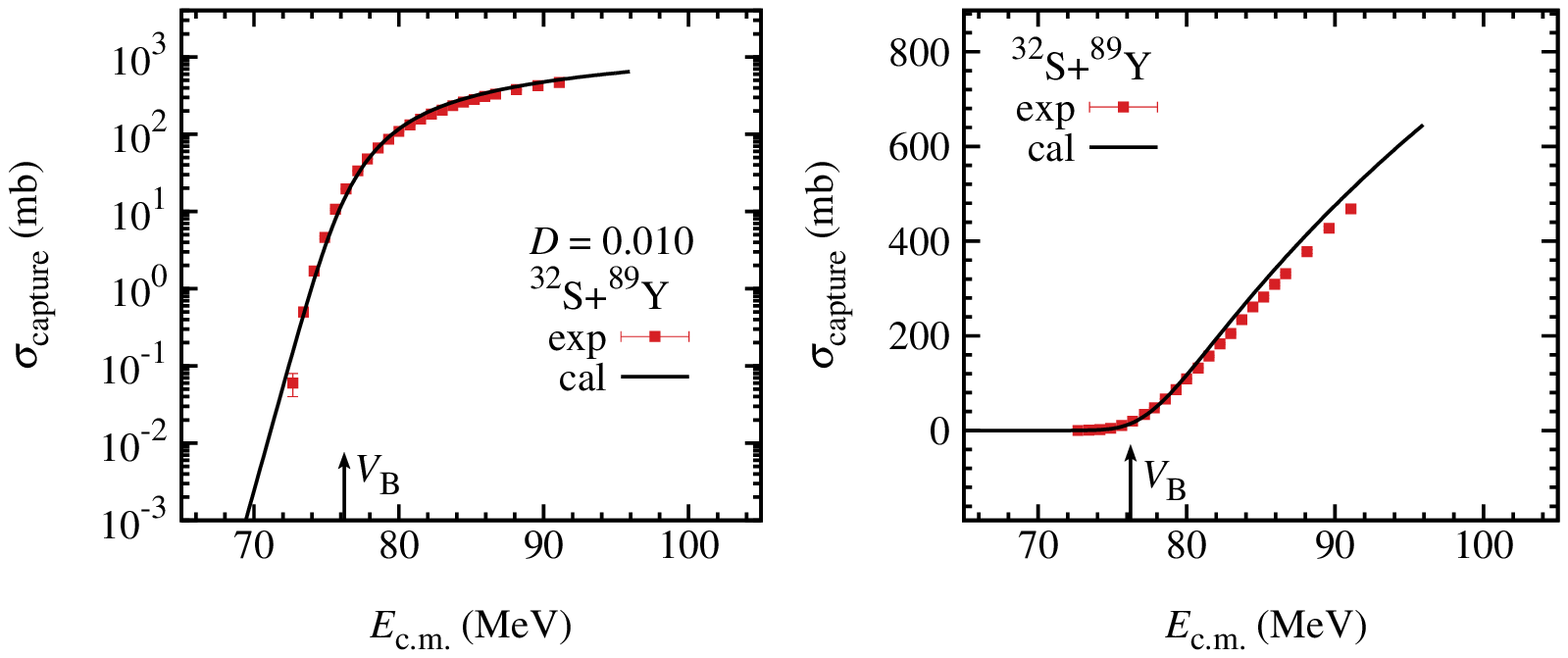}
  \includegraphics[width=0.47\textwidth]{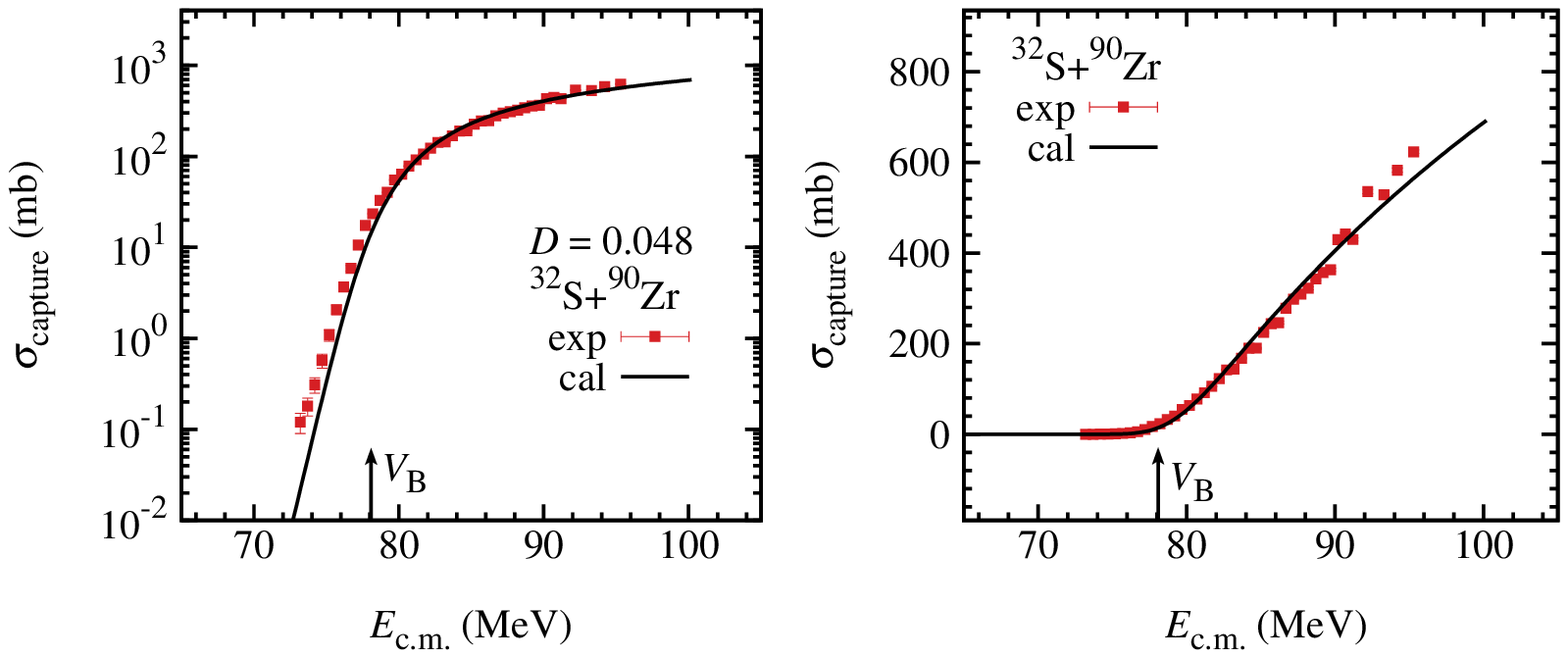}}
  \centerline {Graph 5}
 \end{Dfigures}
 \begin{Dfigures}[!ht]
 \centerline{\includegraphics[width=0.47\textwidth]{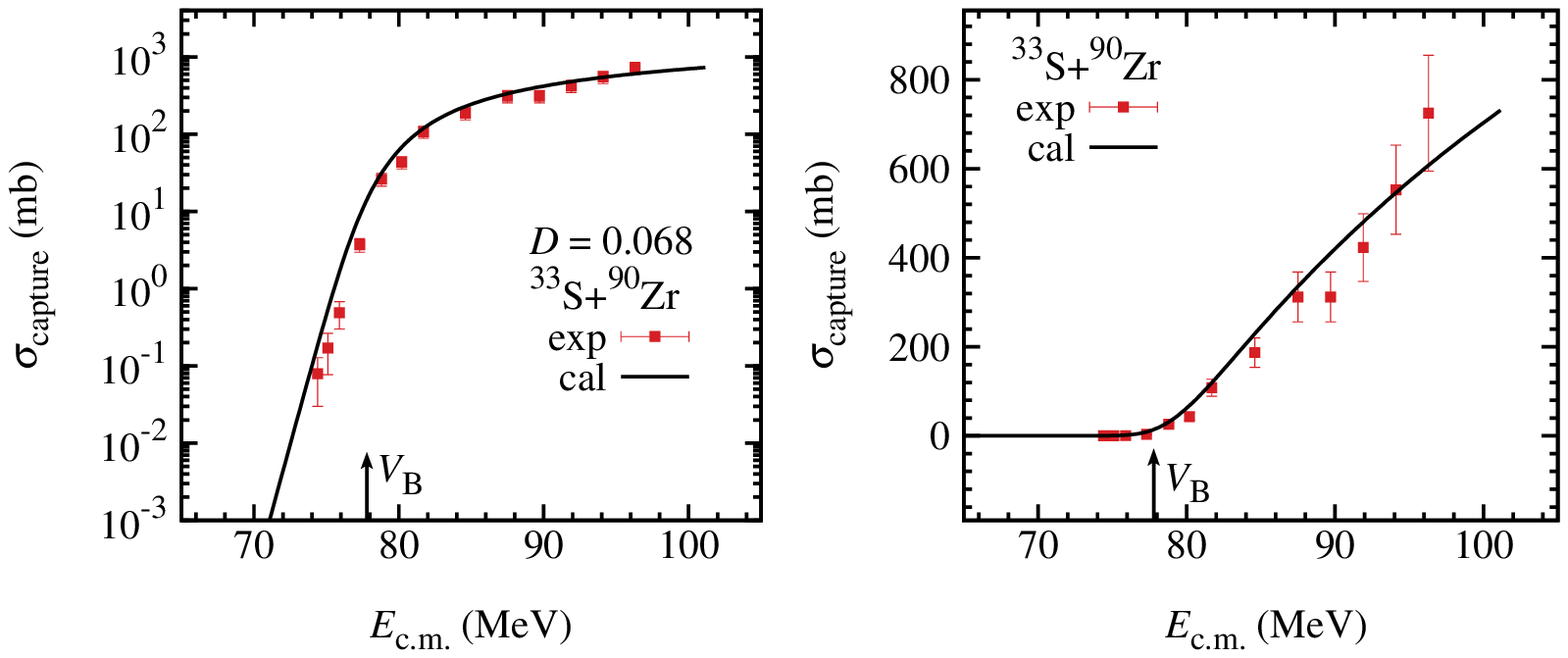}
  \includegraphics[width=0.47\textwidth]{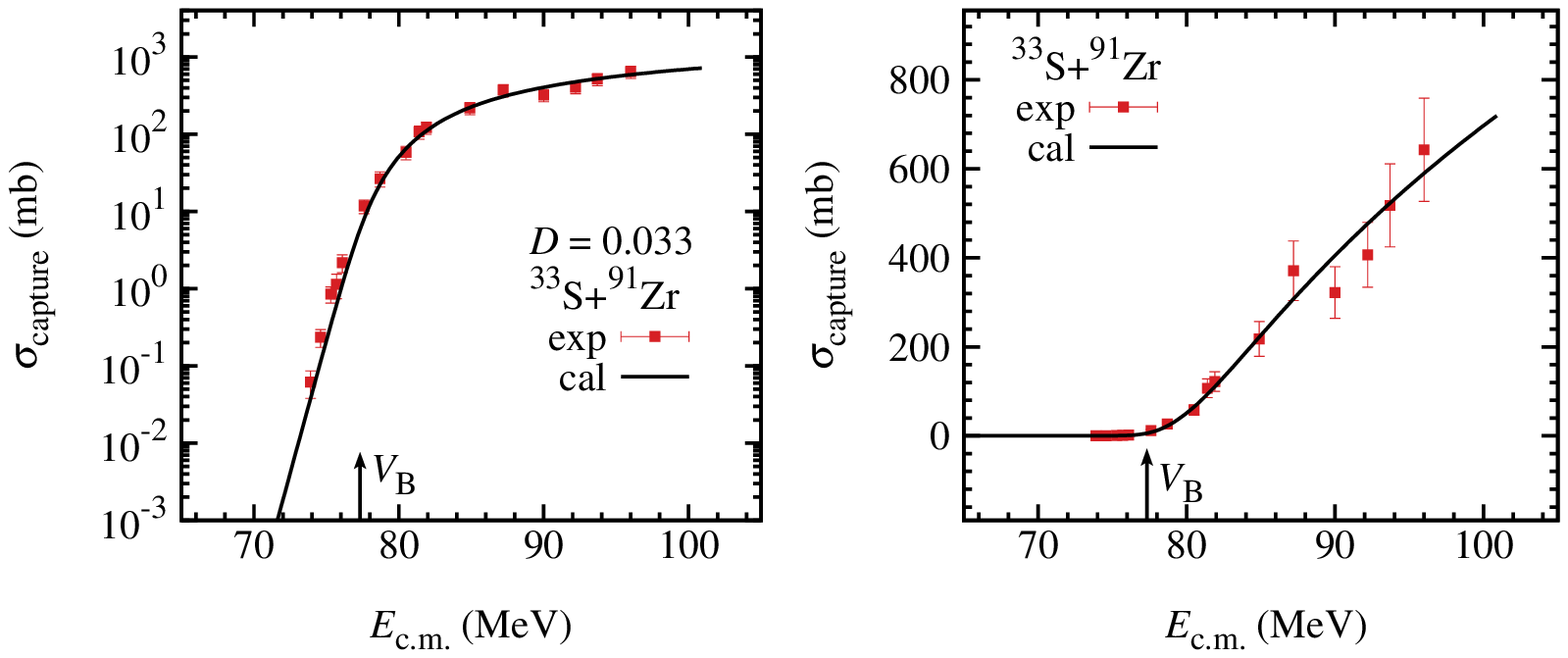}}
 \centerline{\includegraphics[width=0.47\textwidth]{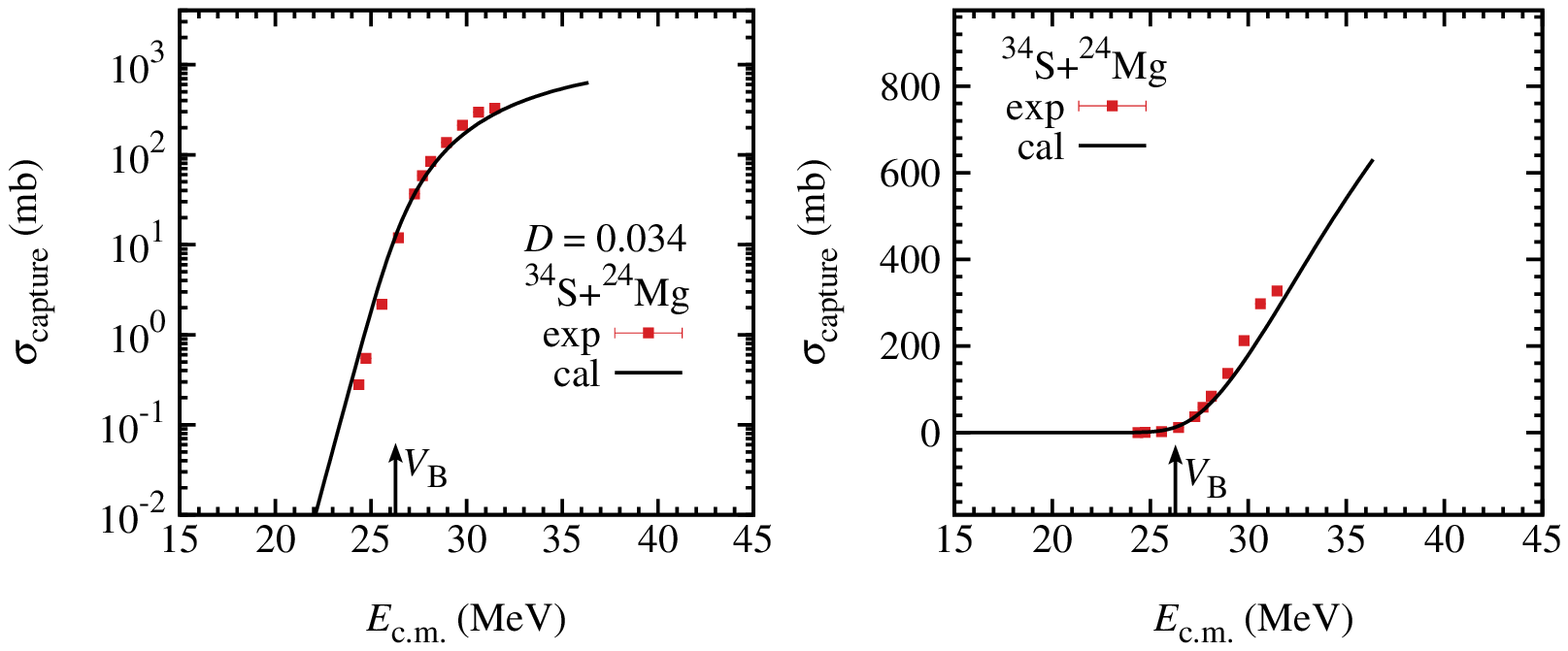}
  \includegraphics[width=0.47\textwidth]{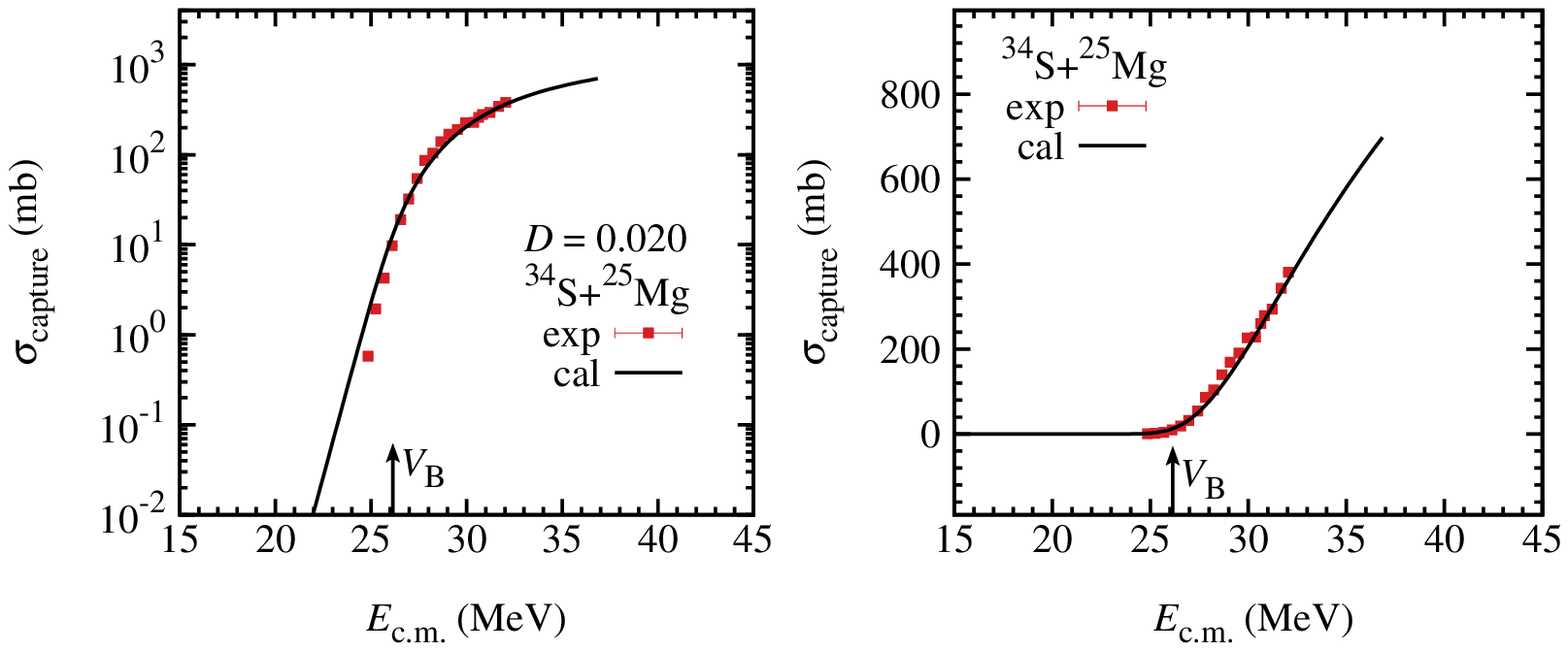}}
 \centerline{\includegraphics[width=0.47\textwidth]{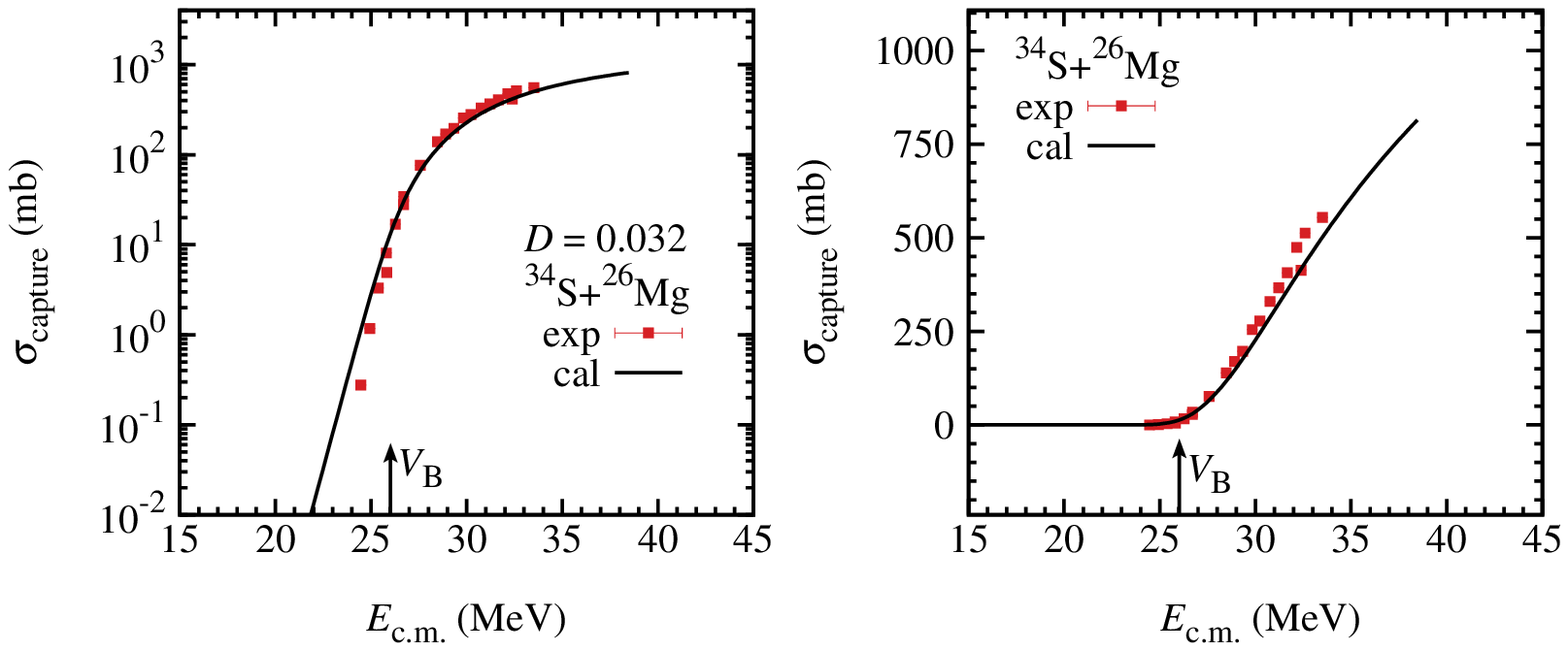}
  \includegraphics[width=0.47\textwidth]{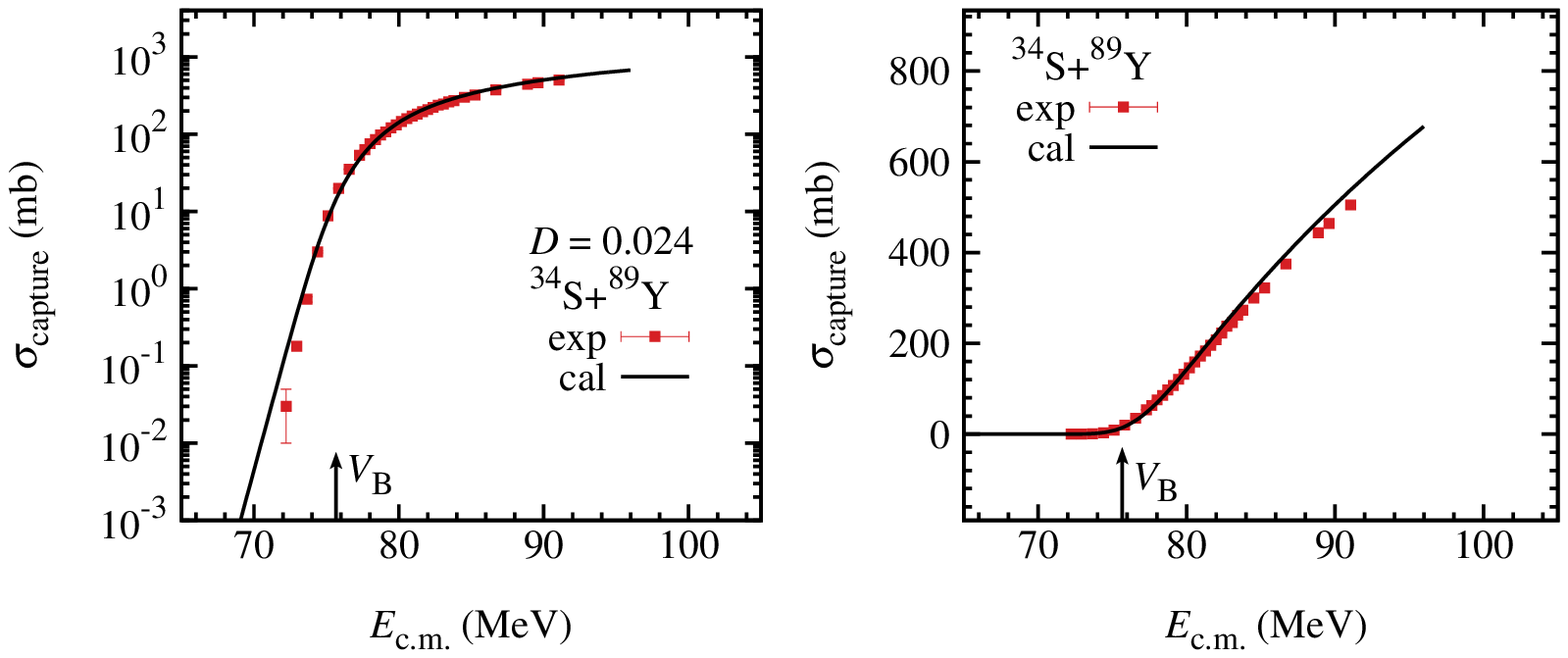}}
 \centerline{\includegraphics[width=0.47\textwidth]{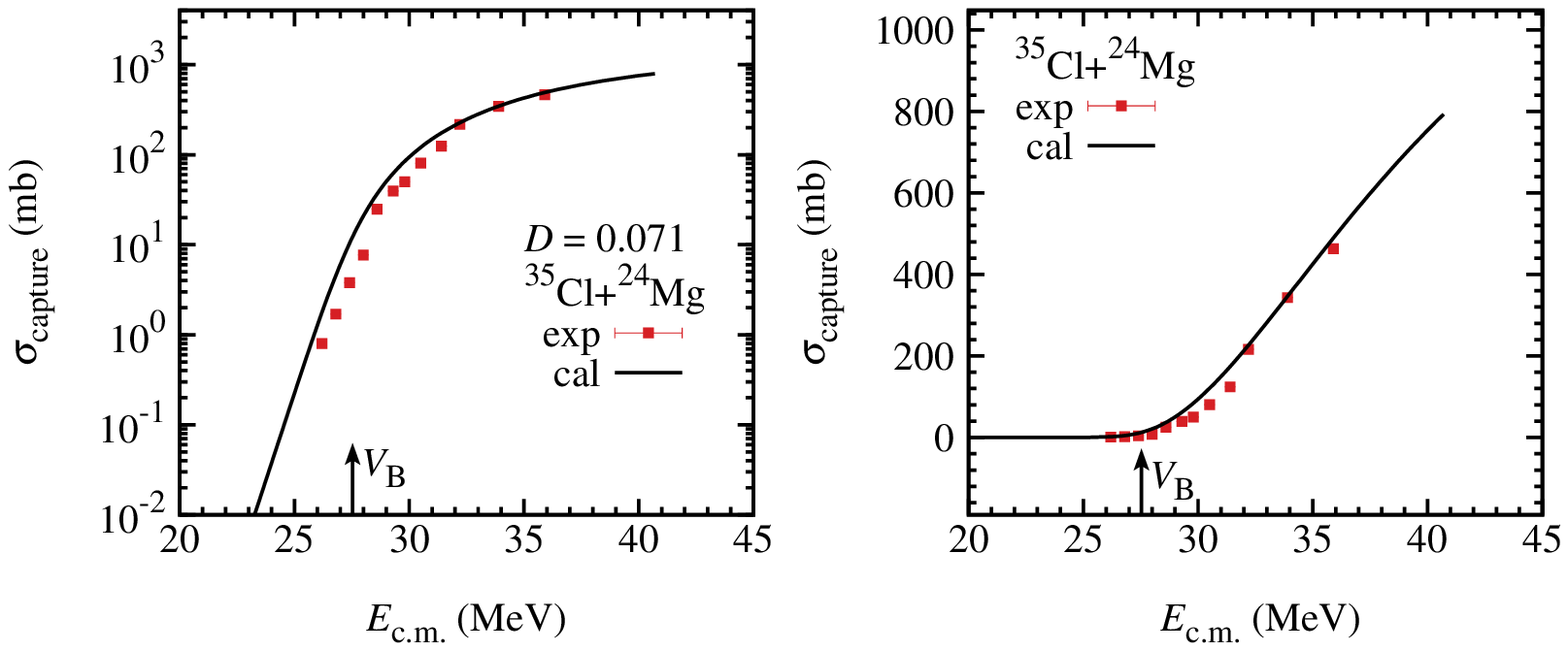}
  \includegraphics[width=0.47\textwidth]{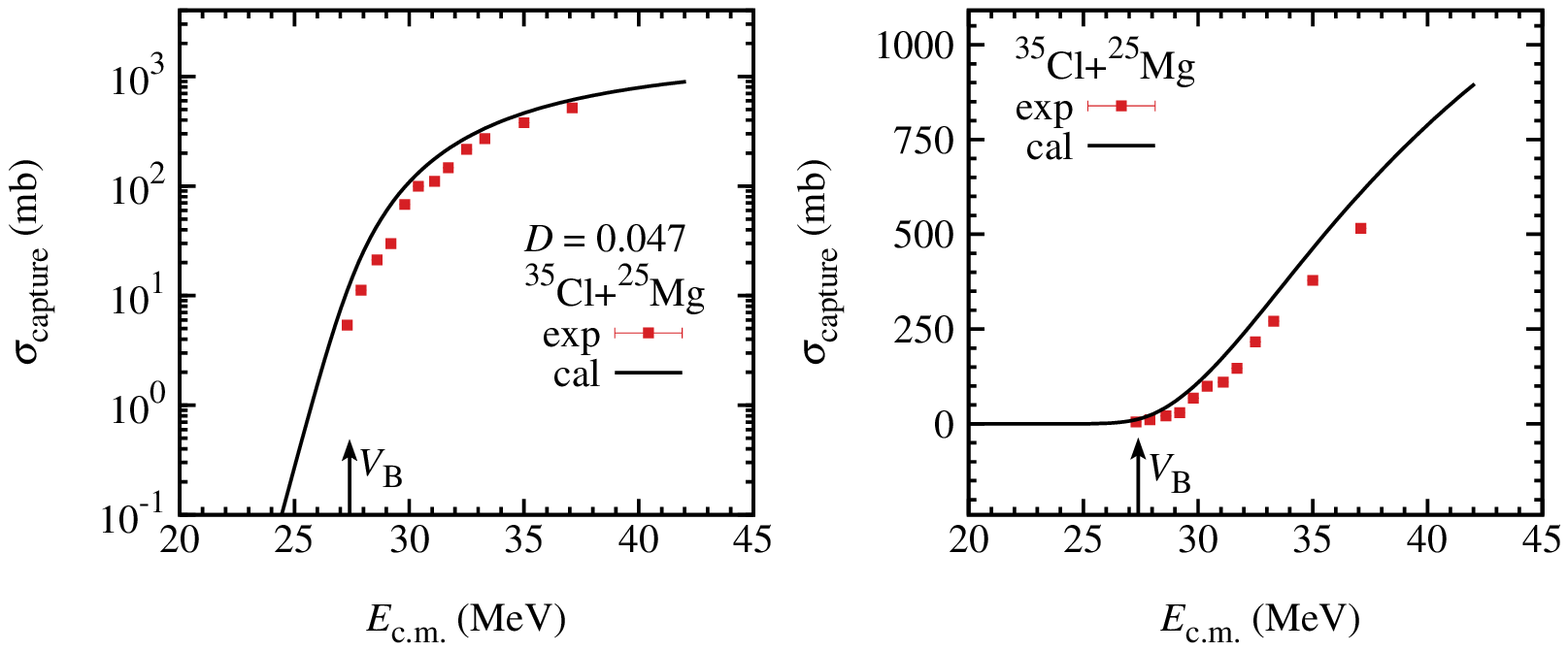}}
 \centerline{\includegraphics[width=0.47\textwidth]{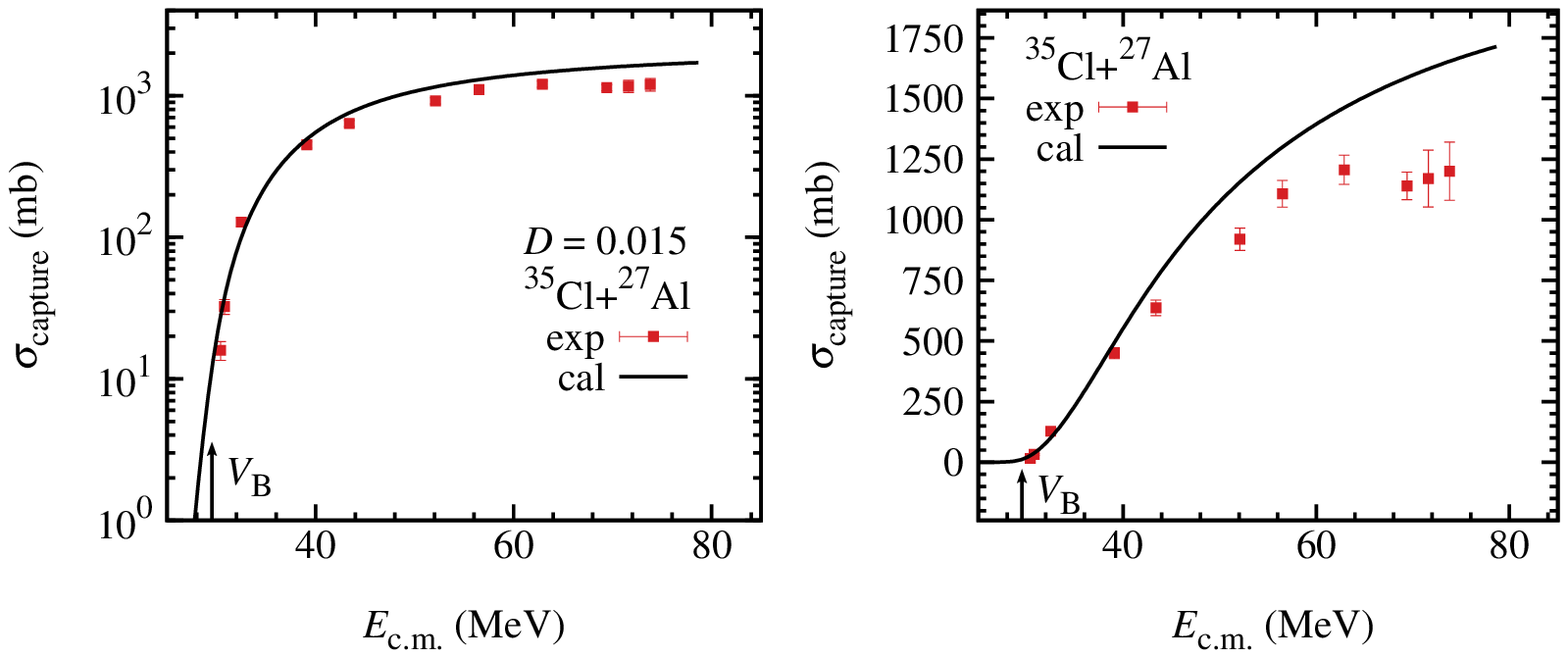}
  \includegraphics[width=0.47\textwidth]{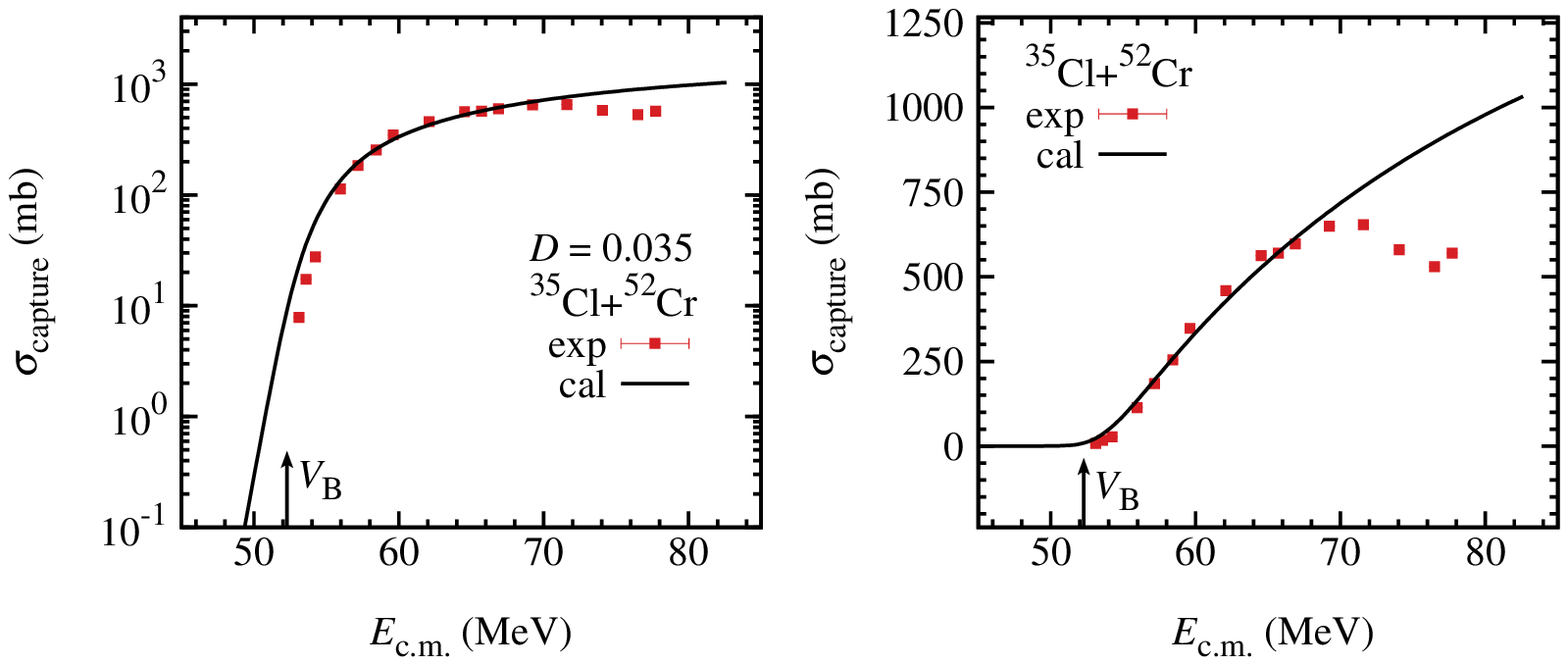}}
 \centerline{\includegraphics[width=0.47\textwidth]{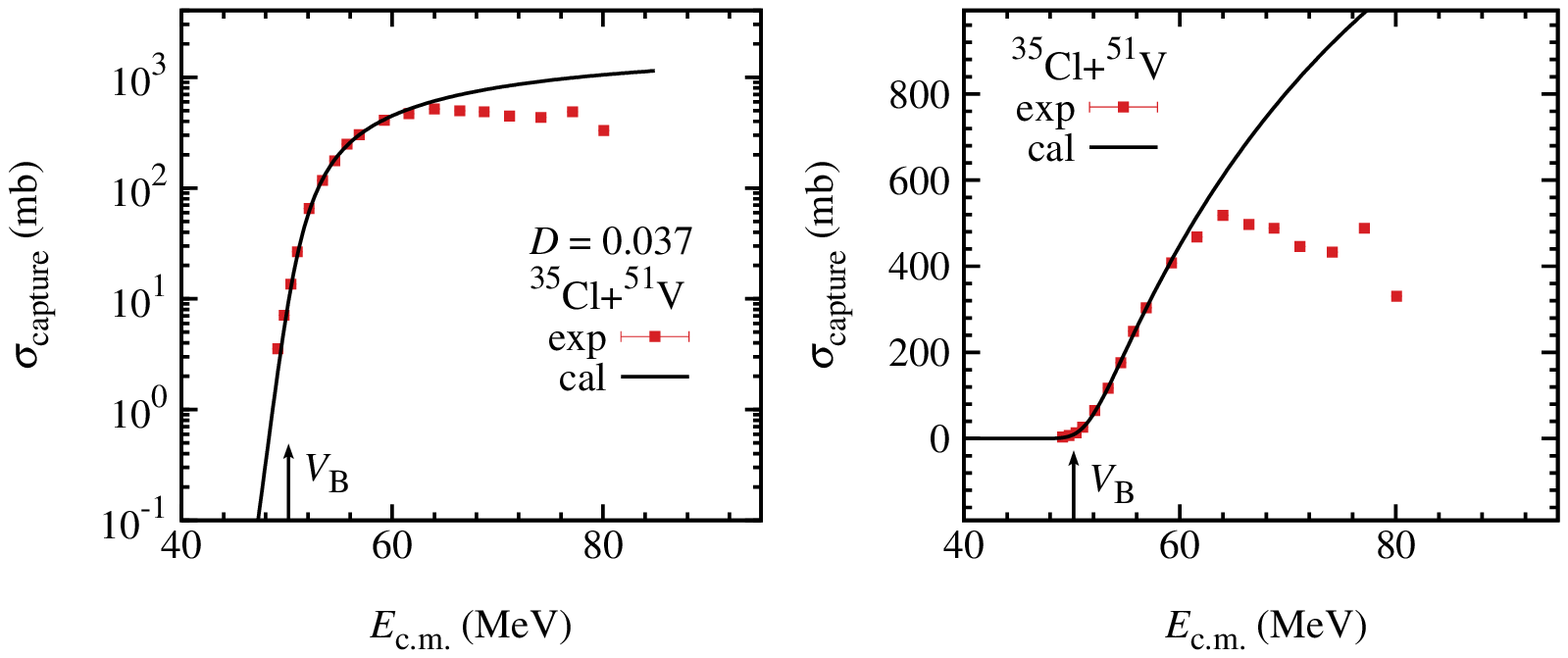}
  \includegraphics[width=0.47\textwidth]{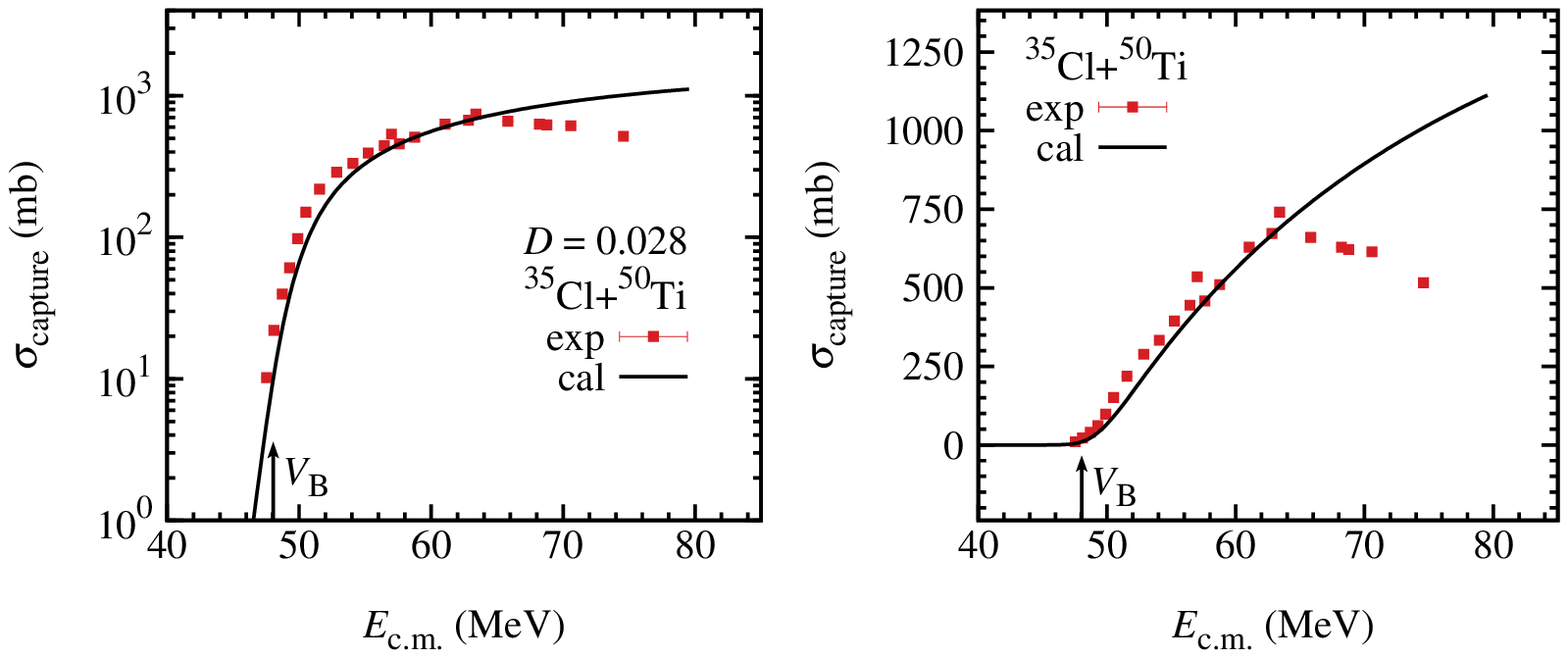}}
  \centerline {Graph 6}
 \end{Dfigures}
 \begin{Dfigures}[!ht]
 \centerline{\includegraphics[width=0.47\textwidth]{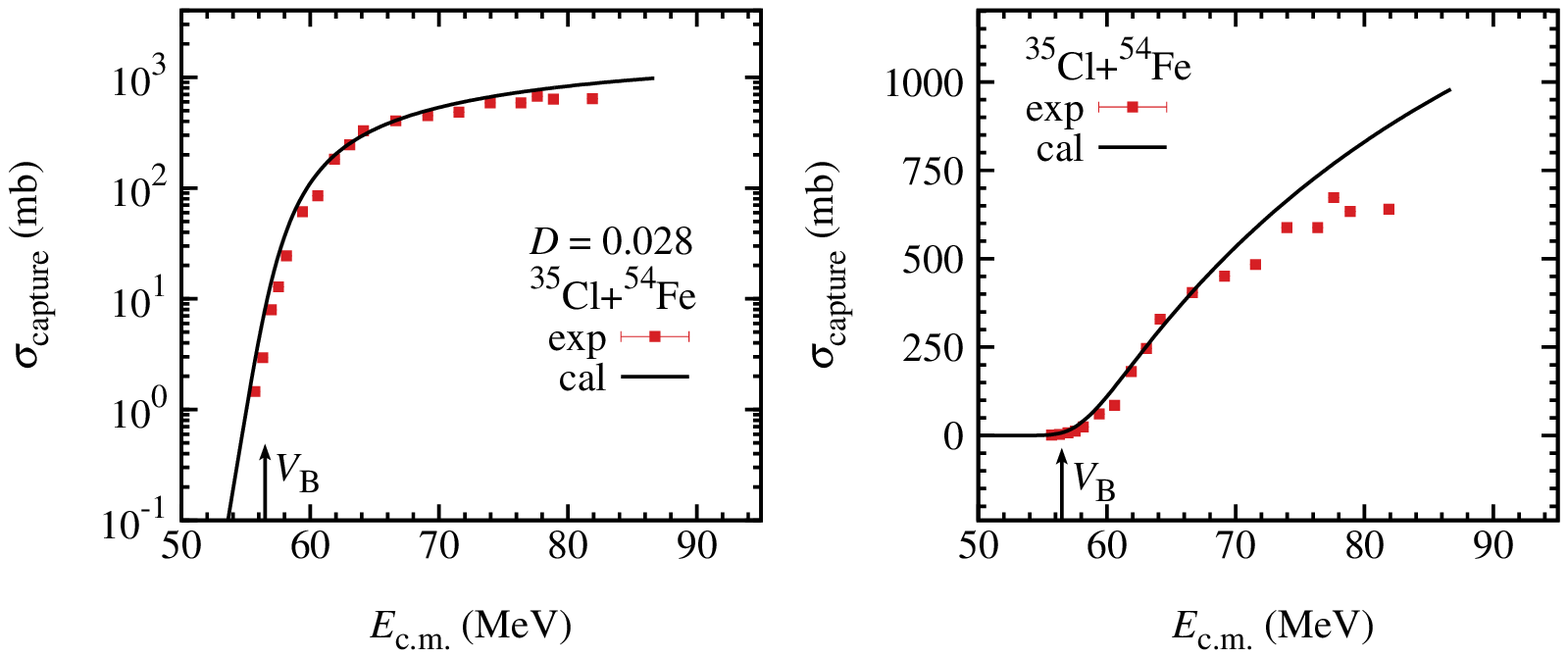}
  \includegraphics[width=0.47\textwidth]{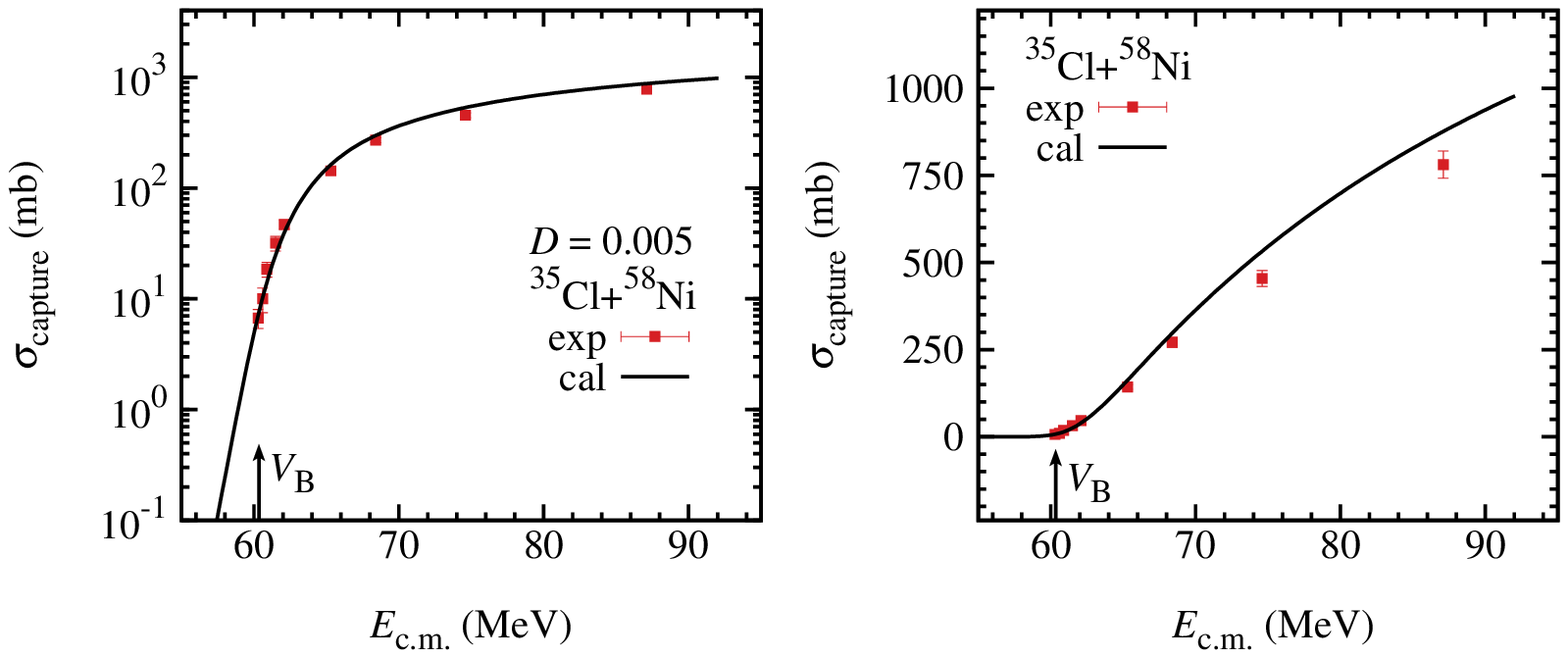}}
 \centerline{\includegraphics[width=0.47\textwidth]{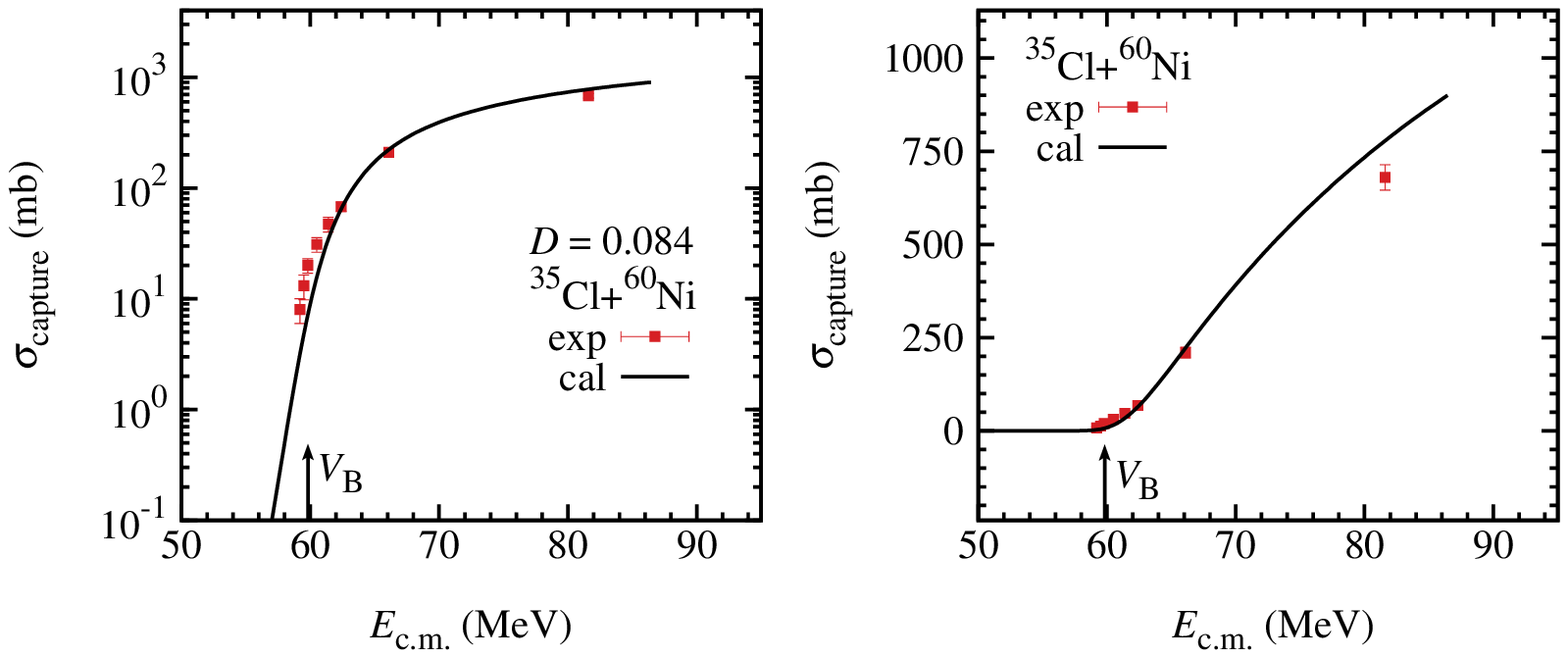}
  \includegraphics[width=0.47\textwidth]{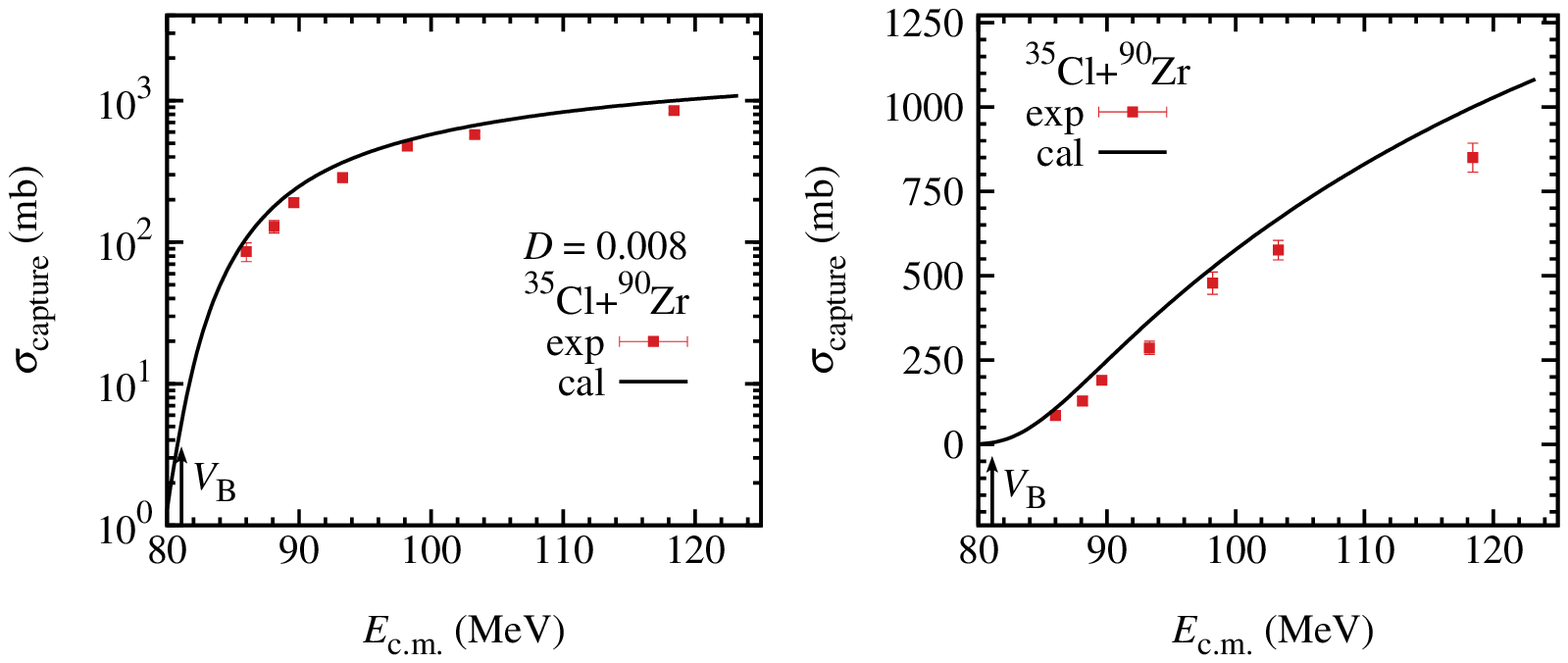}}
 \centerline{\includegraphics[width=0.47\textwidth]{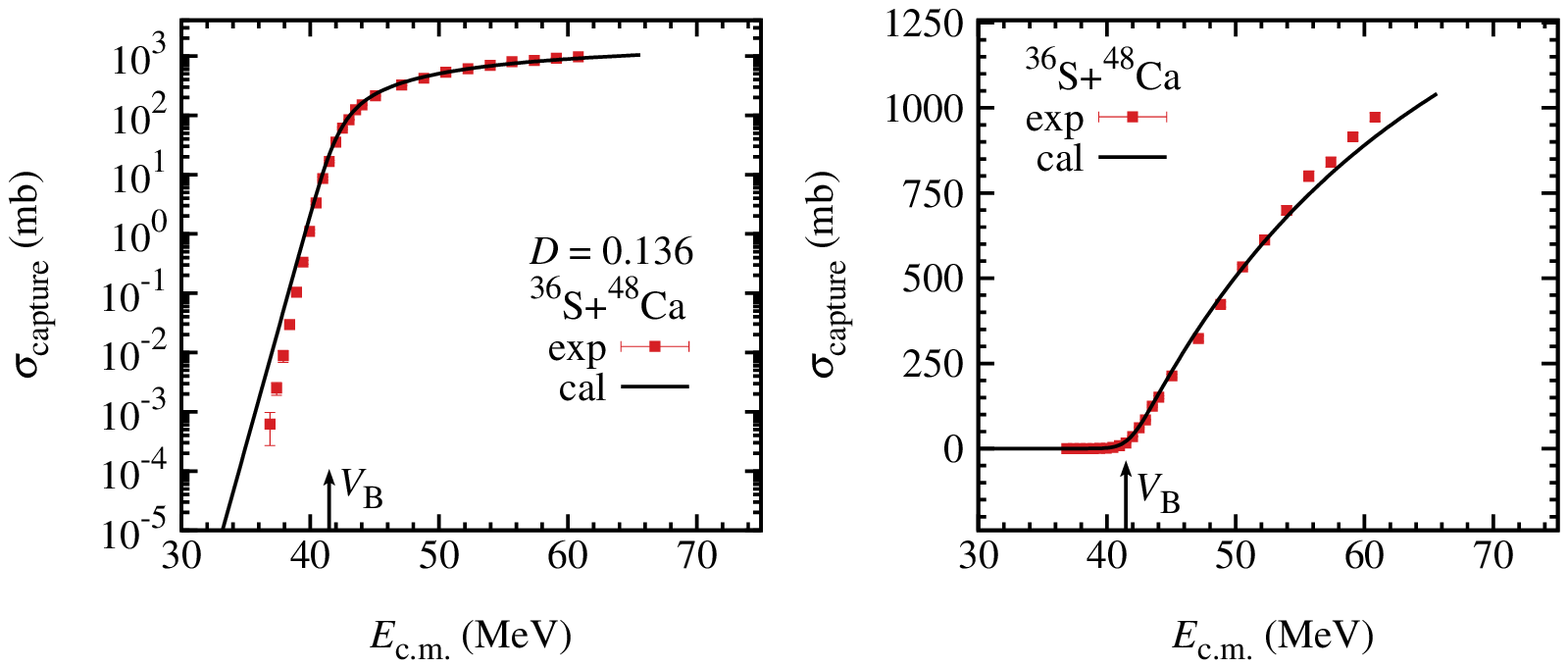}
  \includegraphics[width=0.47\textwidth]{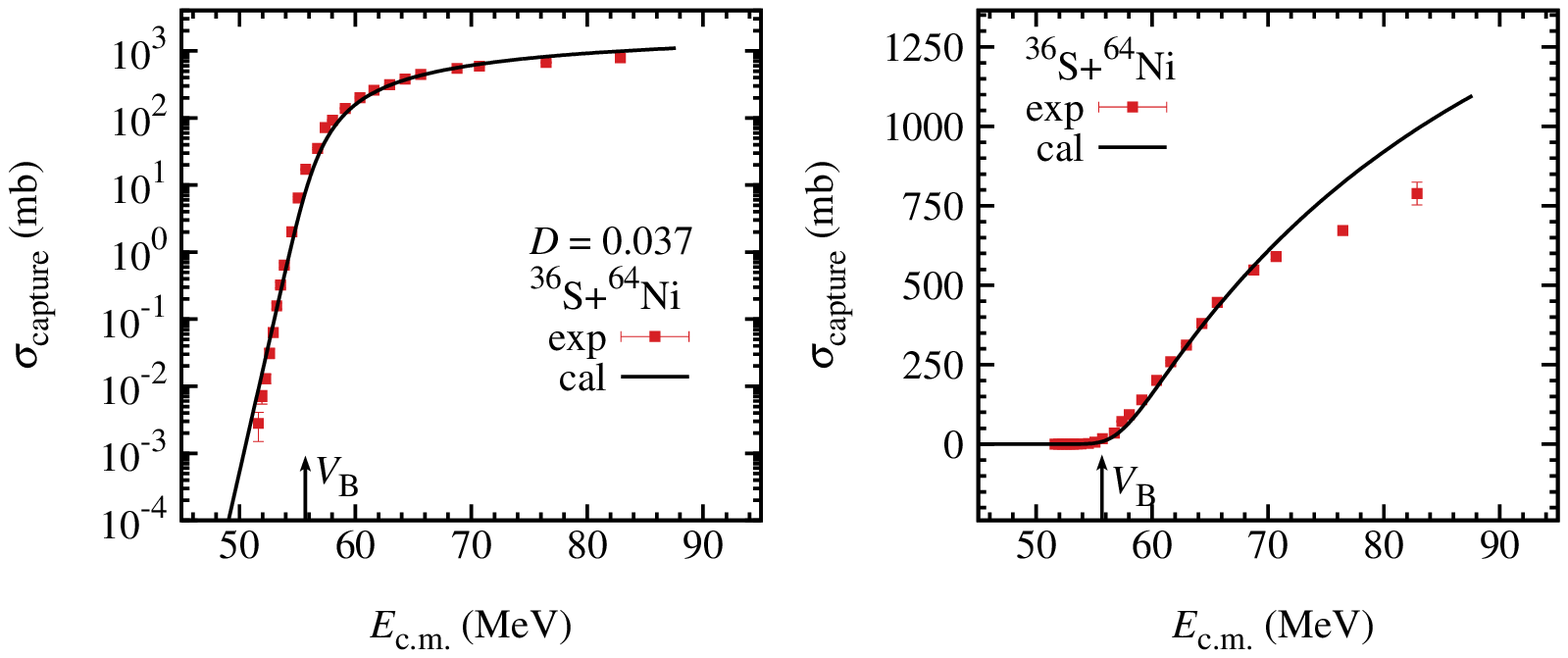}}
 \centerline{\includegraphics[width=0.47\textwidth]{36S90Zr.eps}
  \includegraphics[width=0.47\textwidth]{36S96Zr.eps}}
 \centerline{\includegraphics[width=0.47\textwidth]{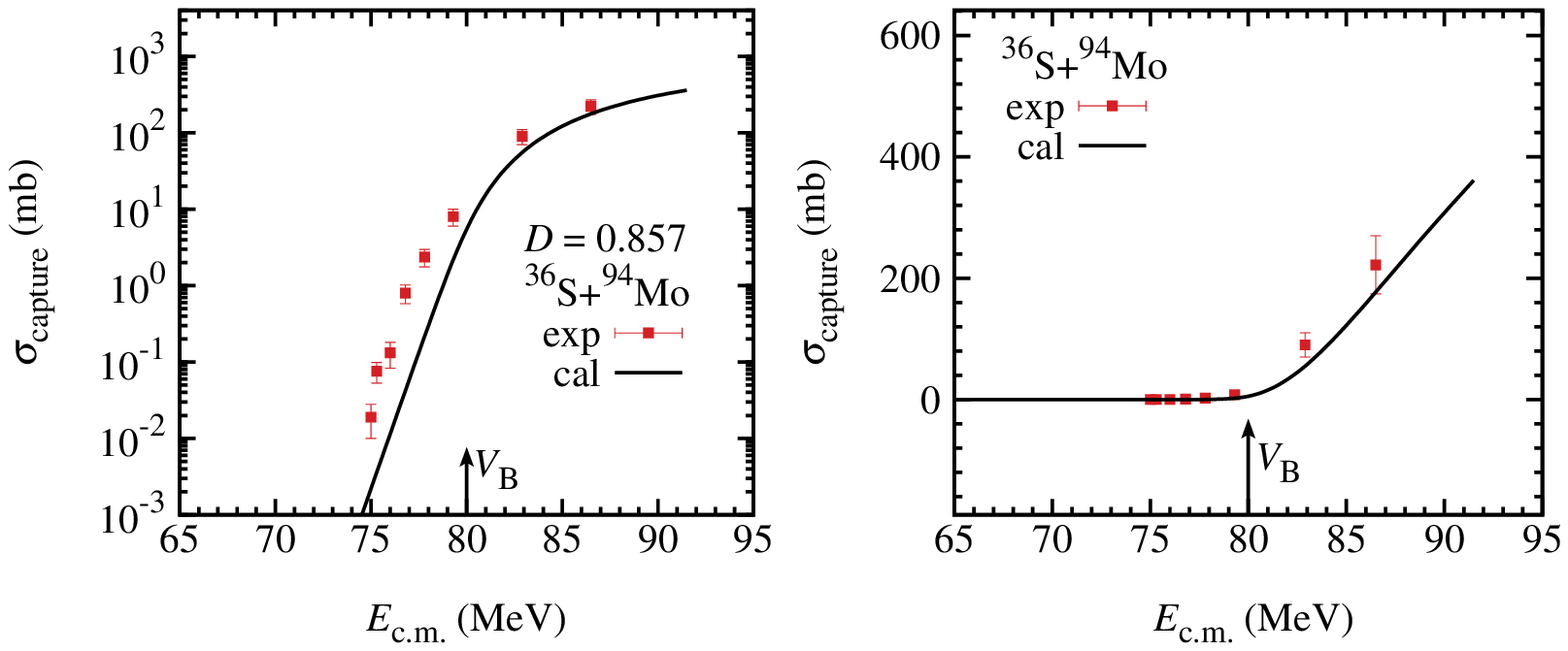}
  \includegraphics[width=0.47\textwidth]{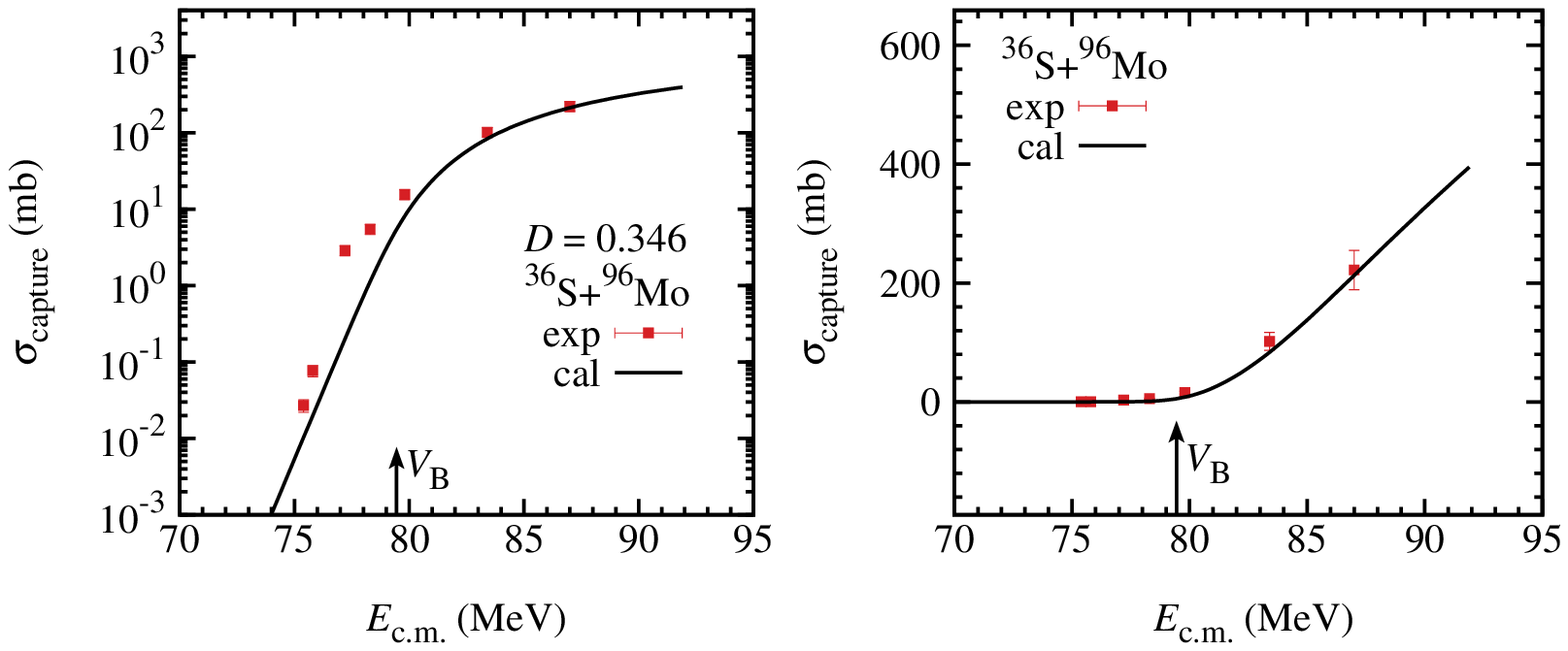}}
 \centerline{\includegraphics[width=0.47\textwidth]{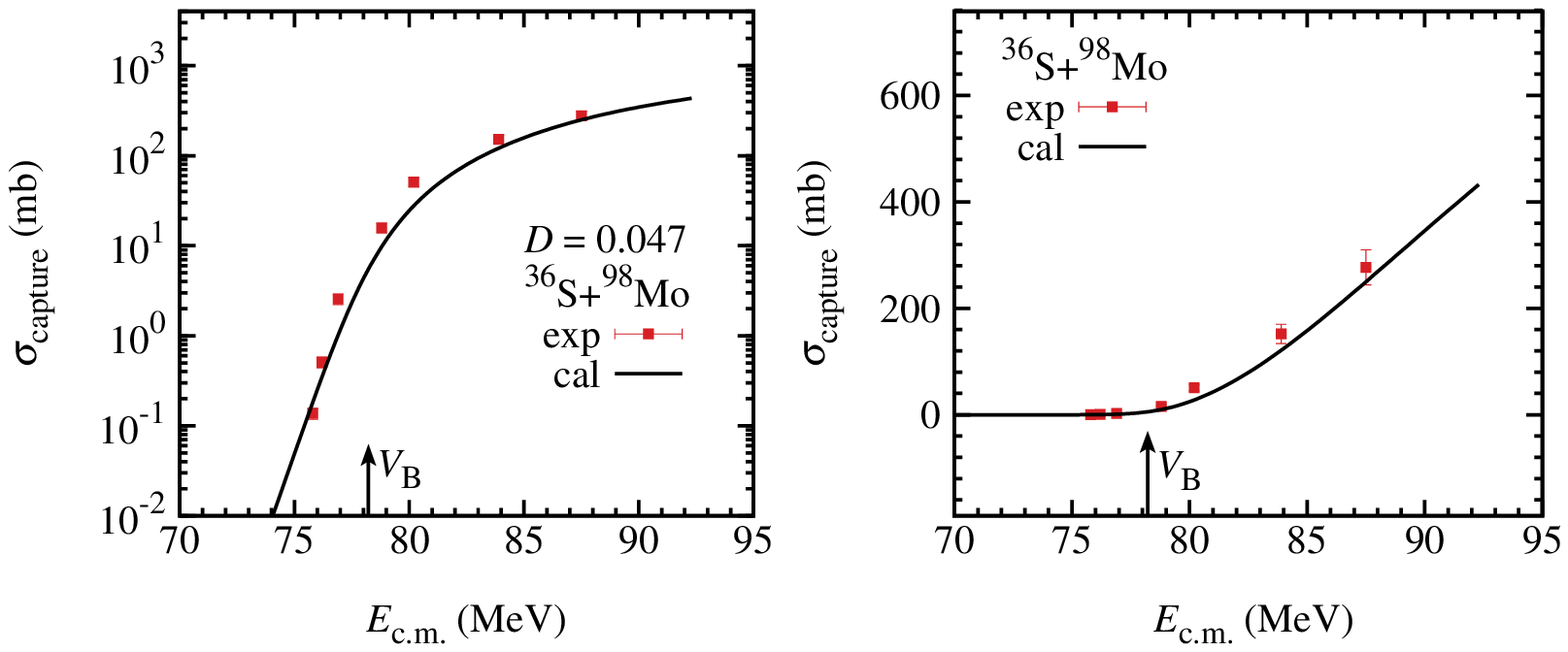}
  \includegraphics[width=0.47\textwidth]{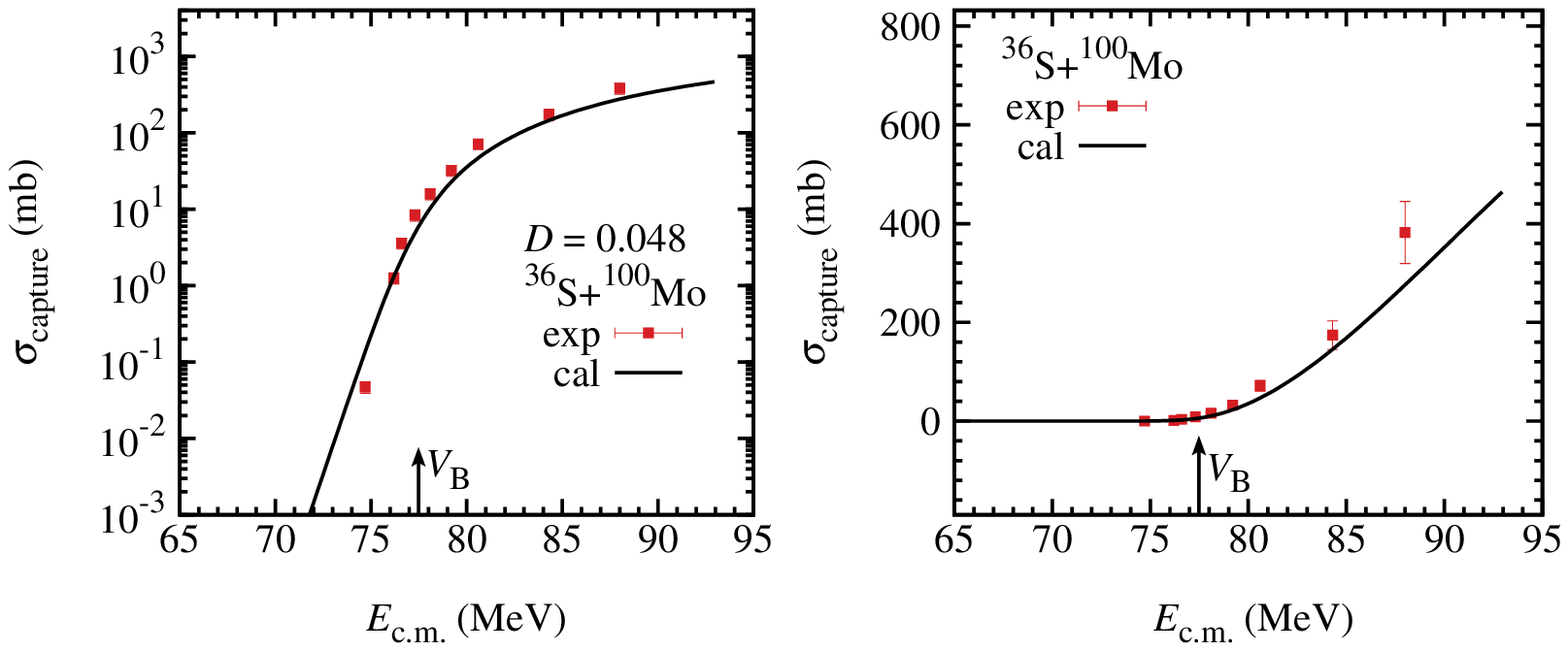}}
  \centerline {Graph 7}
 \end{Dfigures}
 \begin{Dfigures}[!ht]
 \centerline{\includegraphics[width=0.47\textwidth]{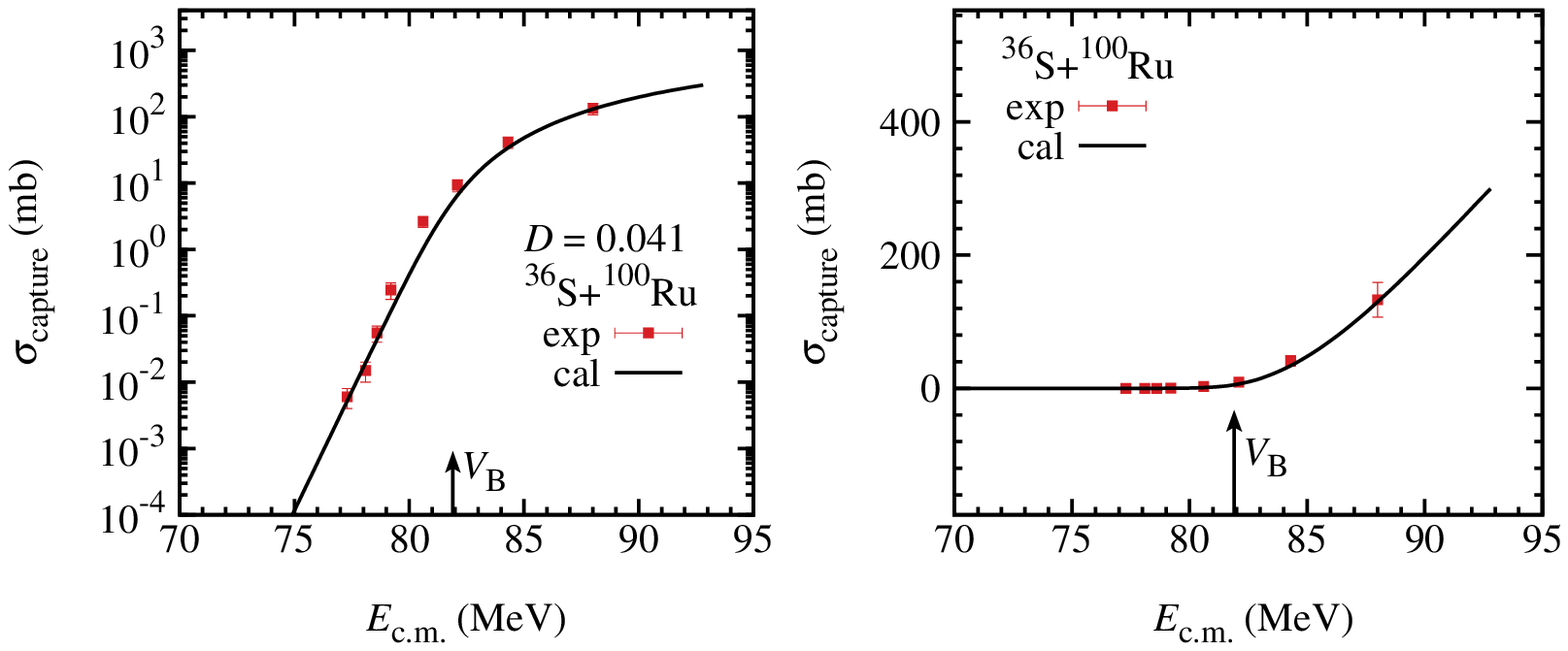}
  \includegraphics[width=0.47\textwidth]{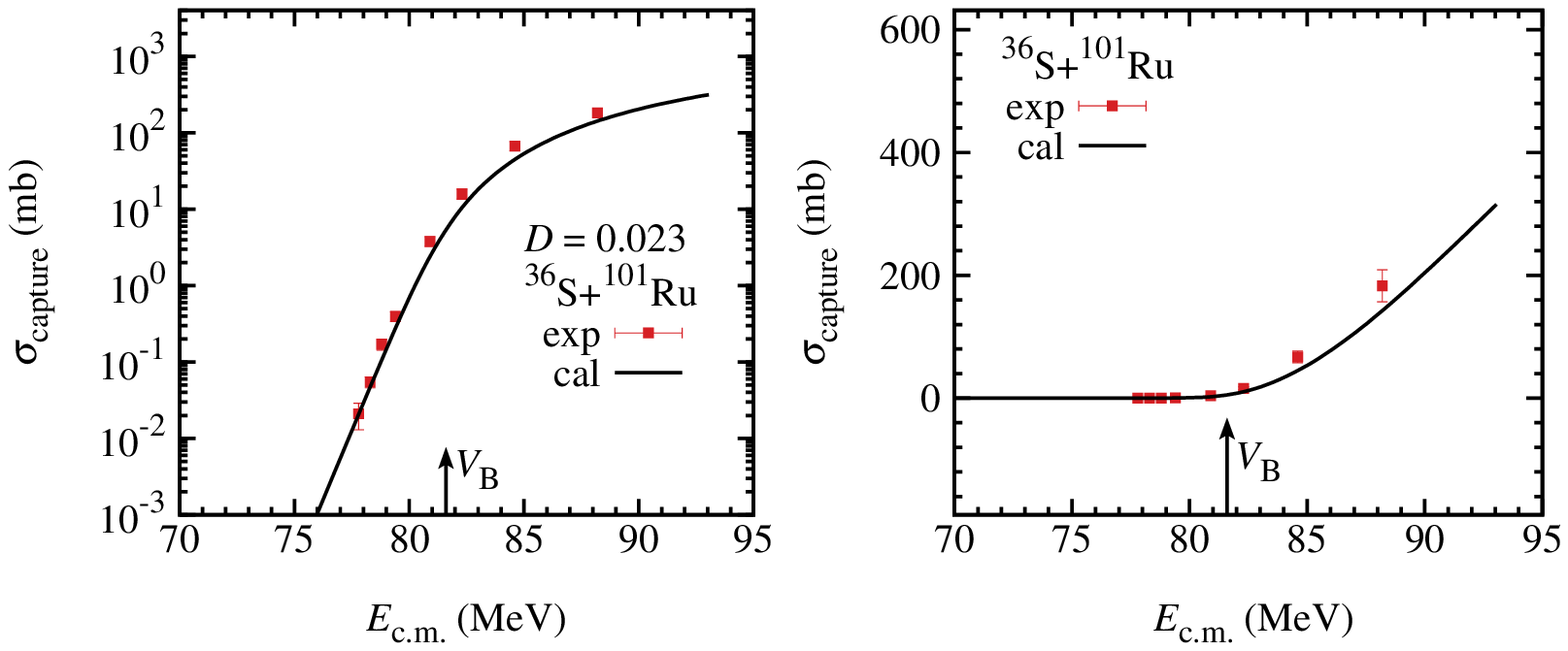}}
 \centerline{\includegraphics[width=0.47\textwidth]{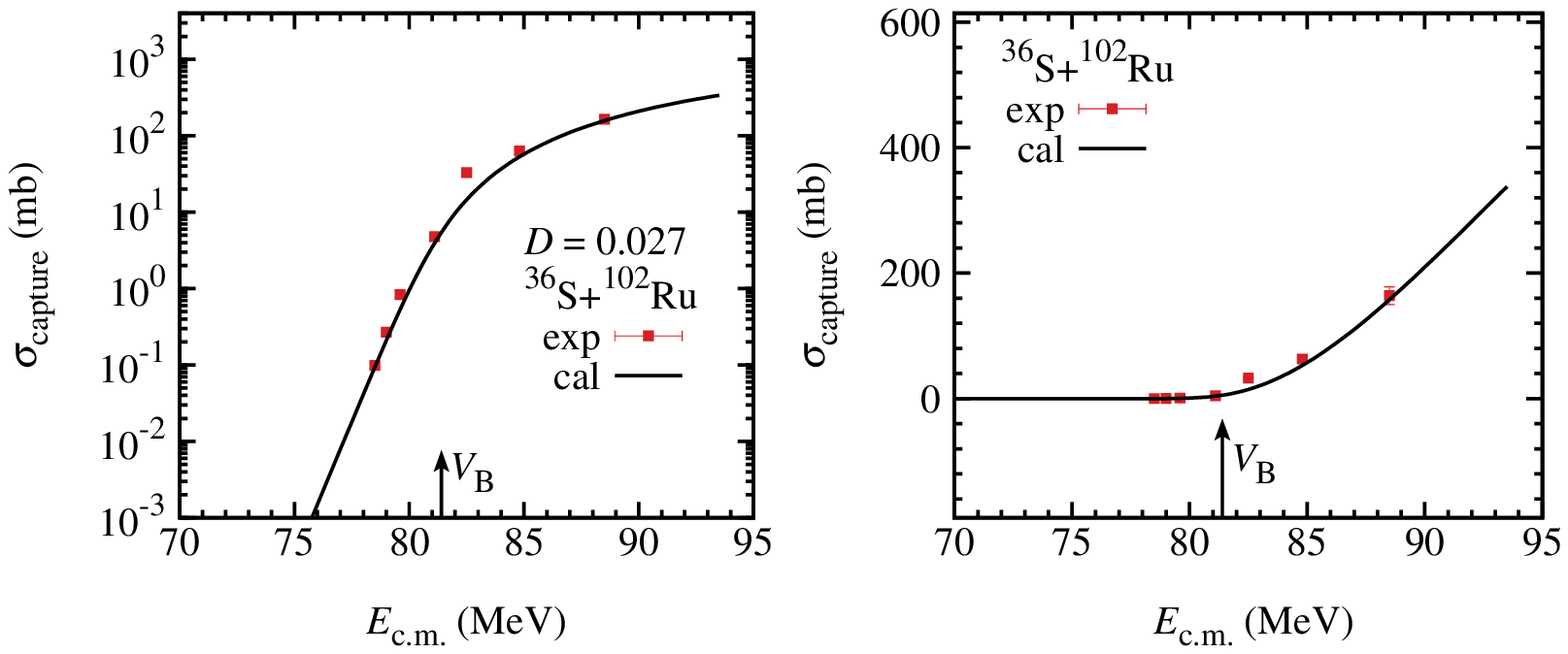}
  \includegraphics[width=0.47\textwidth]{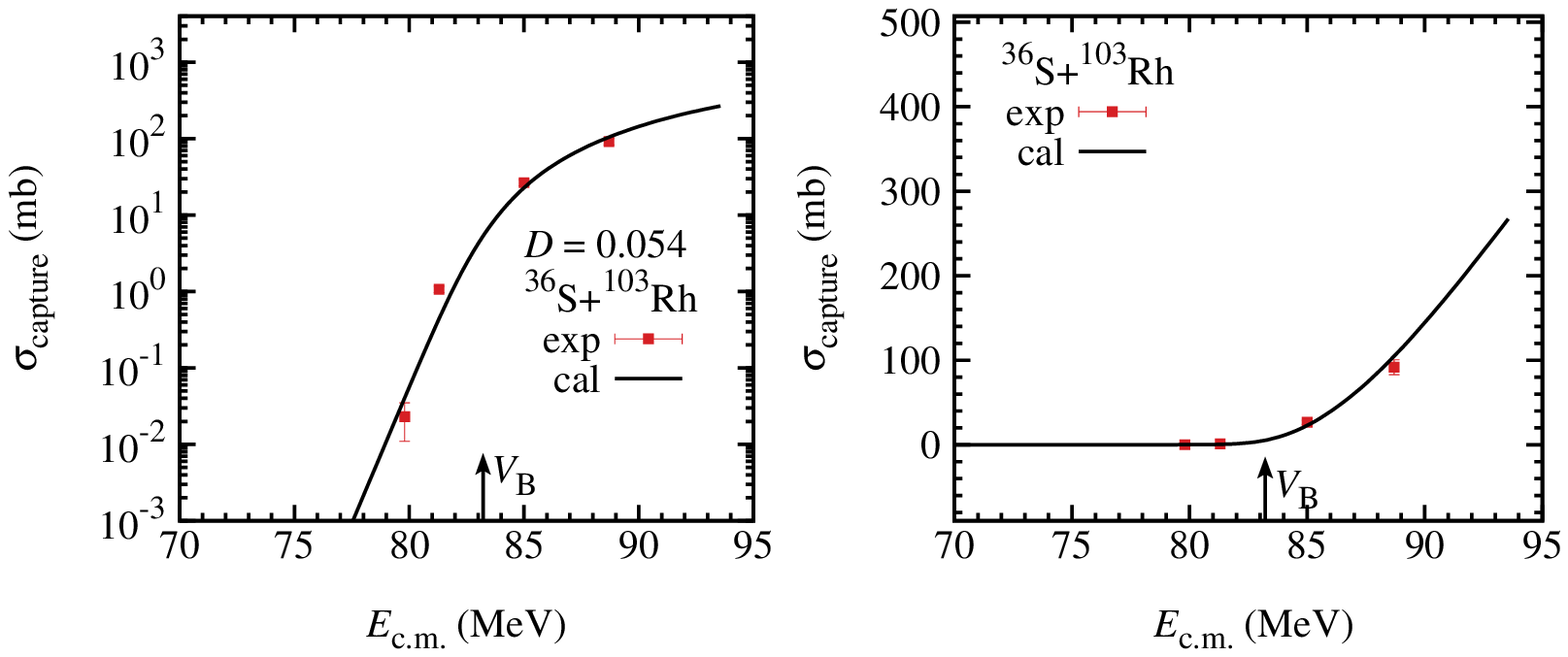}}
 \centerline{\includegraphics[width=0.47\textwidth]{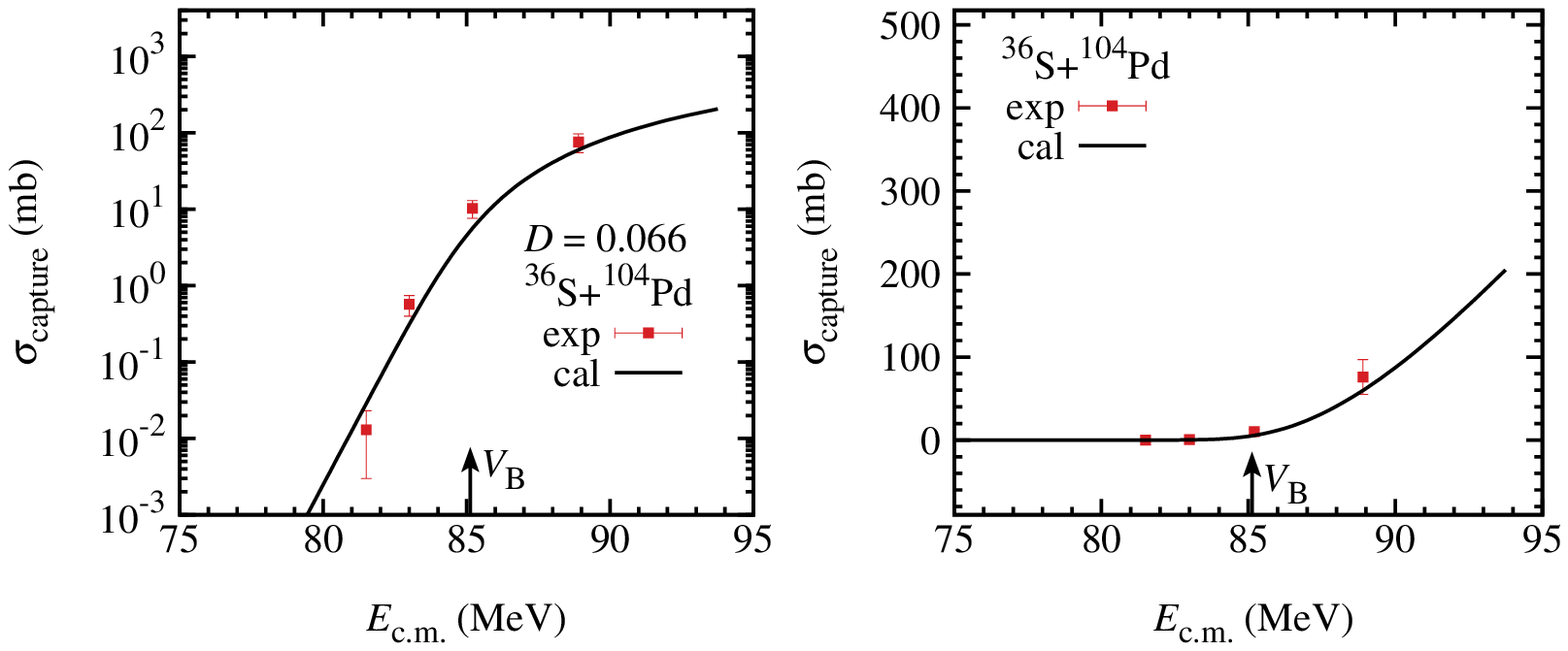}
  \includegraphics[width=0.47\textwidth]{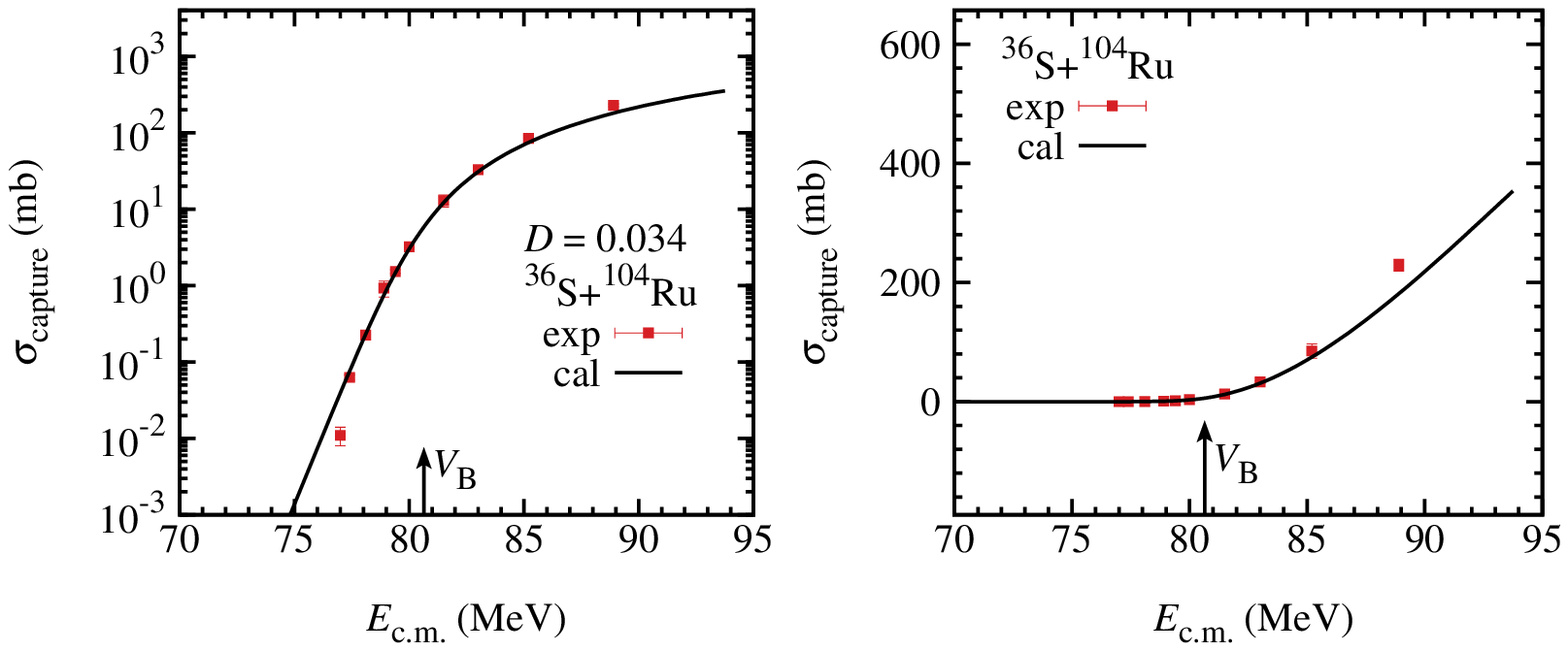}}
 \centerline{\includegraphics[width=0.47\textwidth]{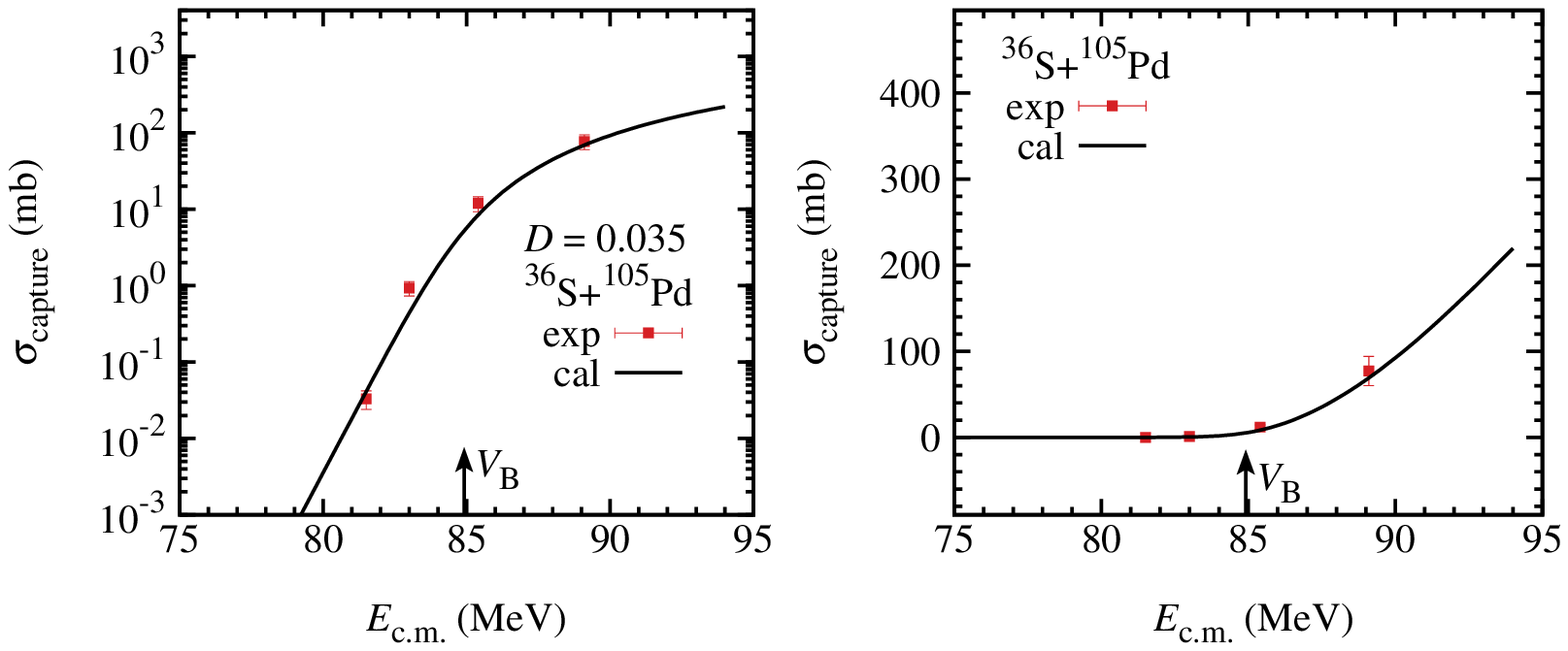}
  \includegraphics[width=0.47\textwidth]{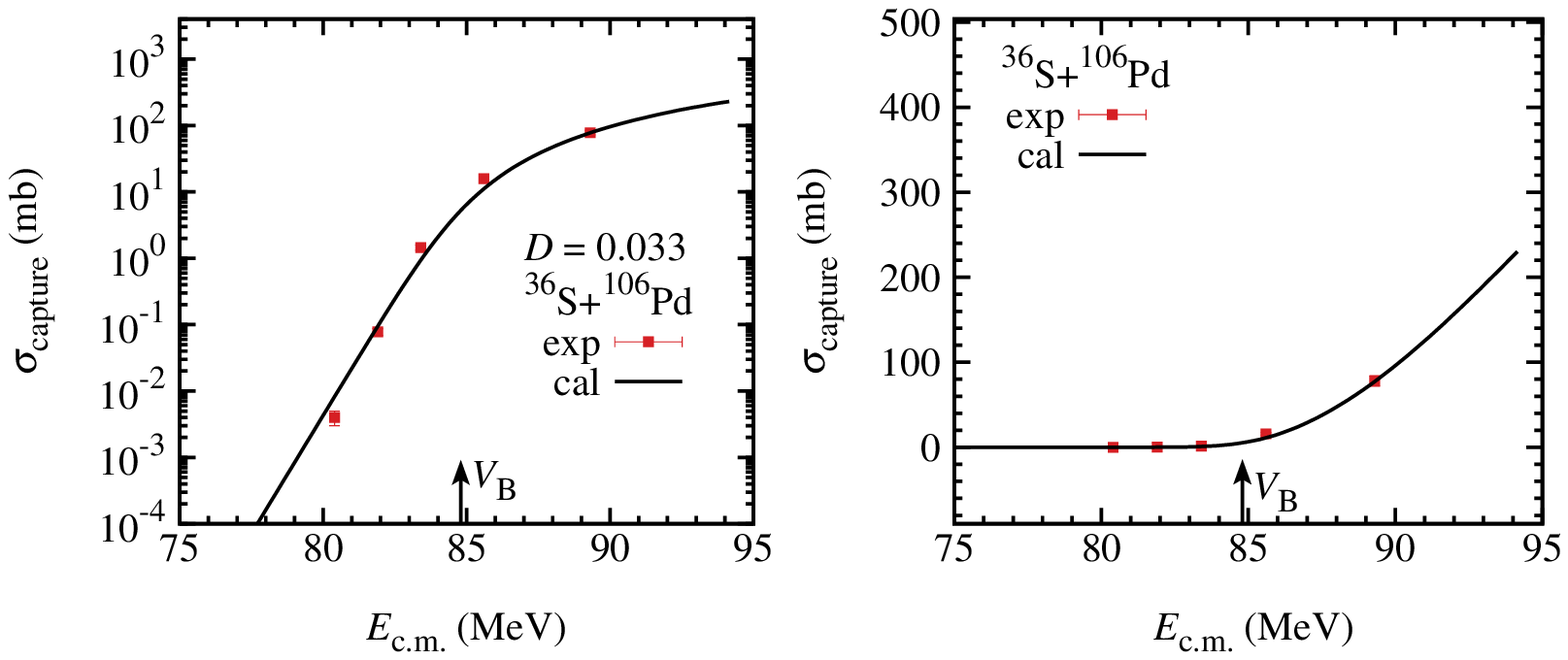}}
 \centerline{\includegraphics[width=0.47\textwidth]{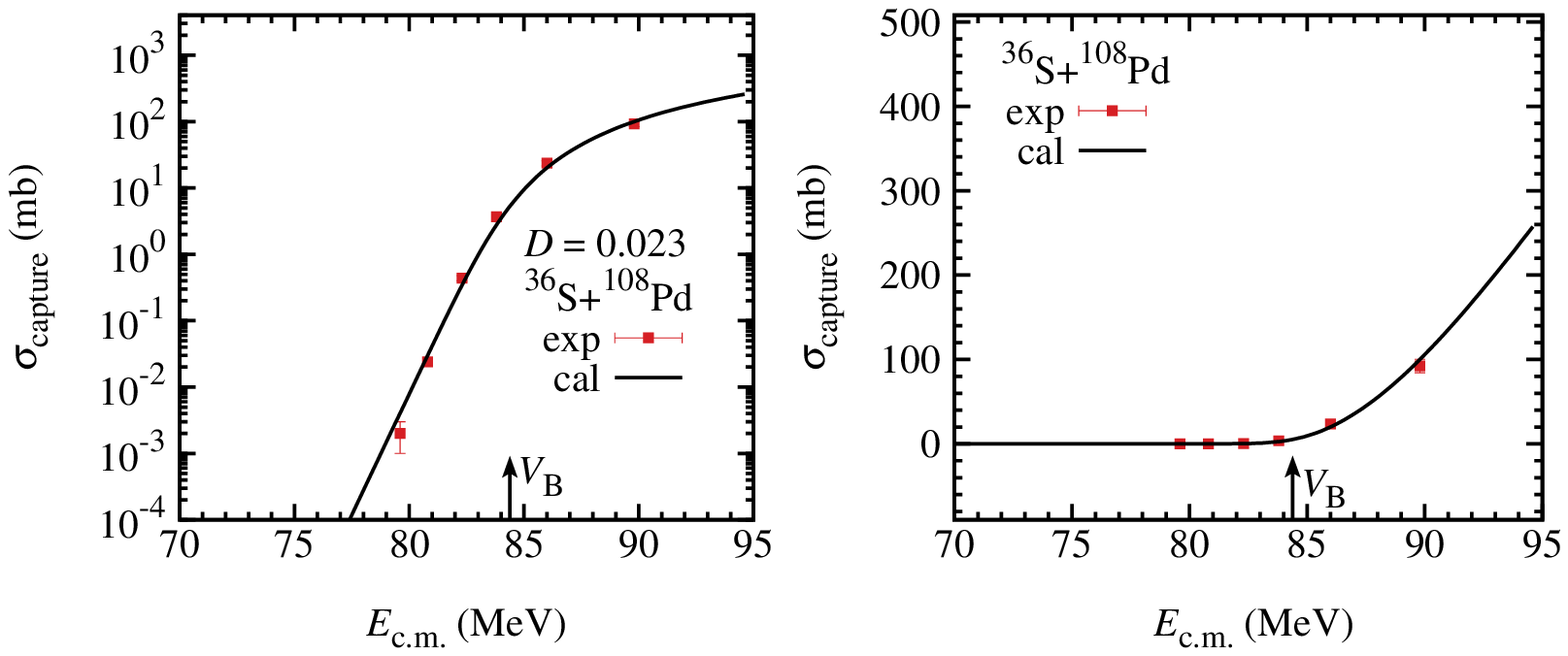}
  \includegraphics[width=0.47\textwidth]{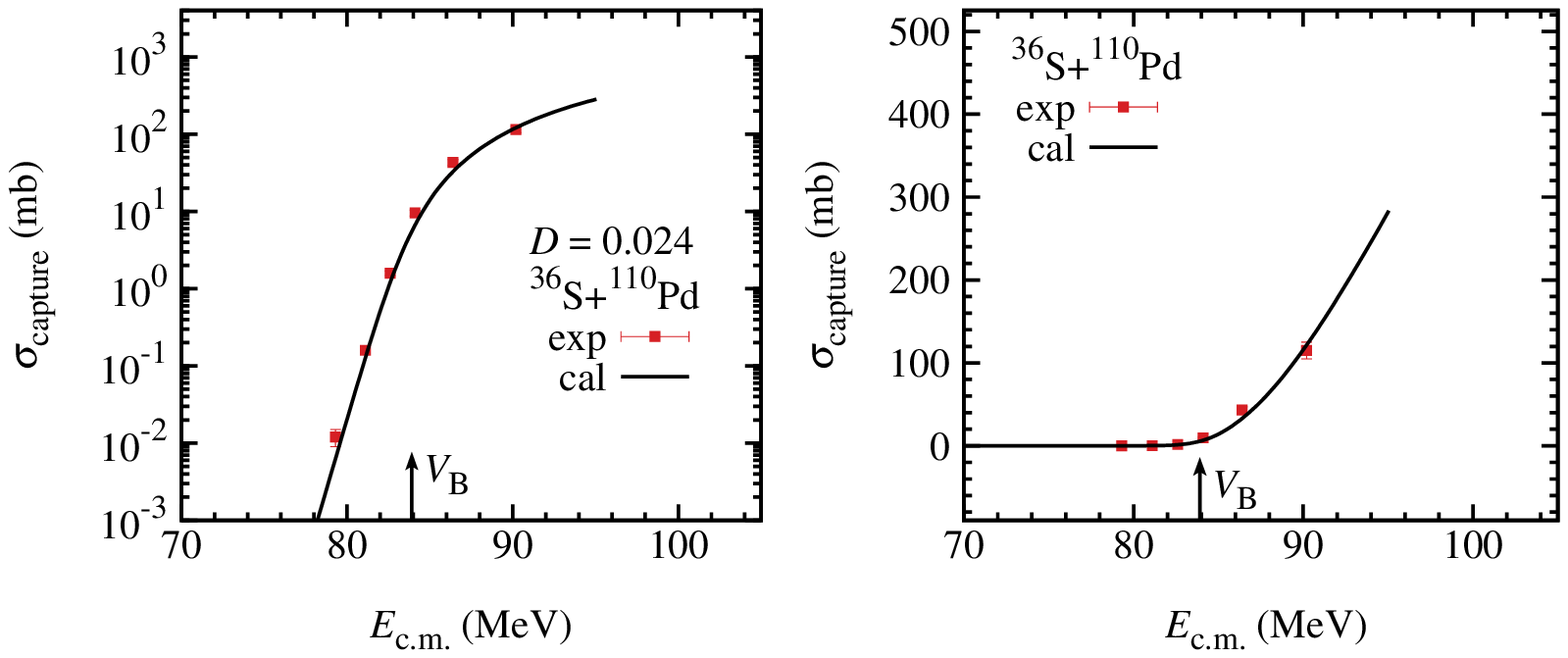}}
 \centerline{\includegraphics[width=0.47\textwidth]{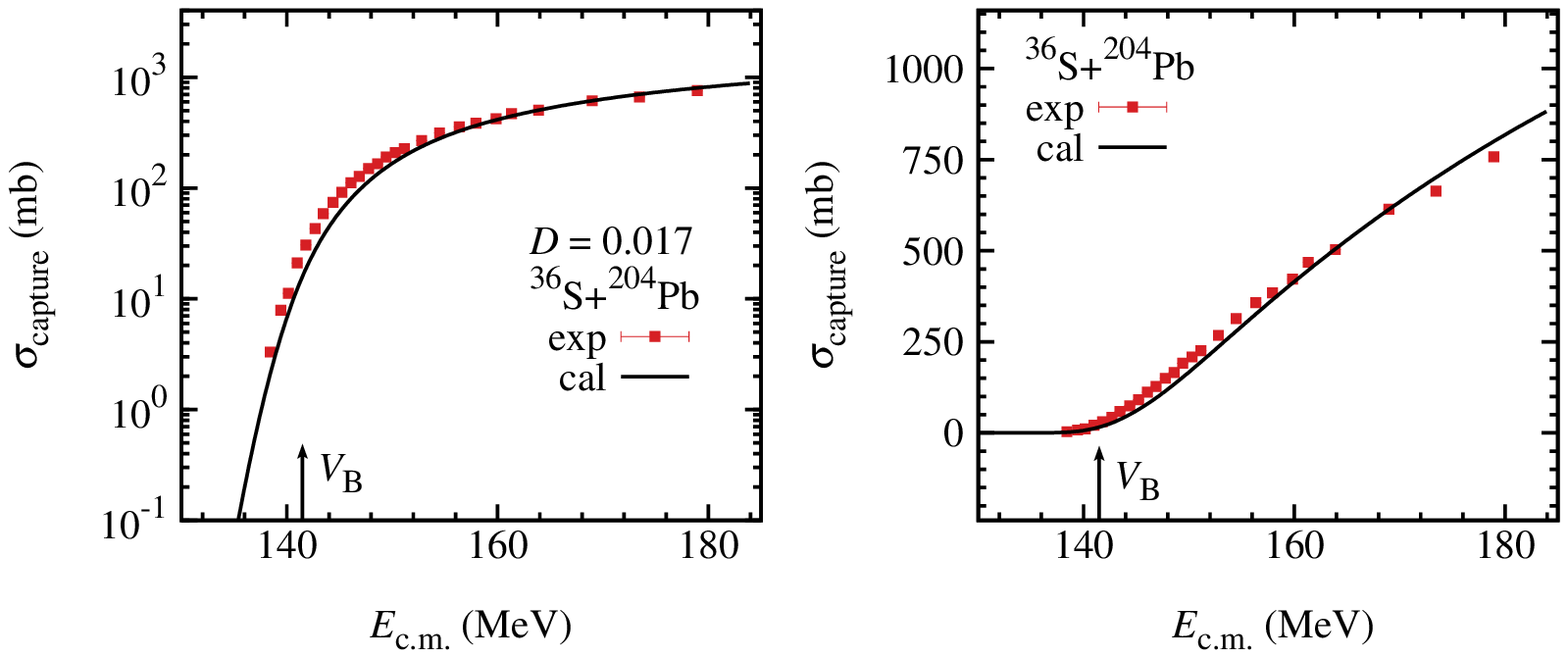}
  \includegraphics[width=0.47\textwidth]{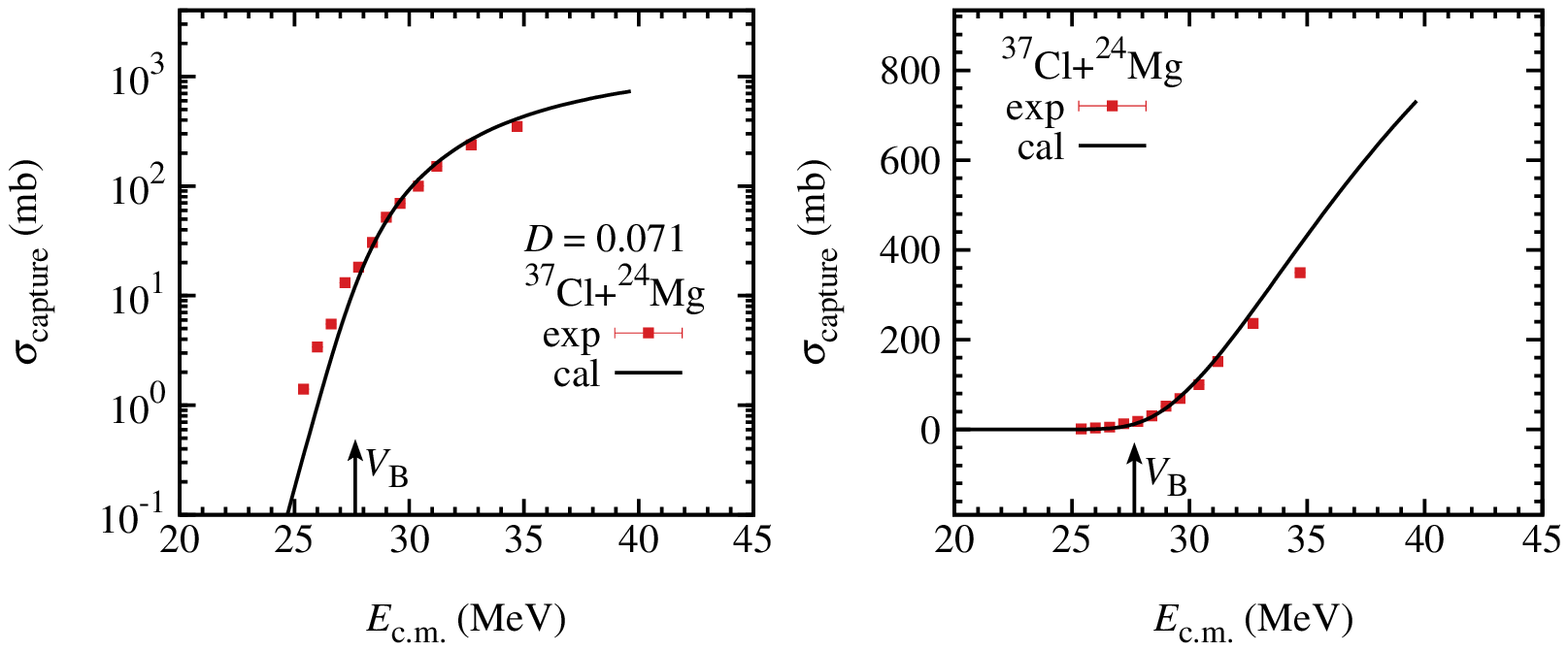}}
  \centerline {Graph 8}
 \end{Dfigures}
 \begin{Dfigures}[!ht]
 \centerline{\includegraphics[width=0.47\textwidth]{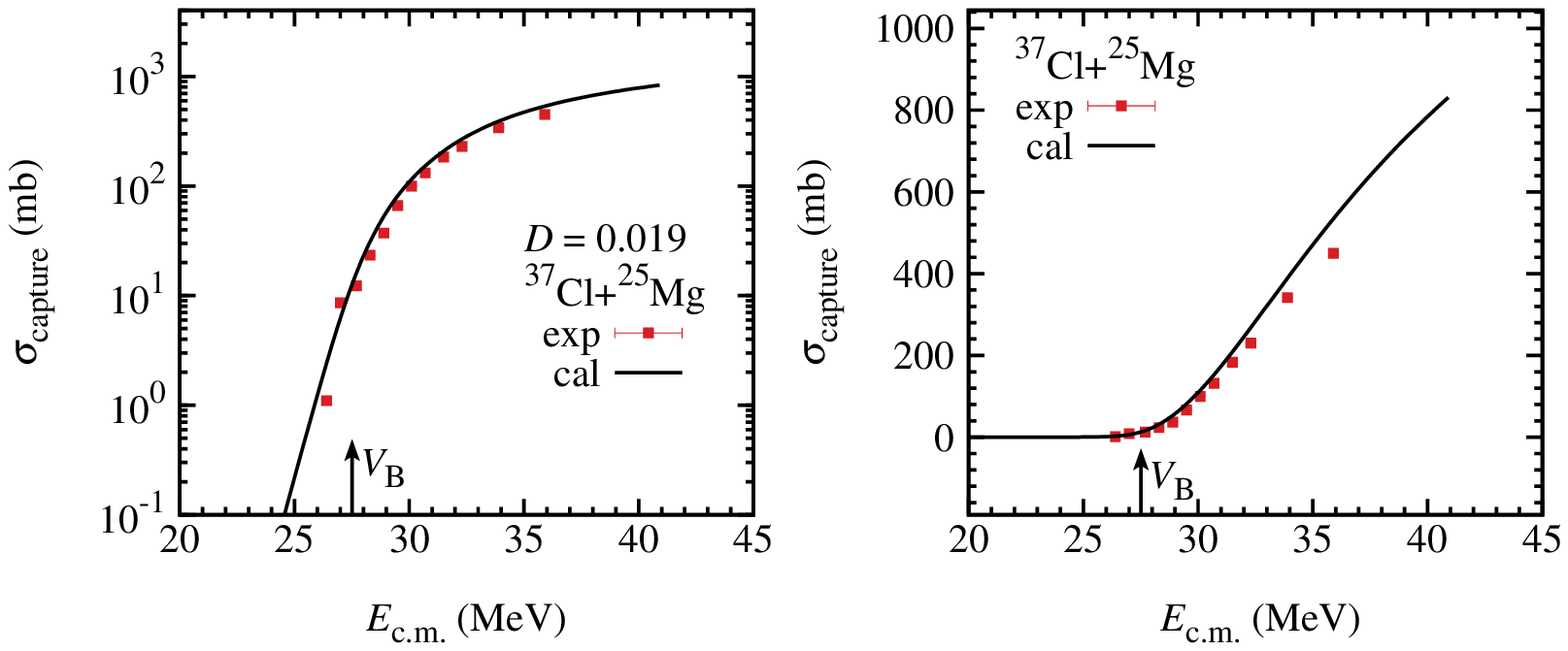}
  \includegraphics[width=0.47\textwidth]{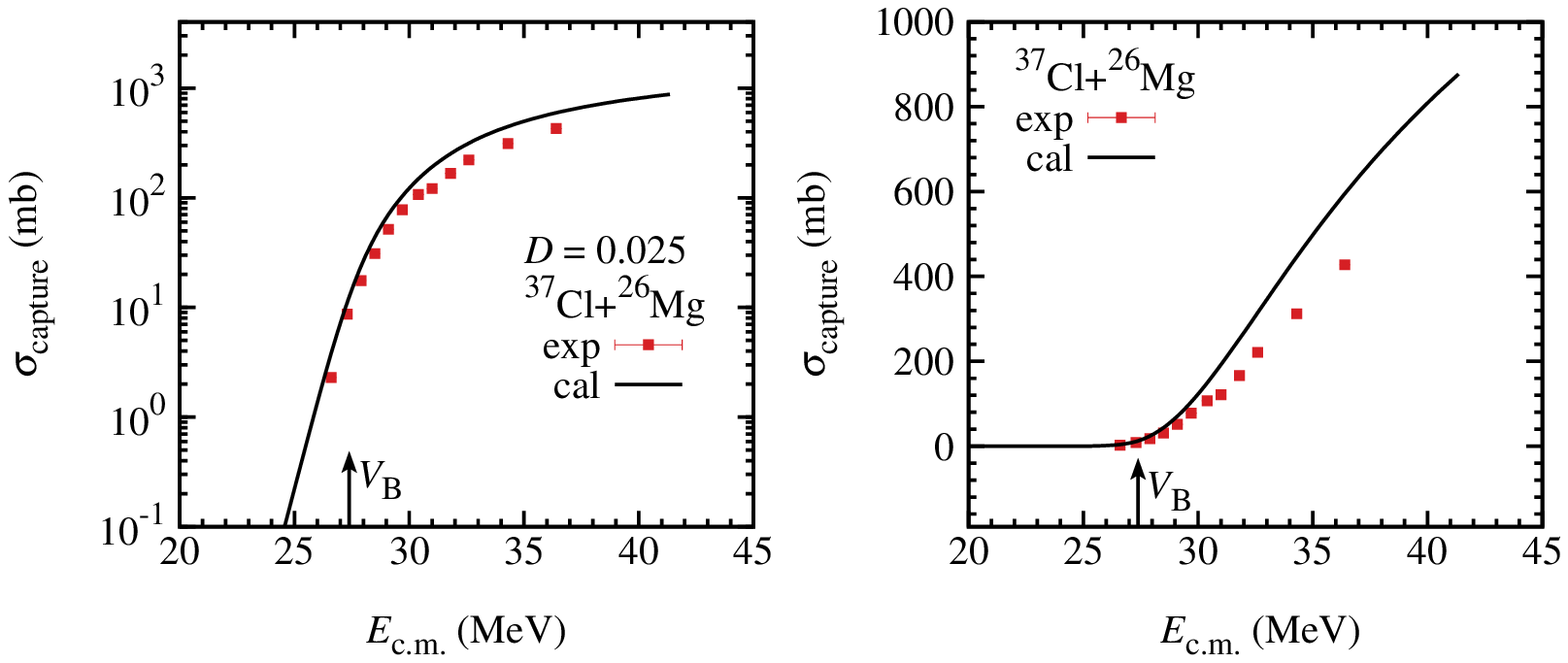}}
 \centerline{\includegraphics[width=0.47\textwidth]{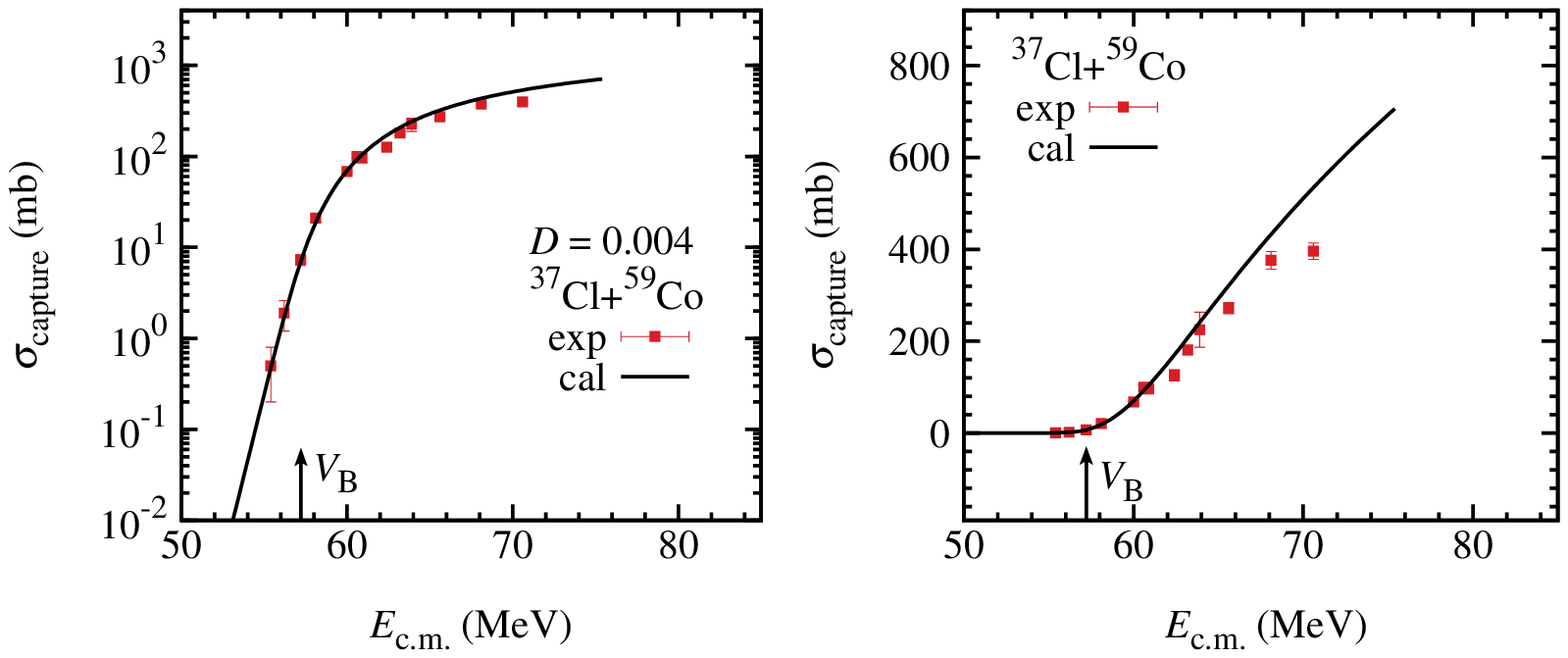}
  \includegraphics[width=0.47\textwidth]{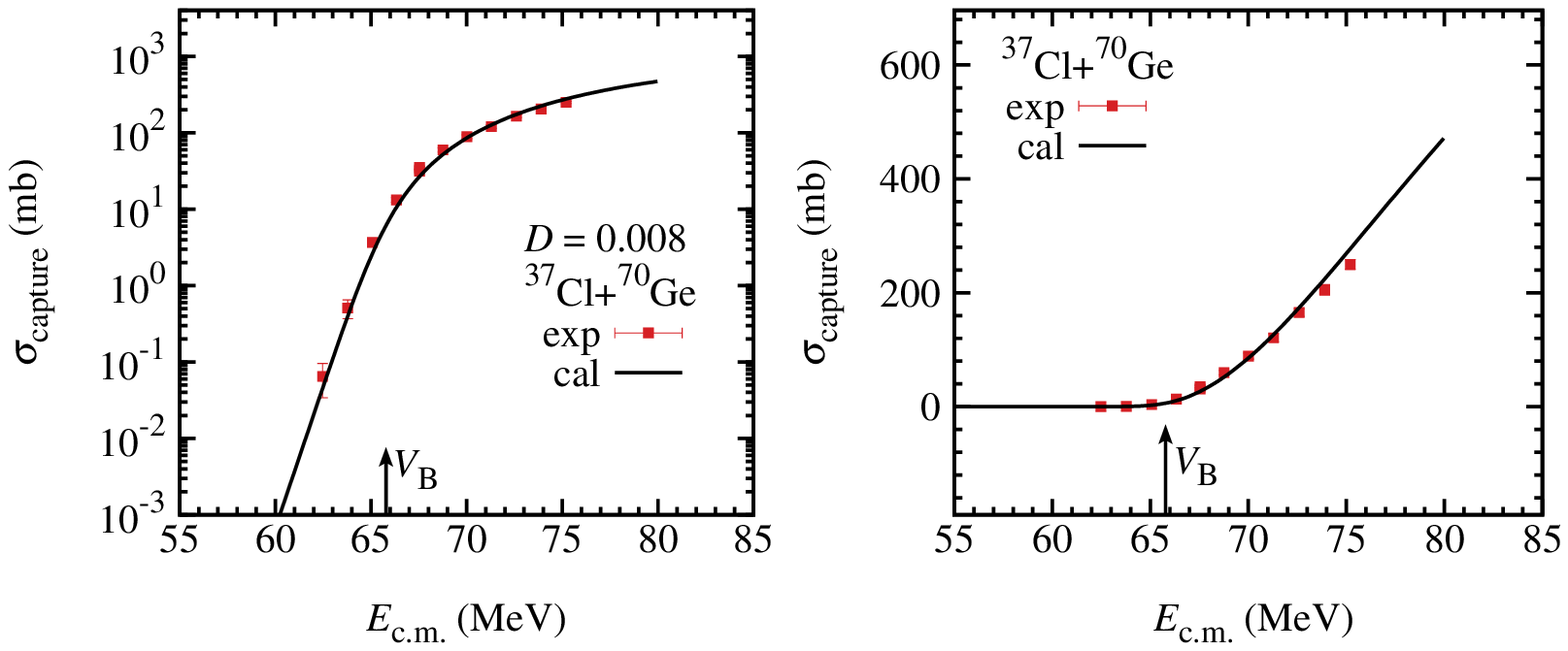}}
 \centerline{\includegraphics[width=0.47\textwidth]{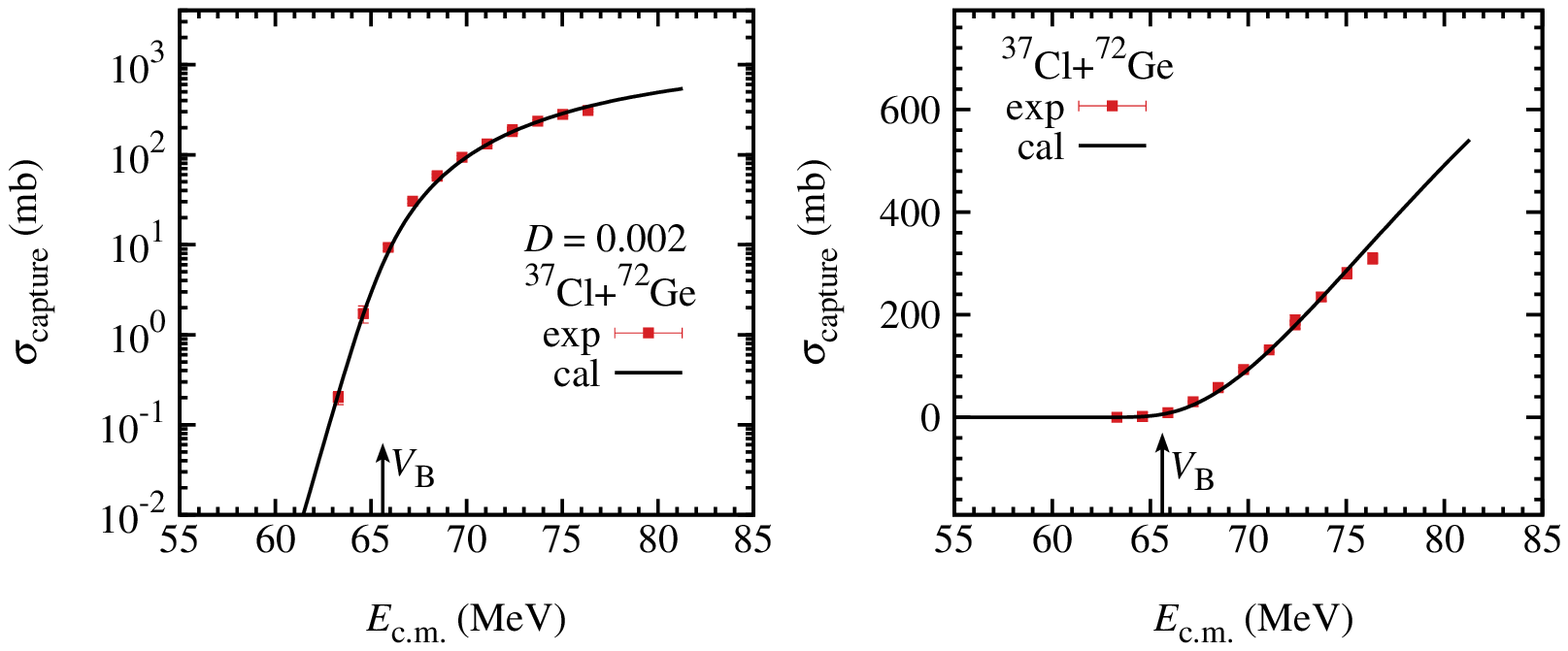}
  \includegraphics[width=0.47\textwidth]{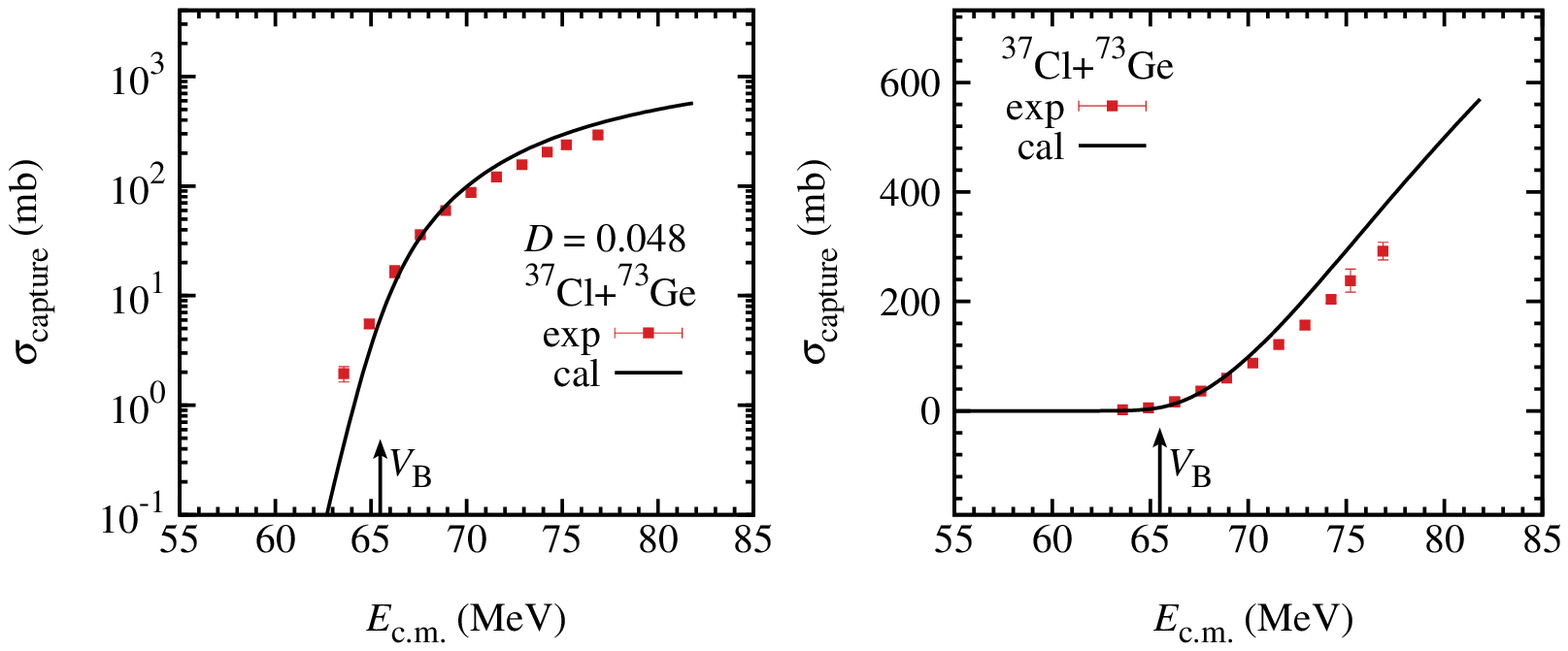}}
 \centerline{\includegraphics[width=0.47\textwidth]{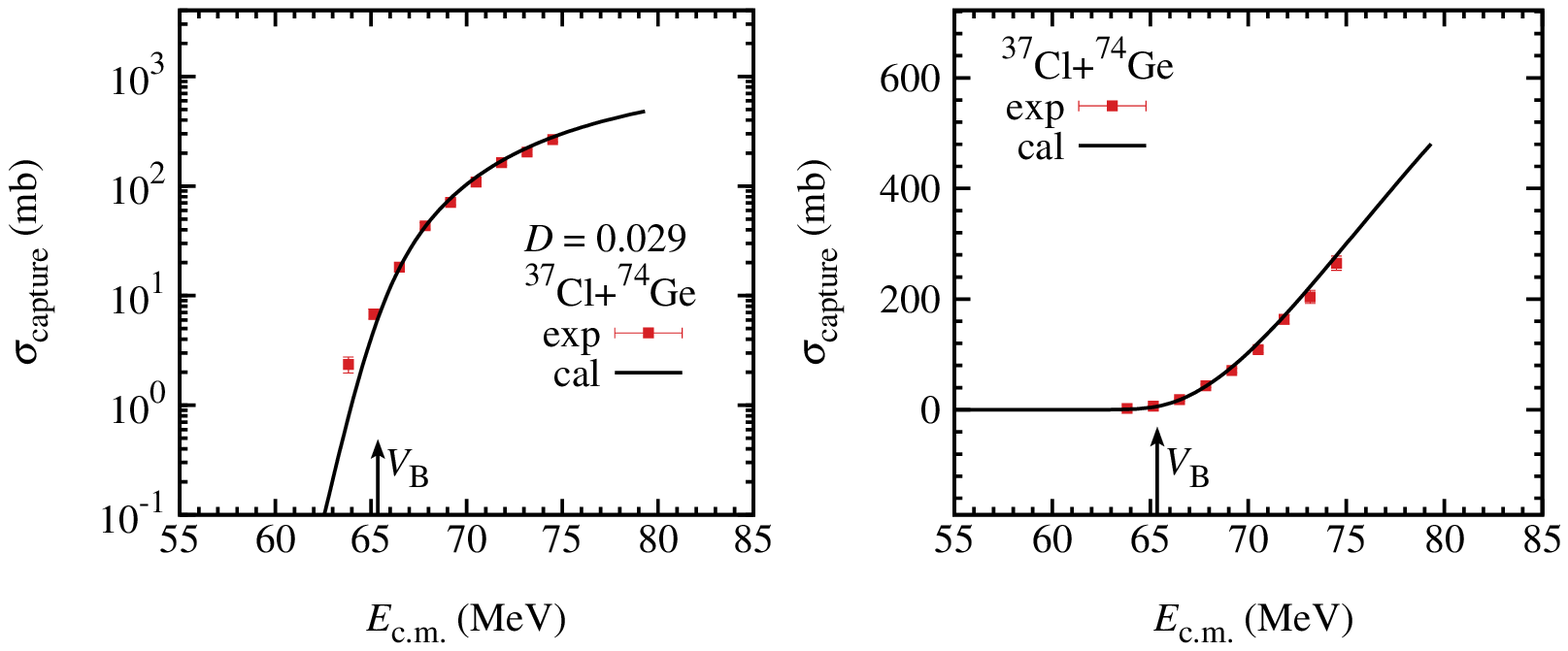}
  \includegraphics[width=0.47\textwidth]{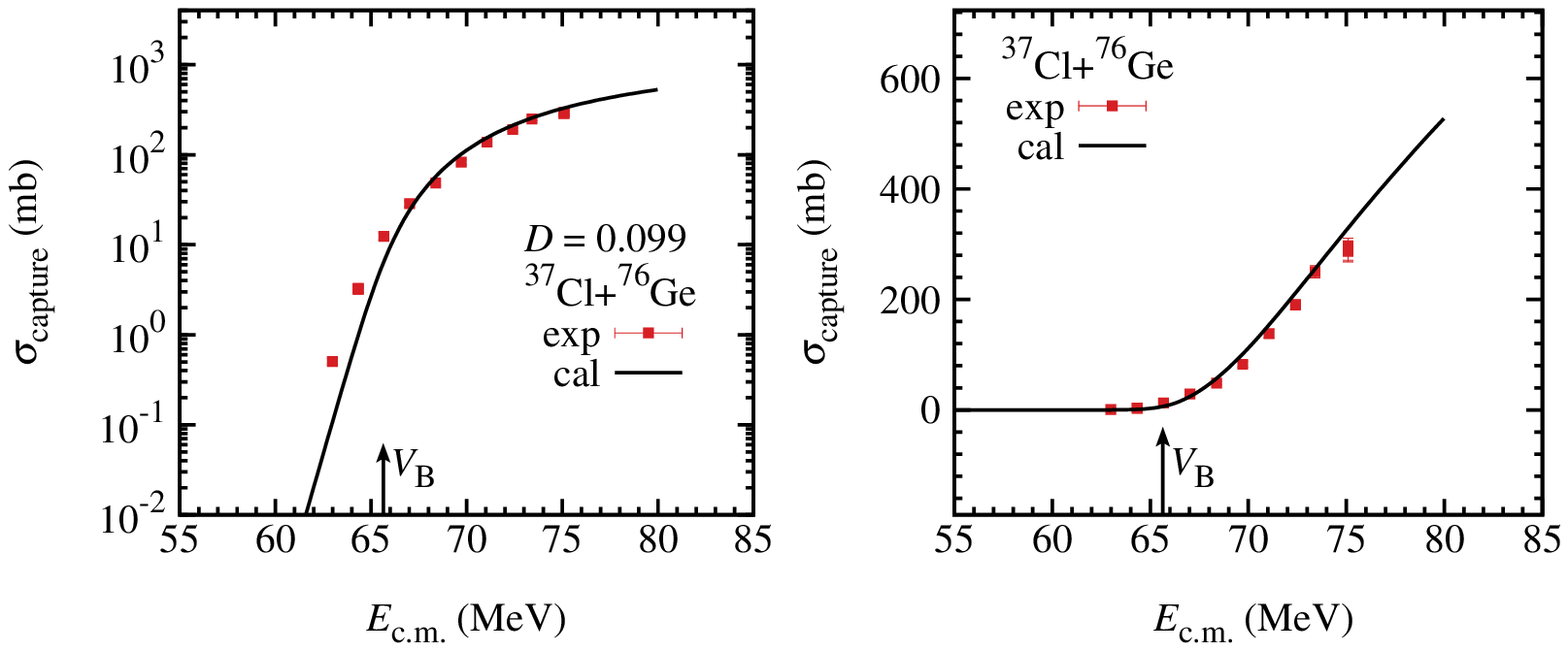}}
 \centerline{\includegraphics[width=0.47\textwidth]{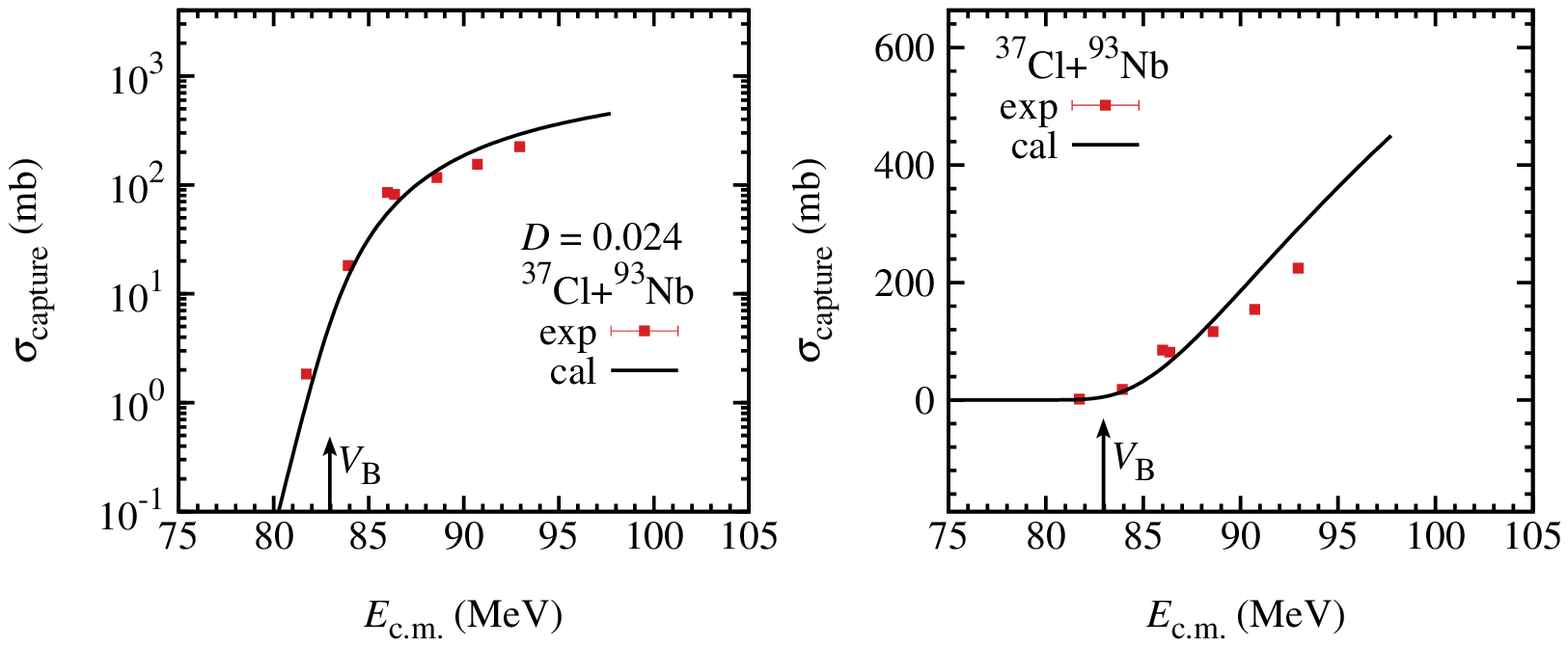}
  \includegraphics[width=0.47\textwidth]{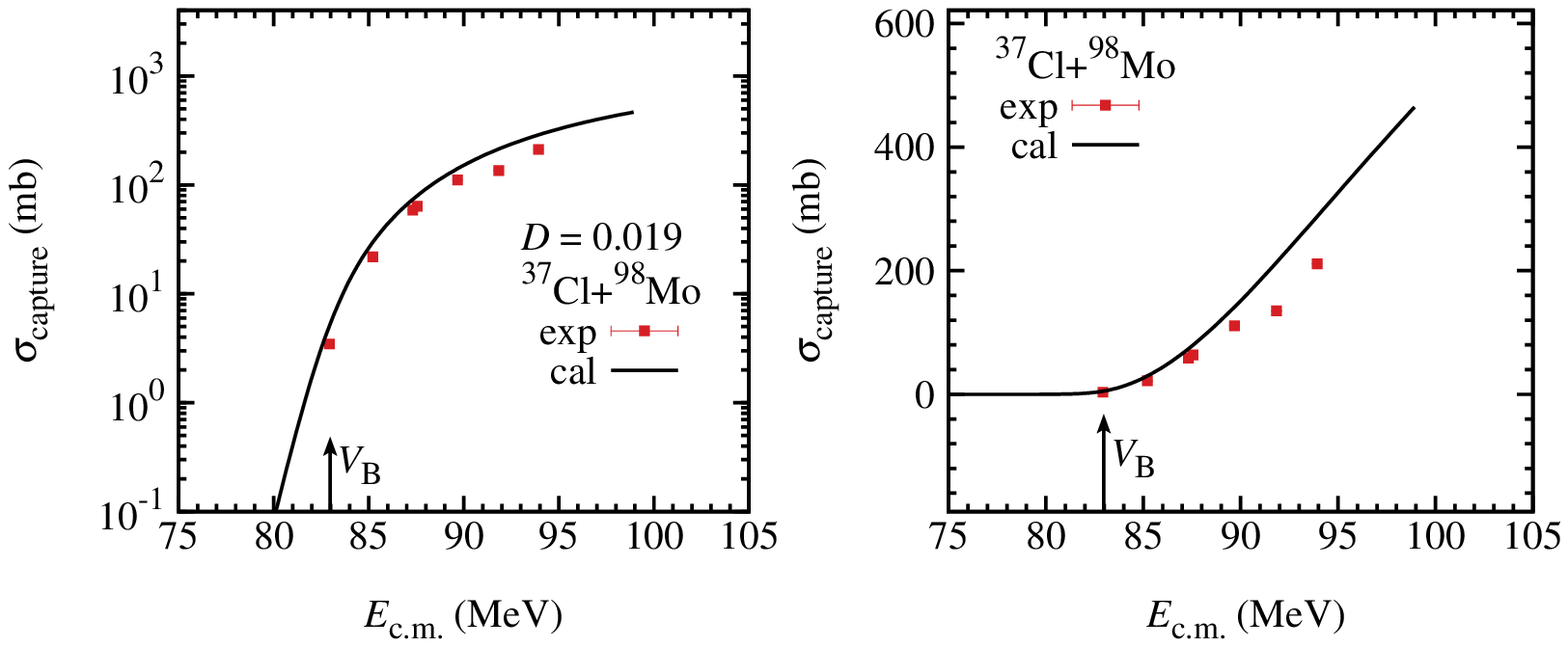}}
 \centerline{\includegraphics[width=0.47\textwidth]{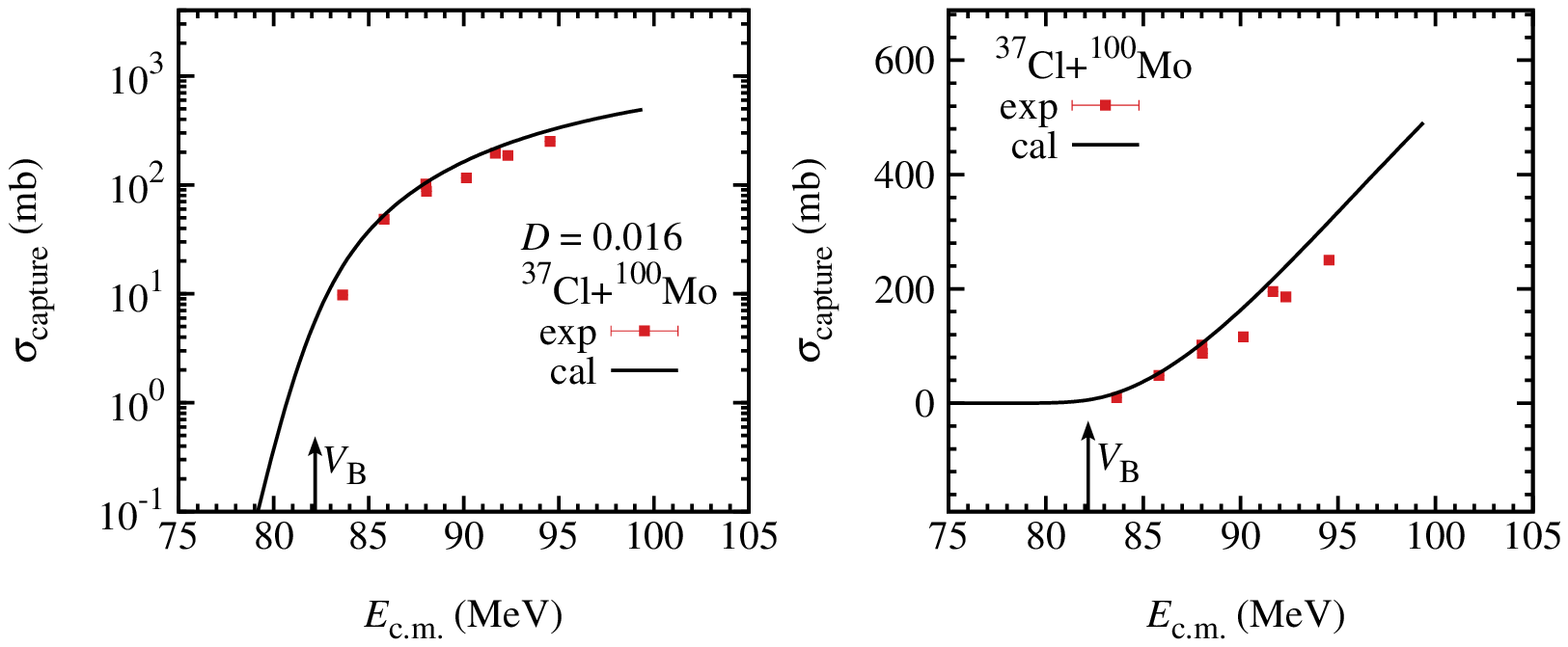}
  \includegraphics[width=0.47\textwidth]{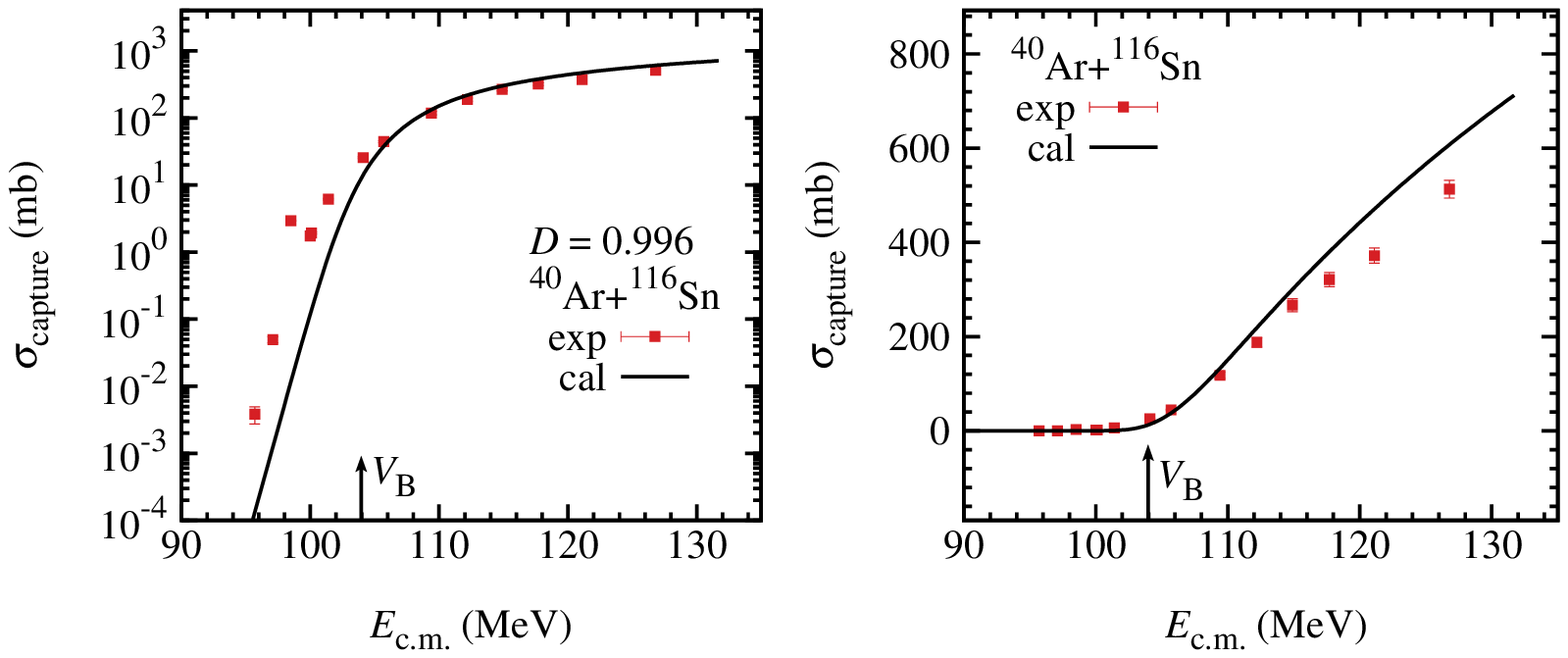}}
  \centerline {Graph 9}
 \end{Dfigures}
 \begin{Dfigures}[!ht]
 \centerline{\includegraphics[width=0.47\textwidth]{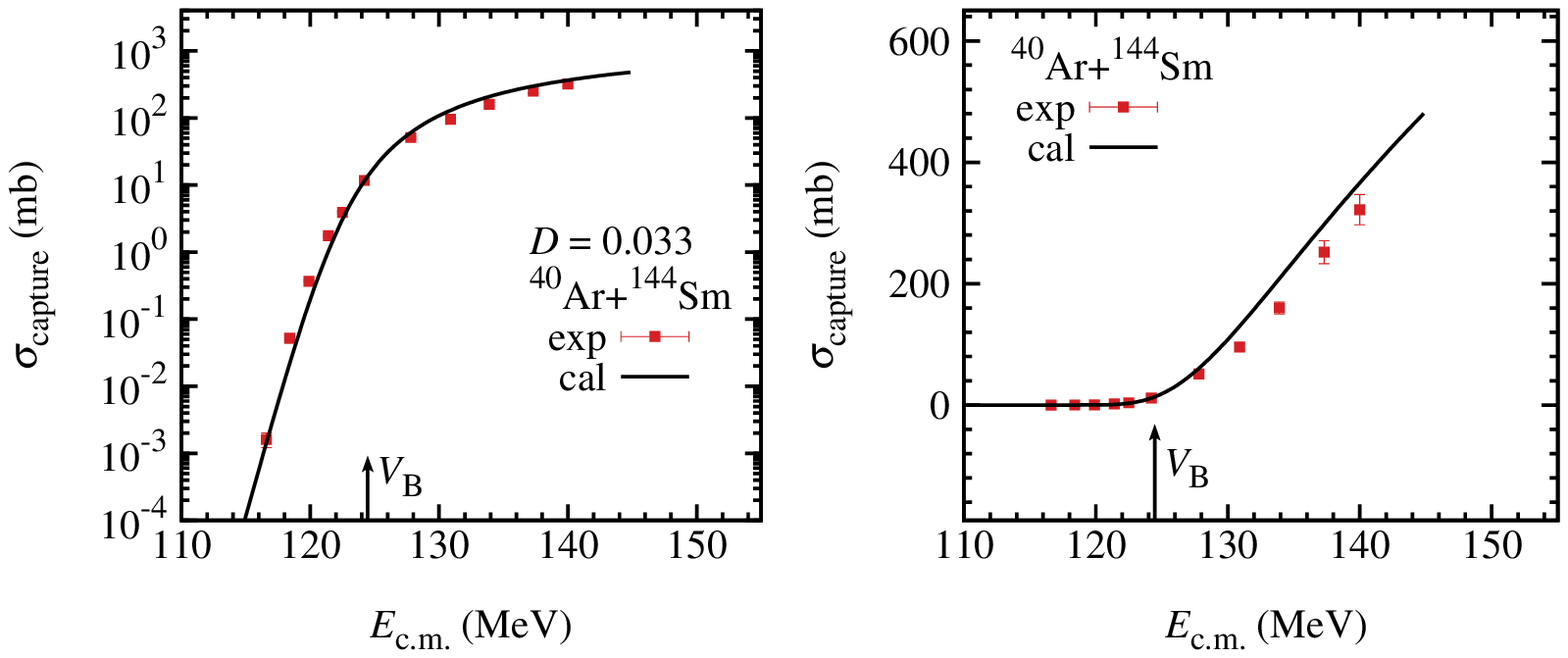}
  \includegraphics[width=0.47\textwidth]{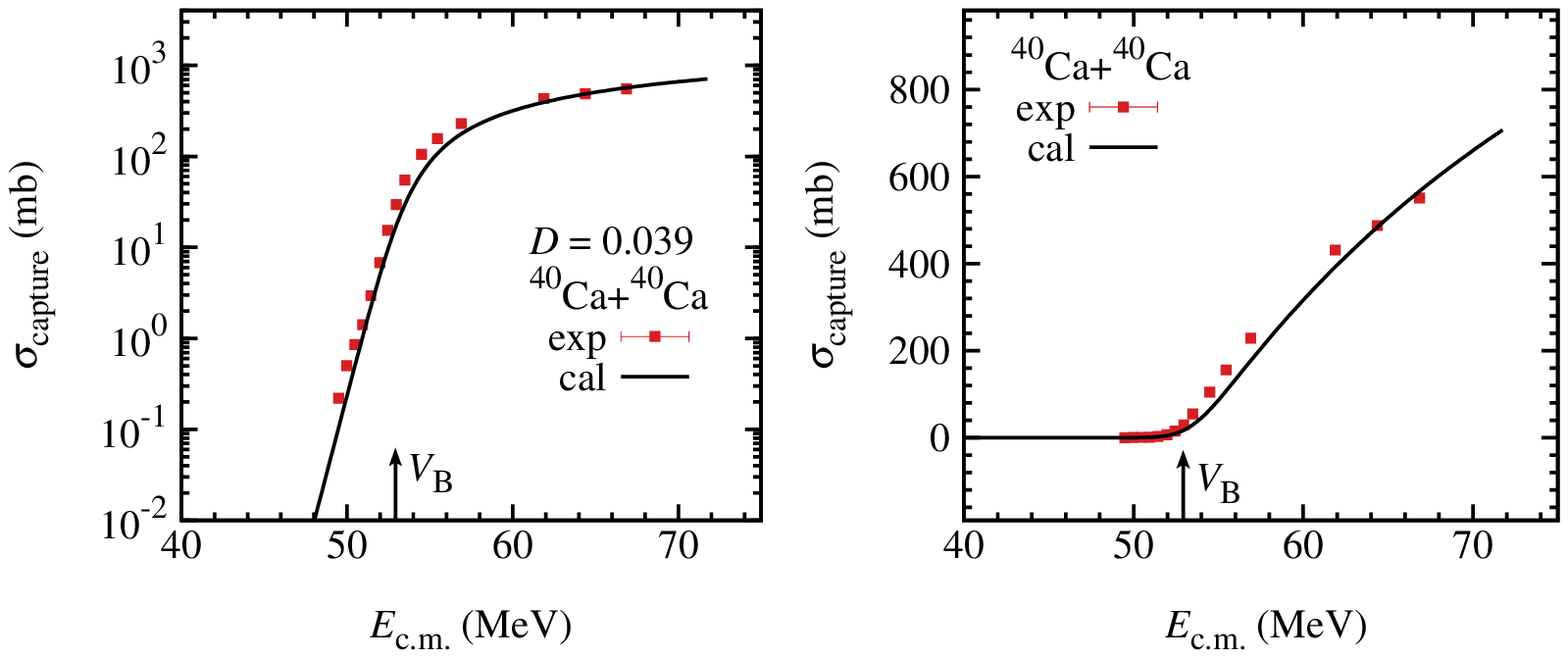}}
 \centerline{\includegraphics[width=0.47\textwidth]{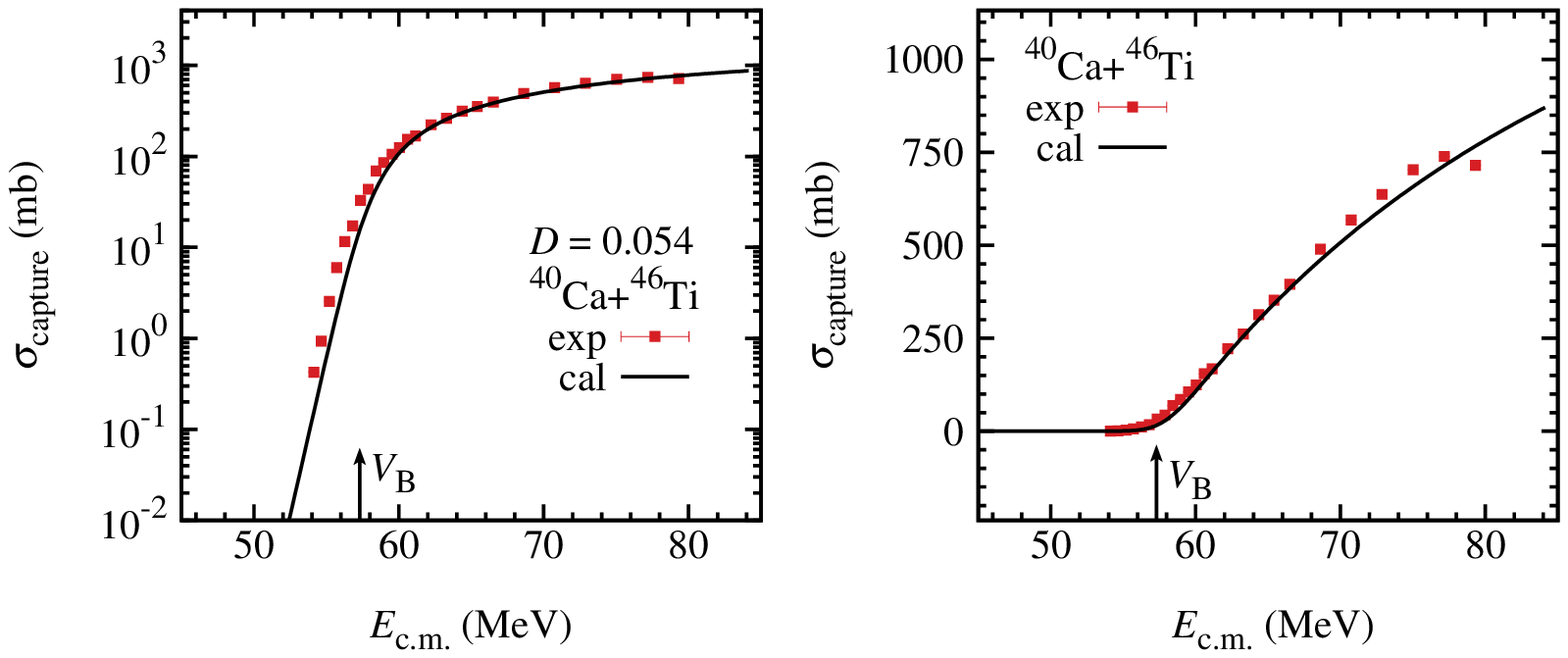}
  \includegraphics[width=0.47\textwidth]{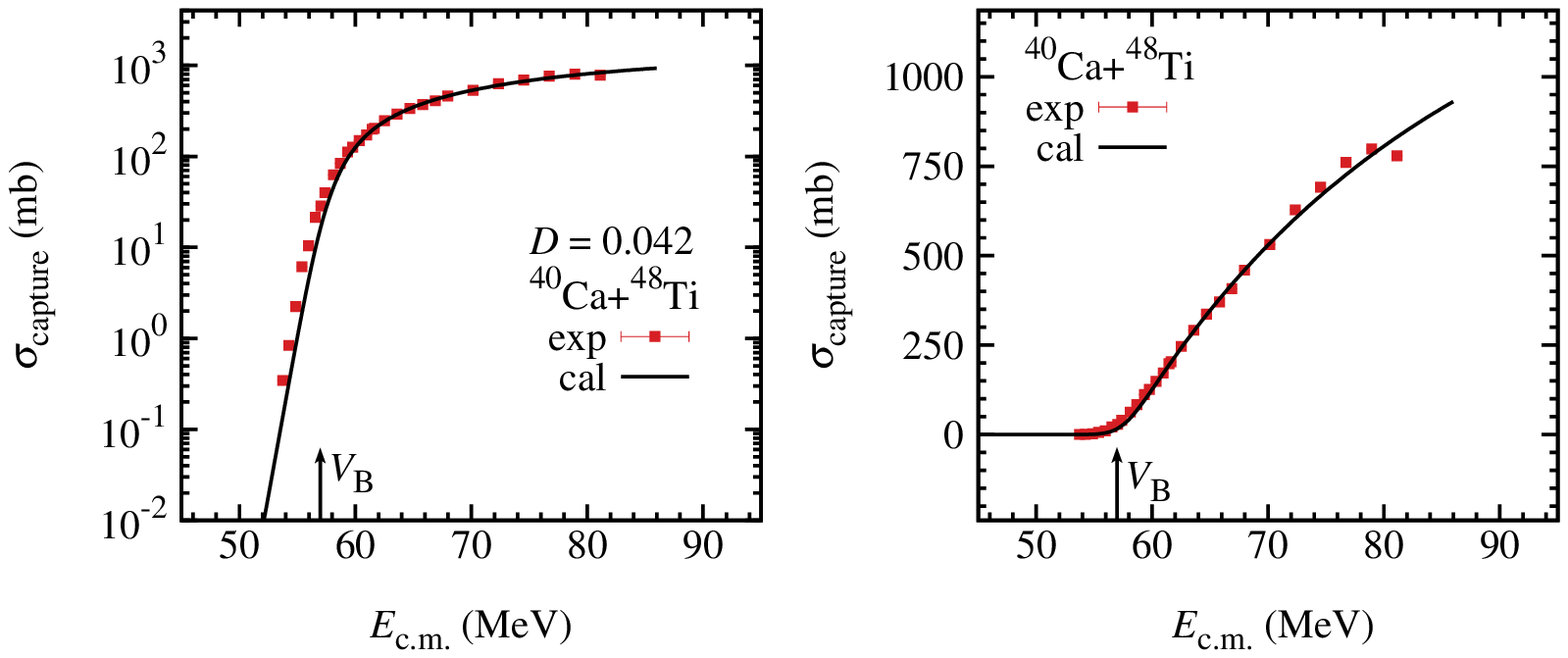}}
 \centerline{\includegraphics[width=0.47\textwidth]{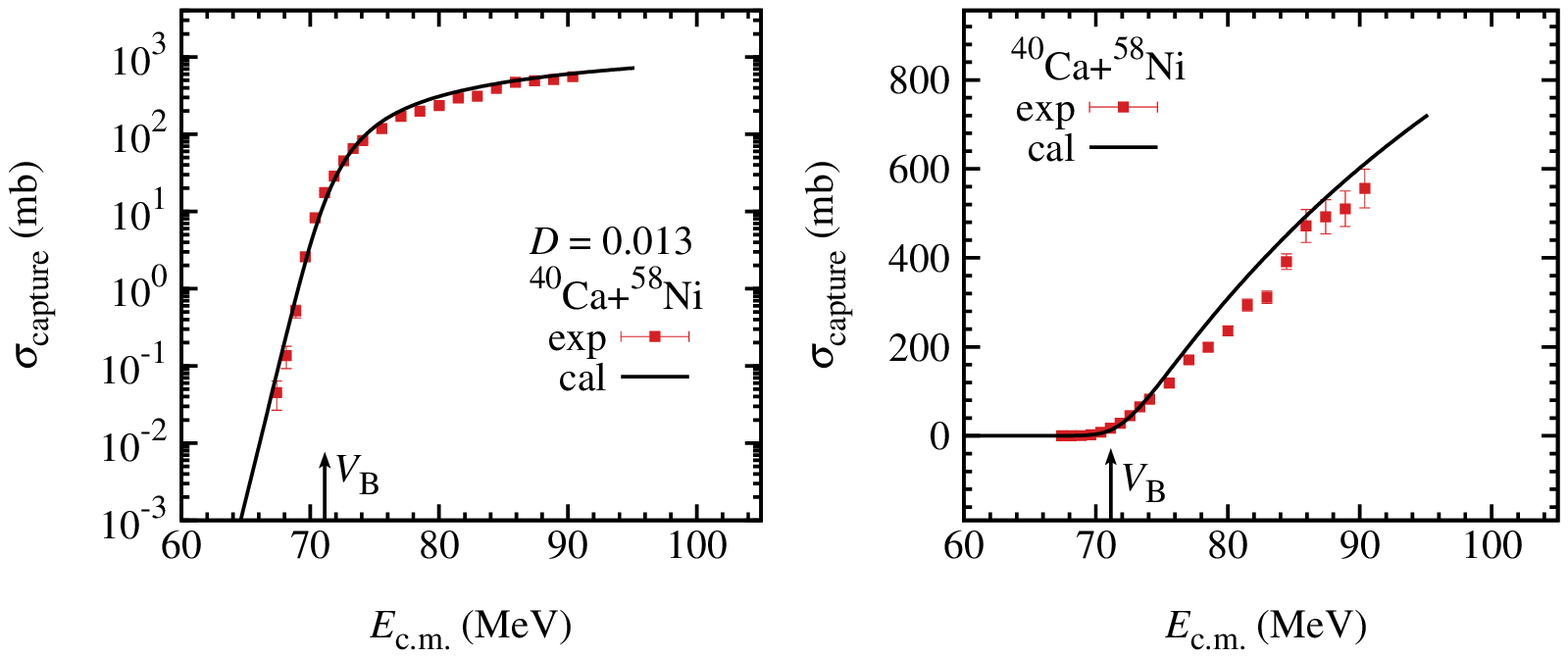}
  \includegraphics[width=0.47\textwidth]{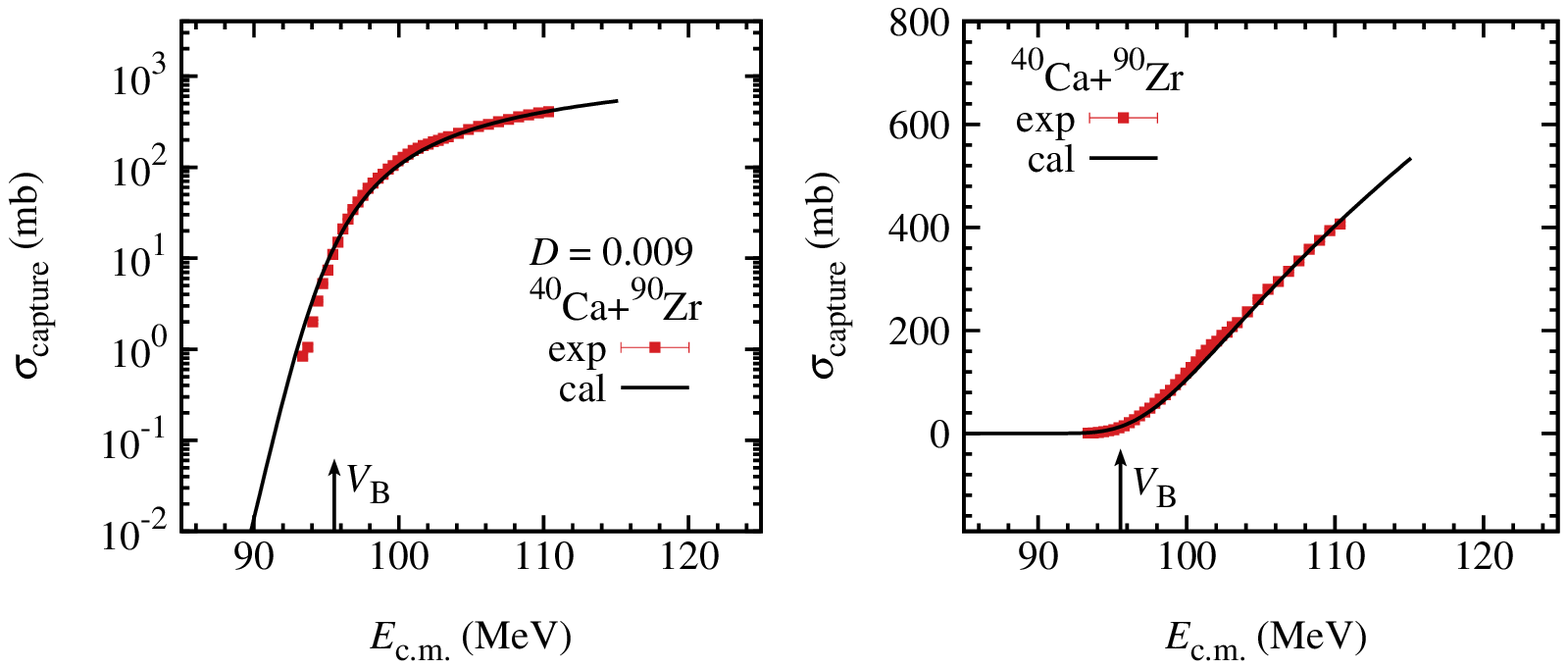}}
 \centerline{\includegraphics[width=0.47\textwidth]{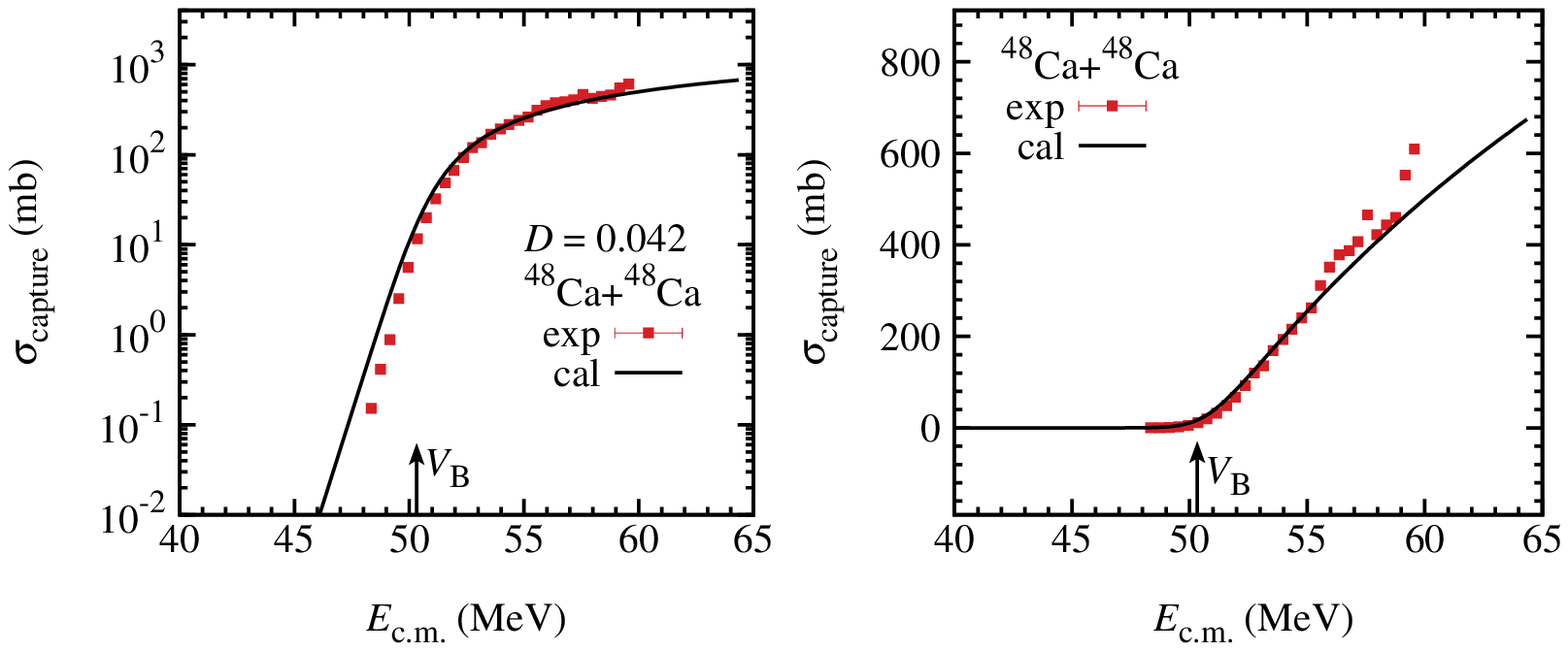}
  \includegraphics[width=0.47\textwidth]{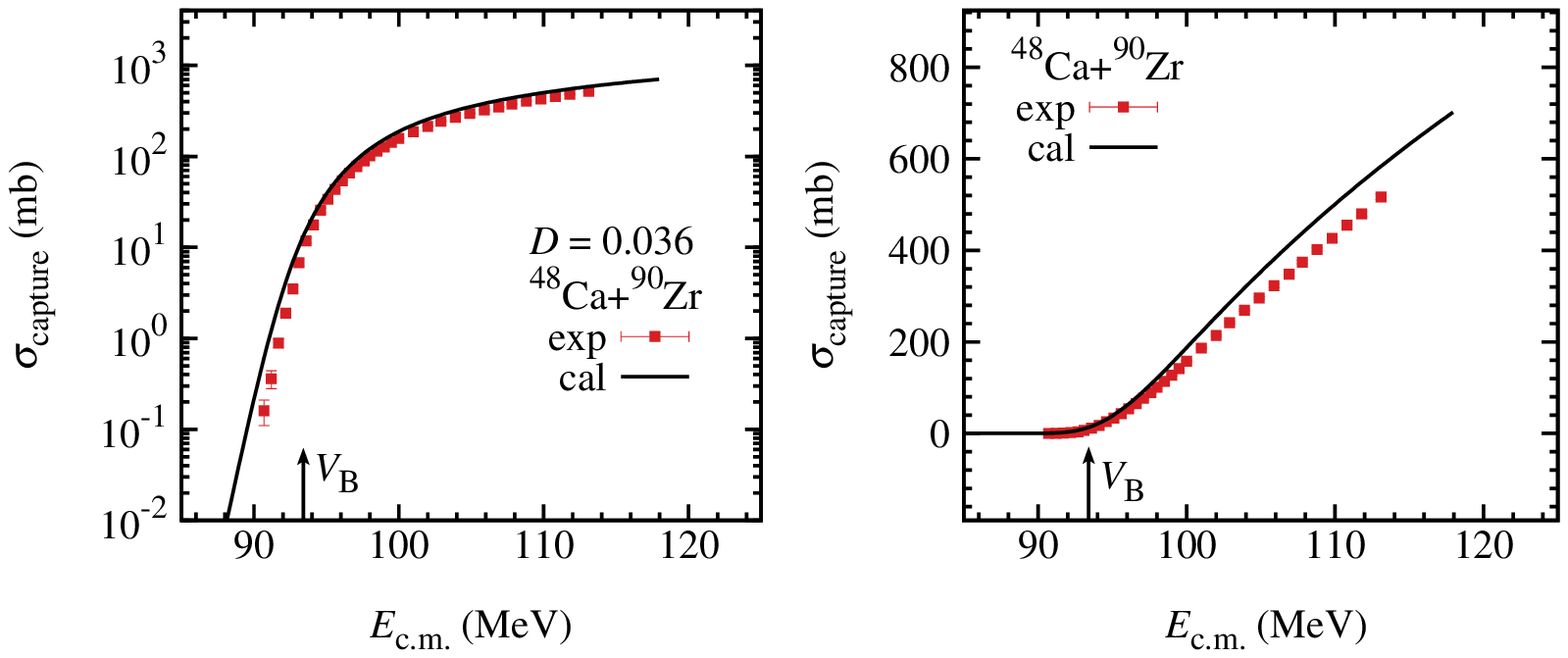}}
 \centerline{\includegraphics[width=0.47\textwidth]{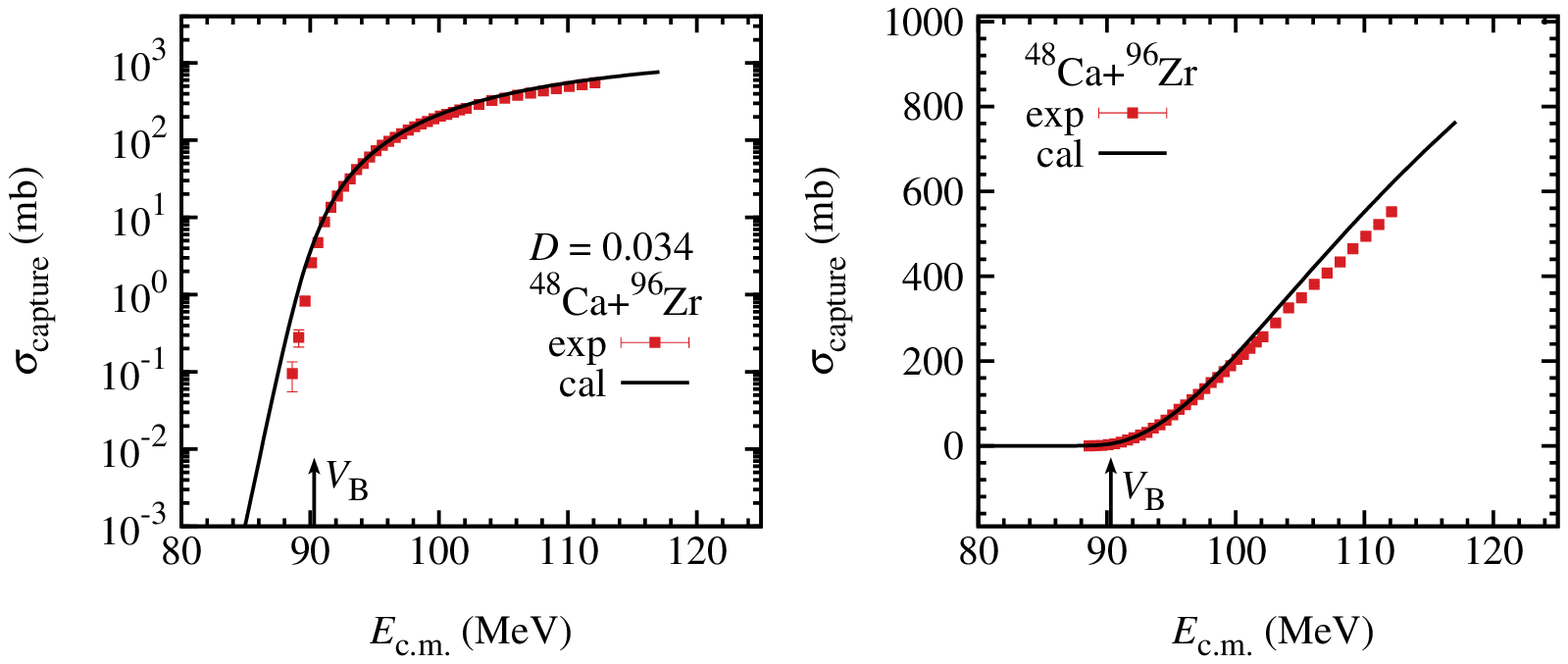}
  \includegraphics[width=0.47\textwidth]{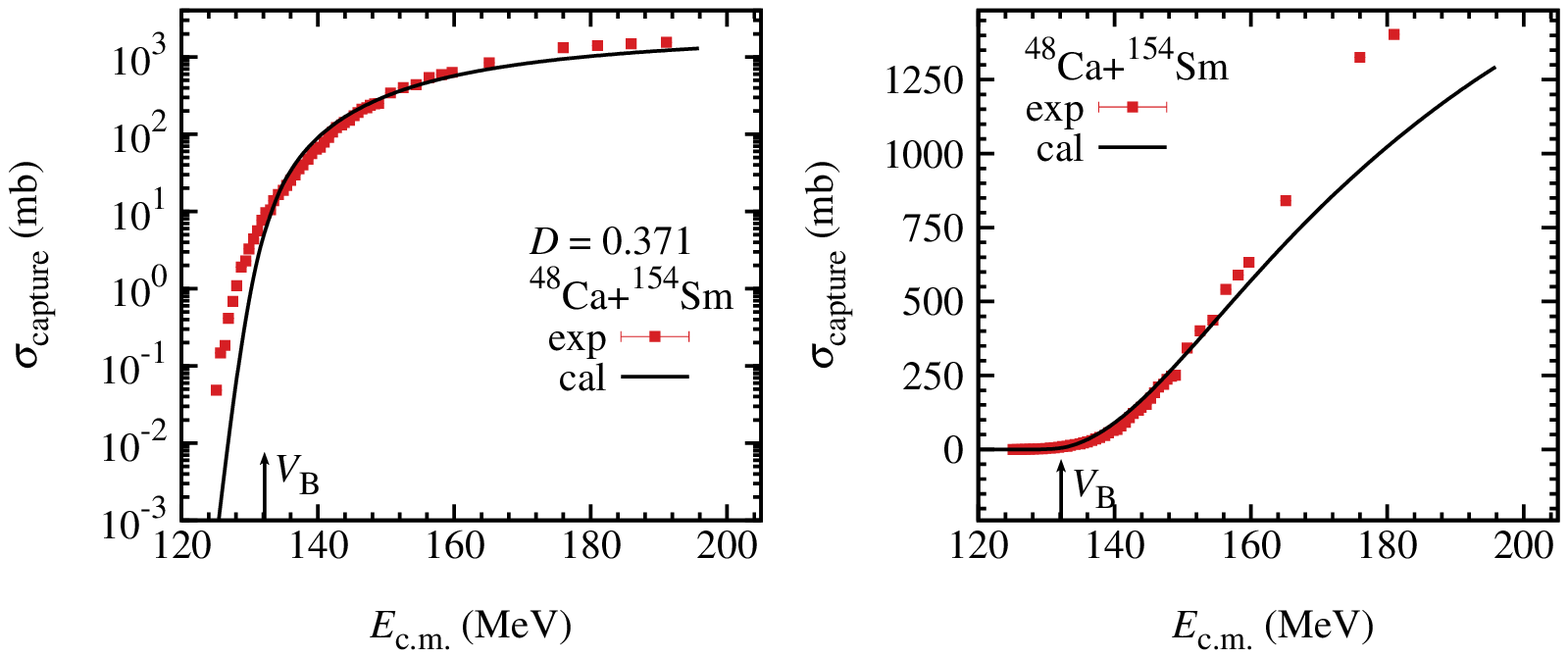}}
 \centerline{\includegraphics[width=0.47\textwidth]{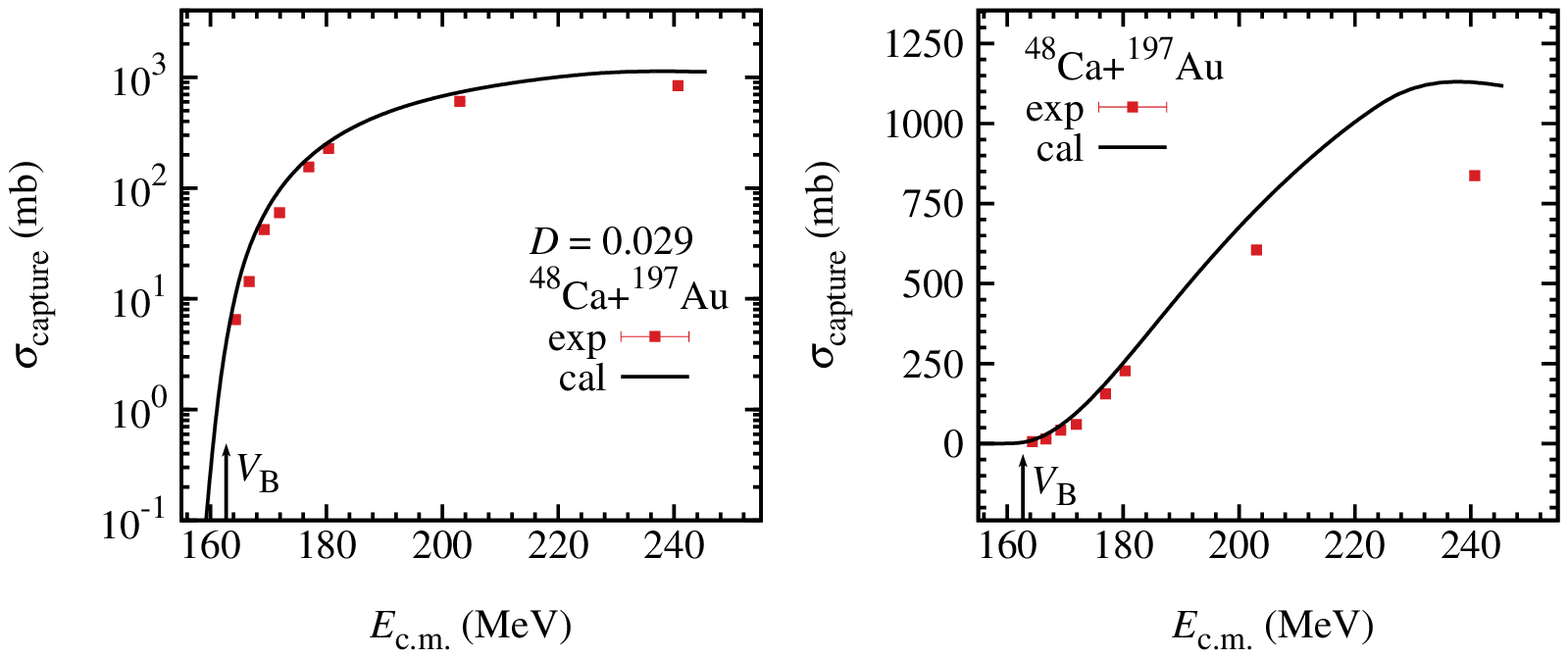}
  \includegraphics[width=0.47\textwidth]{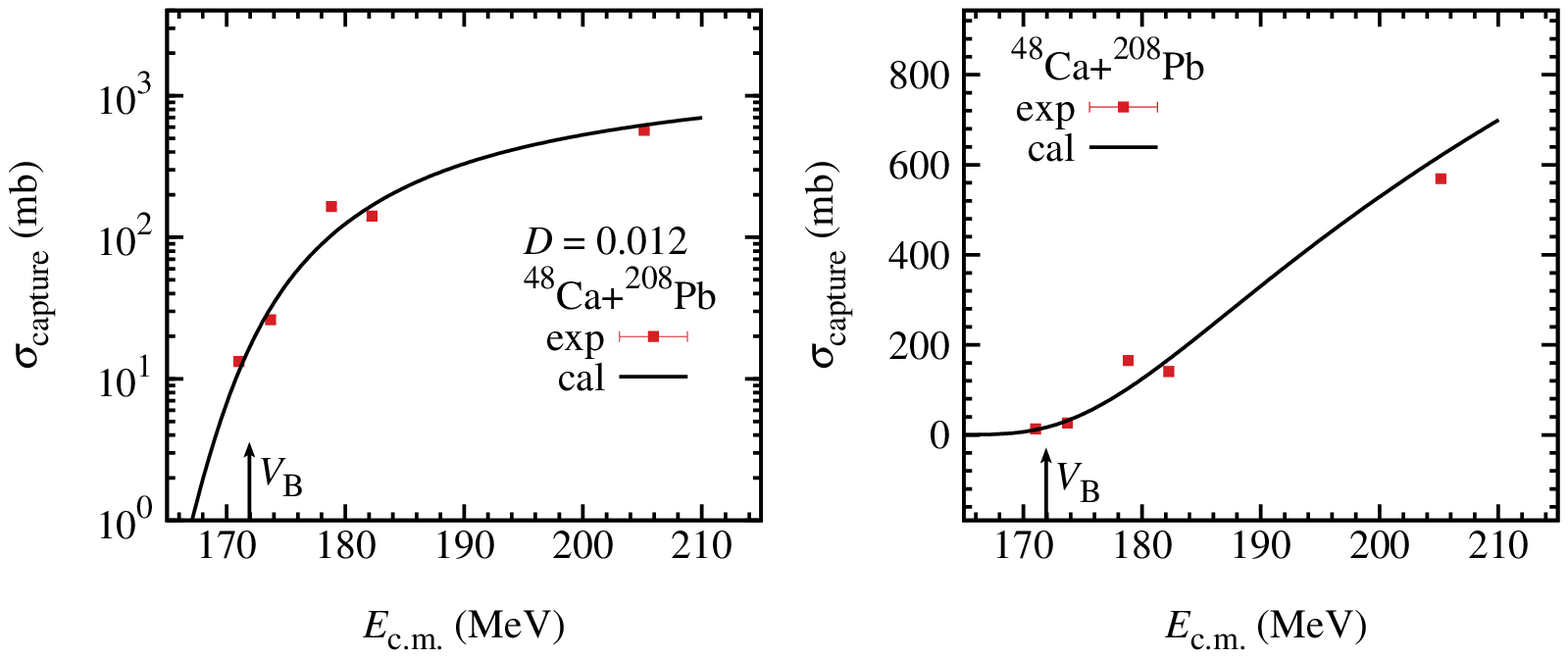}}
  \centerline {Graph 10}
 \end{Dfigures}
 \begin{Dfigures}[!ht]
 \centerline{\includegraphics[width=0.47\textwidth]{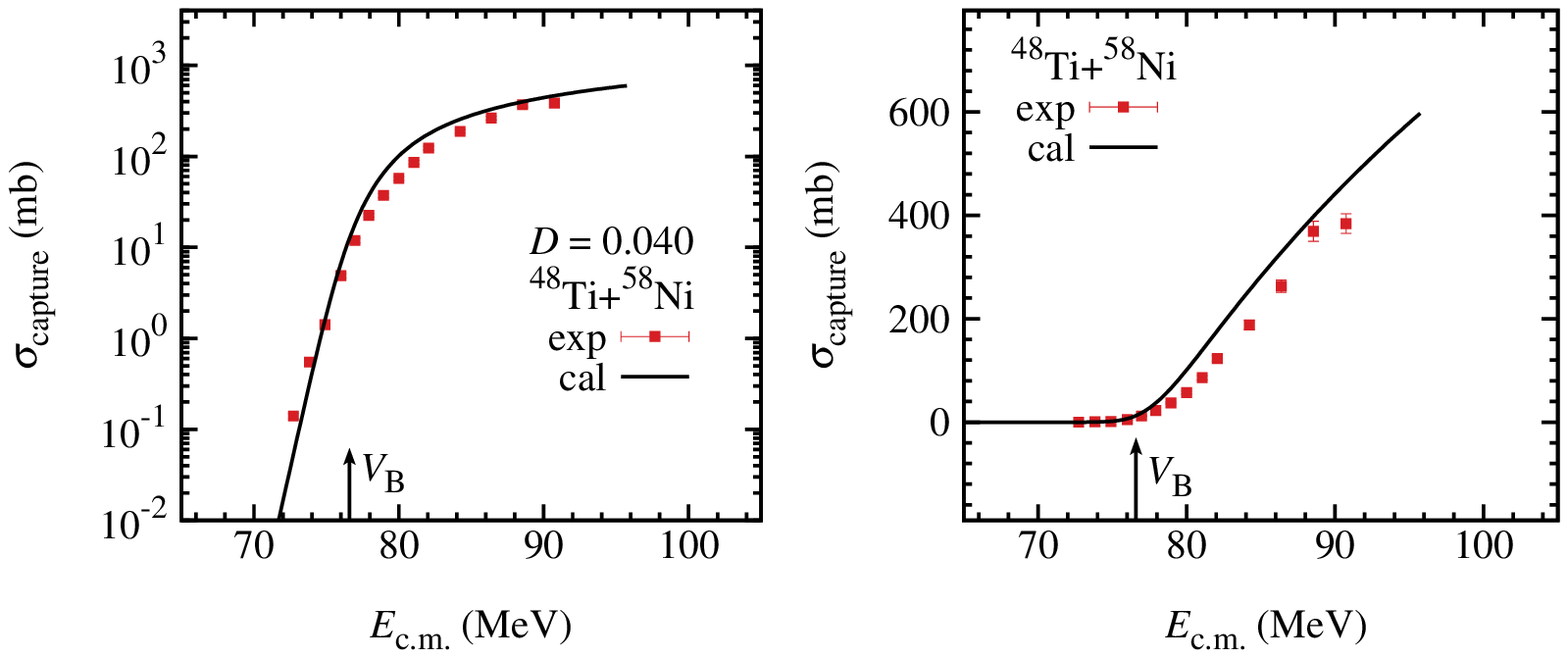}
  \includegraphics[width=0.47\textwidth]{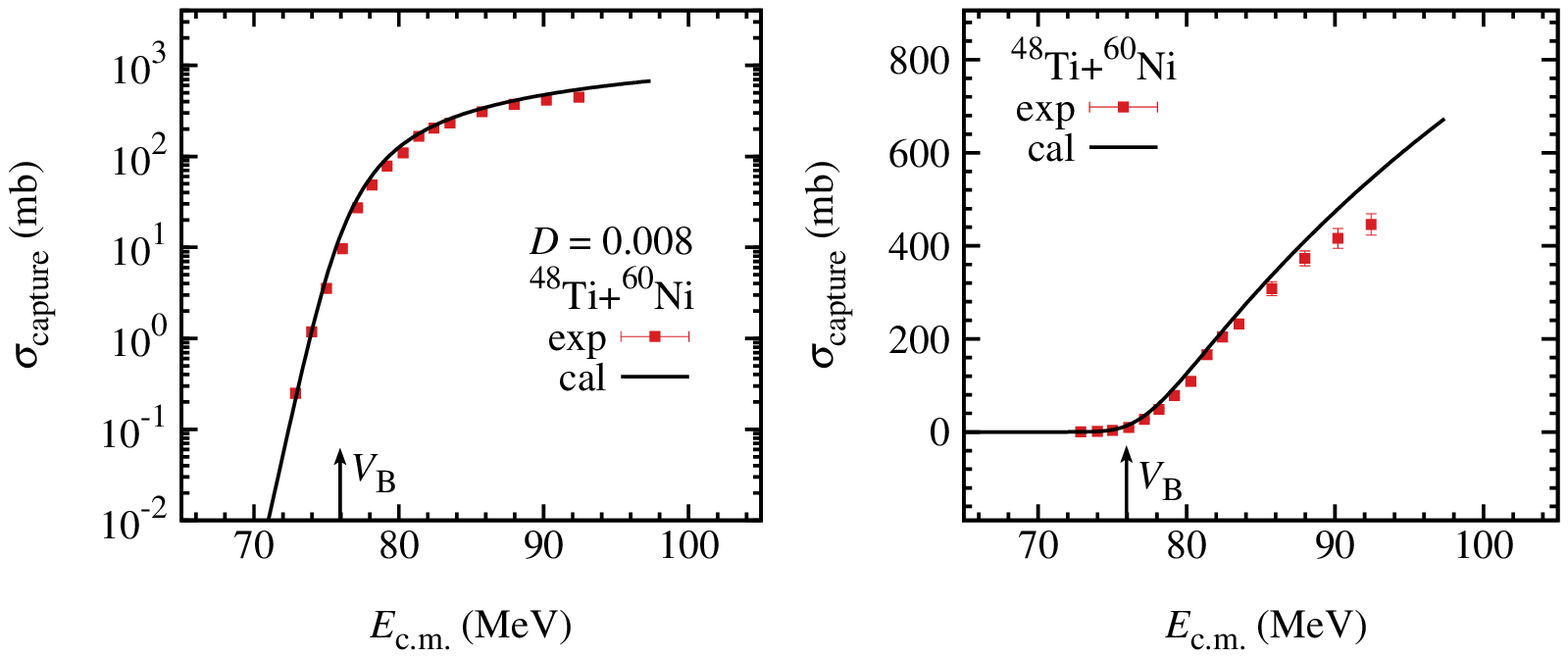}}
 \centerline{\includegraphics[width=0.47\textwidth]{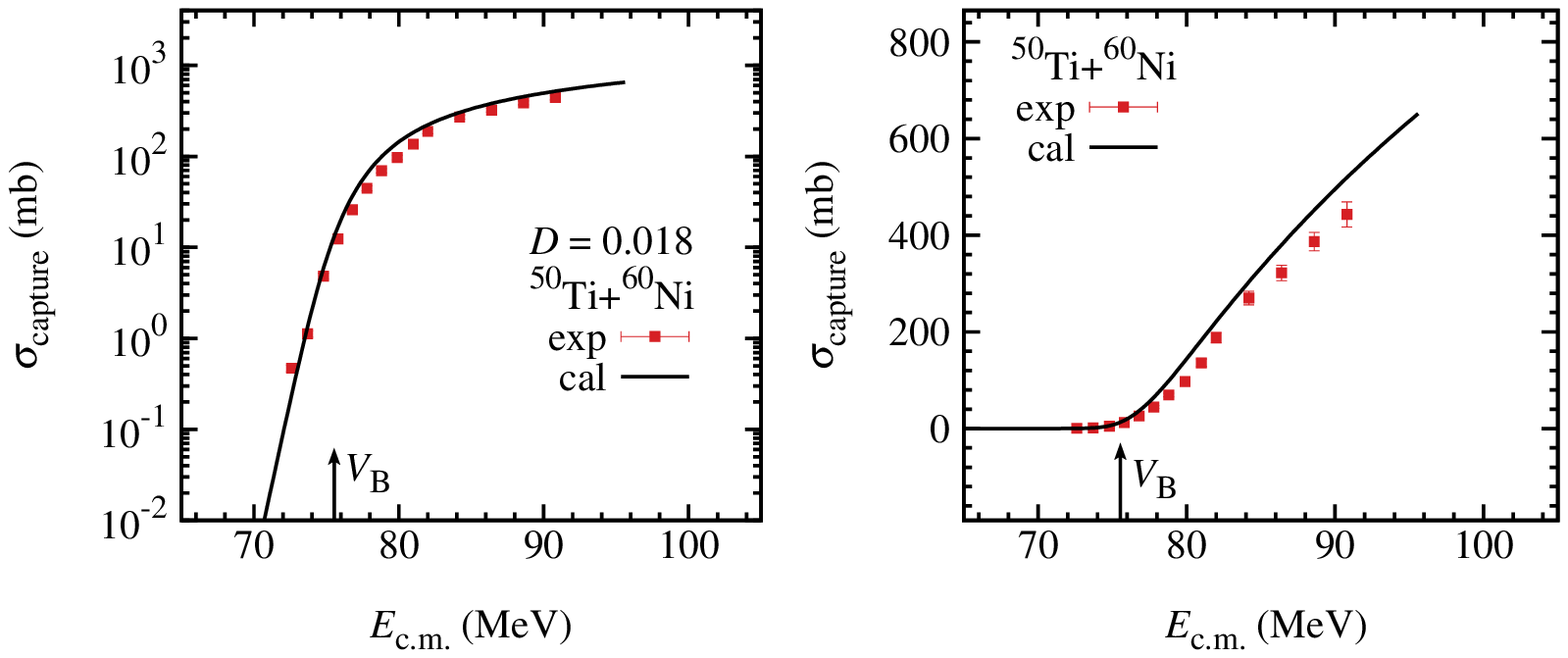}
  \includegraphics[width=0.47\textwidth]{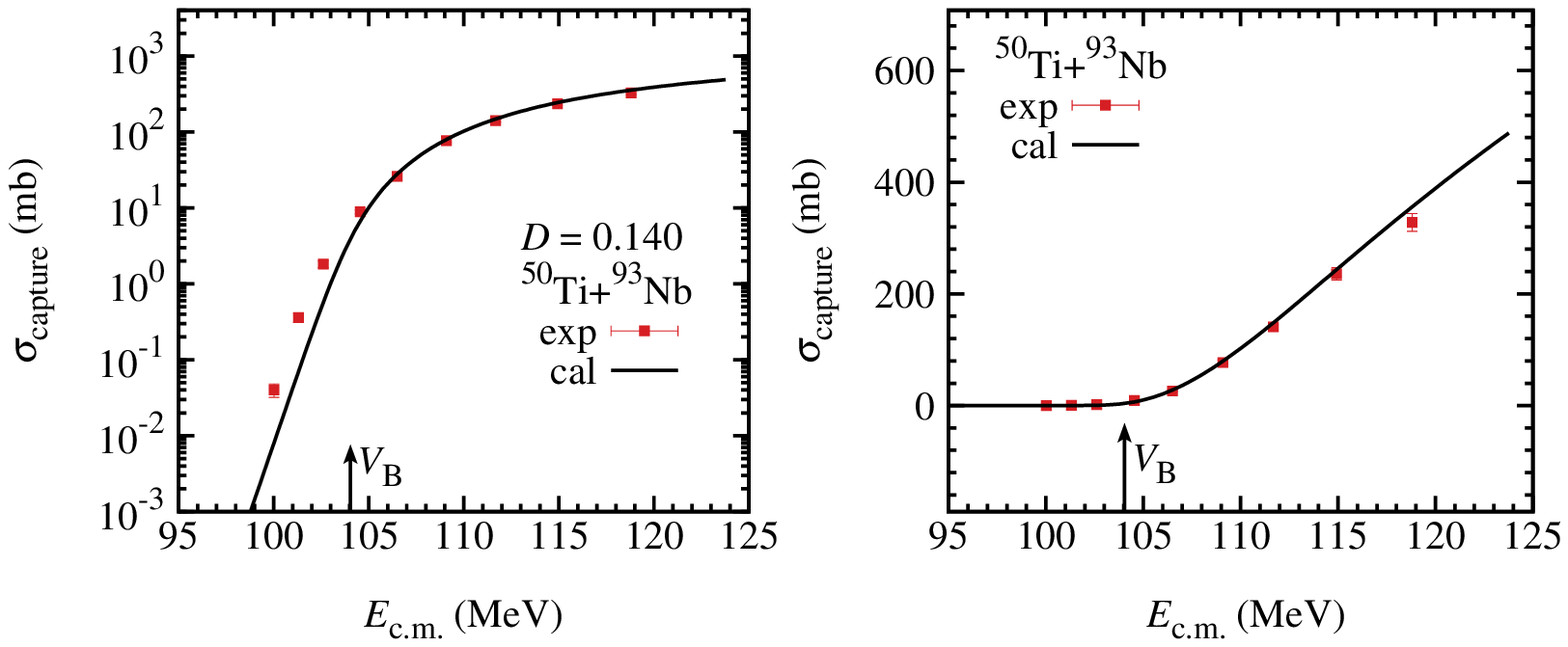}}
 \centerline{\includegraphics[width=0.47\textwidth]{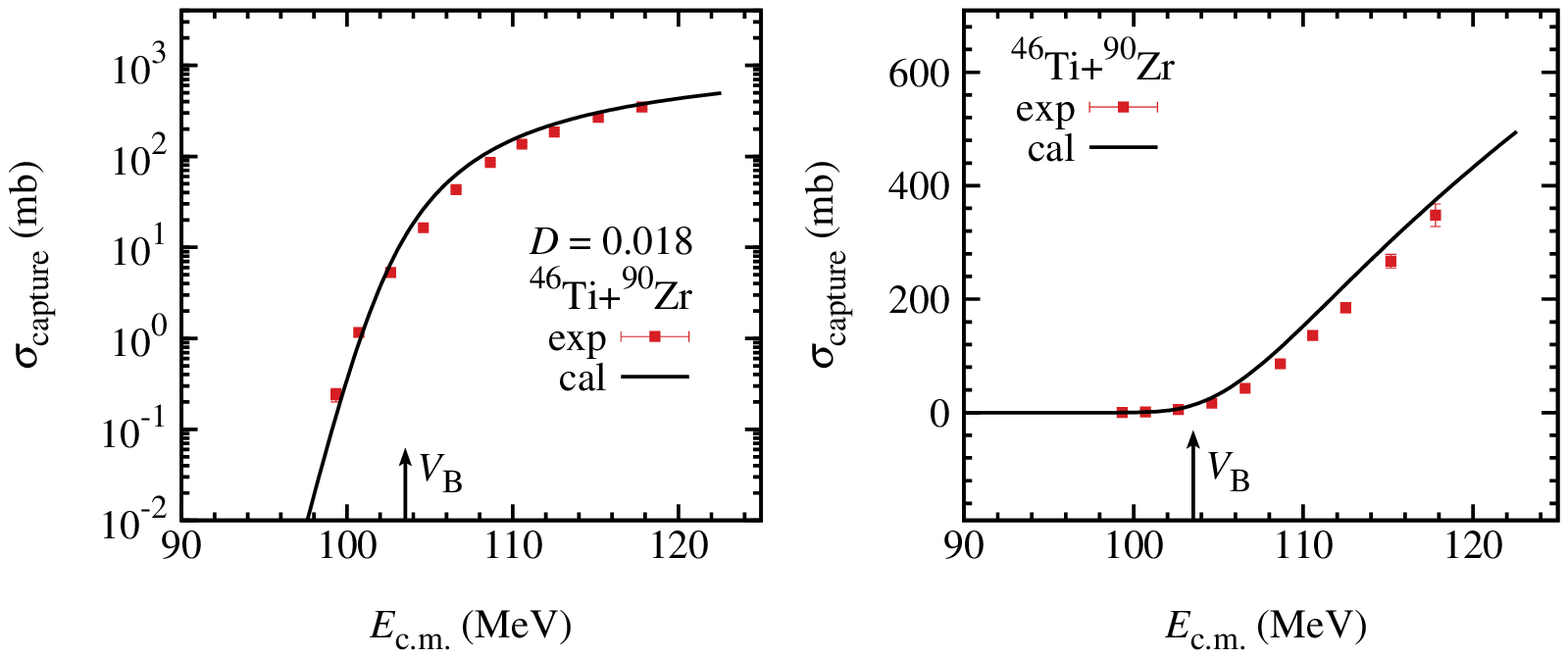}
  \includegraphics[width=0.47\textwidth]{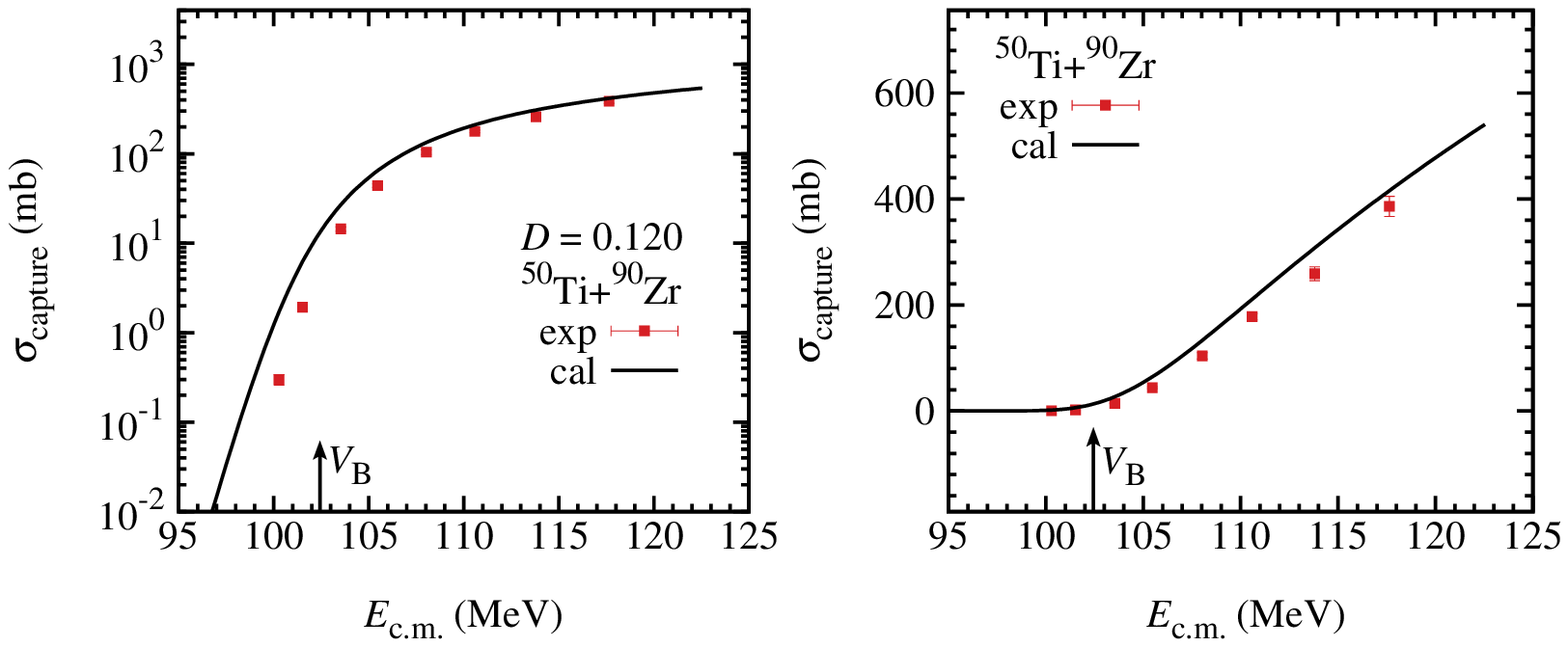}}
 \centerline{\includegraphics[width=0.47\textwidth]{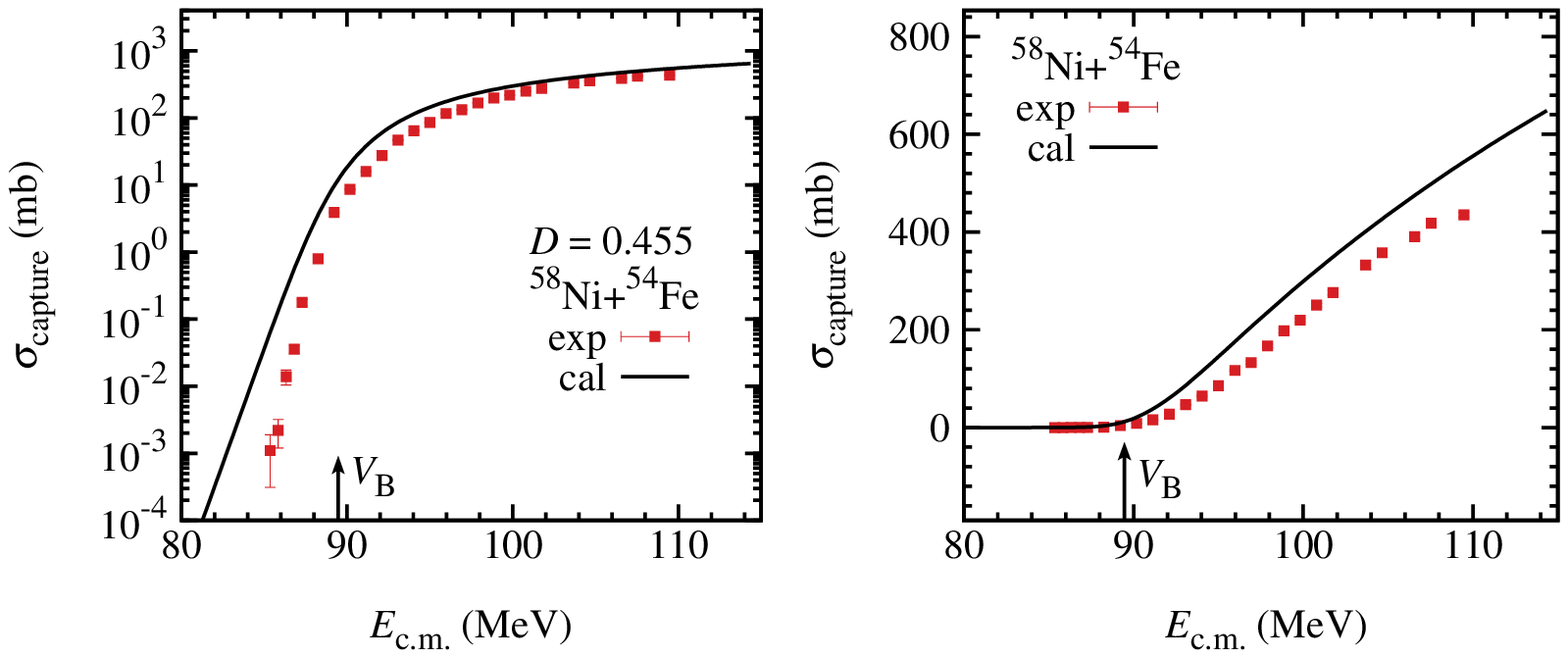}
  \includegraphics[width=0.47\textwidth]{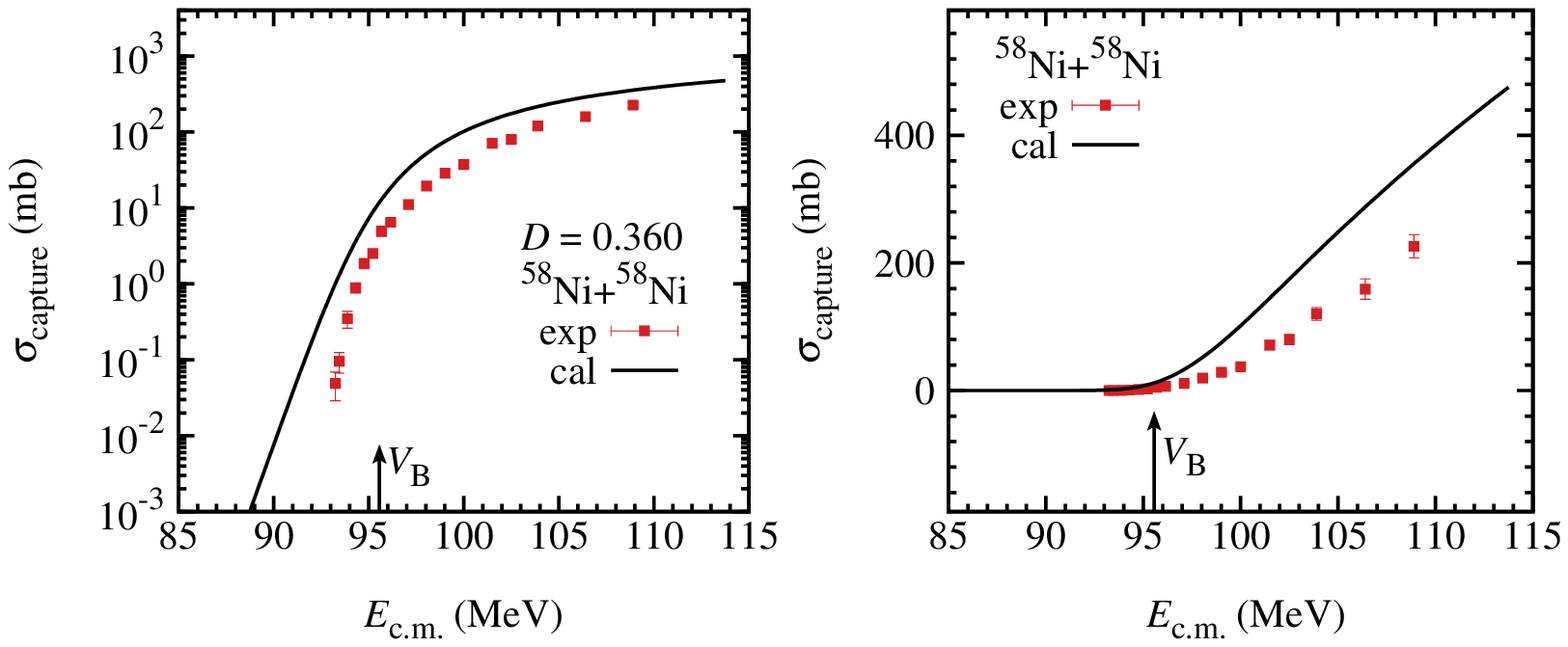}}
 \centerline{\includegraphics[width=0.47\textwidth]{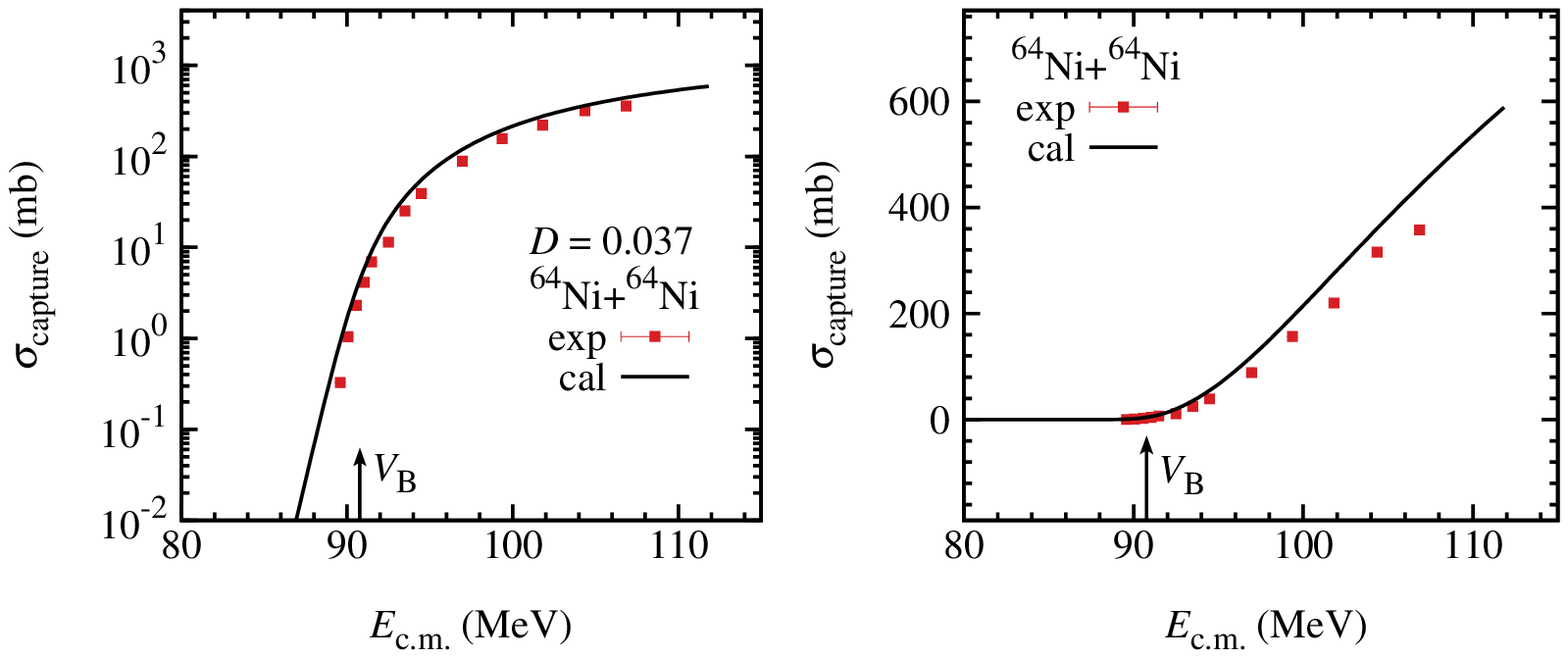}
  \includegraphics[width=0.47\textwidth]{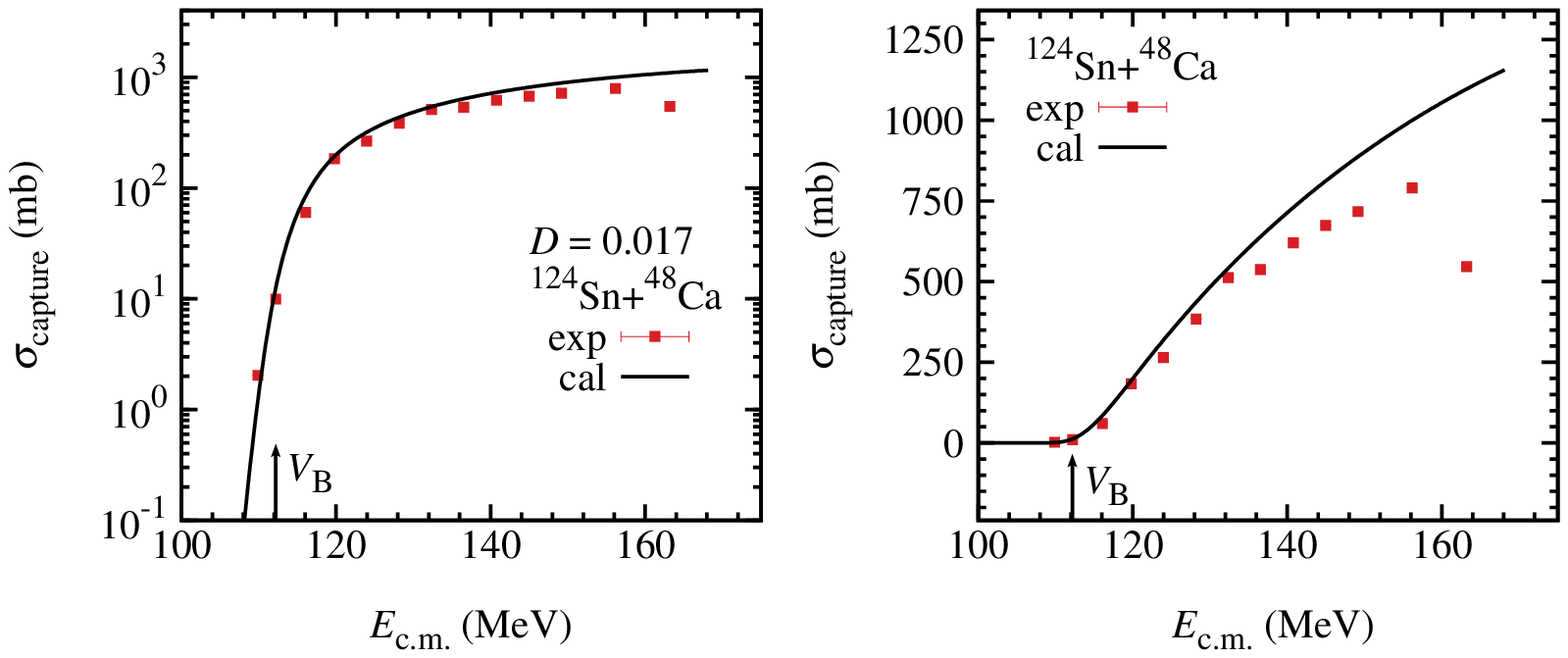}}
 \centerline{\includegraphics[width=0.47\textwidth]{132Sn48Ca.eps}}
   \centerline {Graph 11}
 \end{Dfigures}

\clearpage
  \begin{Dfigures}[!ht]
 \centerline{\includegraphics[width=0.47\textwidth]{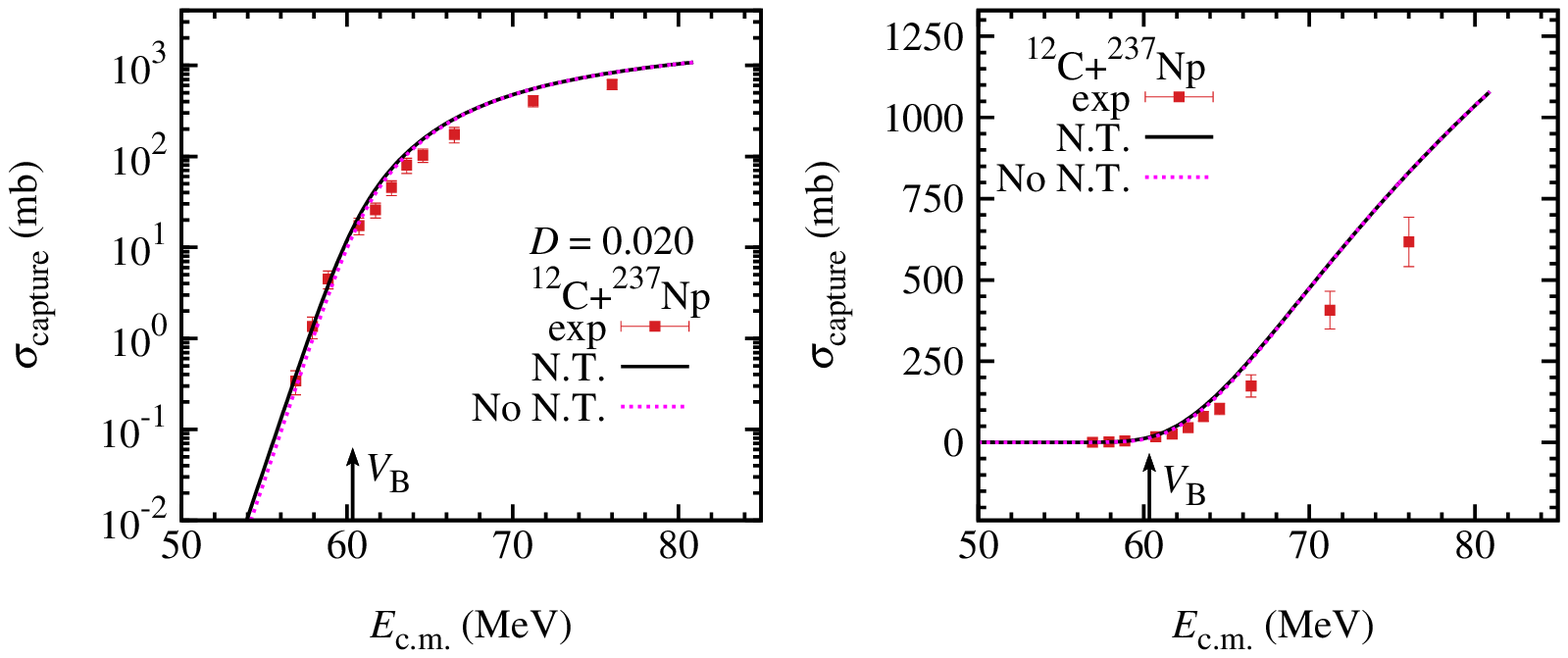}
  \includegraphics[width=0.47\textwidth]{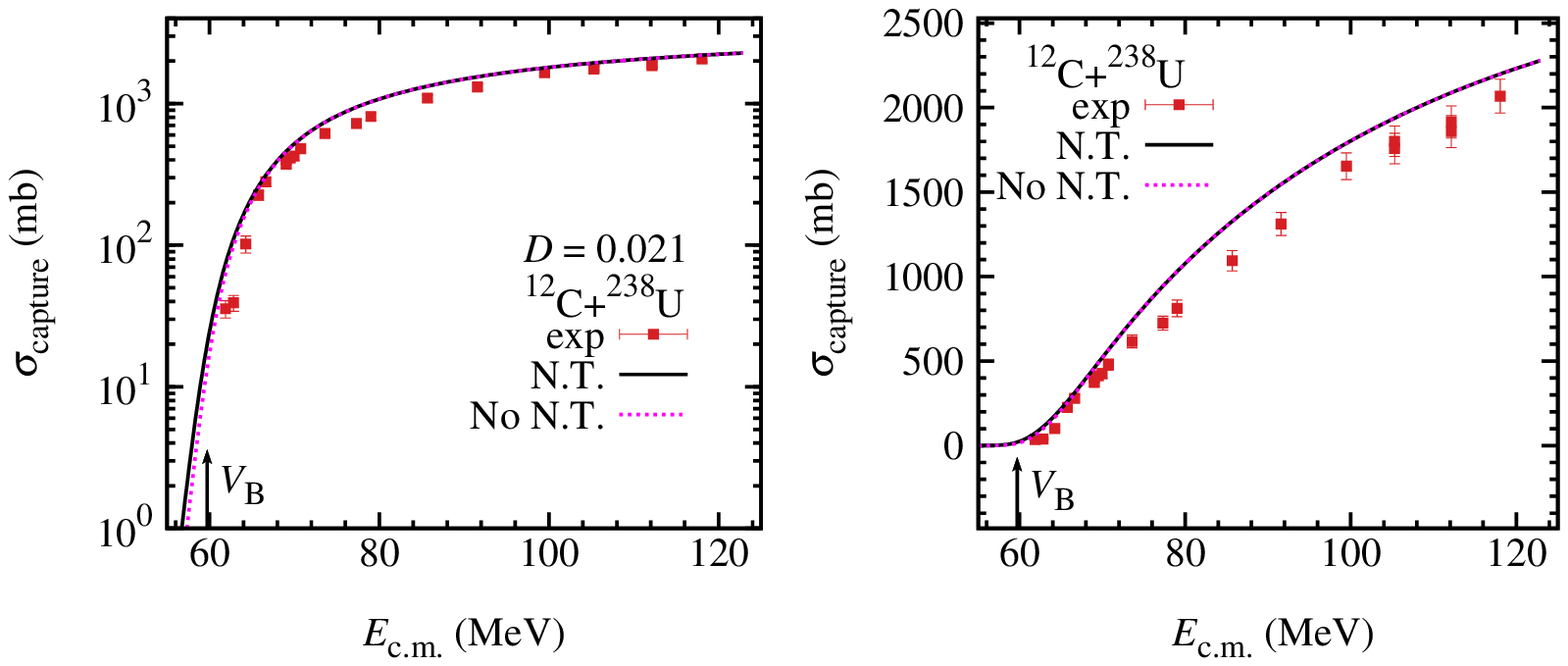}}
 \centerline{\includegraphics[width=0.47\textwidth]{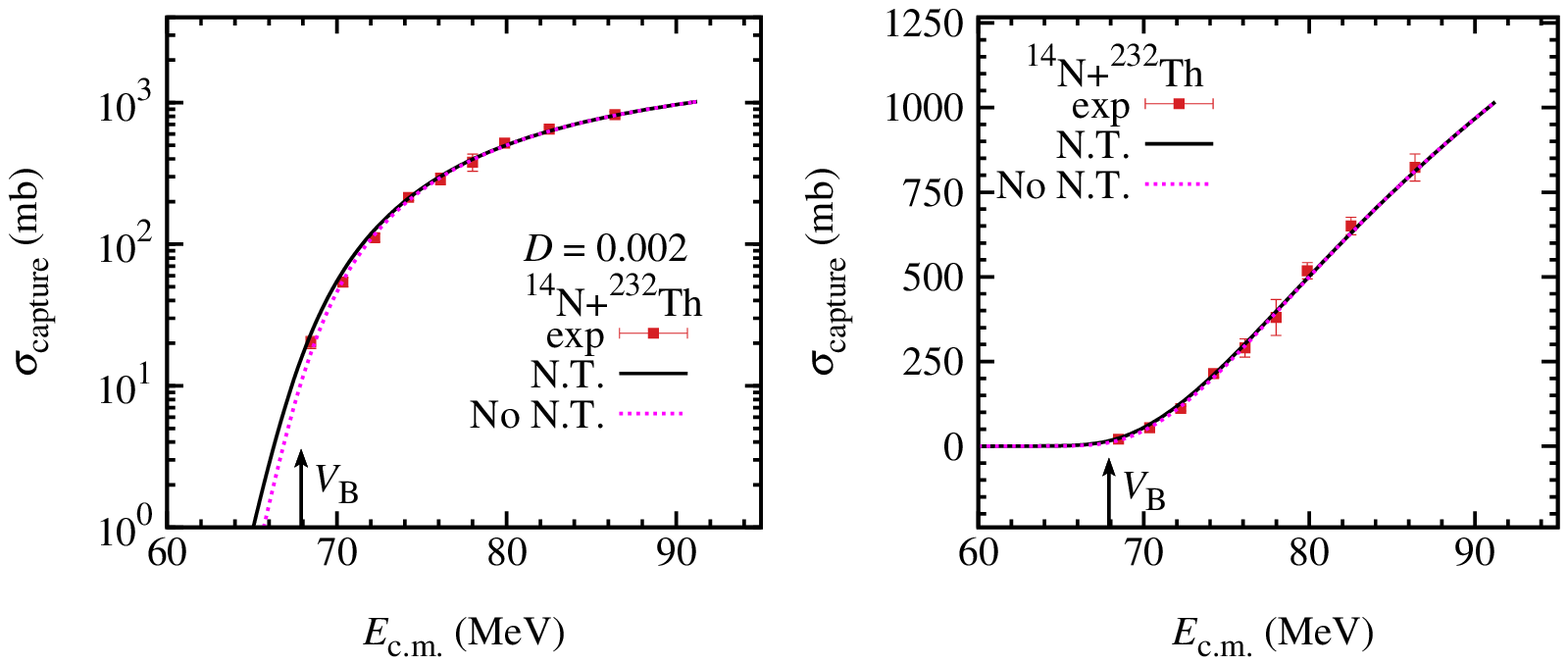}
  \includegraphics[width=0.47\textwidth]{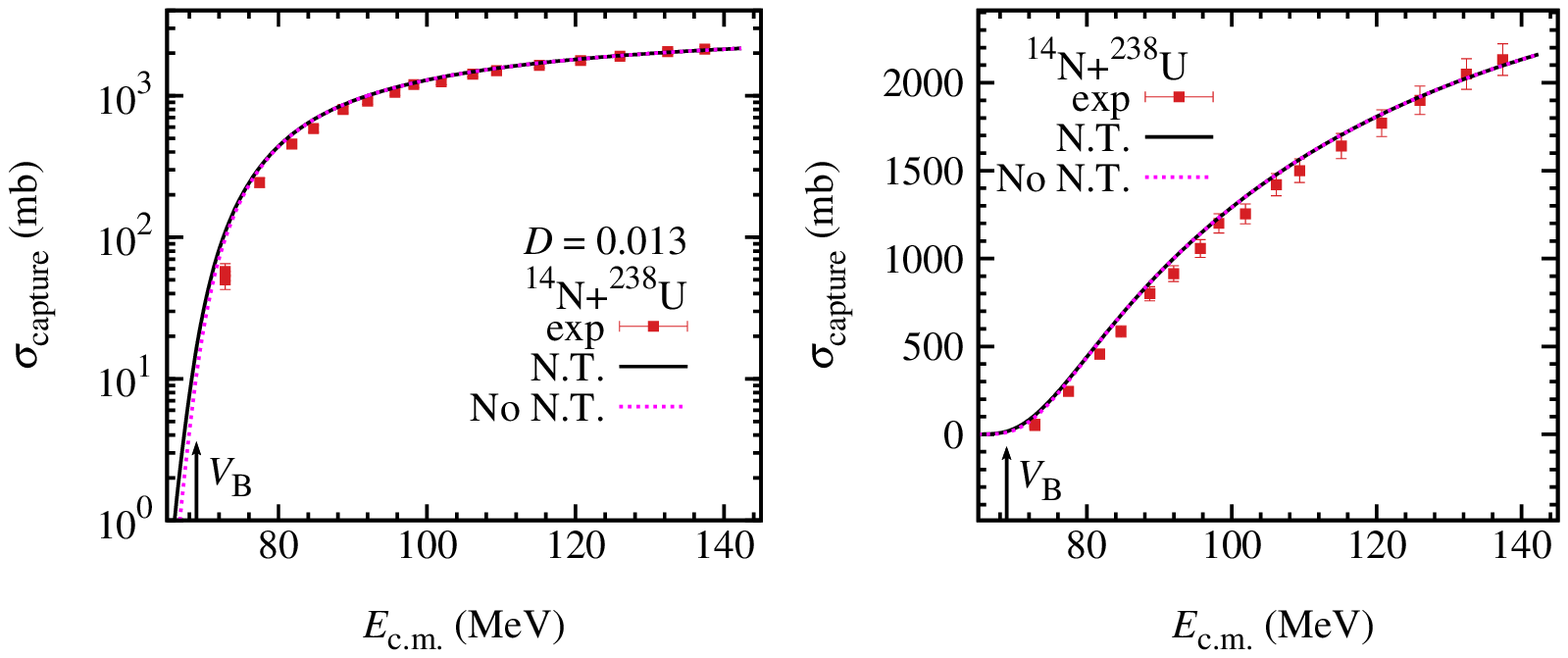}}
 \centerline{\includegraphics[width=0.47\textwidth]{16O232Th.eps}
  \includegraphics[width=0.47\textwidth]{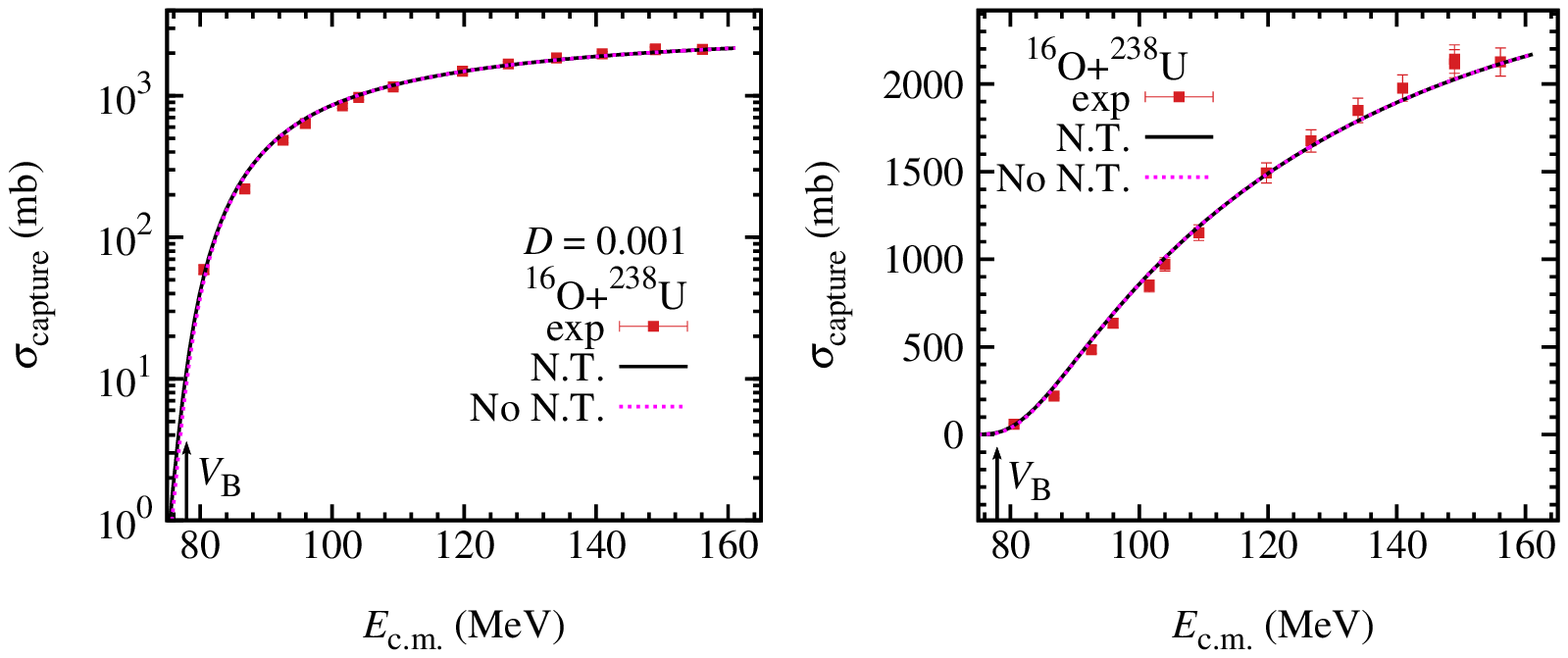}}
 \centerline{\includegraphics[width=0.47\textwidth]{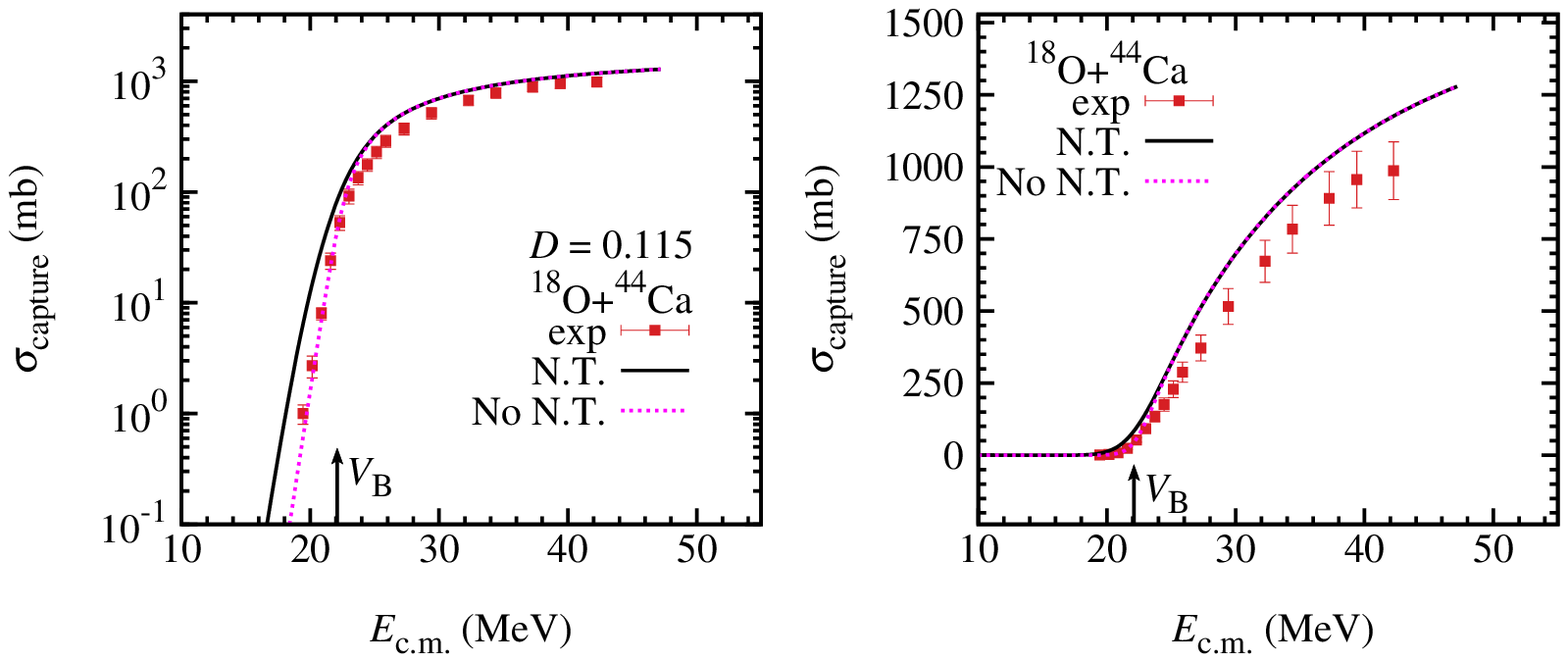}
  \includegraphics[width=0.47\textwidth]{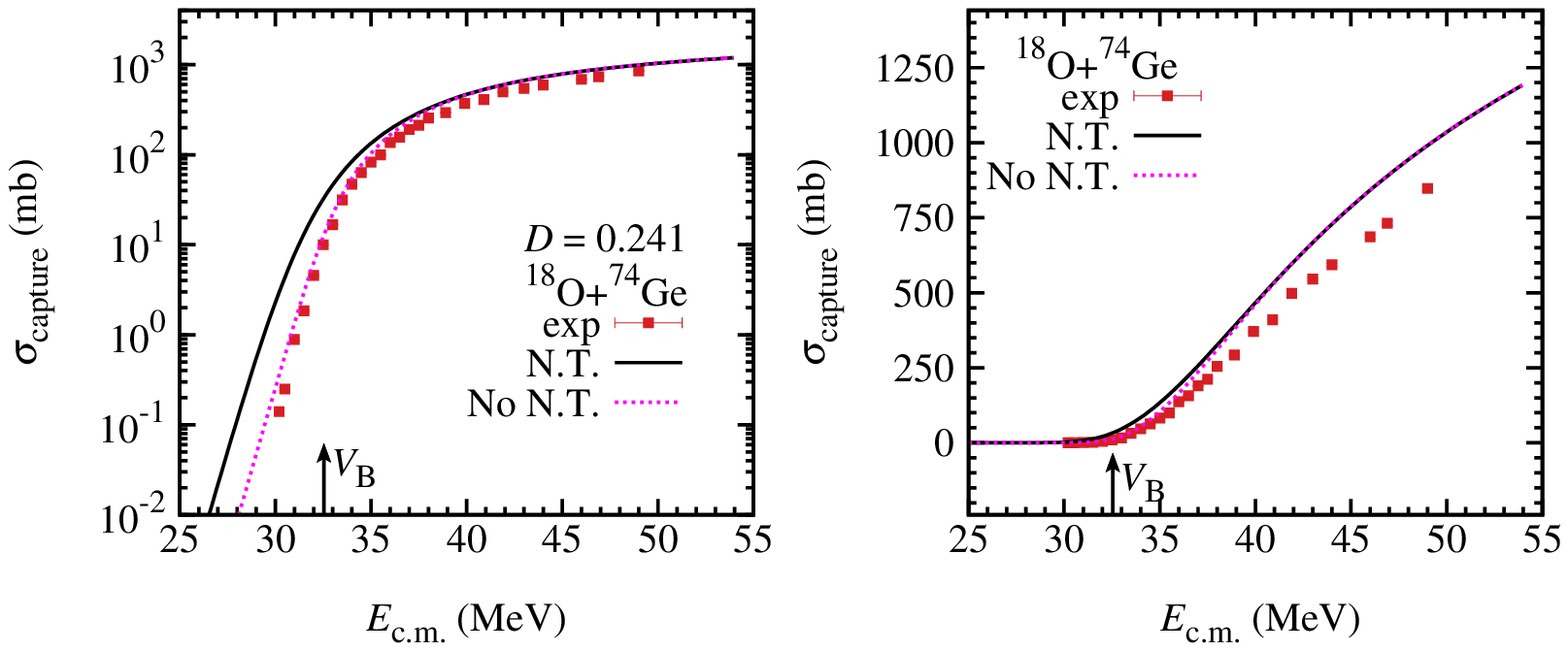}}
 \centerline{\includegraphics[width=0.47\textwidth]{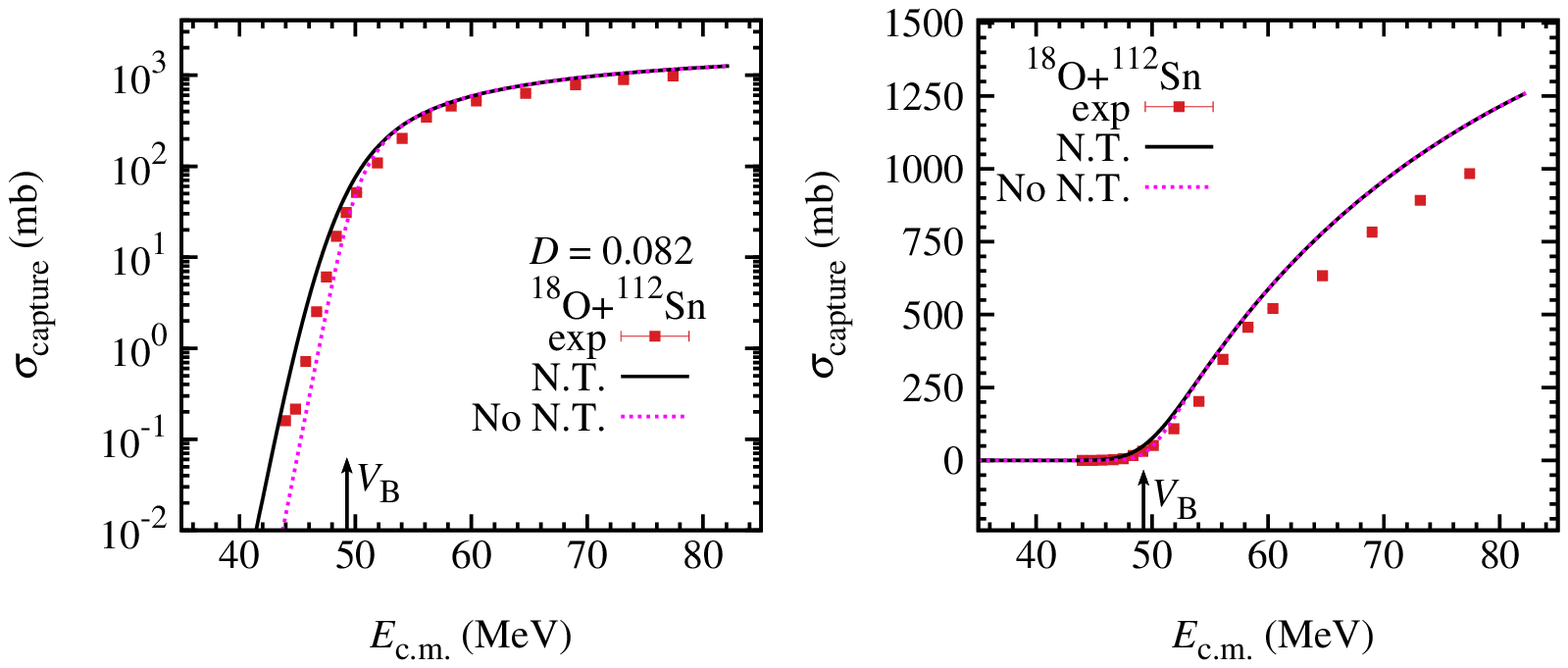}
  \includegraphics[width=0.47\textwidth]{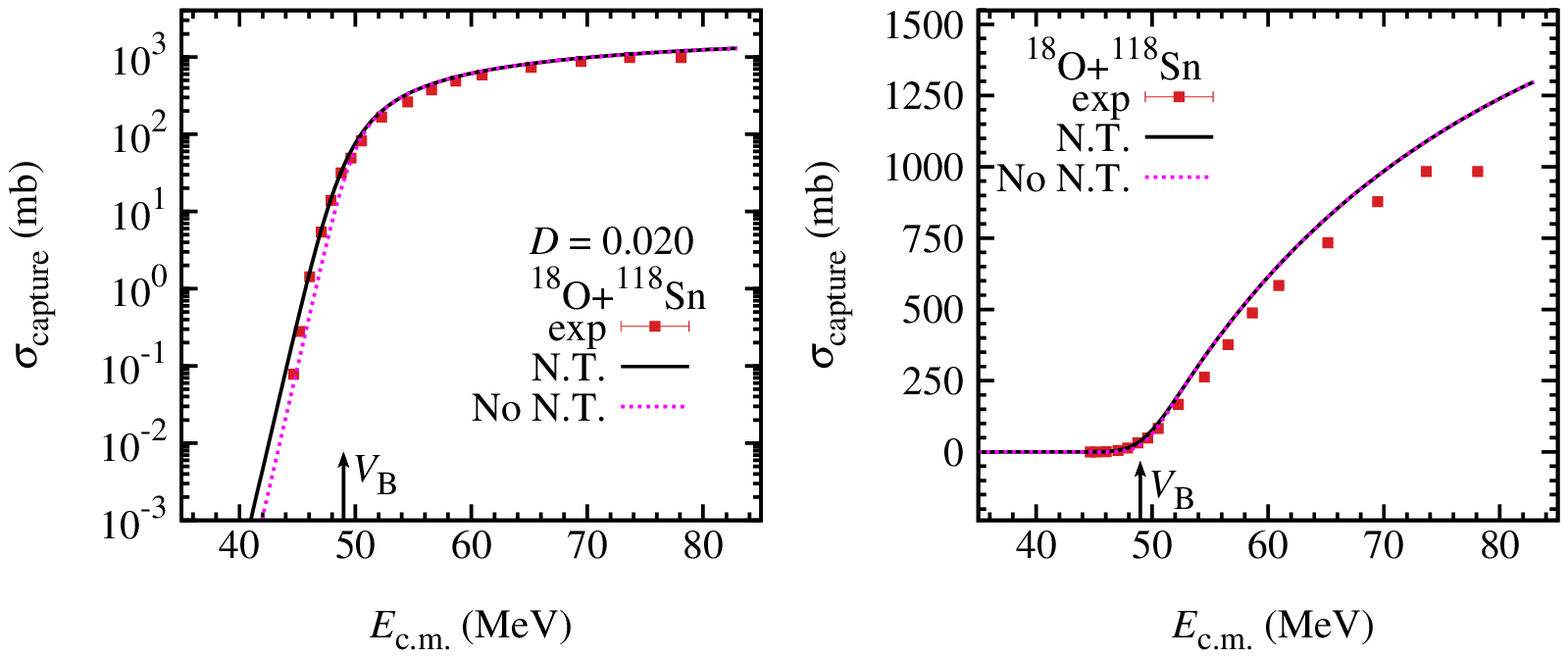}}
 \centerline{\includegraphics[width=0.47\textwidth]{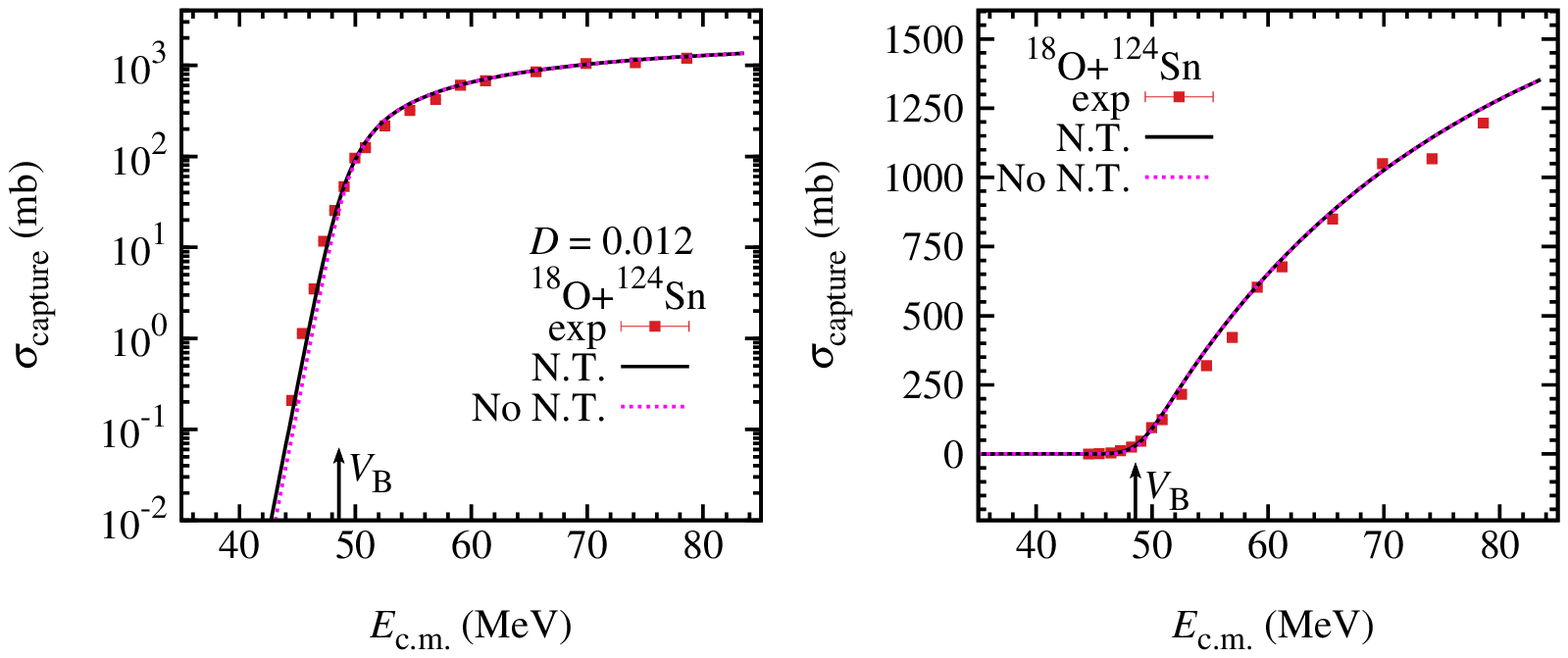}
  \includegraphics[width=0.47\textwidth]{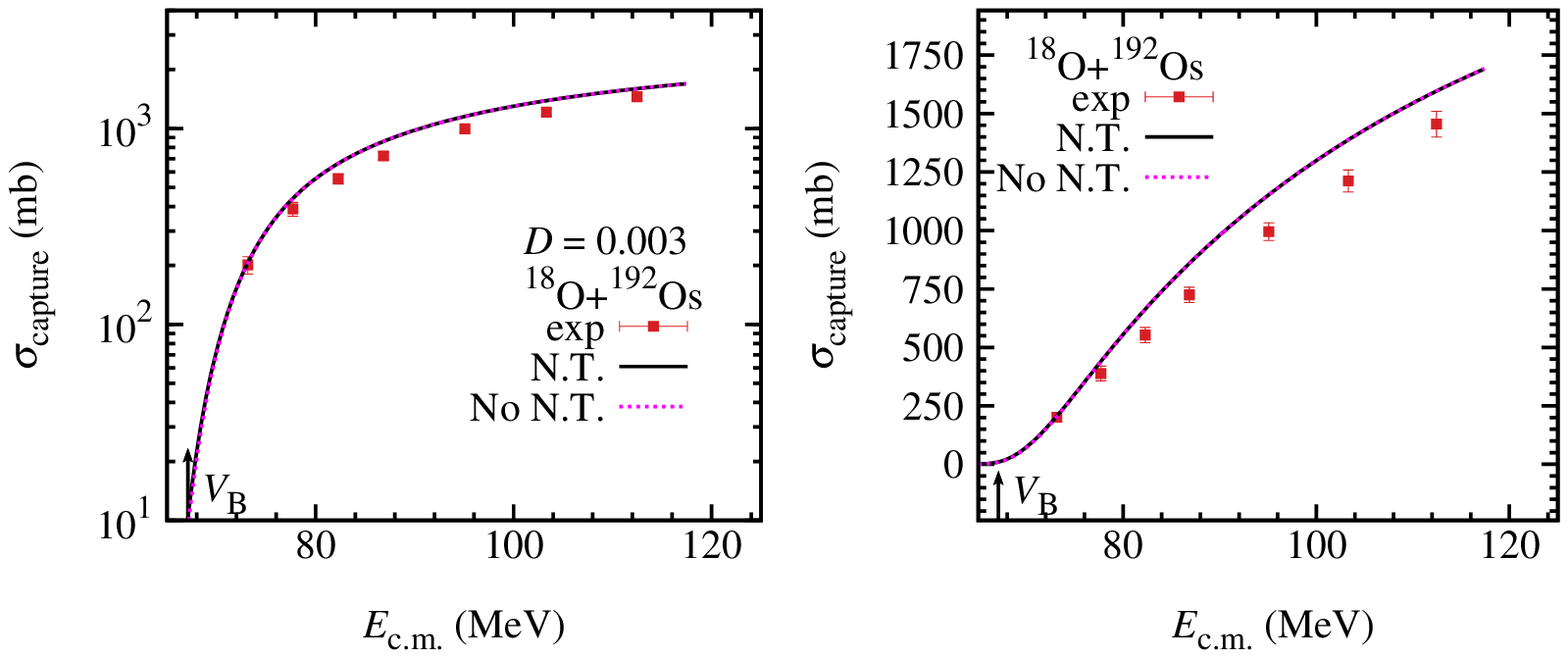}}
  \centerline{Graph 12}
 \end{Dfigures}
 \begin{Dfigures}[!ht]
 \centerline{\includegraphics[width=0.47\textwidth]{19F208Pb.eps}
  \includegraphics[width=0.47\textwidth]{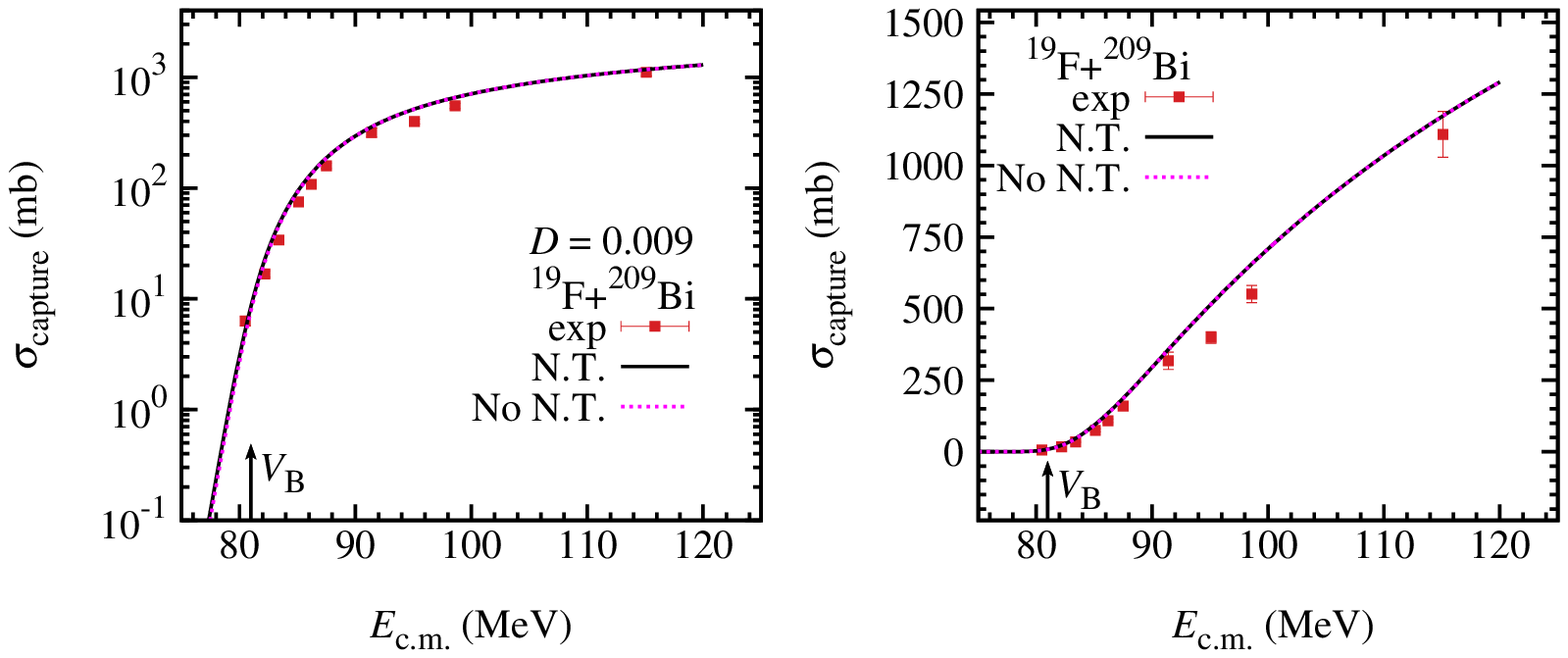}}
 \centerline{\includegraphics[width=0.47\textwidth]{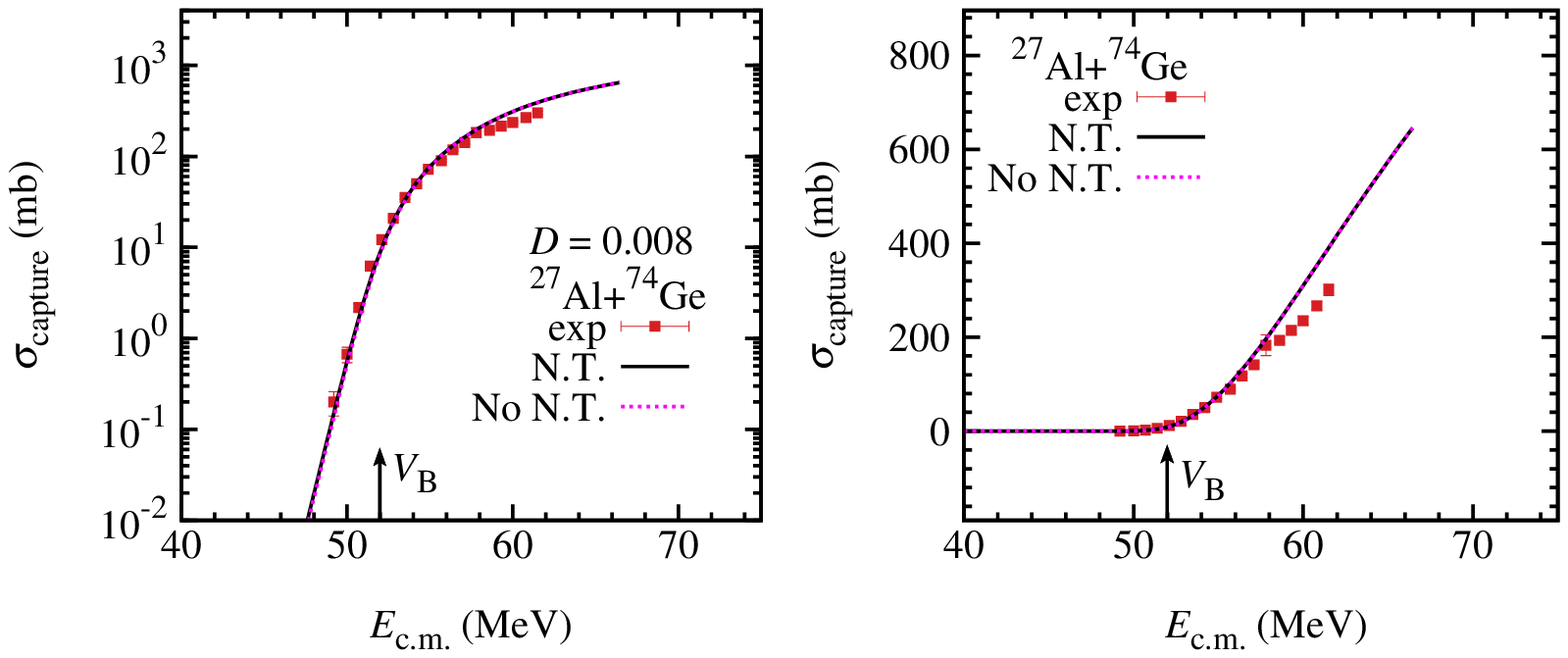}
  \includegraphics[width=0.47\textwidth]{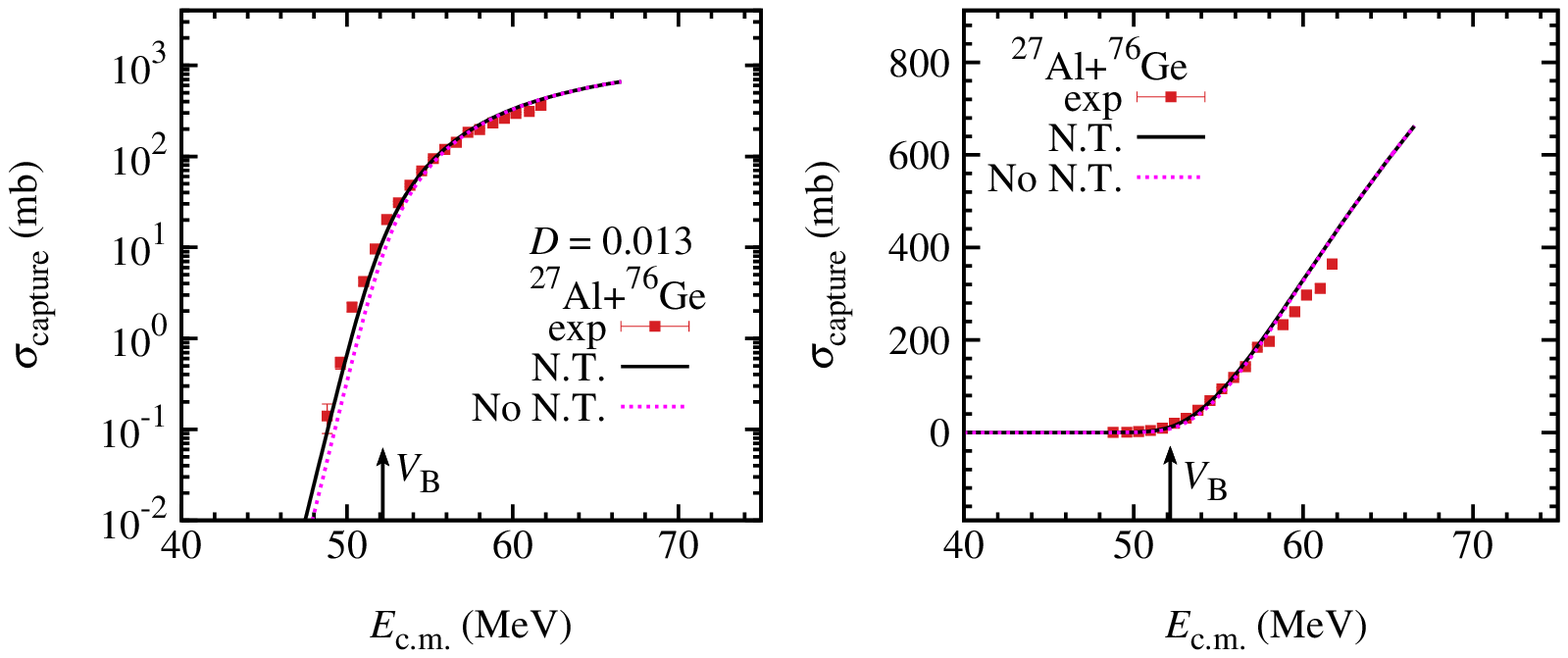}}
 \centerline{\includegraphics[width=0.47\textwidth]{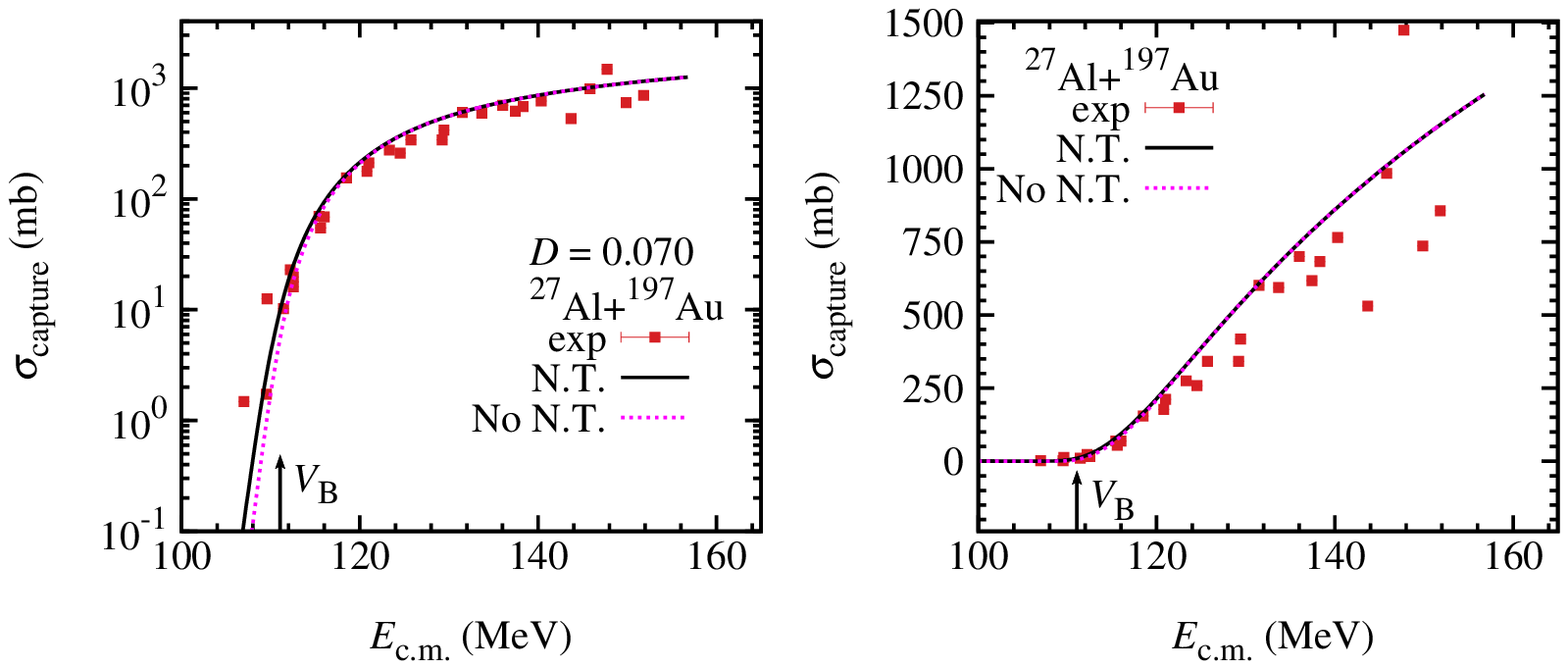}
  \includegraphics[width=0.47\textwidth]{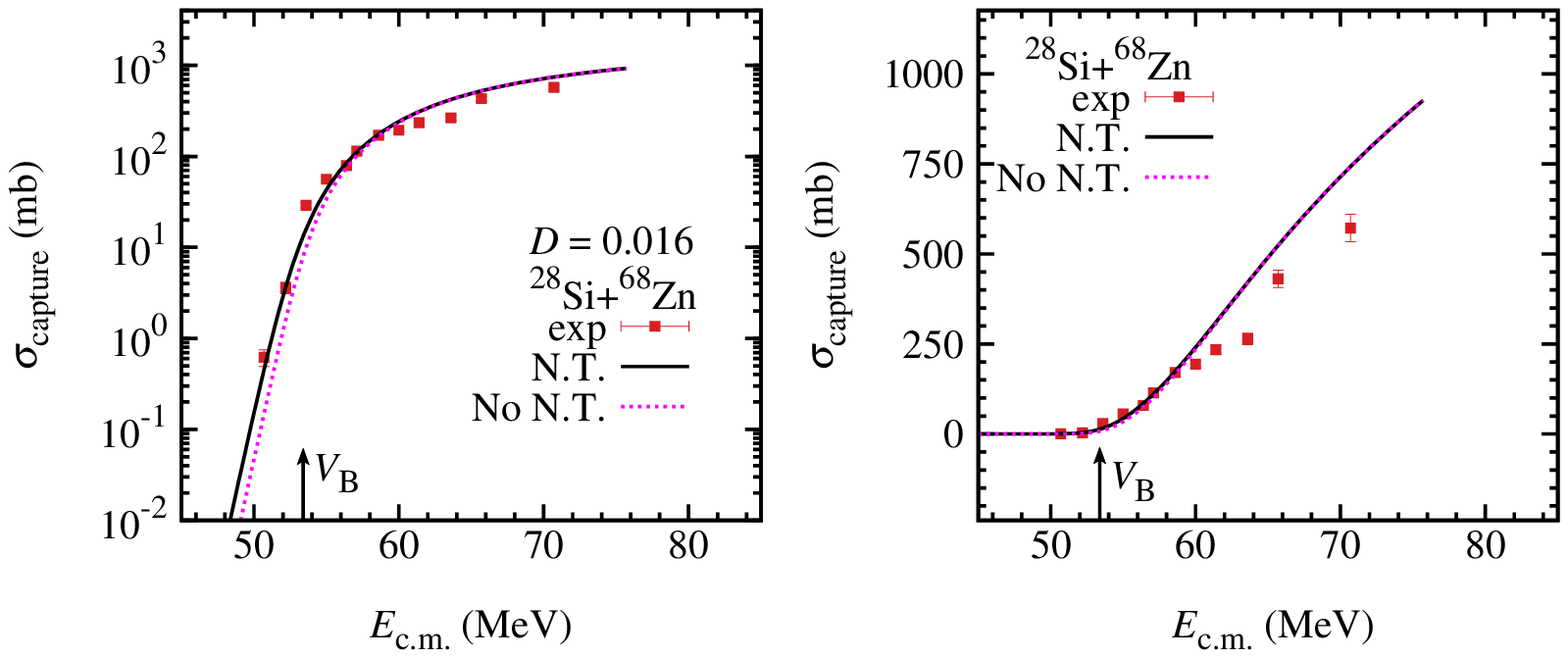}}
 \centerline{\includegraphics[width=0.47\textwidth]{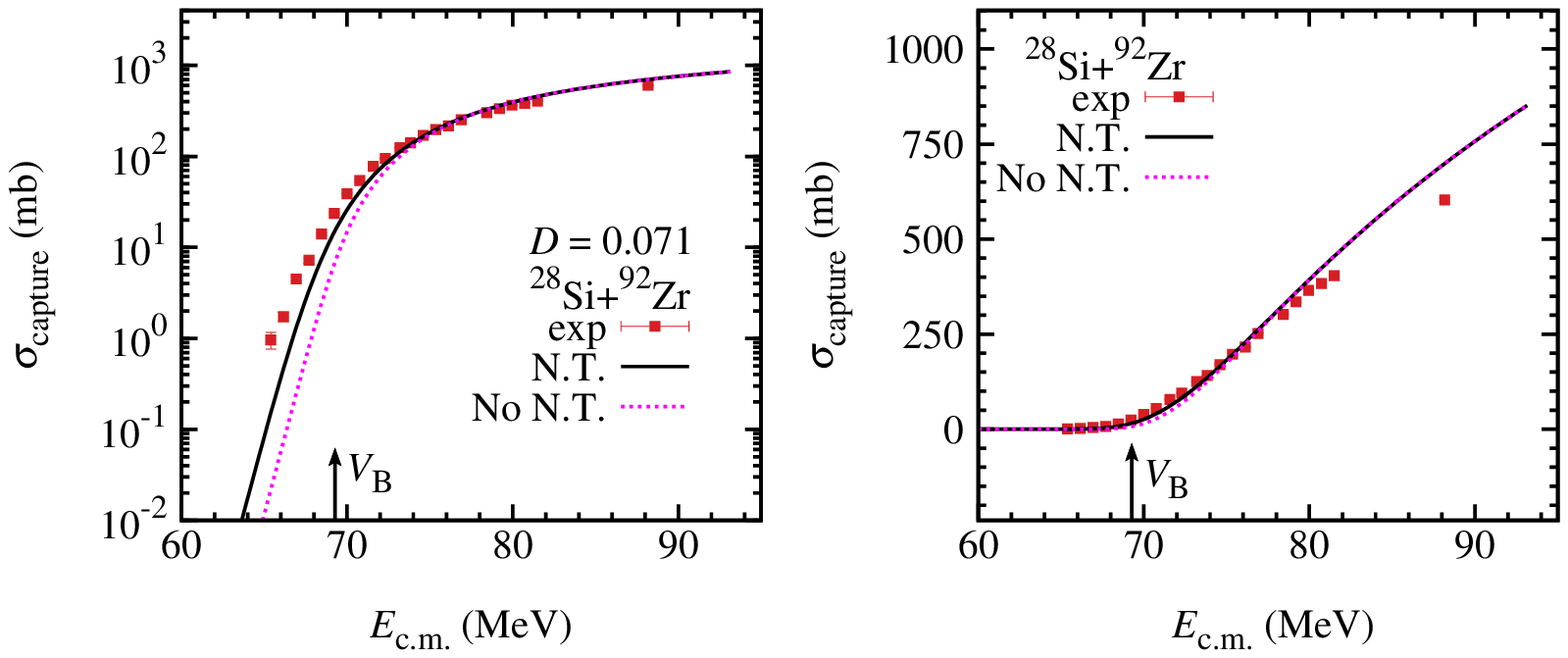}
  \includegraphics[width=0.47\textwidth]{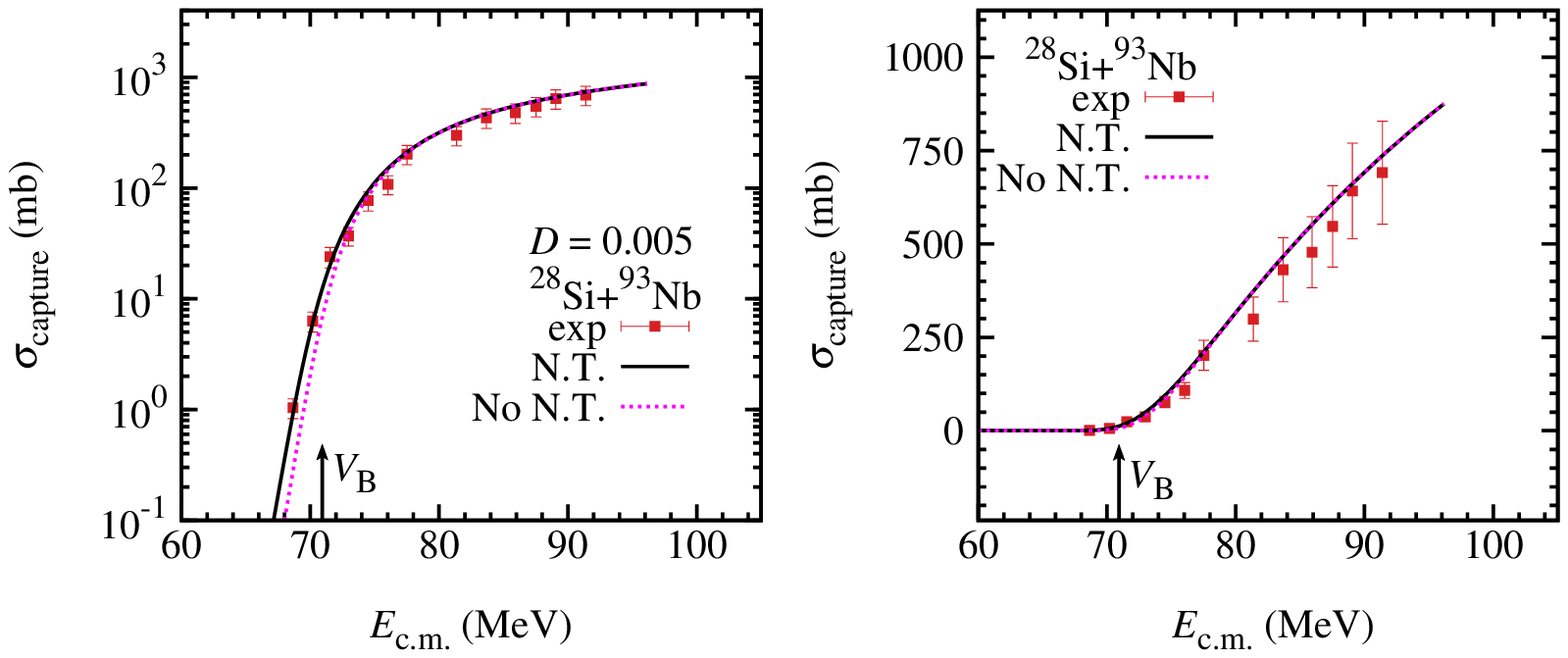}}
 \centerline{\includegraphics[width=0.47\textwidth]{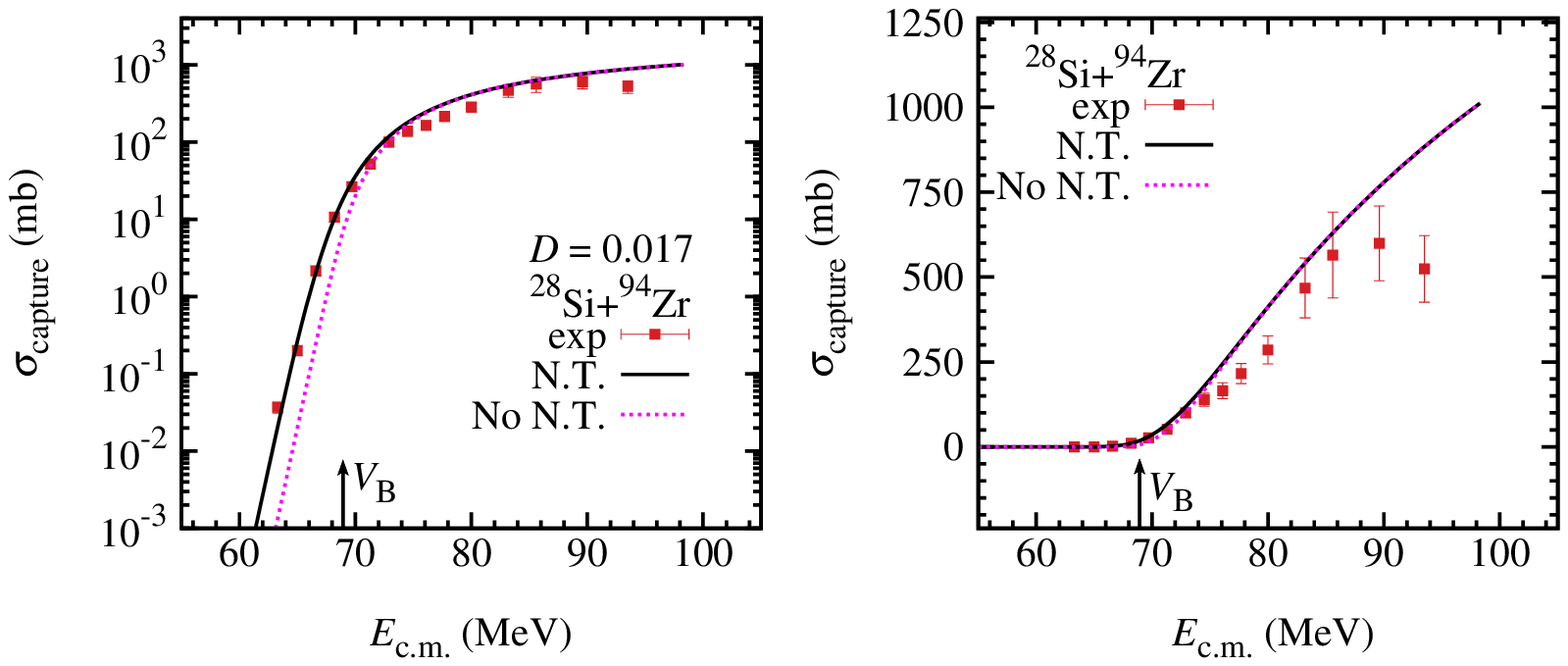}
  \includegraphics[width=0.47\textwidth]{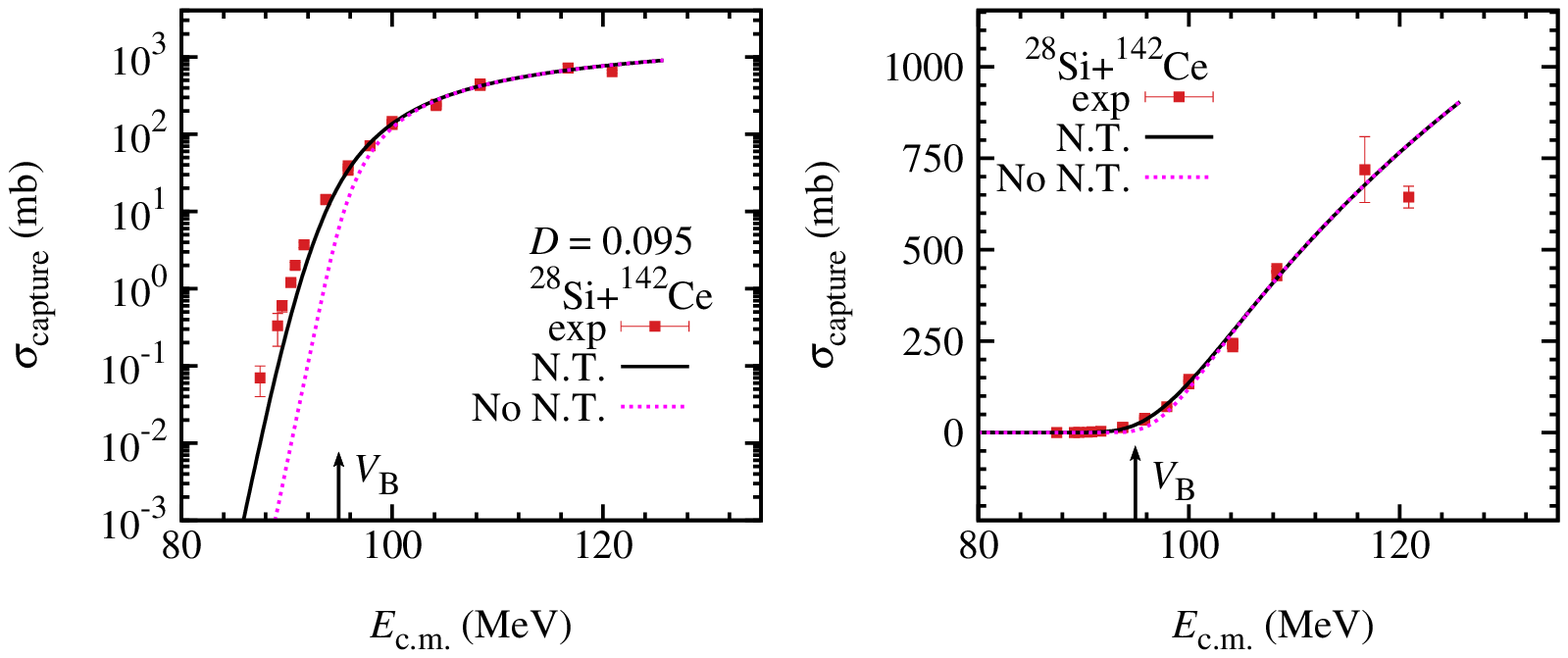}}
 \centerline{\includegraphics[width=0.47\textwidth]{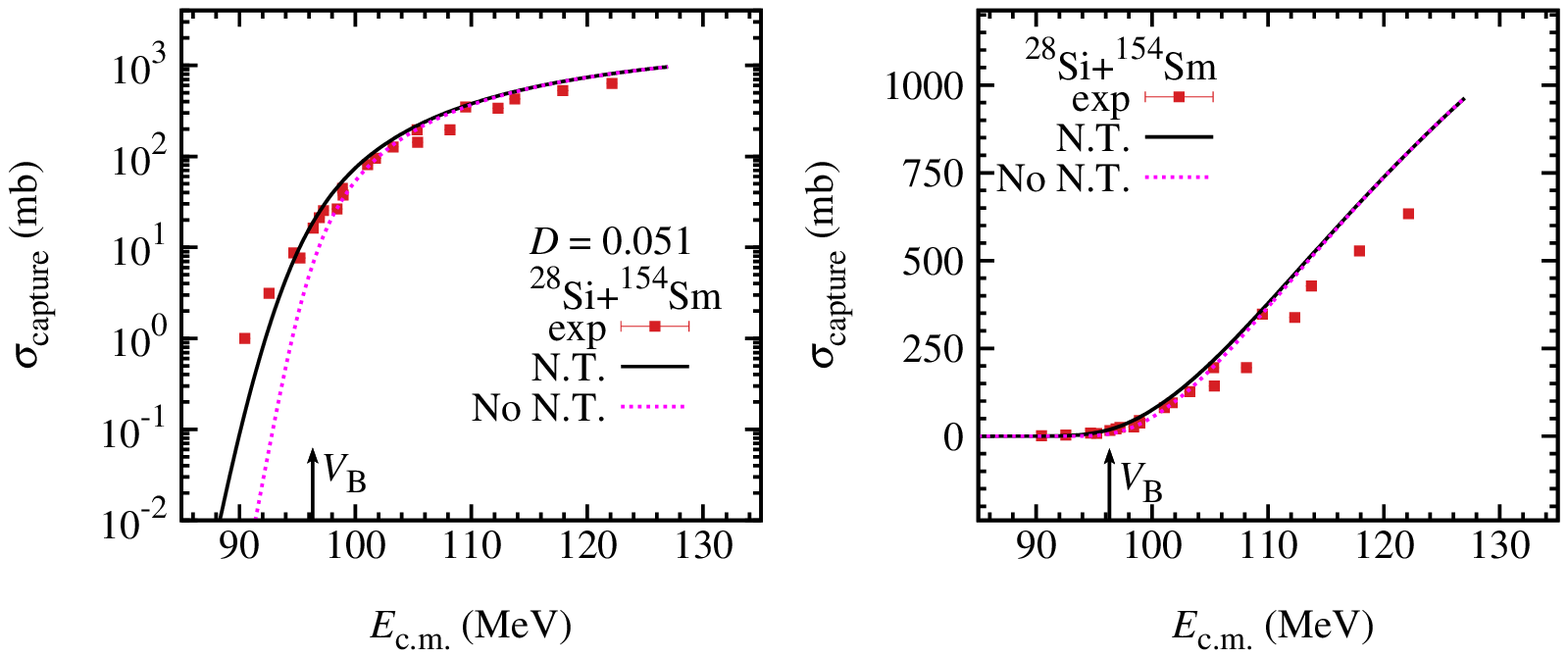}
  \includegraphics[width=0.47\textwidth]{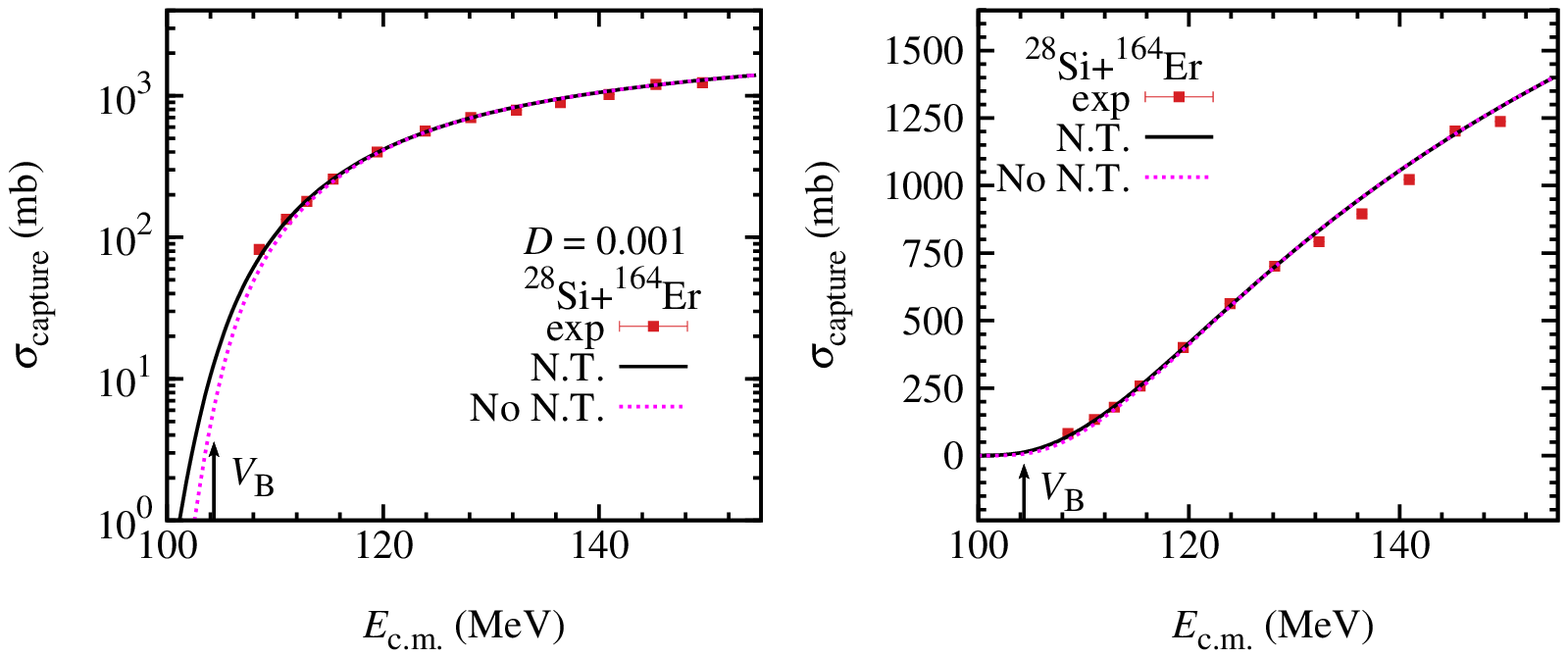}}
  \centerline{Graph 13}
 \end{Dfigures}
 \begin{Dfigures}[!ht]
 \centerline{\includegraphics[width=0.47\textwidth]{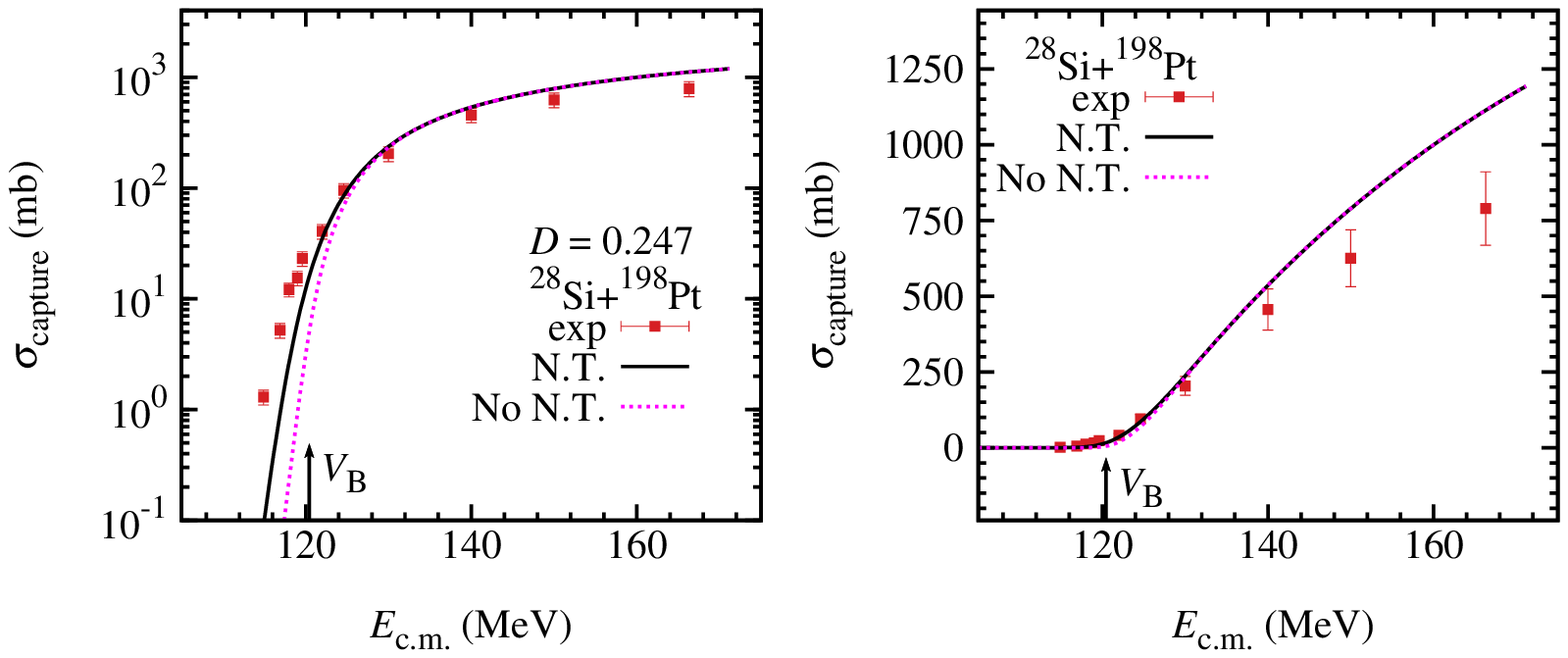}
  \includegraphics[width=0.47\textwidth]{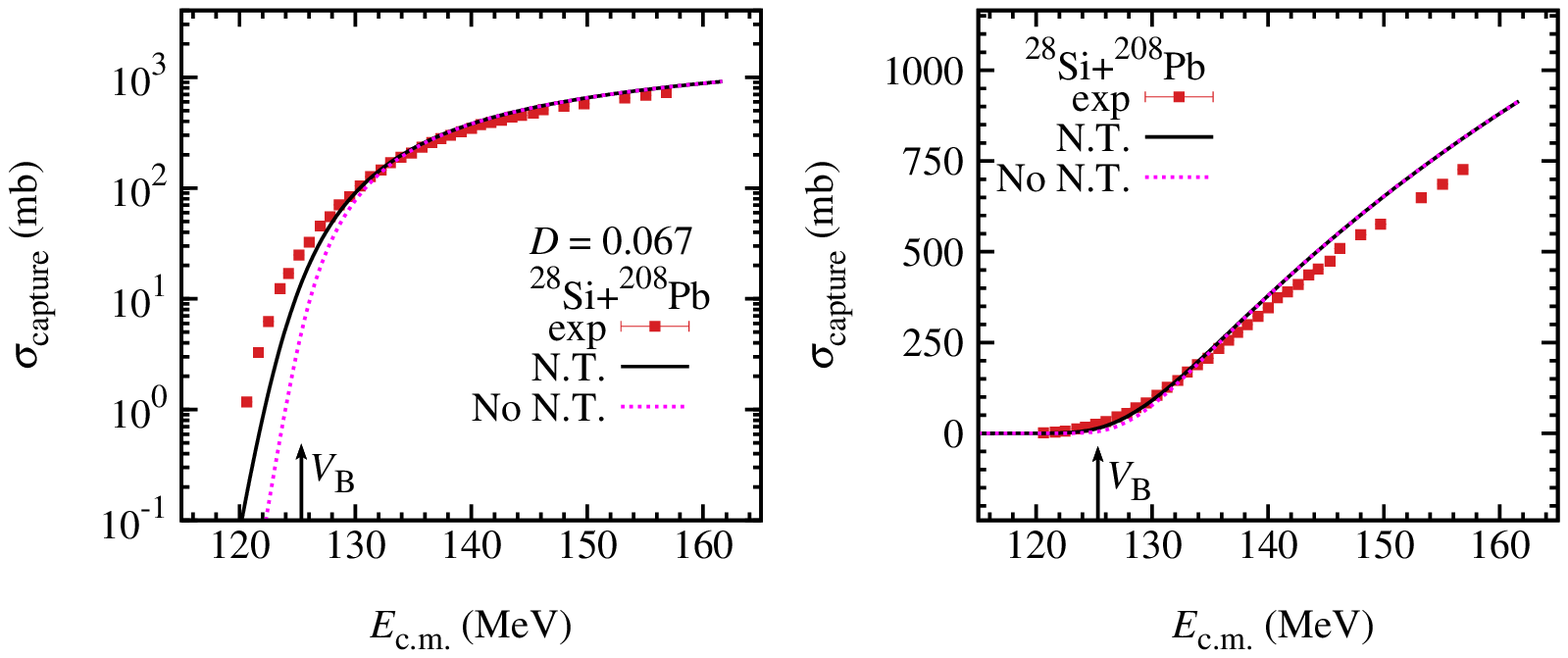}}
 \centerline{\includegraphics[width=0.47\textwidth]{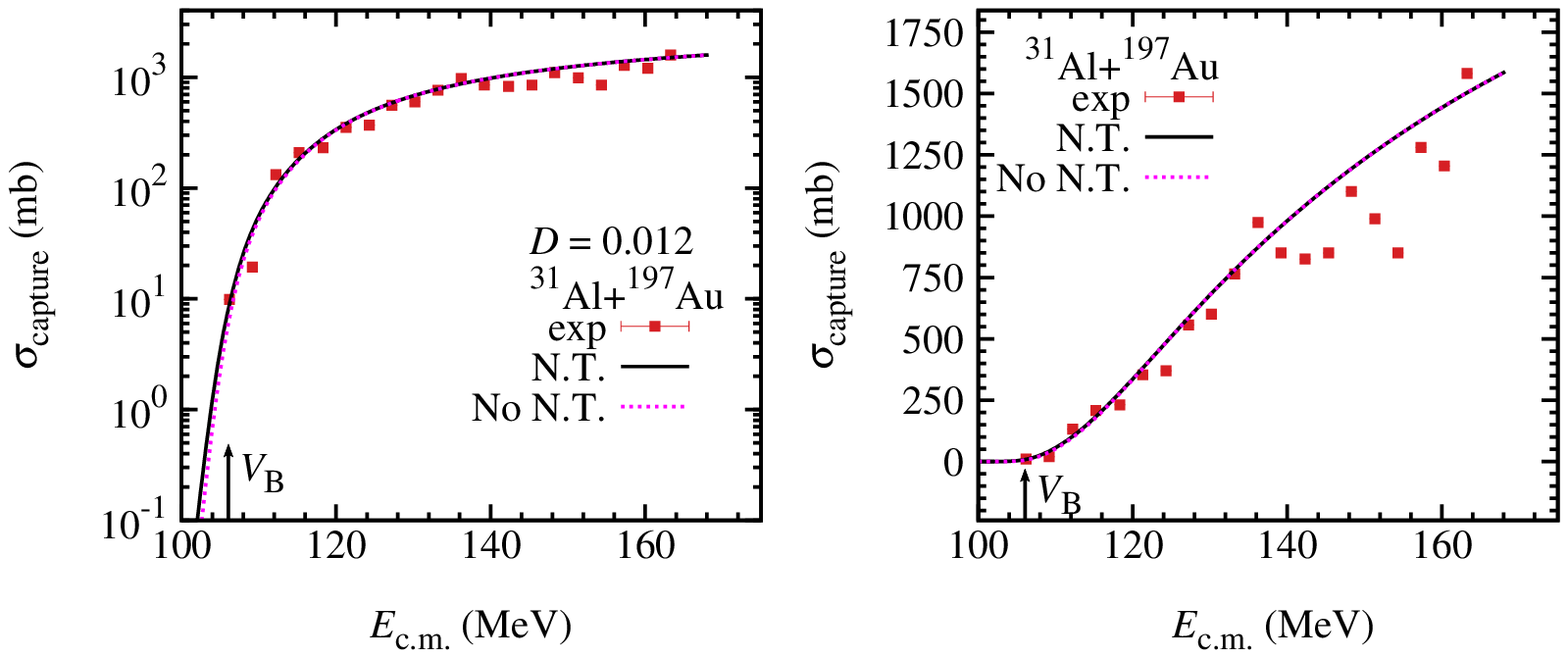}
  \includegraphics[width=0.47\textwidth]{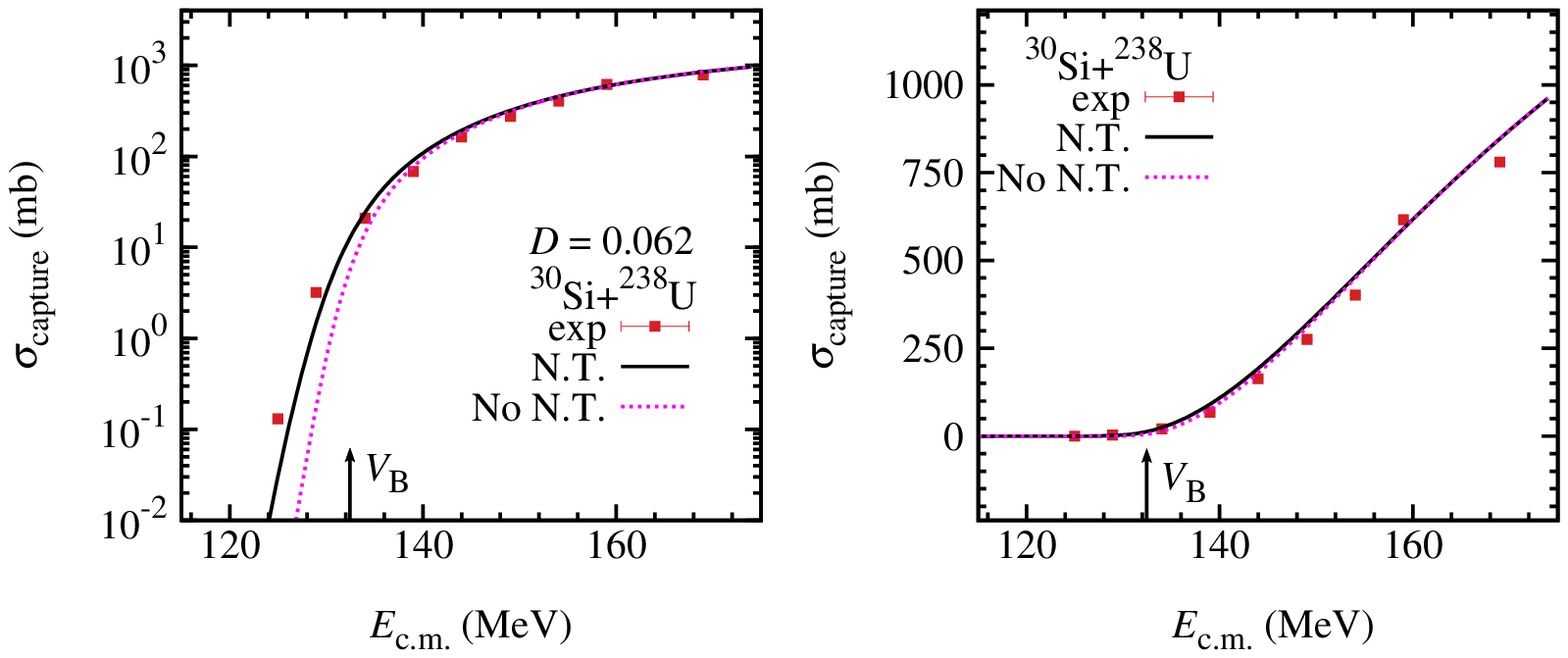}}
 \centerline{\includegraphics[width=0.47\textwidth]{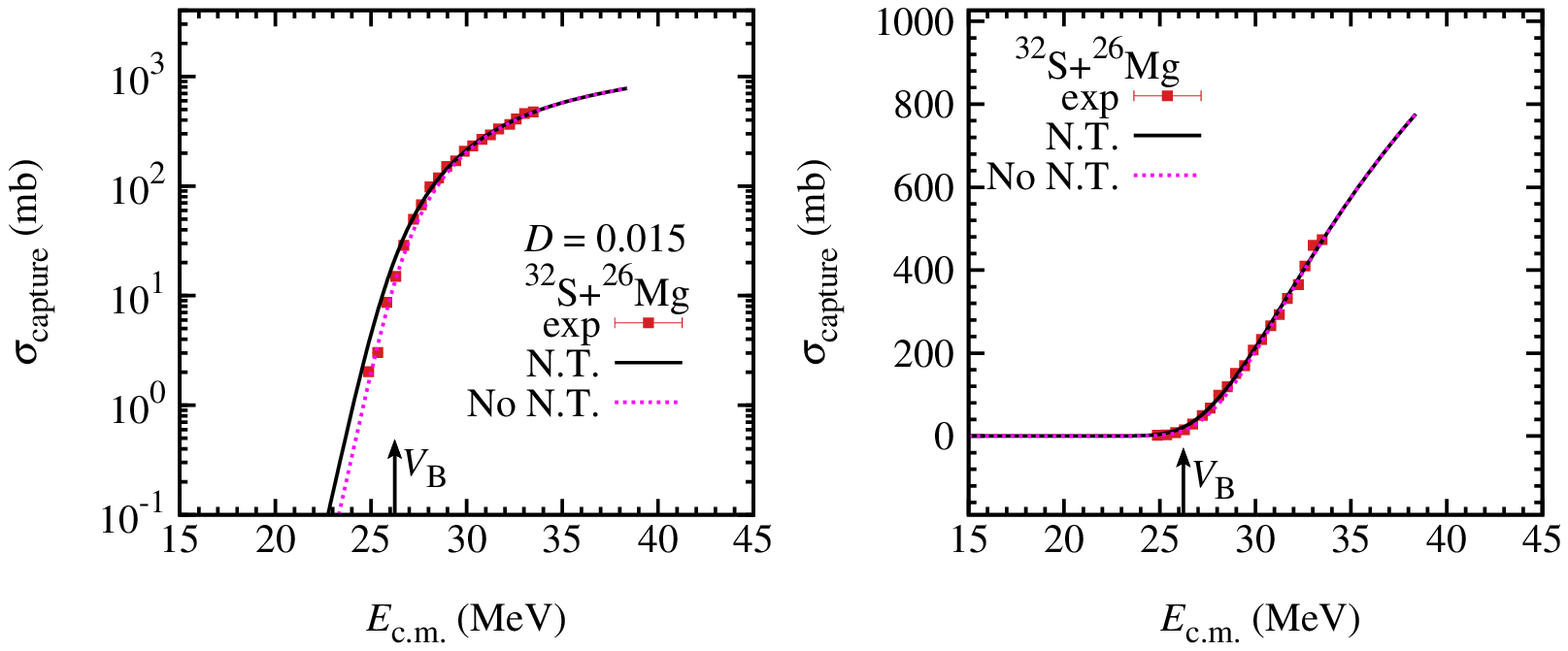}
  \includegraphics[width=0.47\textwidth]{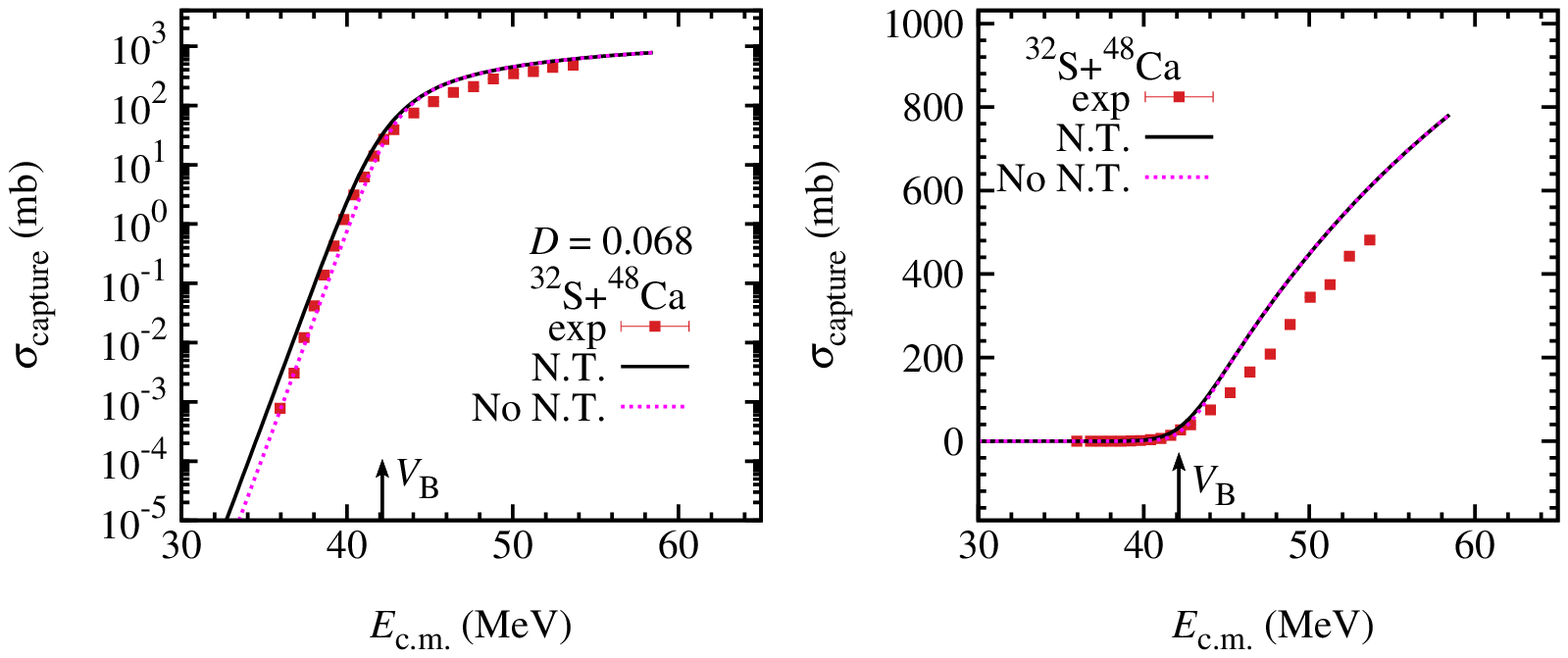}}
 \centerline{\includegraphics[width=0.47\textwidth]{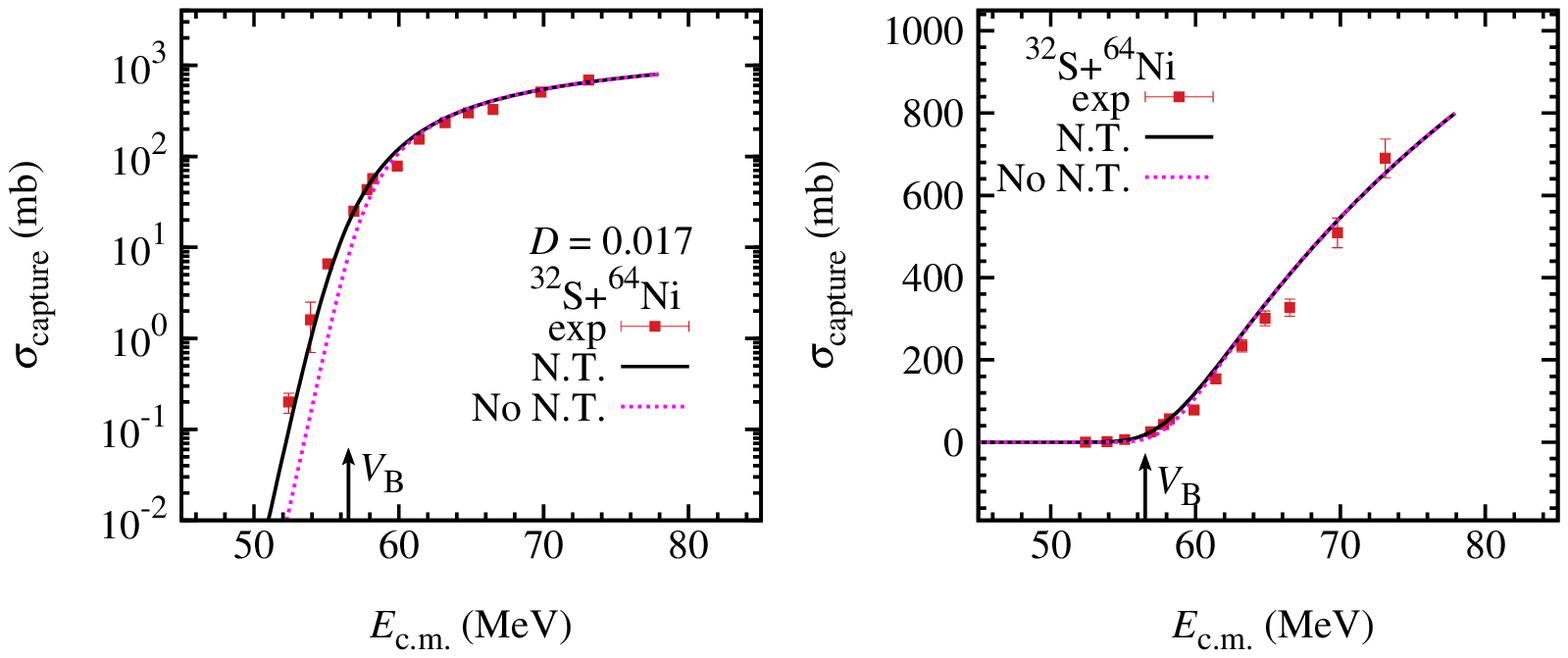}
  \includegraphics[width=0.47\textwidth]{32S94Zr.eps}}
 \centerline{\includegraphics[width=0.47\textwidth]{32S96Zr.eps}
  \includegraphics[width=0.47\textwidth]{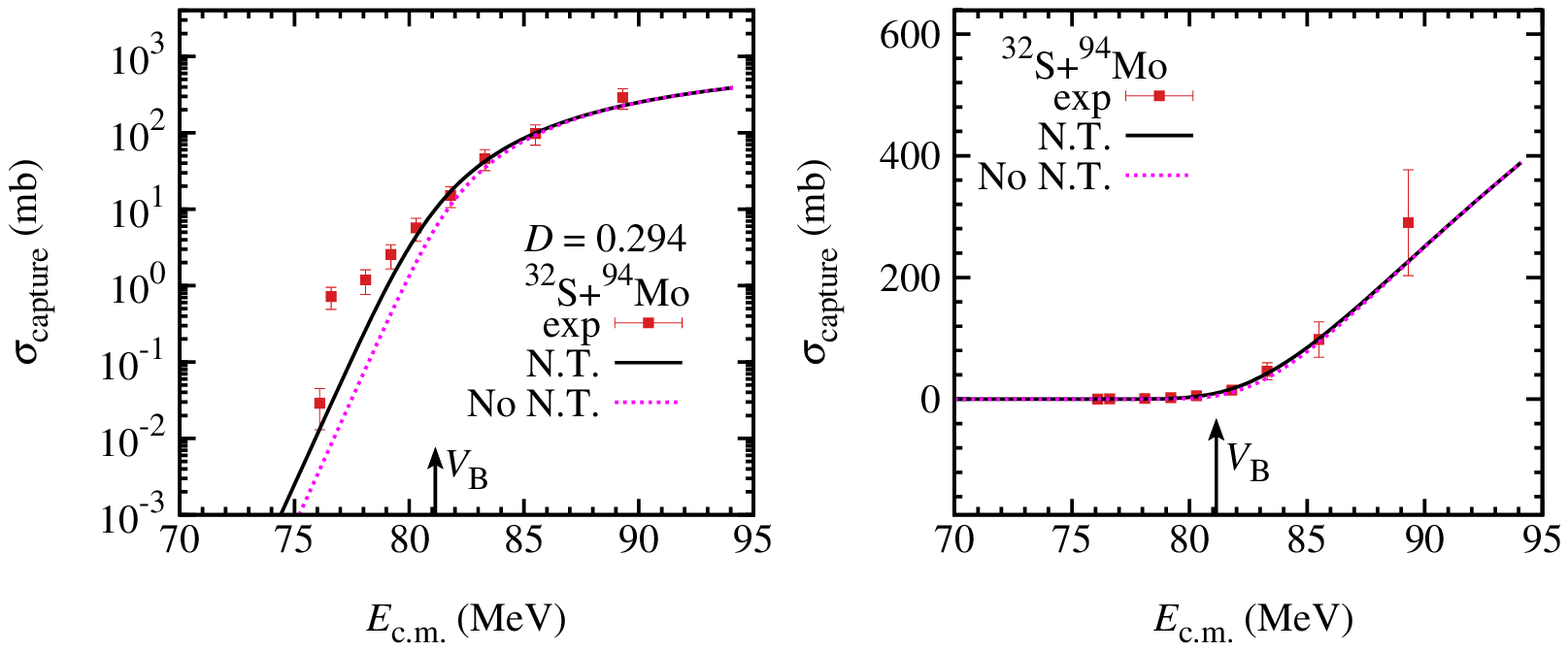}}
 \centerline{\includegraphics[width=0.47\textwidth]{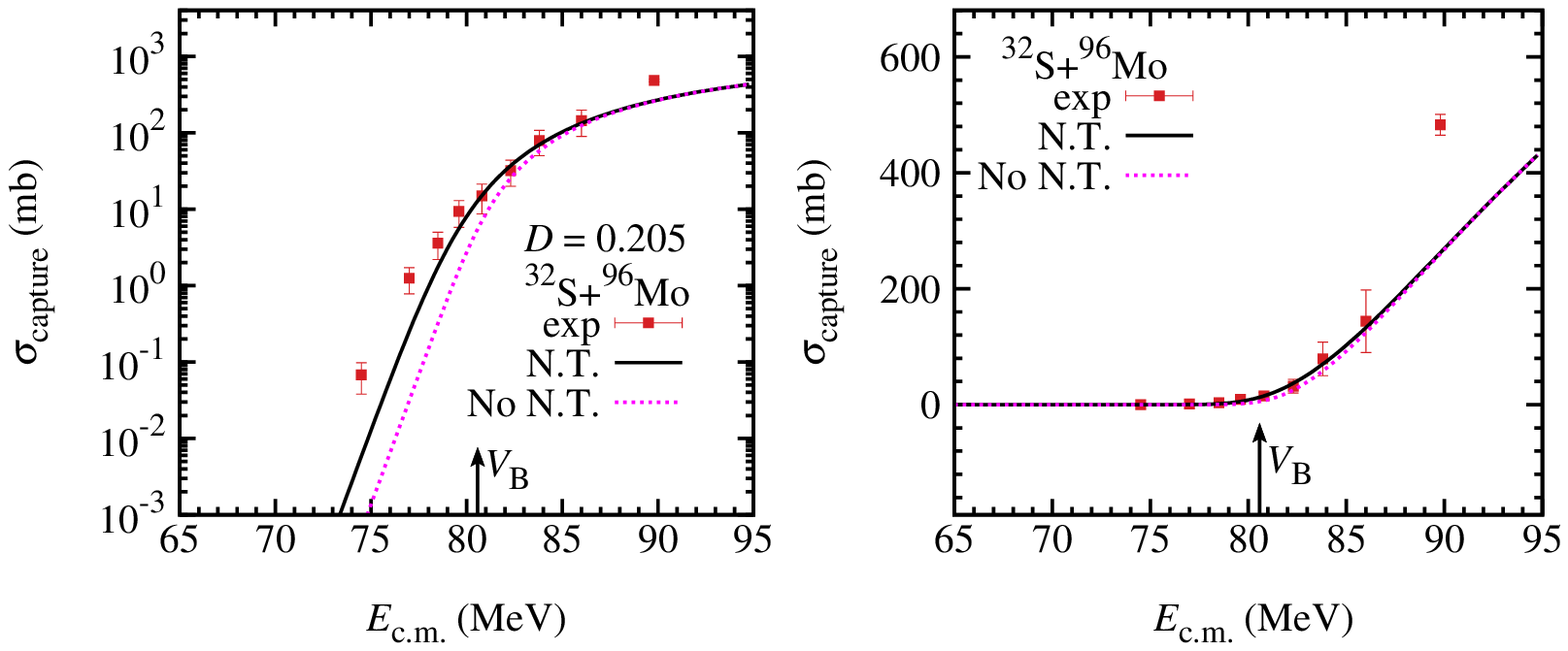}
  \includegraphics[width=0.47\textwidth]{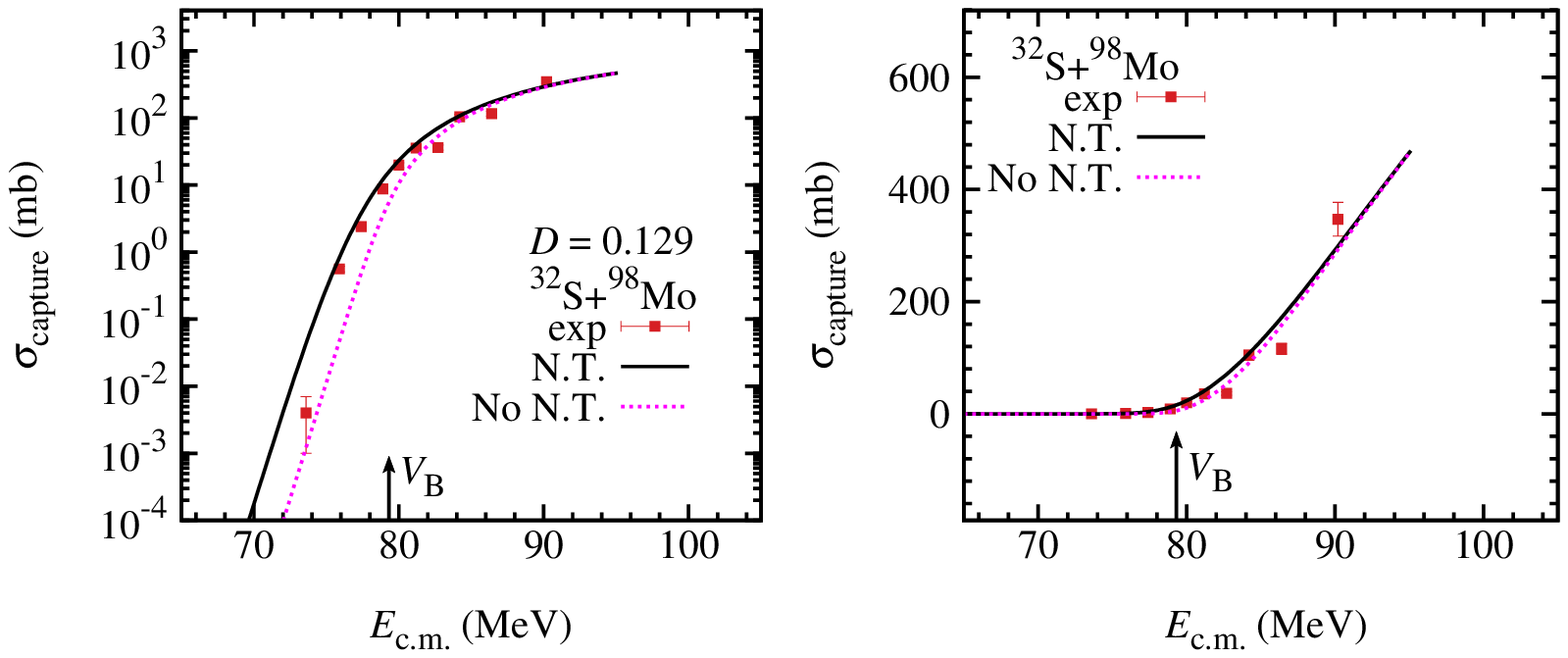}}
  \centerline{Graph 14}
 \end{Dfigures}
 \begin{Dfigures}[!ht]
 \centerline{\includegraphics[width=0.47\textwidth]{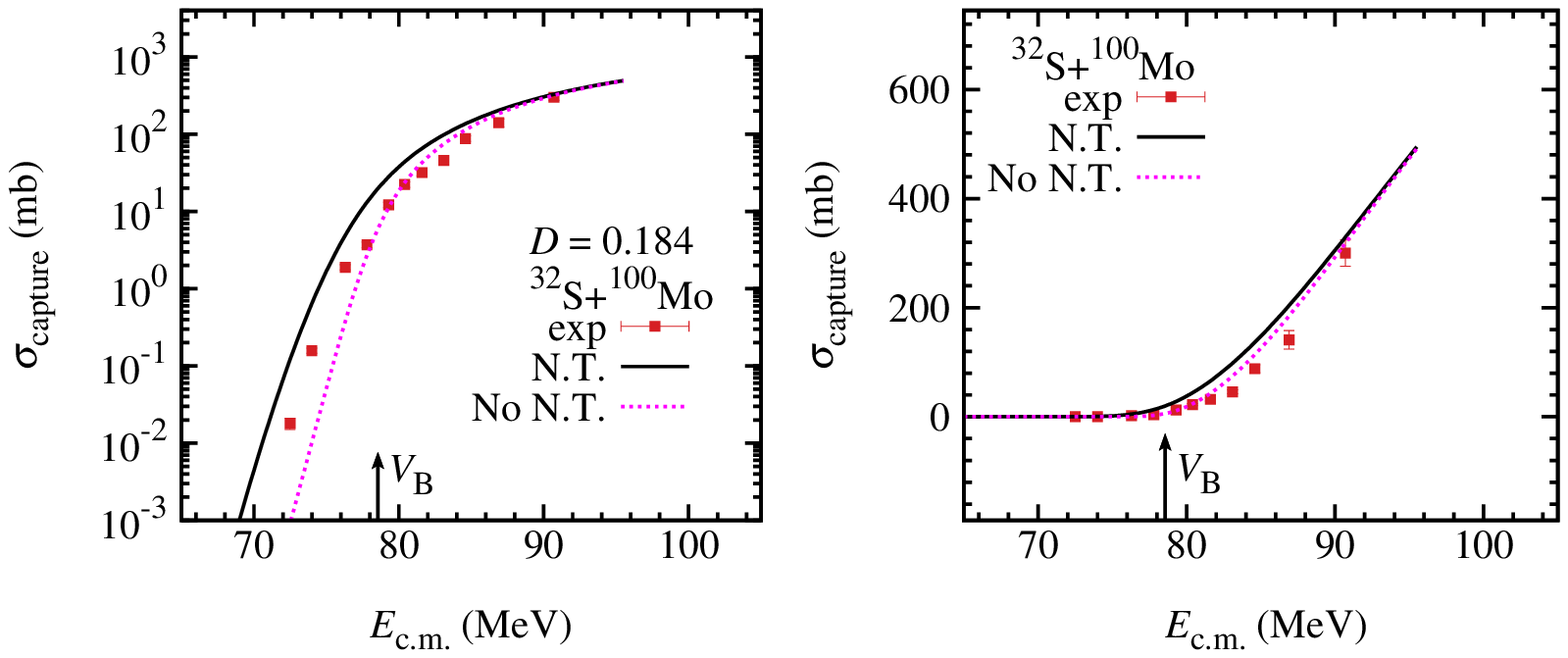}
  \includegraphics[width=0.47\textwidth]{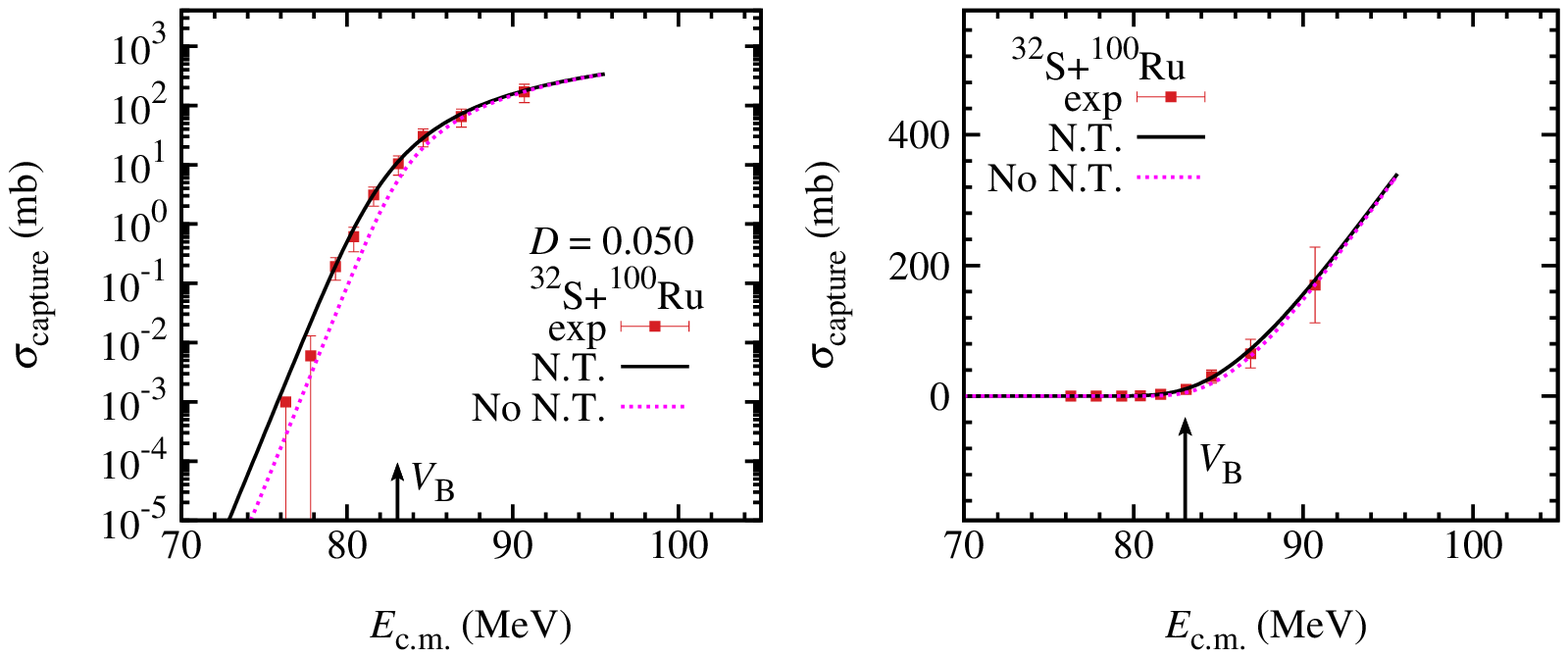}}
 \centerline{\includegraphics[width=0.47\textwidth]{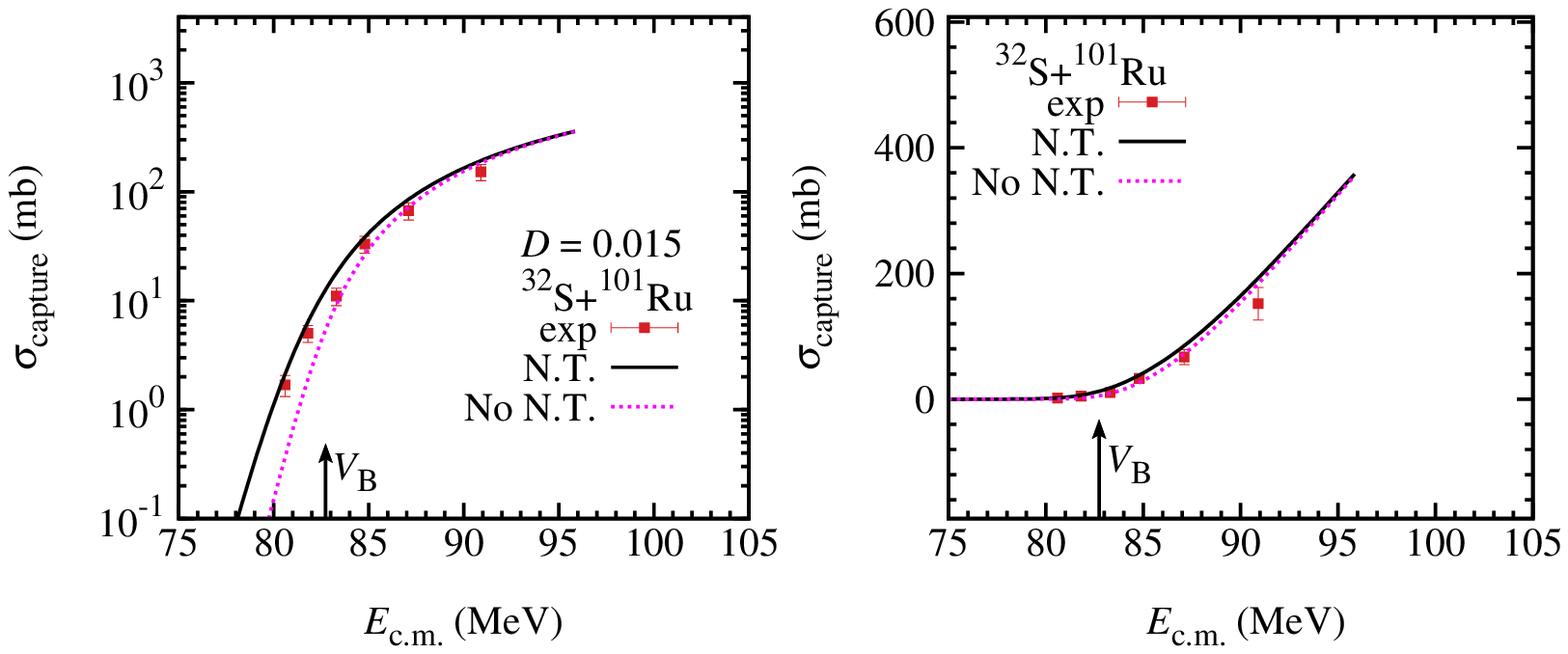}
  \includegraphics[width=0.47\textwidth]{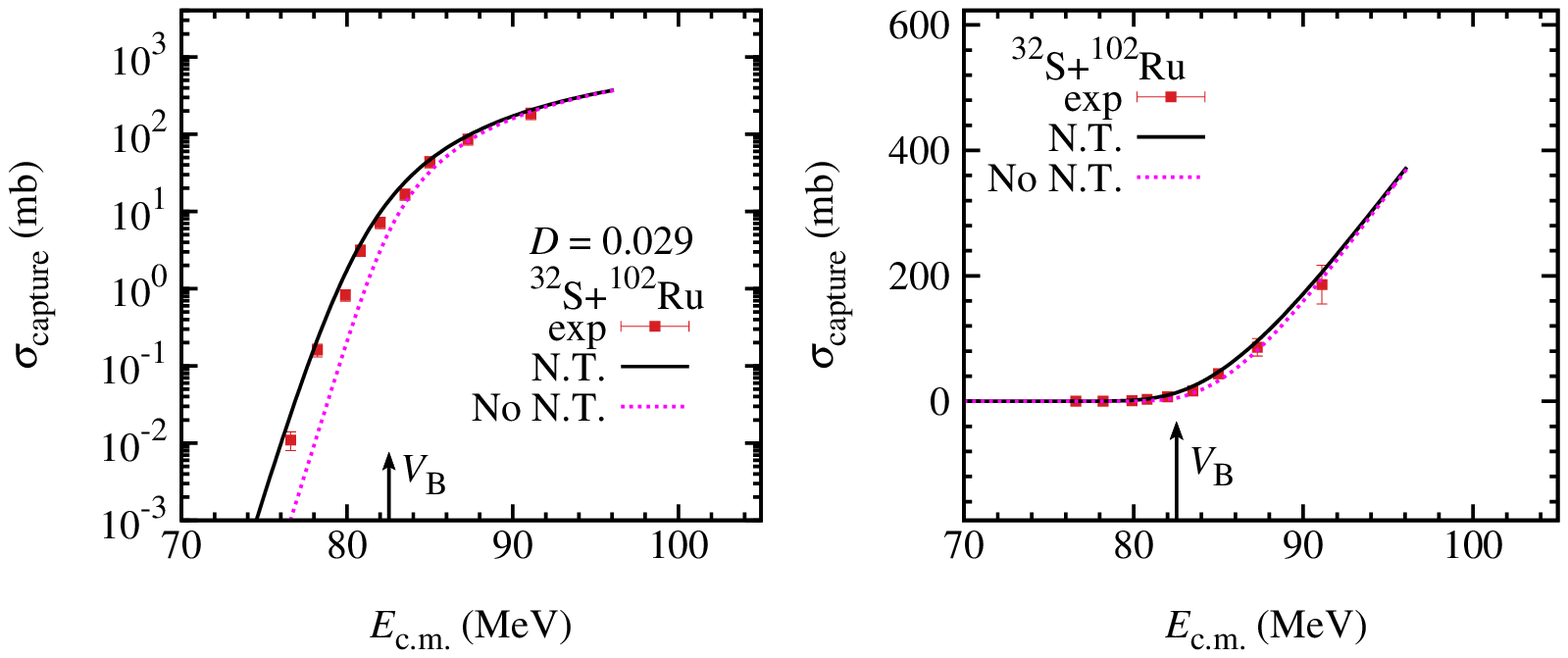}}
 \centerline{\includegraphics[width=0.47\textwidth]{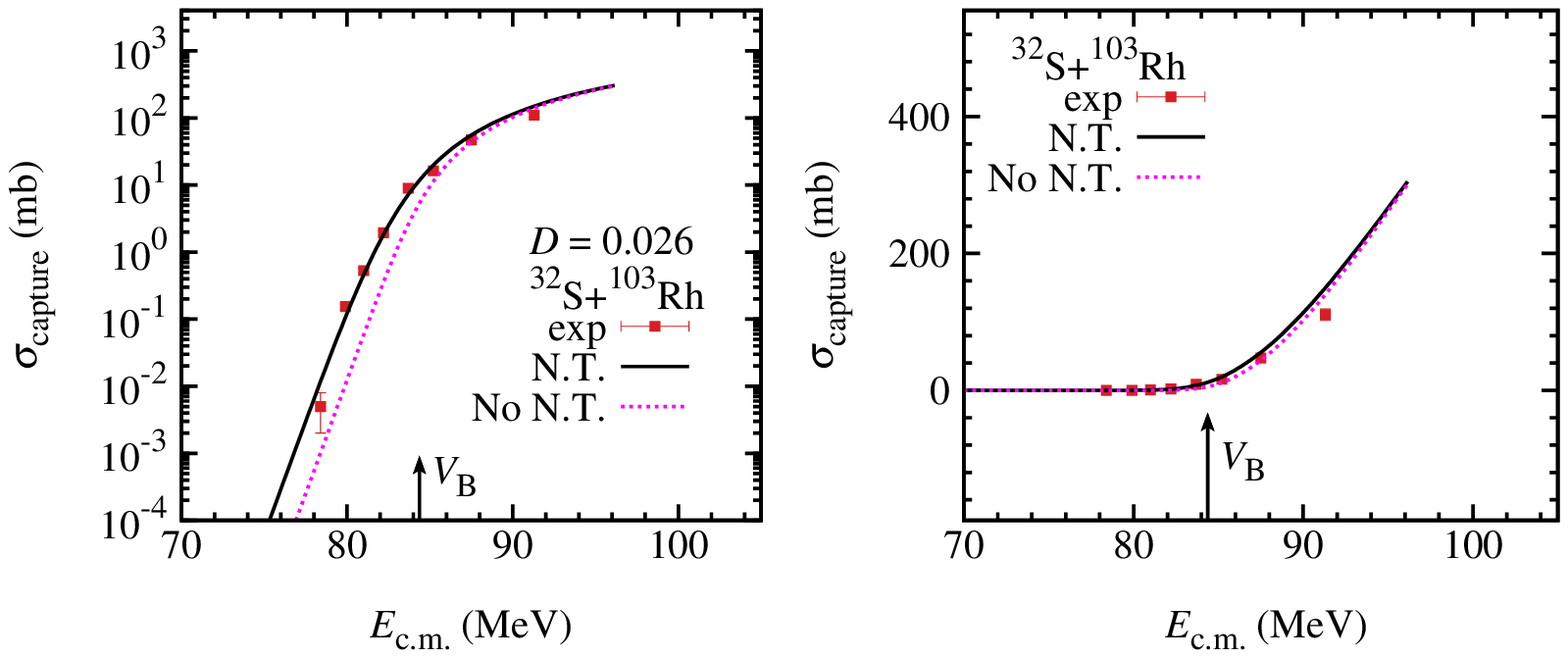}
  \includegraphics[width=0.47\textwidth]{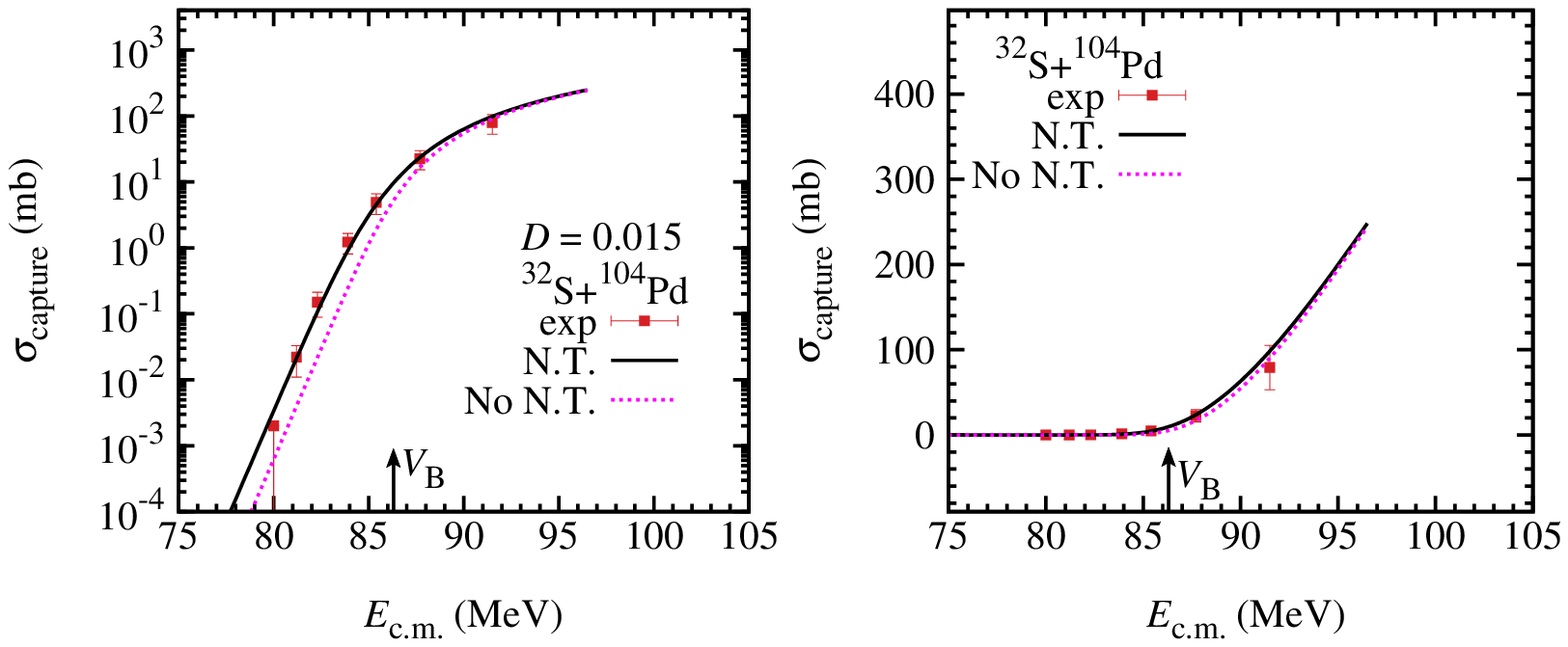}}
 \centerline{\includegraphics[width=0.47\textwidth]{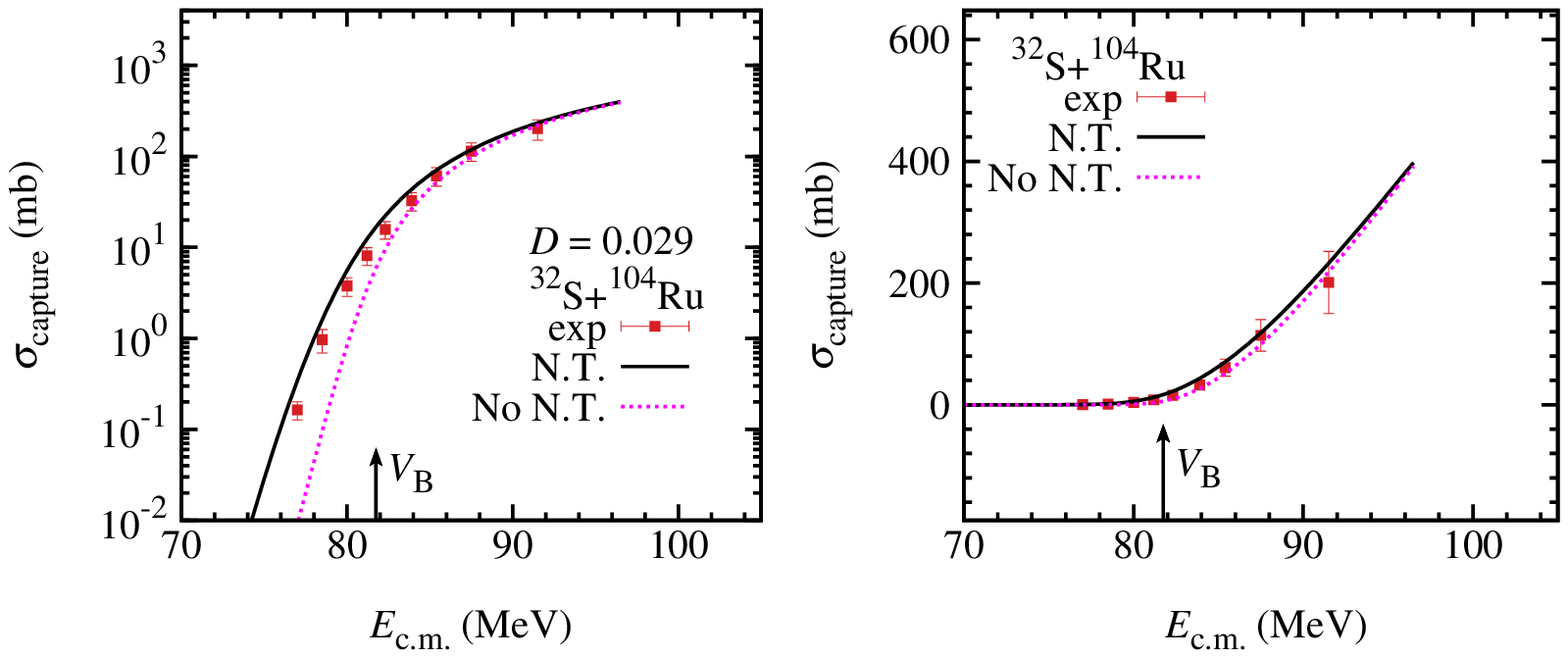}
  \includegraphics[width=0.47\textwidth]{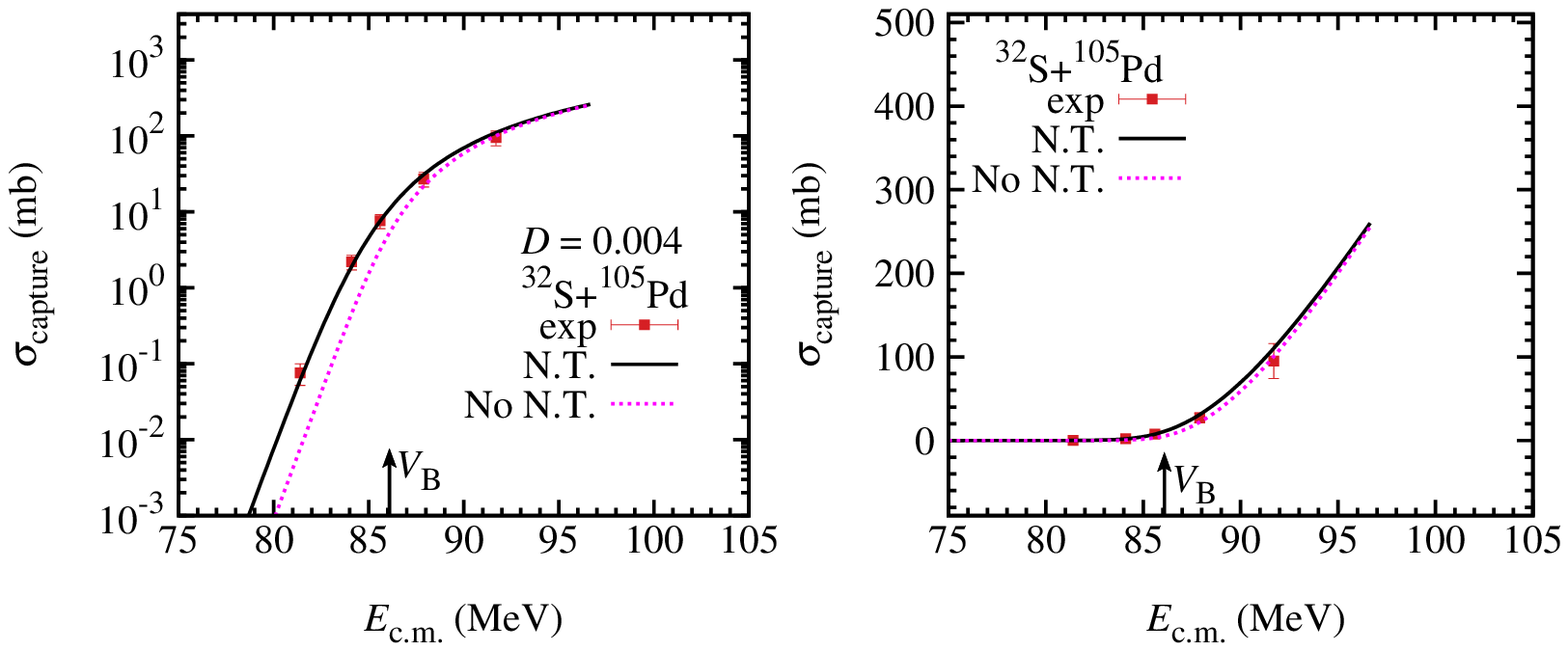}}
 \centerline{\includegraphics[width=0.47\textwidth]{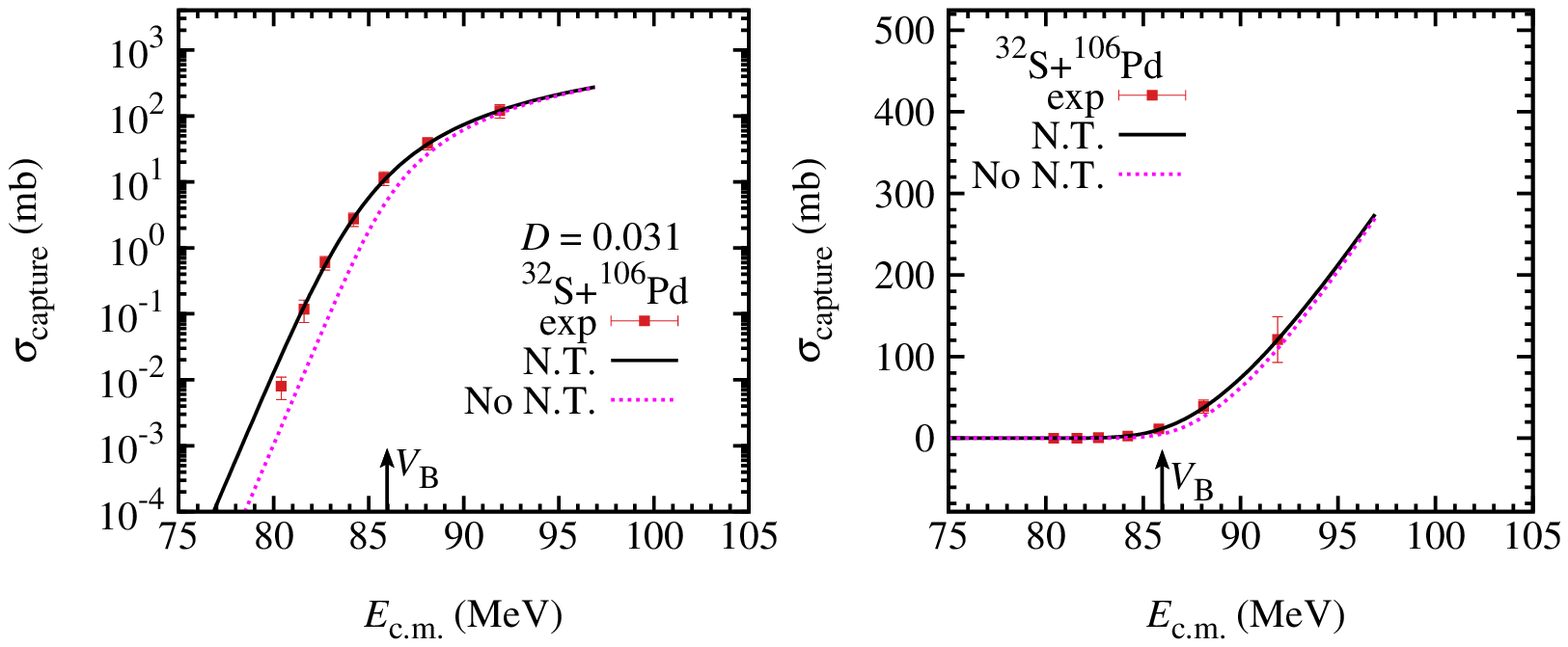}
  \includegraphics[width=0.47\textwidth]{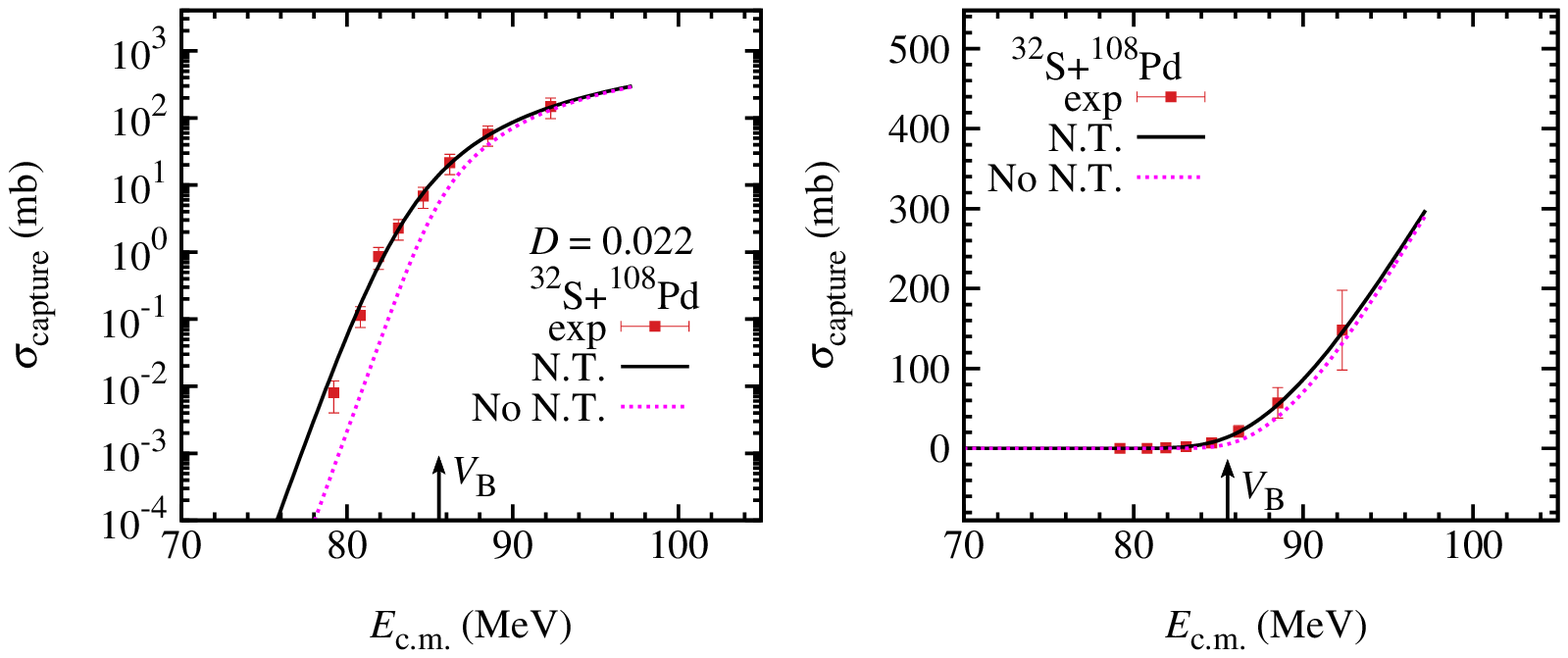}}
 \centerline{\includegraphics[width=0.47\textwidth]{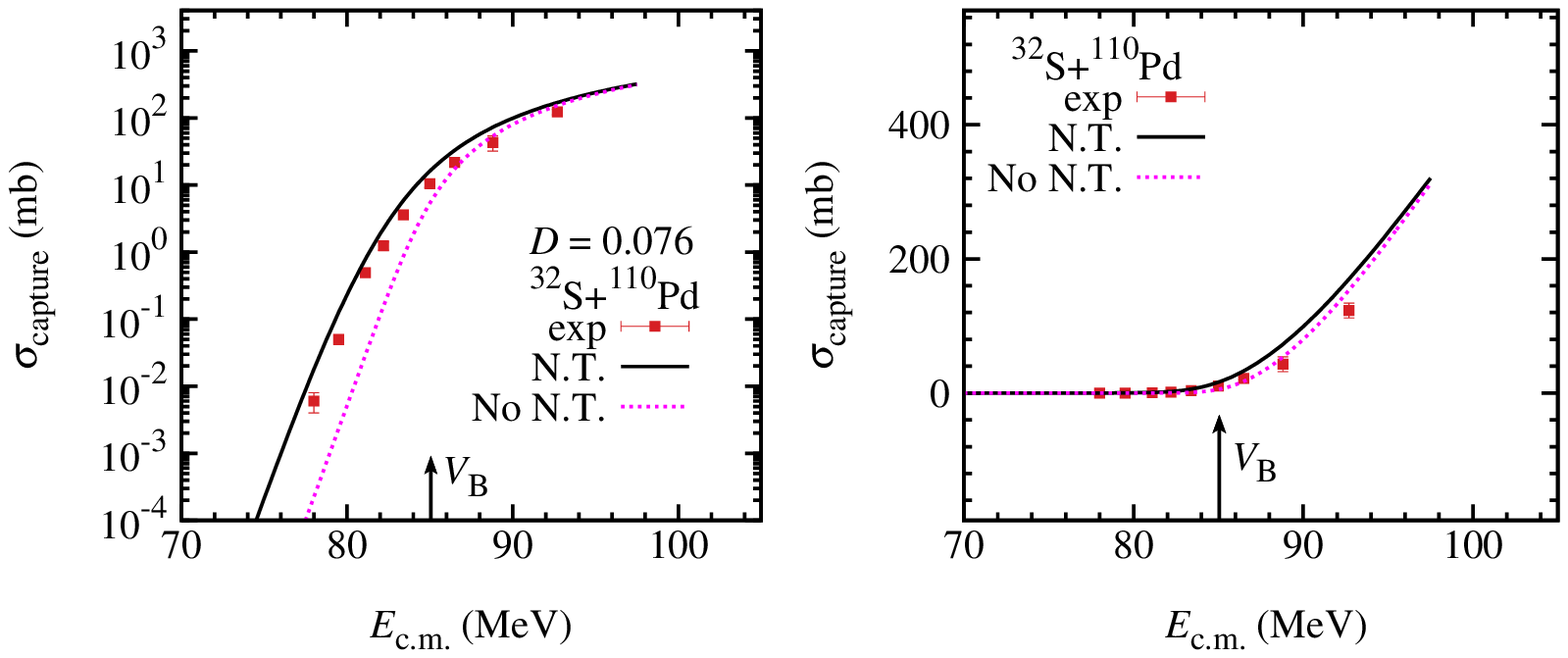}
  \includegraphics[width=0.47\textwidth]{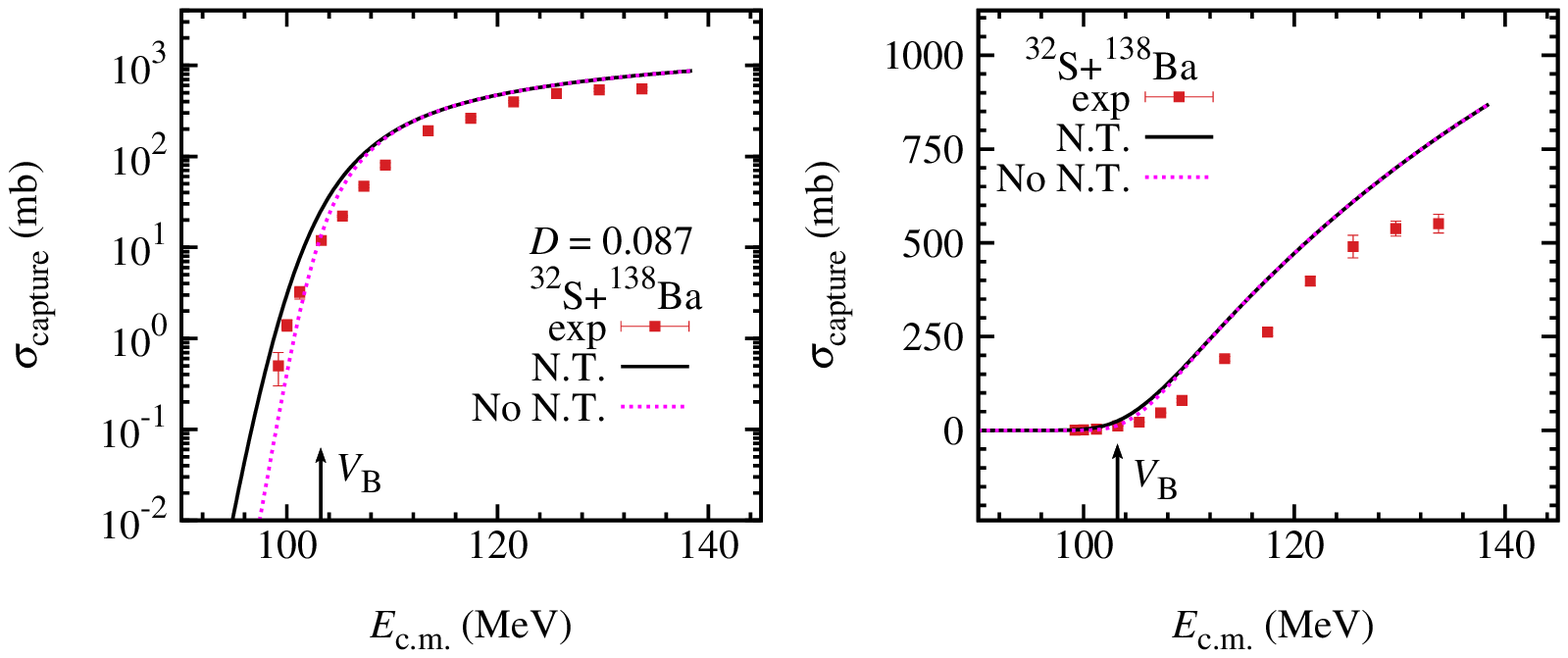}}
  \centerline{Graph 15}
 \end{Dfigures}
 \begin{Dfigures}[!ht]
 \centerline{\includegraphics[width=0.47\textwidth]{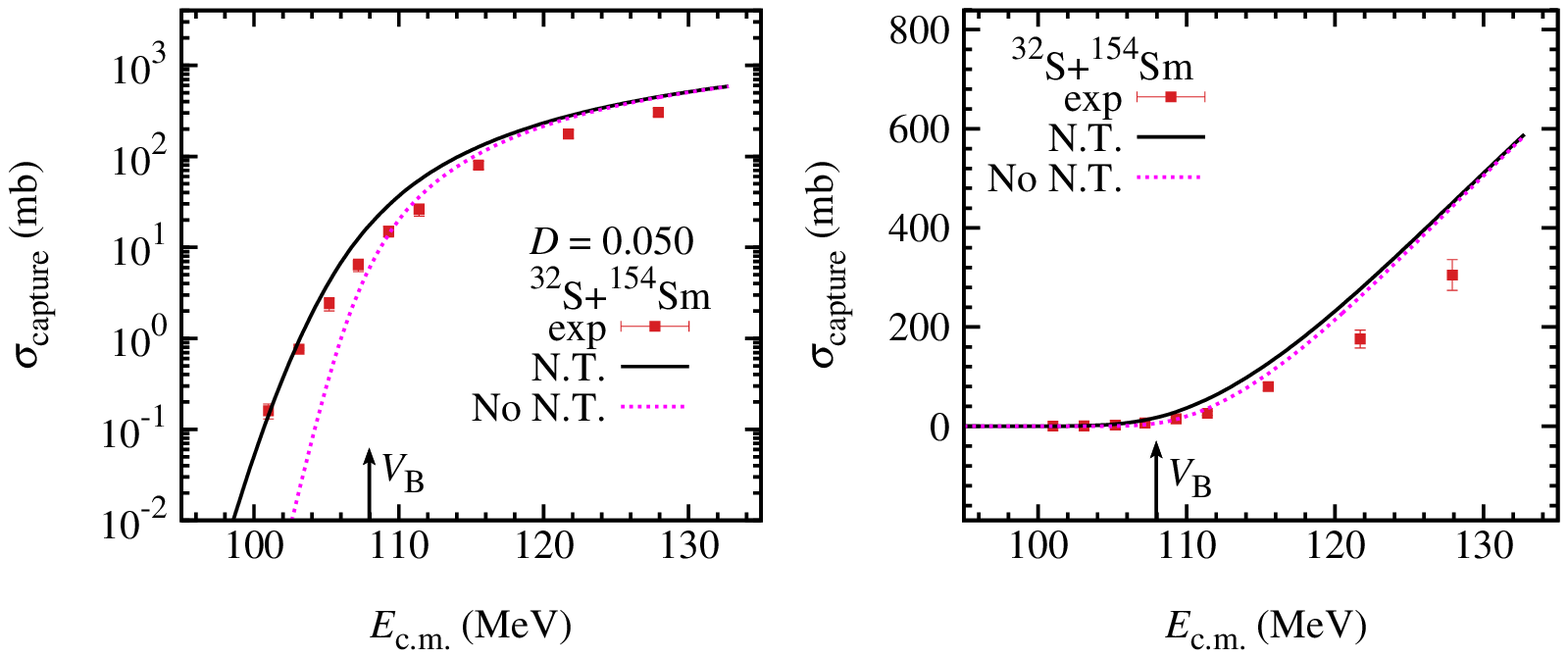}
  \includegraphics[width=0.47\textwidth]{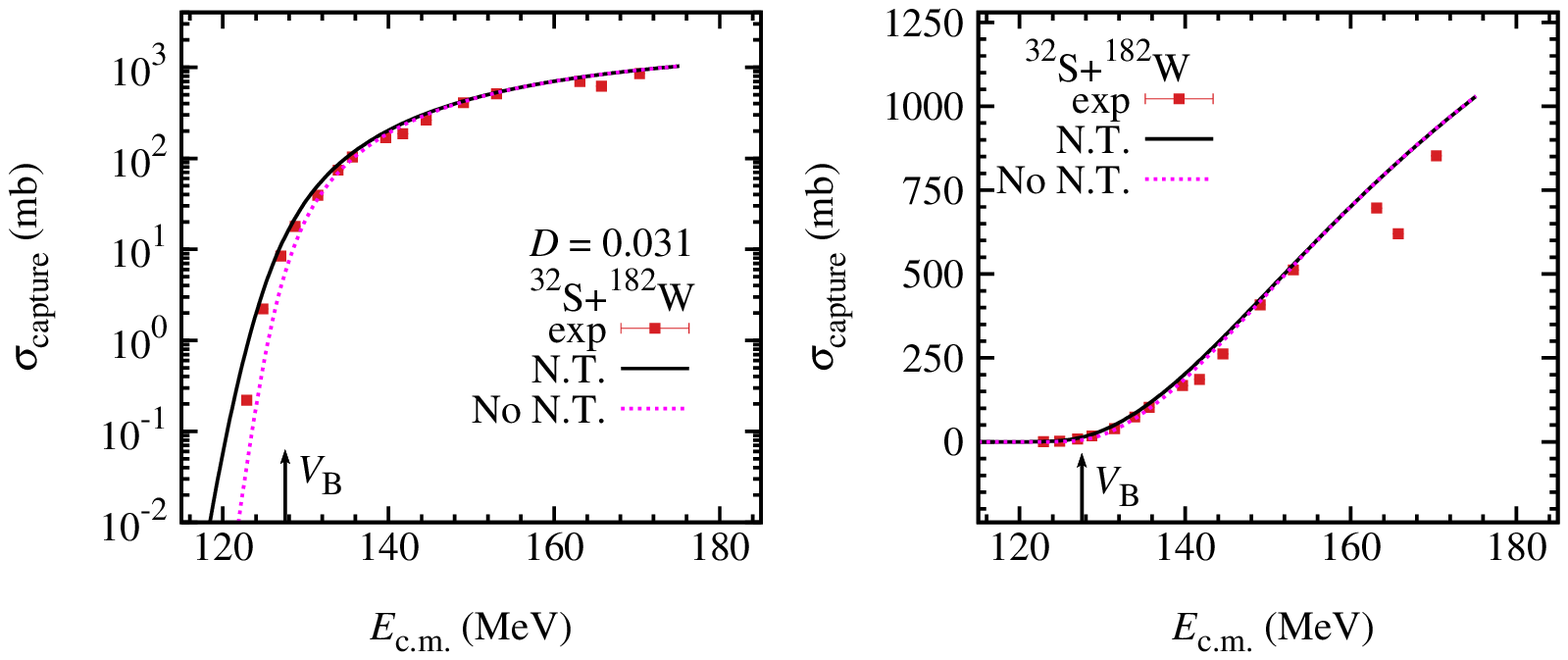}}
 \centerline{\includegraphics[width=0.47\textwidth]{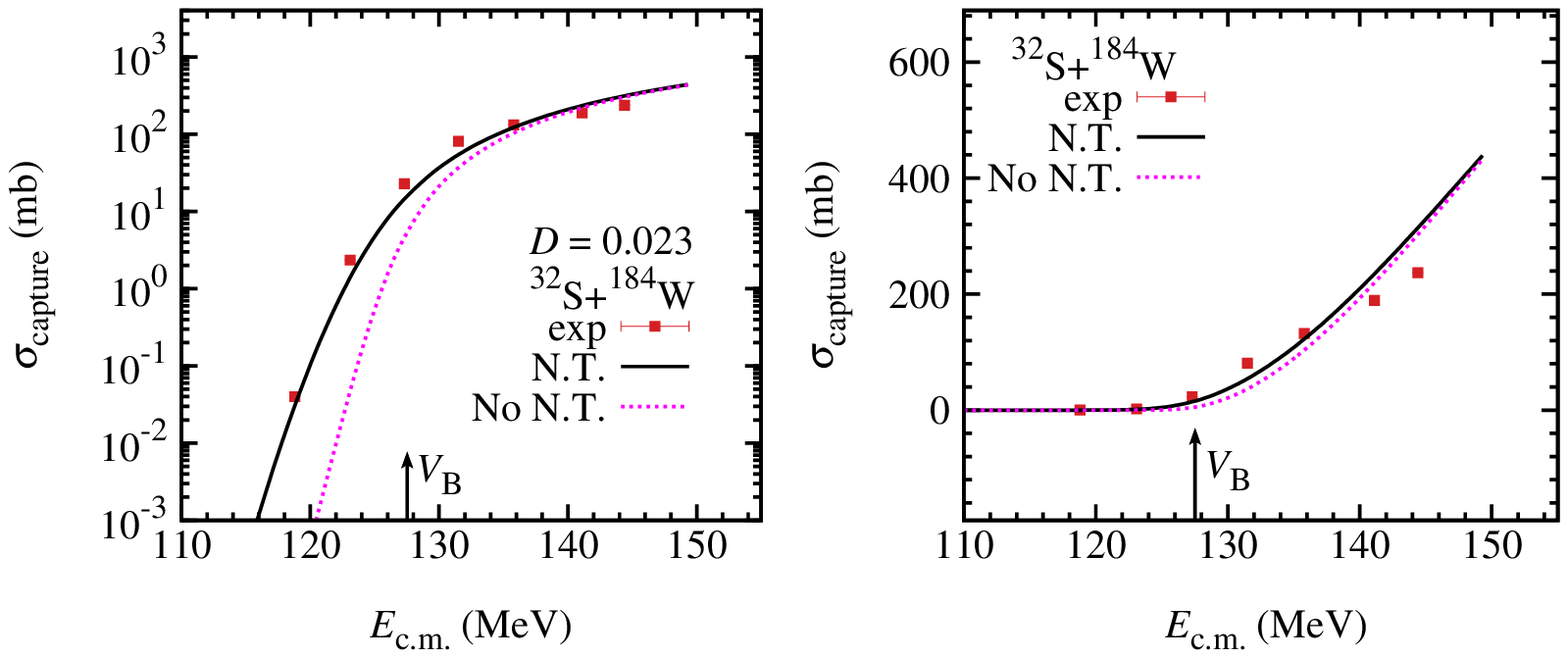}
  \includegraphics[width=0.47\textwidth]{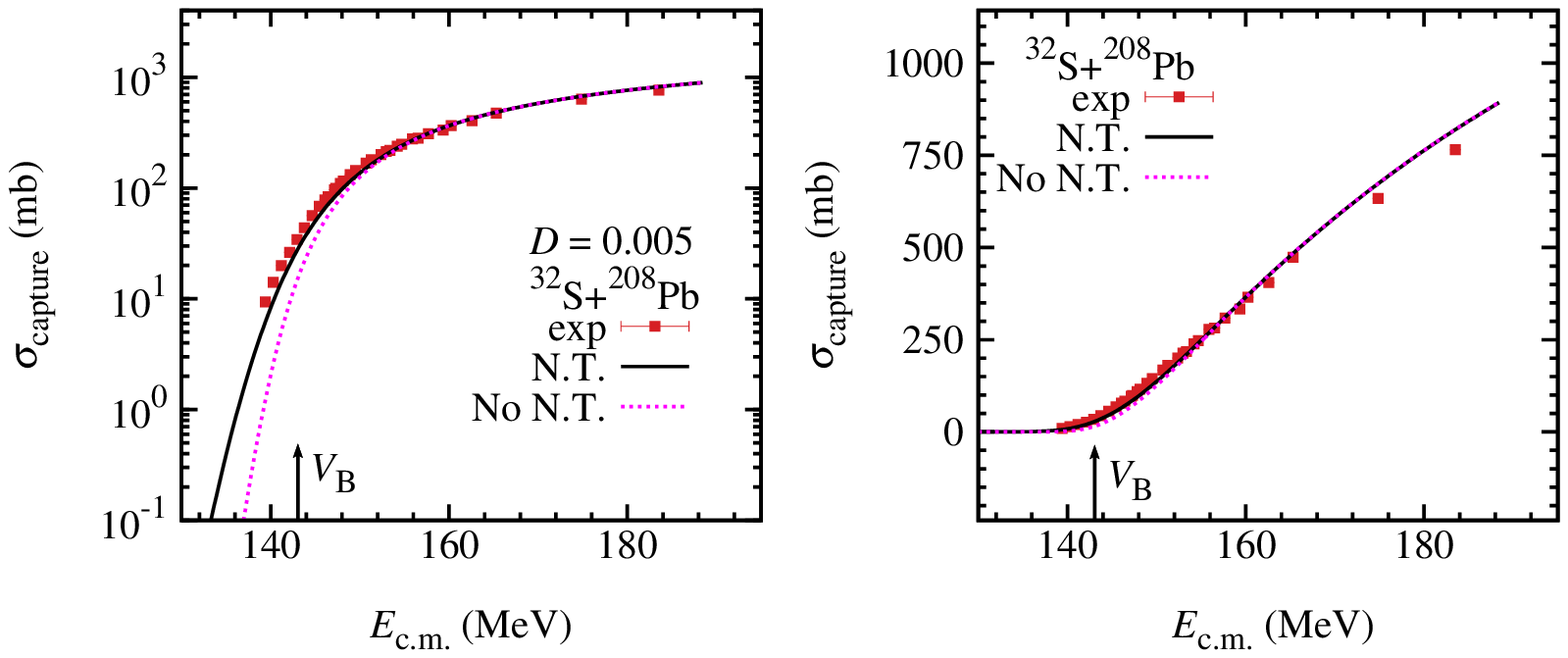}}
 \centerline{\includegraphics[width=0.47\textwidth]{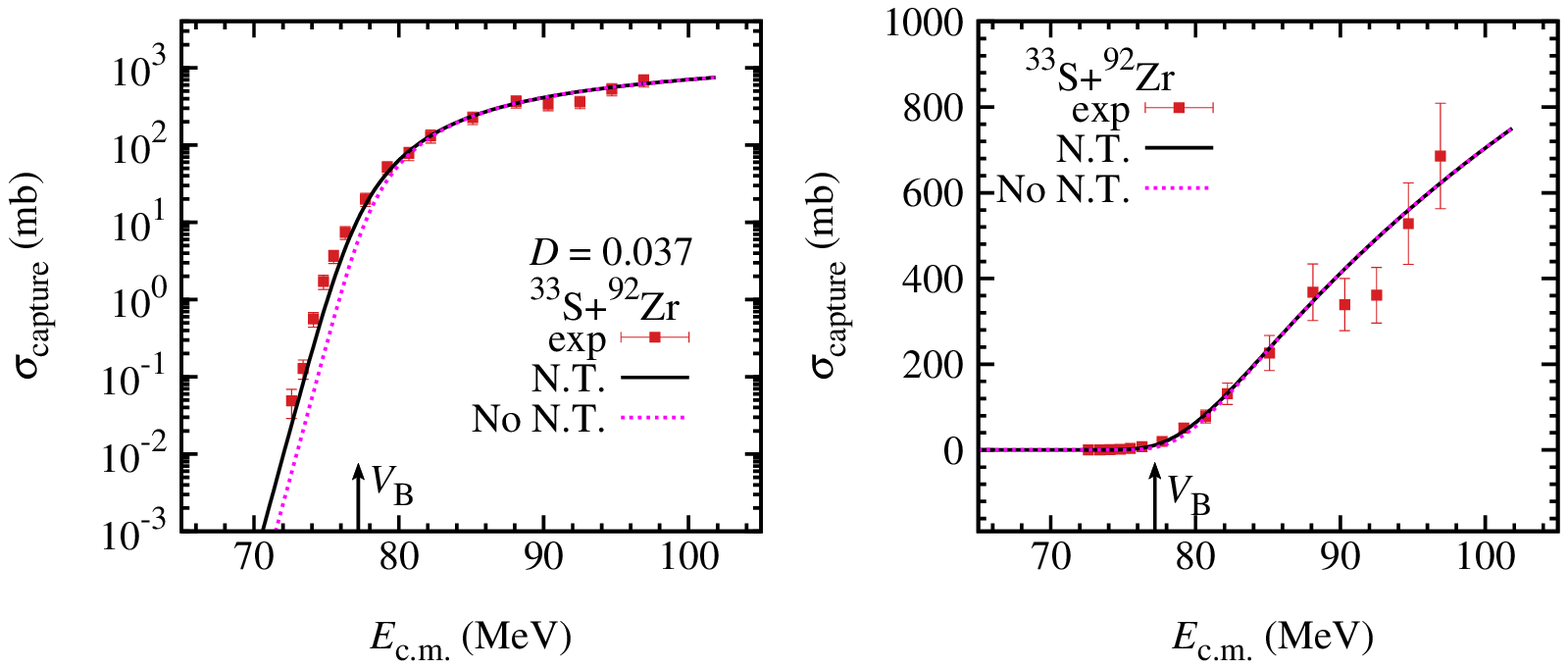}
  \includegraphics[width=0.47\textwidth]{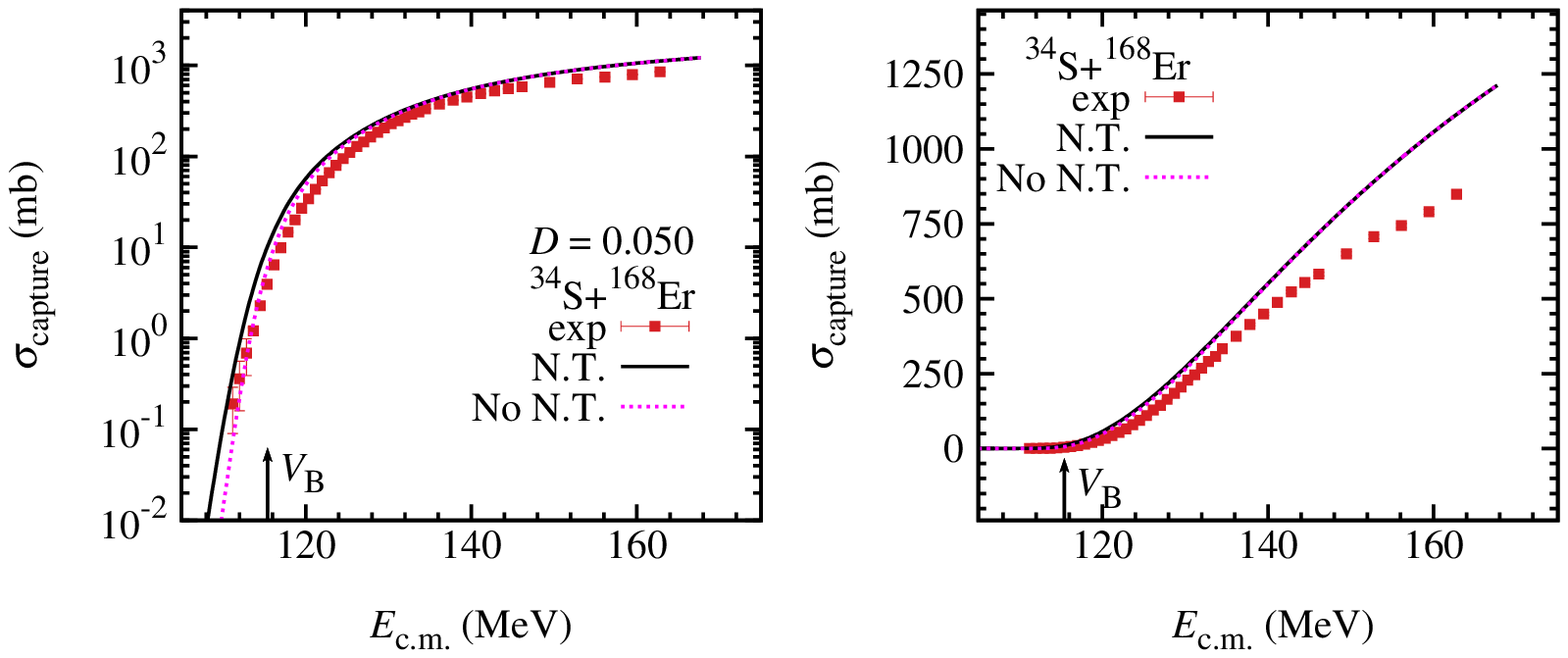}}
 \centerline{\includegraphics[width=0.47\textwidth]{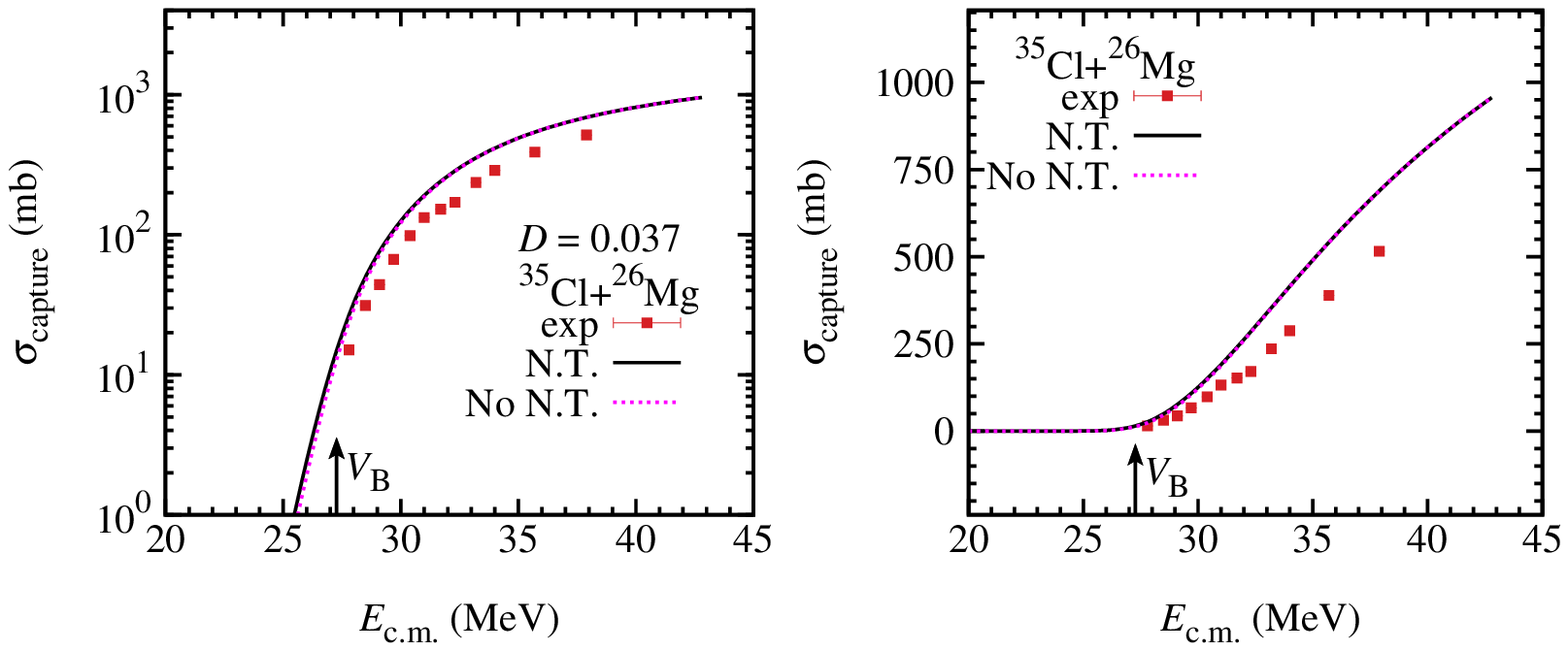}
  \includegraphics[width=0.47\textwidth]{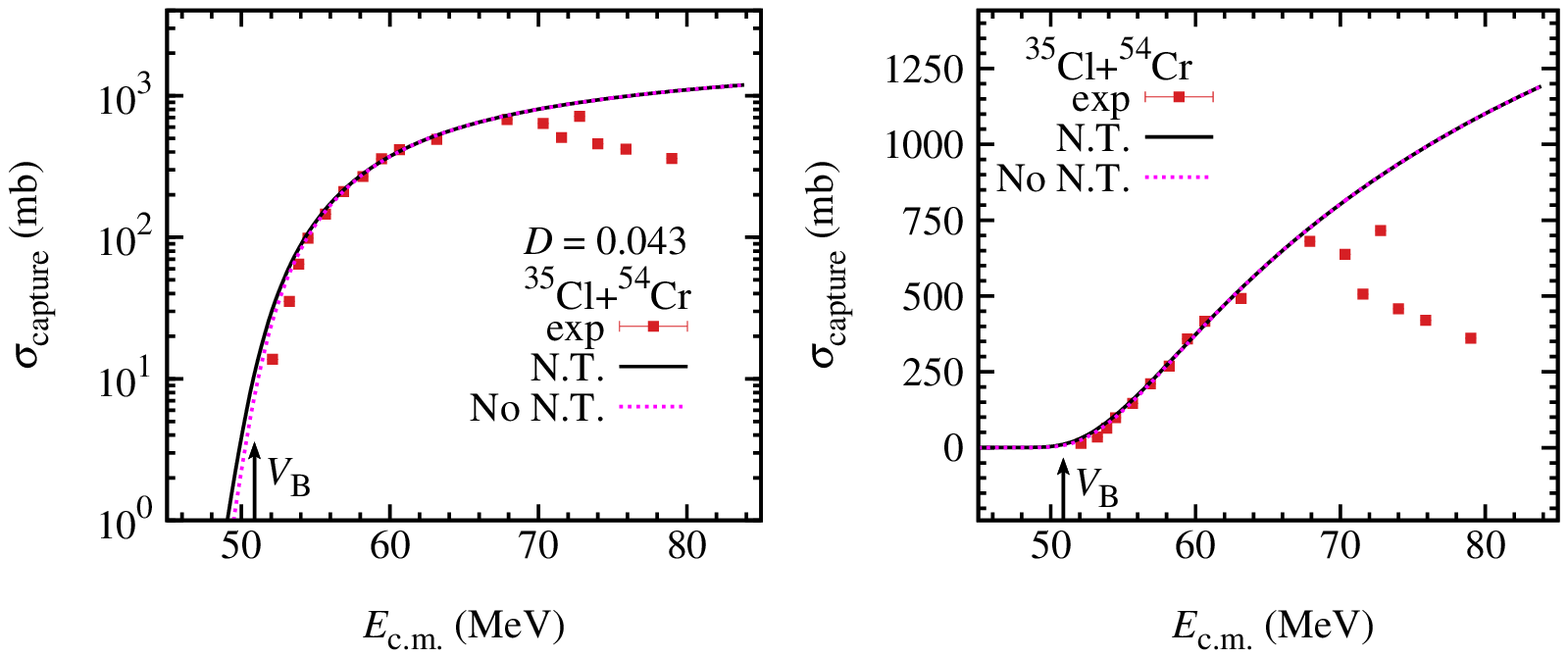}}
 \centerline{\includegraphics[width=0.47\textwidth]{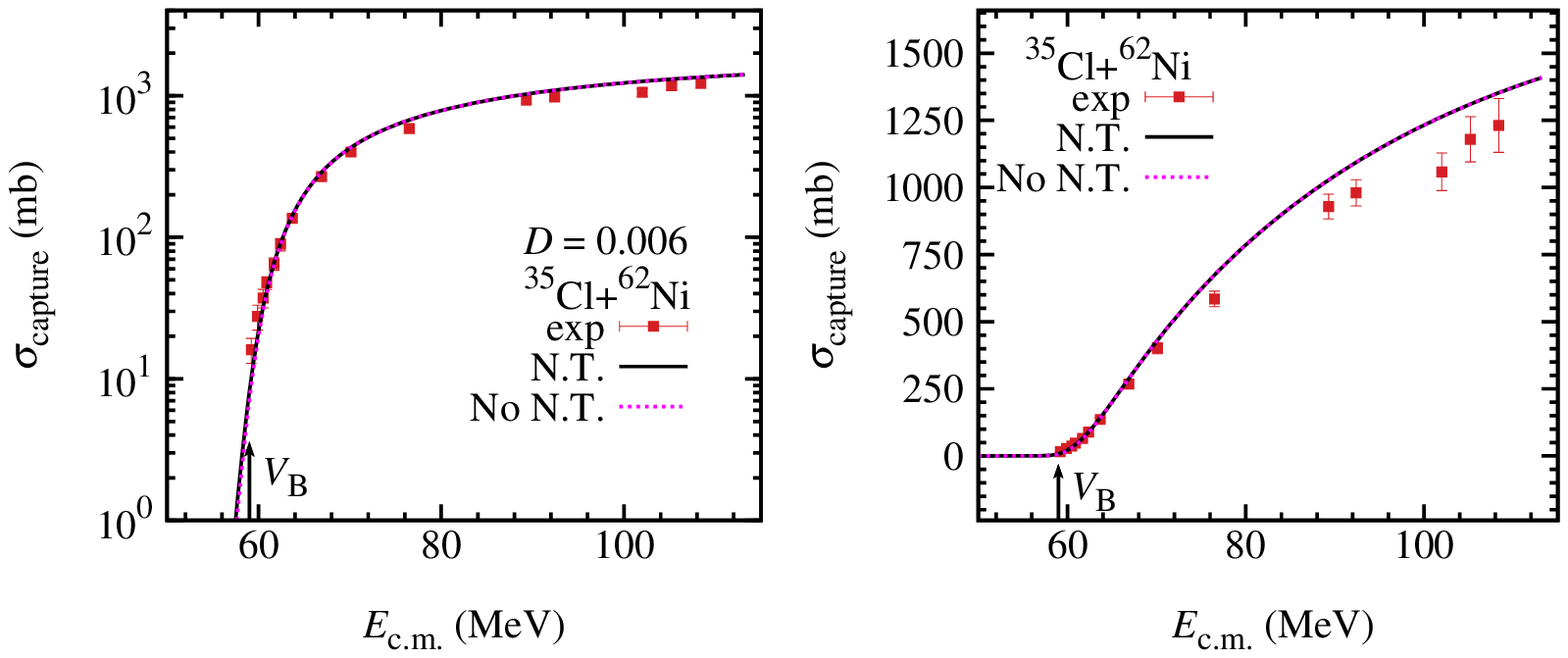}
  \includegraphics[width=0.47\textwidth]{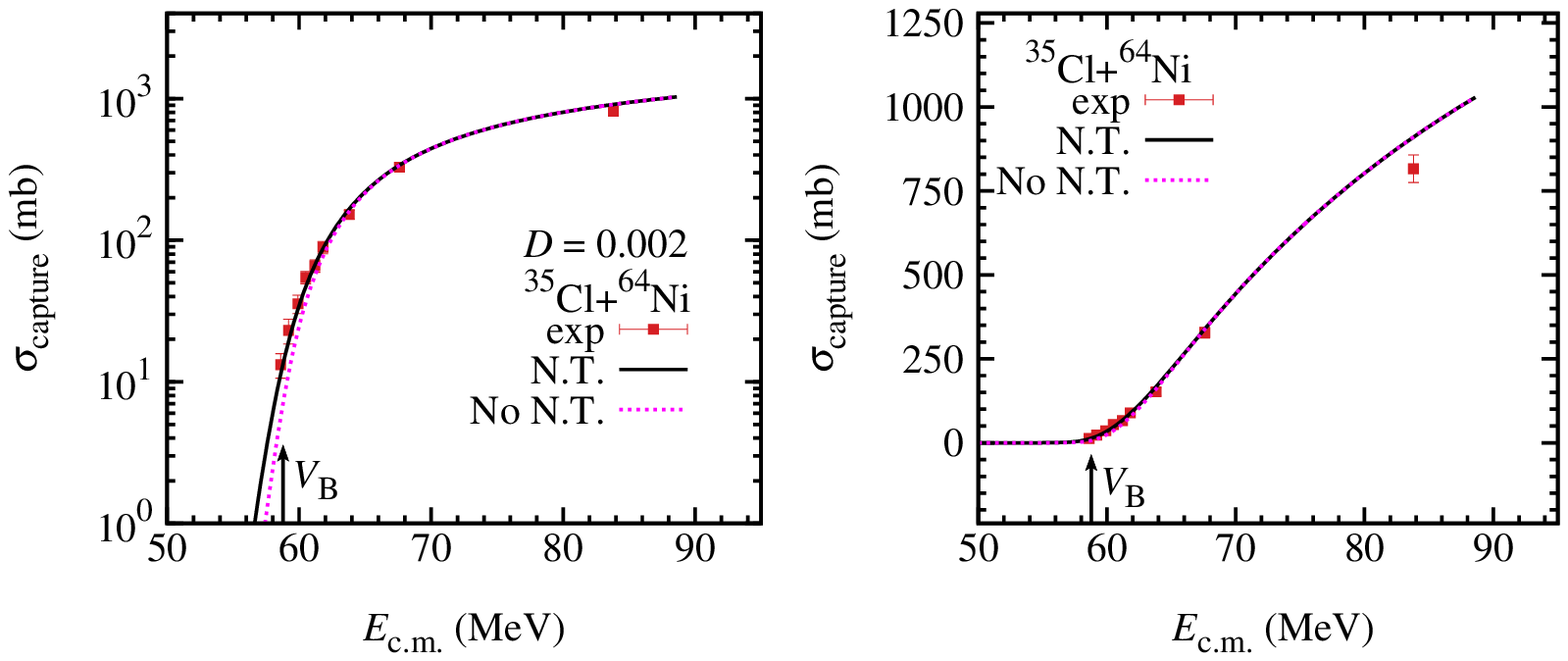}}
 \centerline{\includegraphics[width=0.47\textwidth]{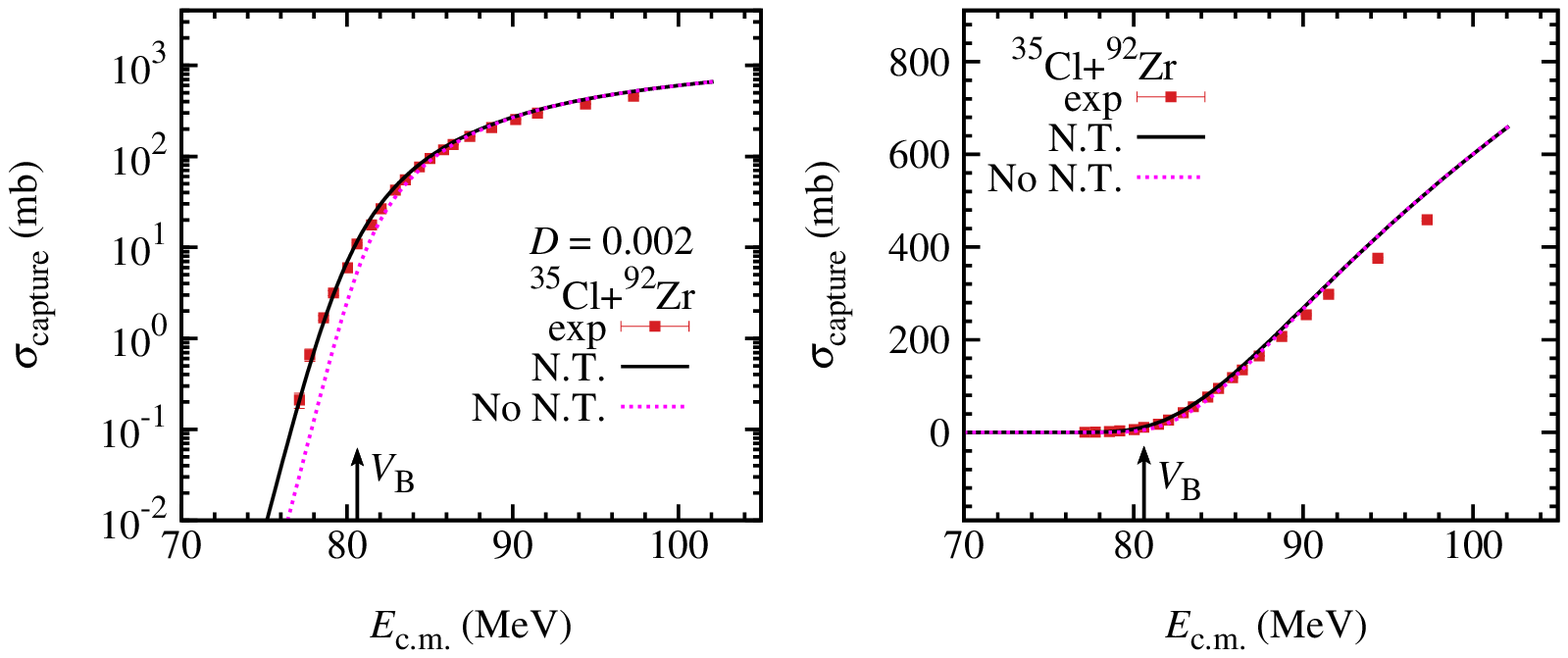}
  \includegraphics[width=0.47\textwidth]{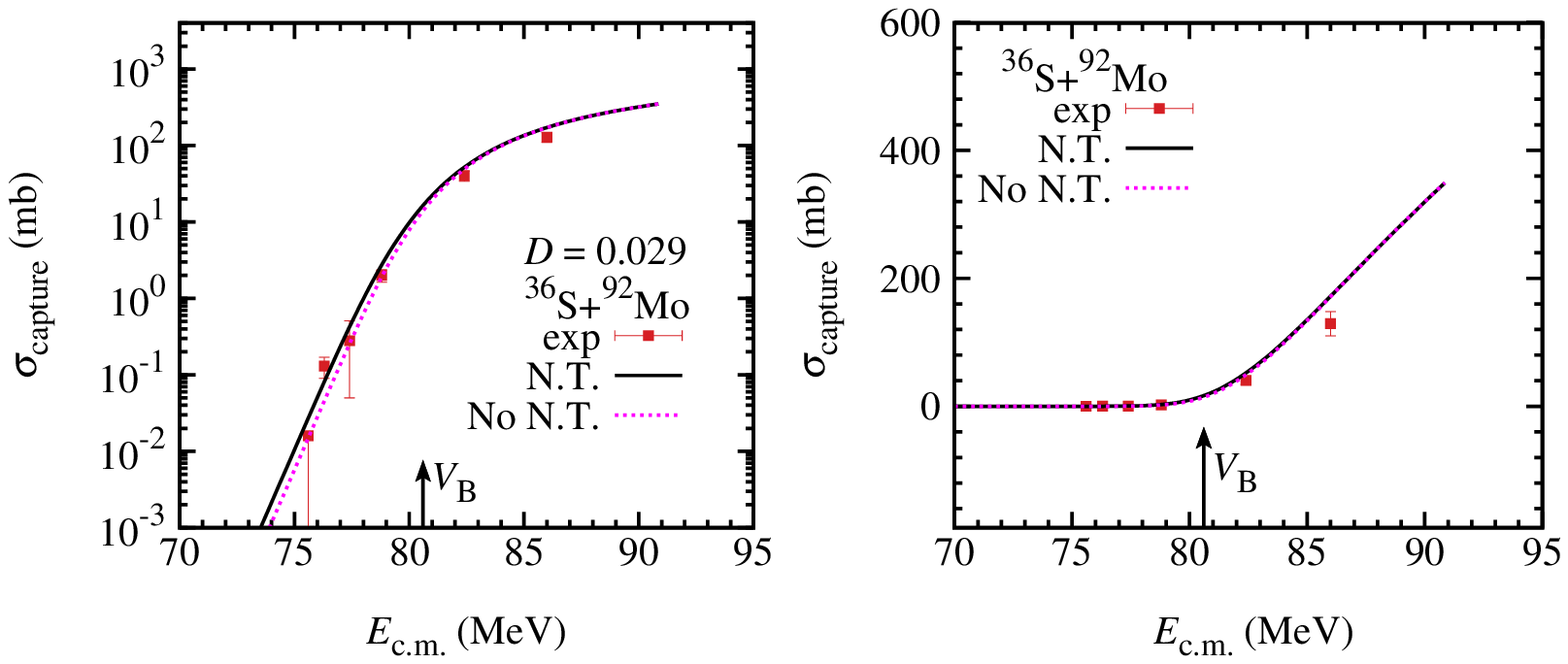}}
  \centerline{Graph 16}
 \end{Dfigures}
 \begin{Dfigures}[!ht]
 \centerline{\includegraphics[width=0.47\textwidth]{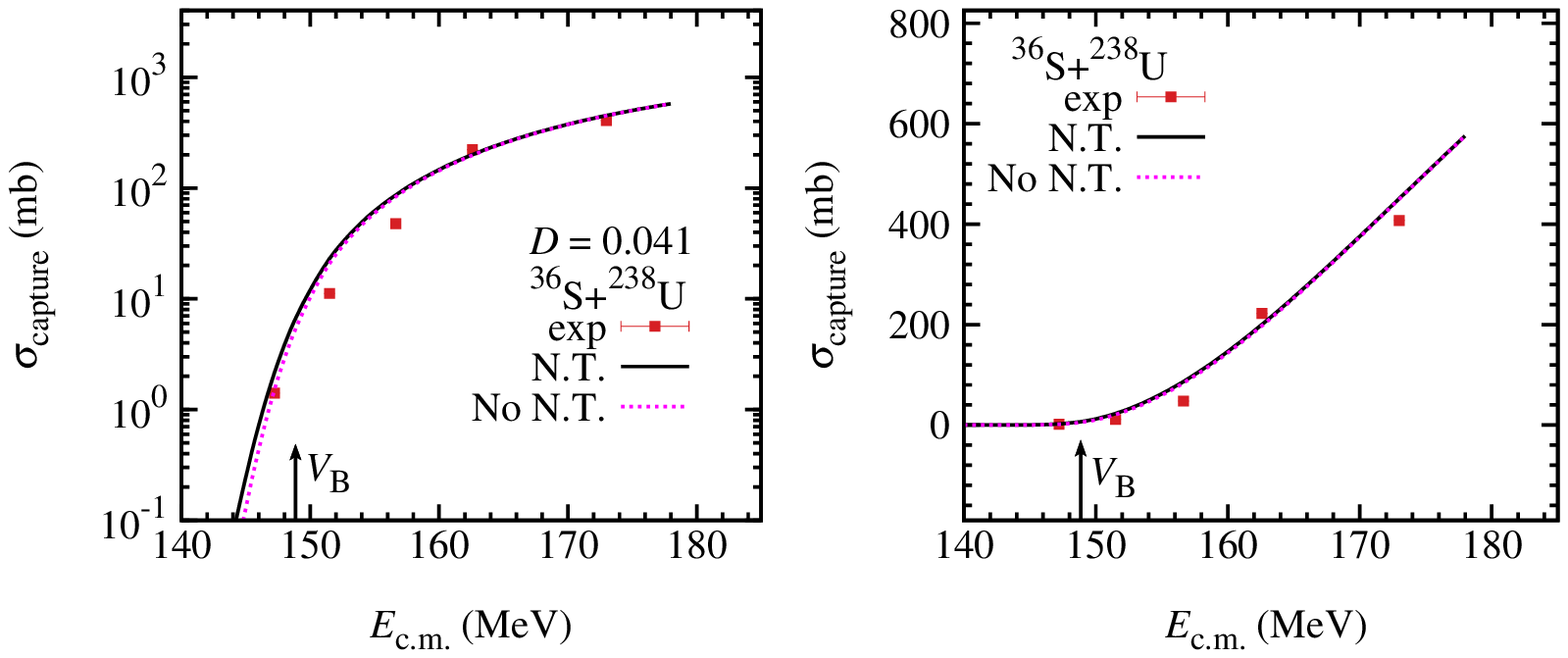}
  \includegraphics[width=0.47\textwidth]{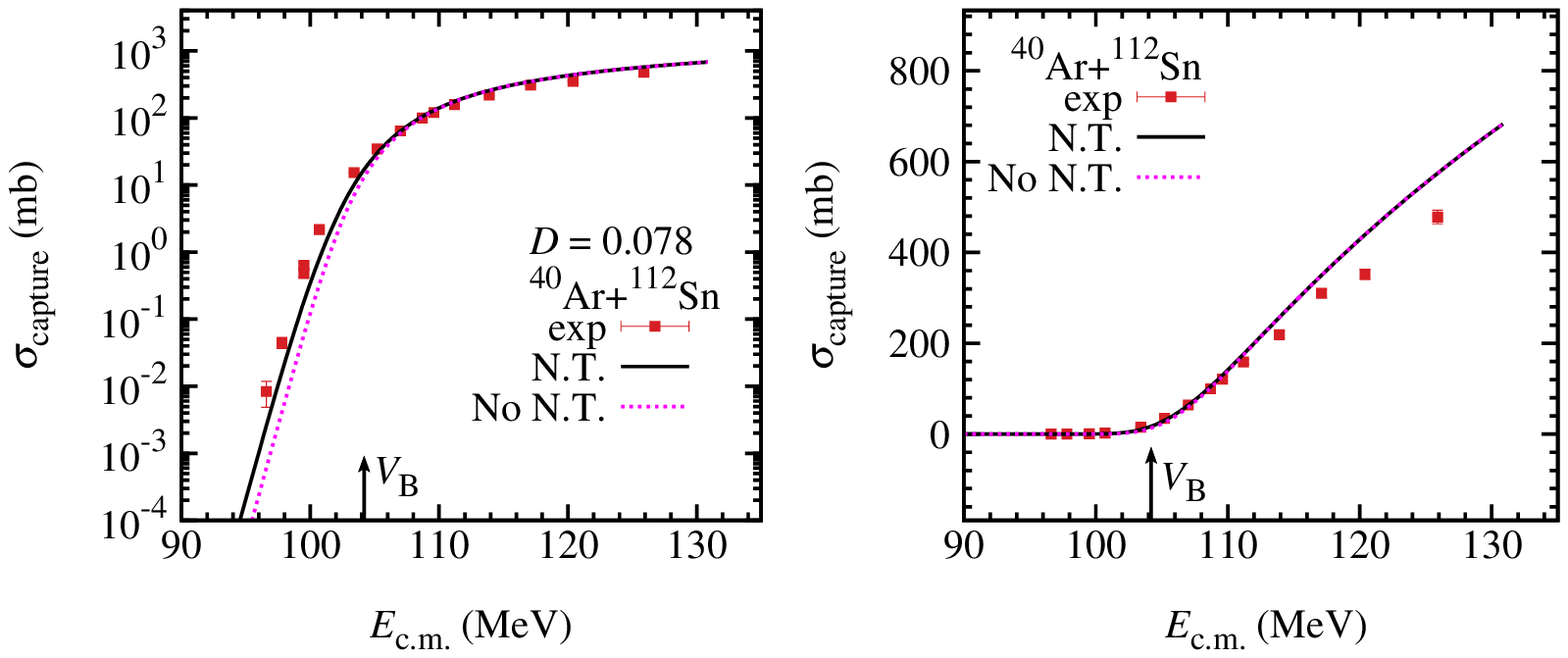}}
 \centerline{\includegraphics[width=0.47\textwidth]{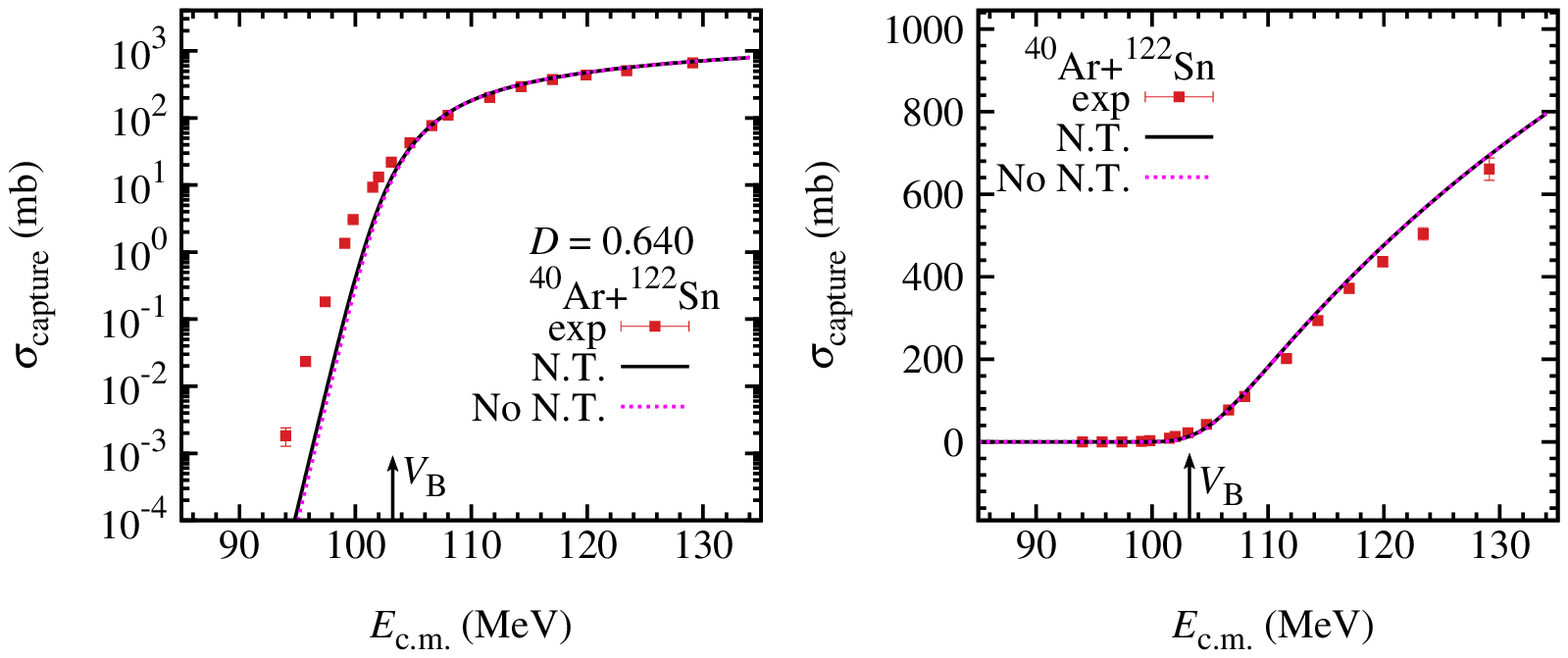}
  \includegraphics[width=0.47\textwidth]{40Ar148Sm.eps}}
 \centerline{\includegraphics[width=0.47\textwidth]{40Ar154Sm.eps}
  \includegraphics[width=0.47\textwidth]{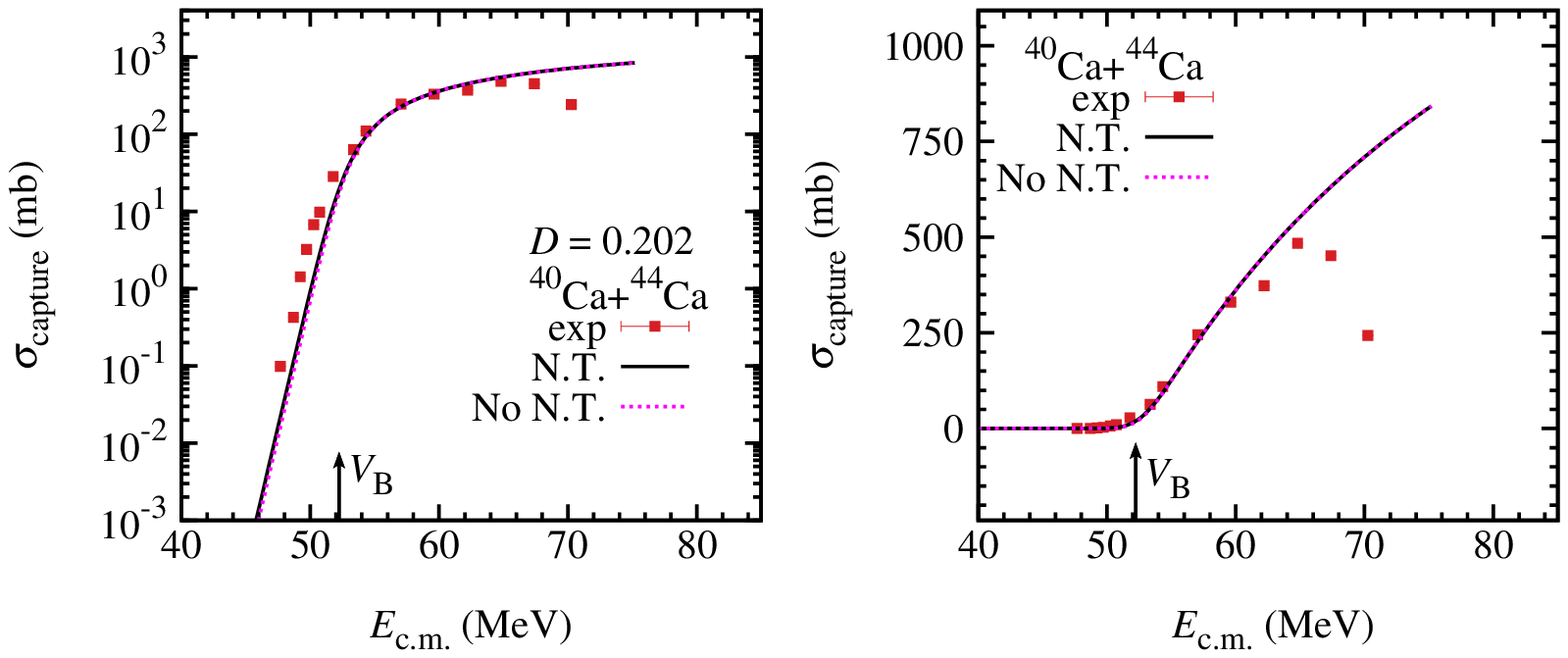}}
 \centerline{\includegraphics[width=0.47\textwidth]{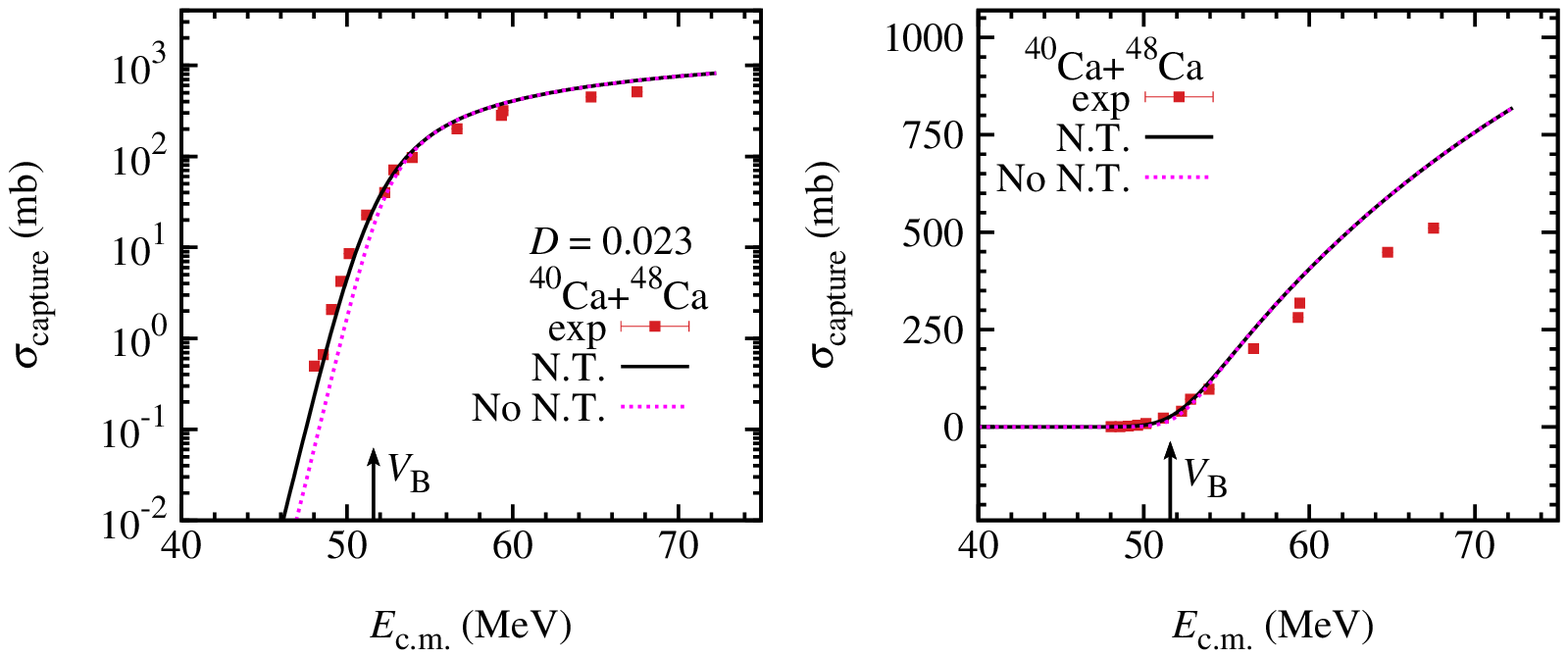}
  \includegraphics[width=0.47\textwidth]{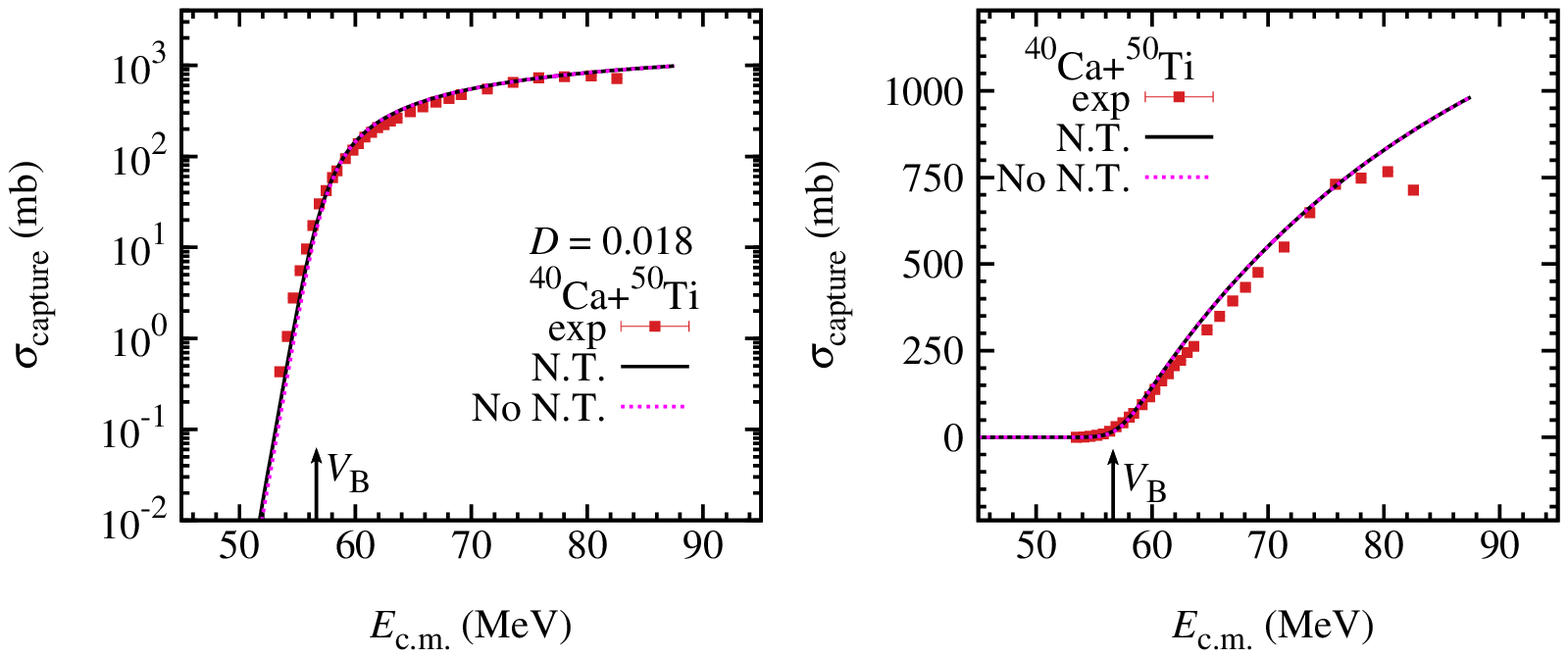}}
 \centerline{\includegraphics[width=0.47\textwidth]{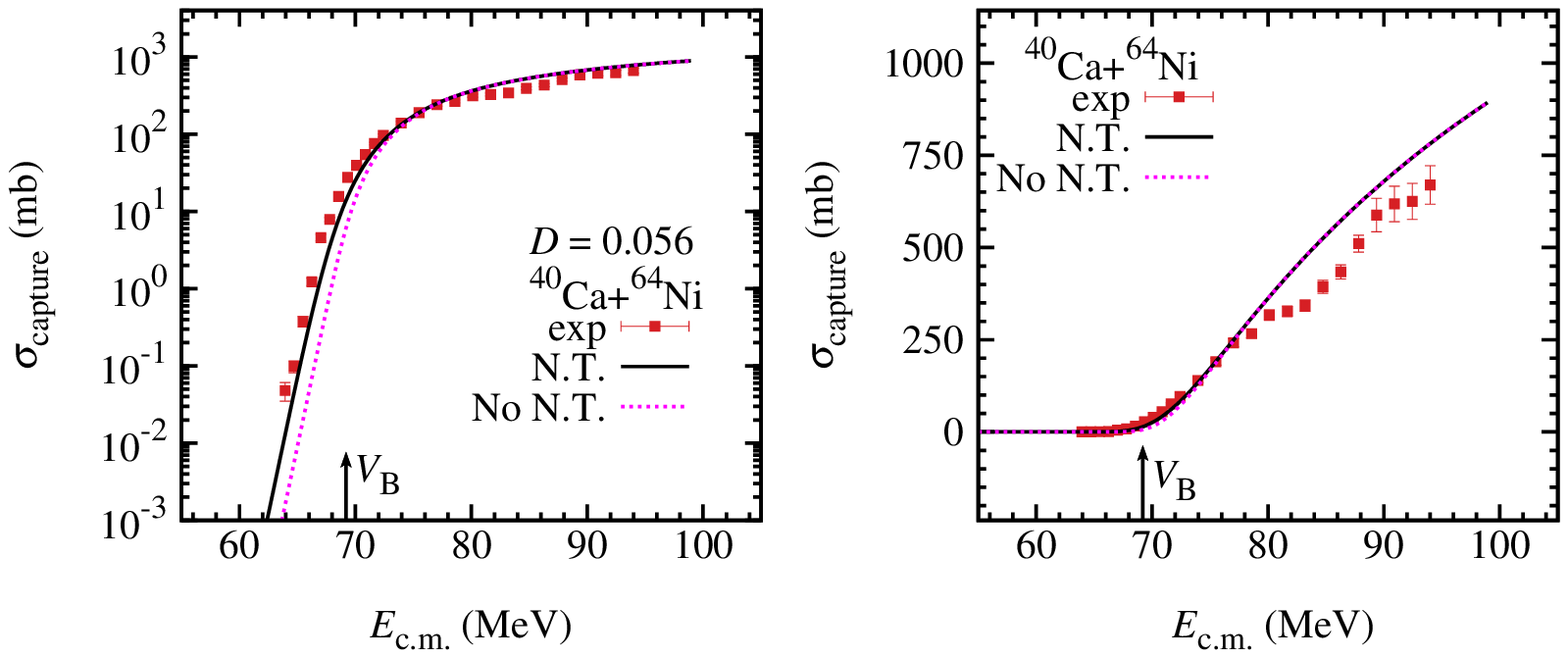}
  \includegraphics[width=0.47\textwidth]{40Ca94Zr.eps}}
 \centerline{\includegraphics[width=0.47\textwidth]{40Ca96Zr.eps}
  \includegraphics[width=0.47\textwidth]{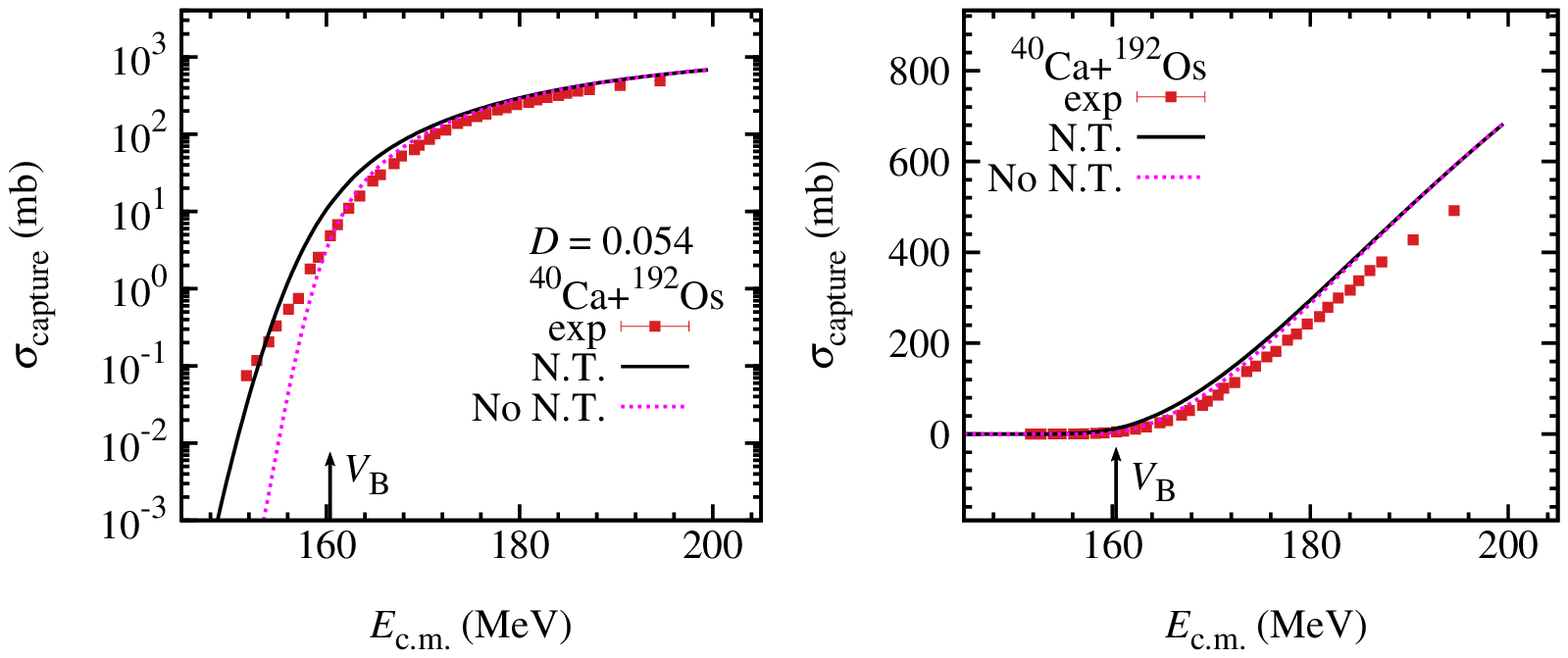}}
  \centerline{Graph 17}
 \end{Dfigures}
 \begin{Dfigures}[!ht]
 \centerline{\includegraphics[width=0.47\textwidth]{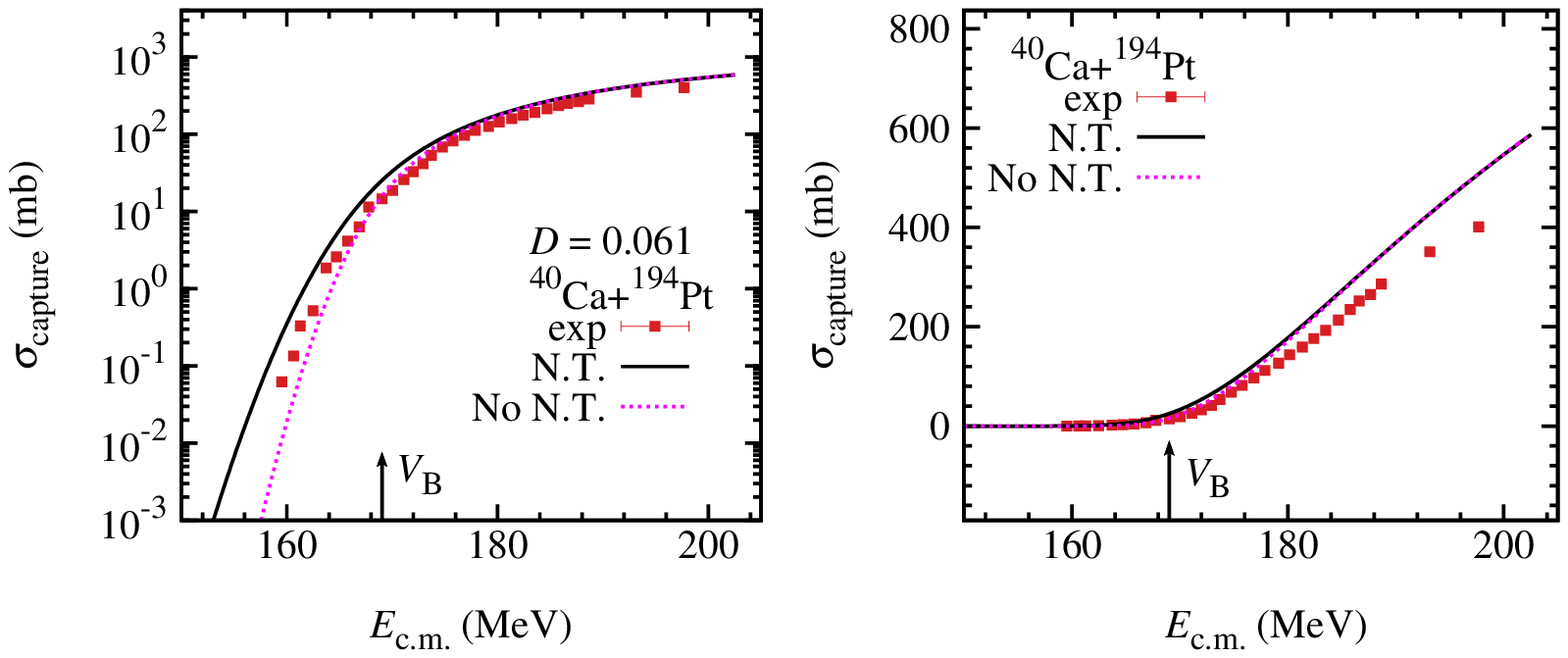}
  \includegraphics[width=0.47\textwidth]{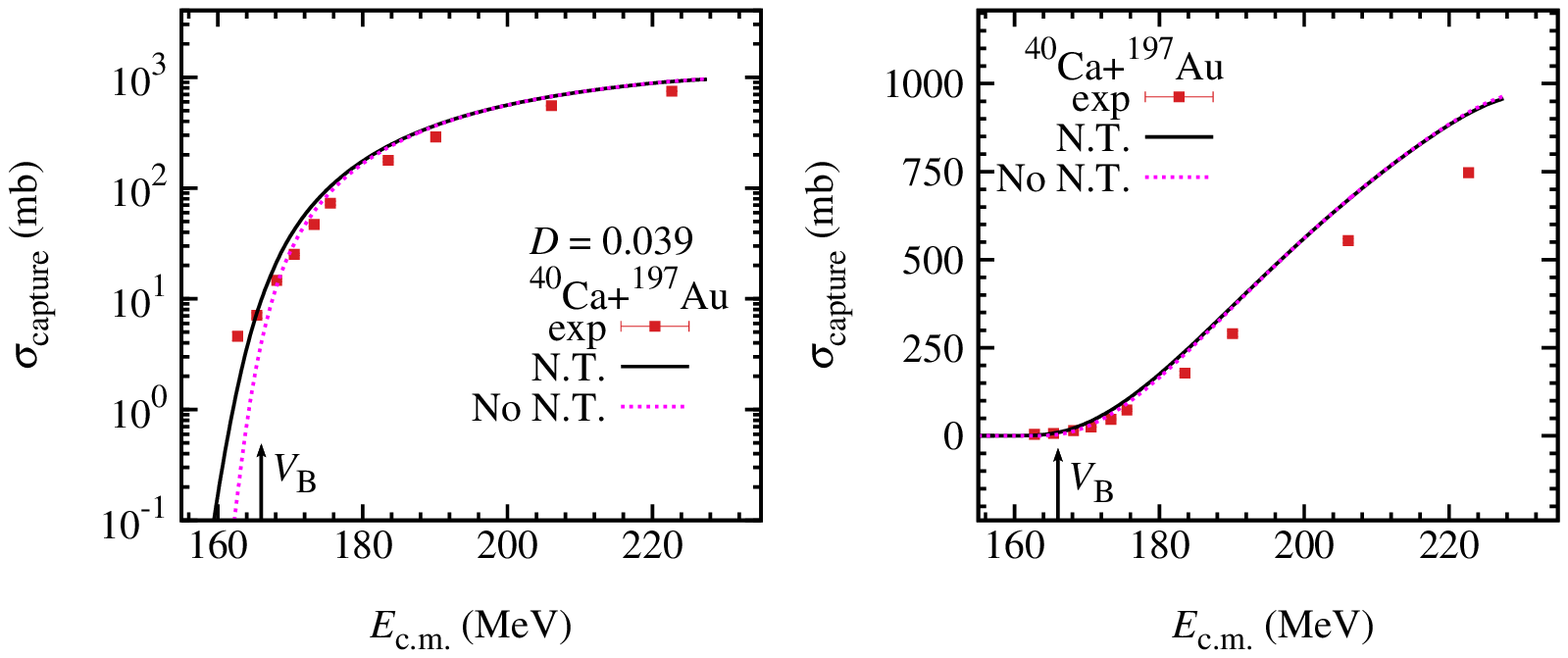}}
 \centerline{\includegraphics[width=0.47\textwidth]{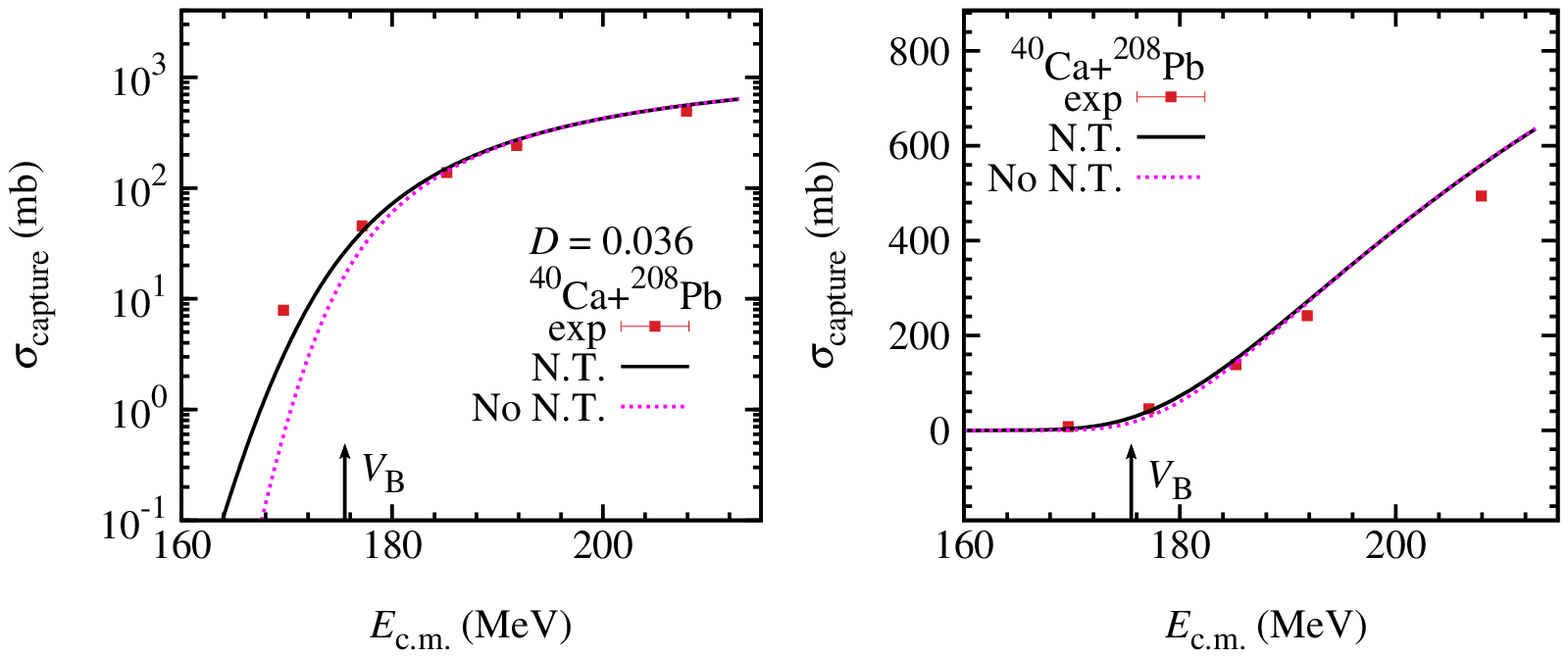}
 \includegraphics[width=0.47\textwidth]{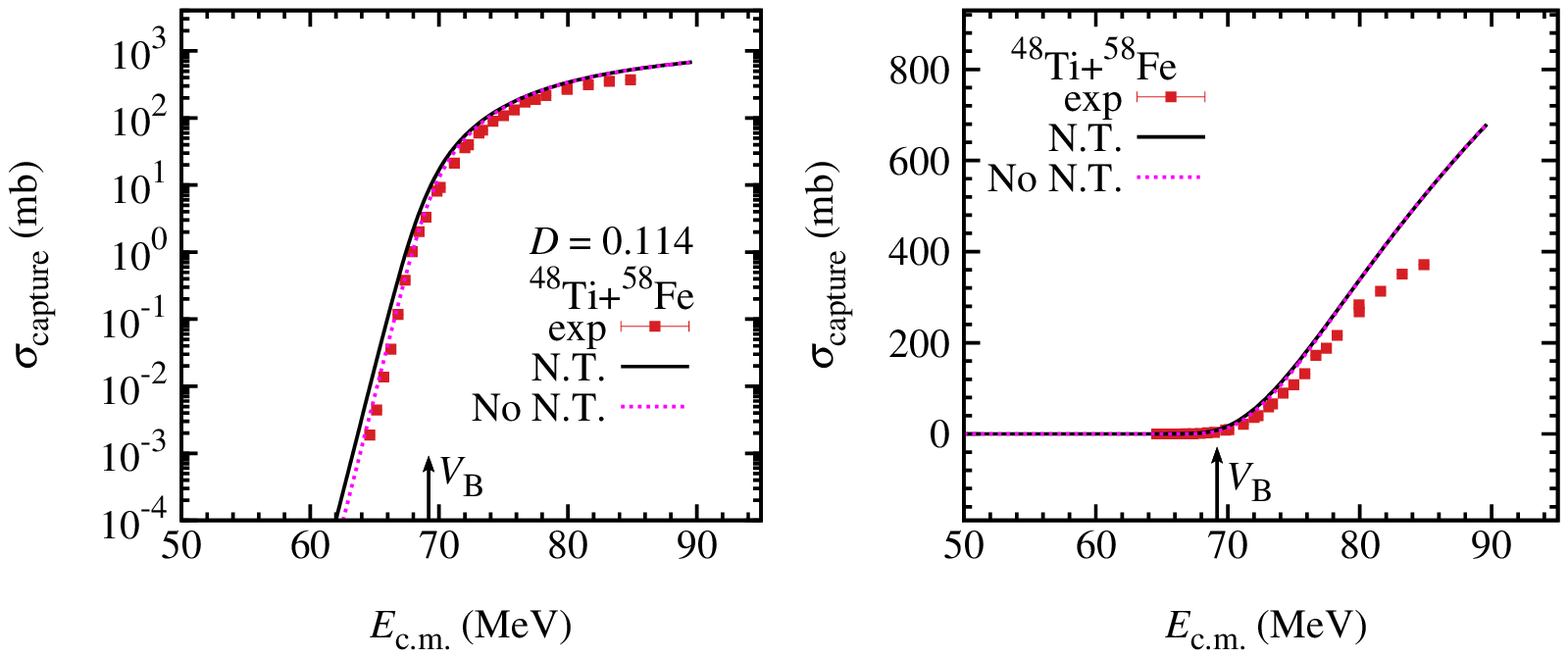}}
\centerline{\includegraphics[width=0.47\textwidth]{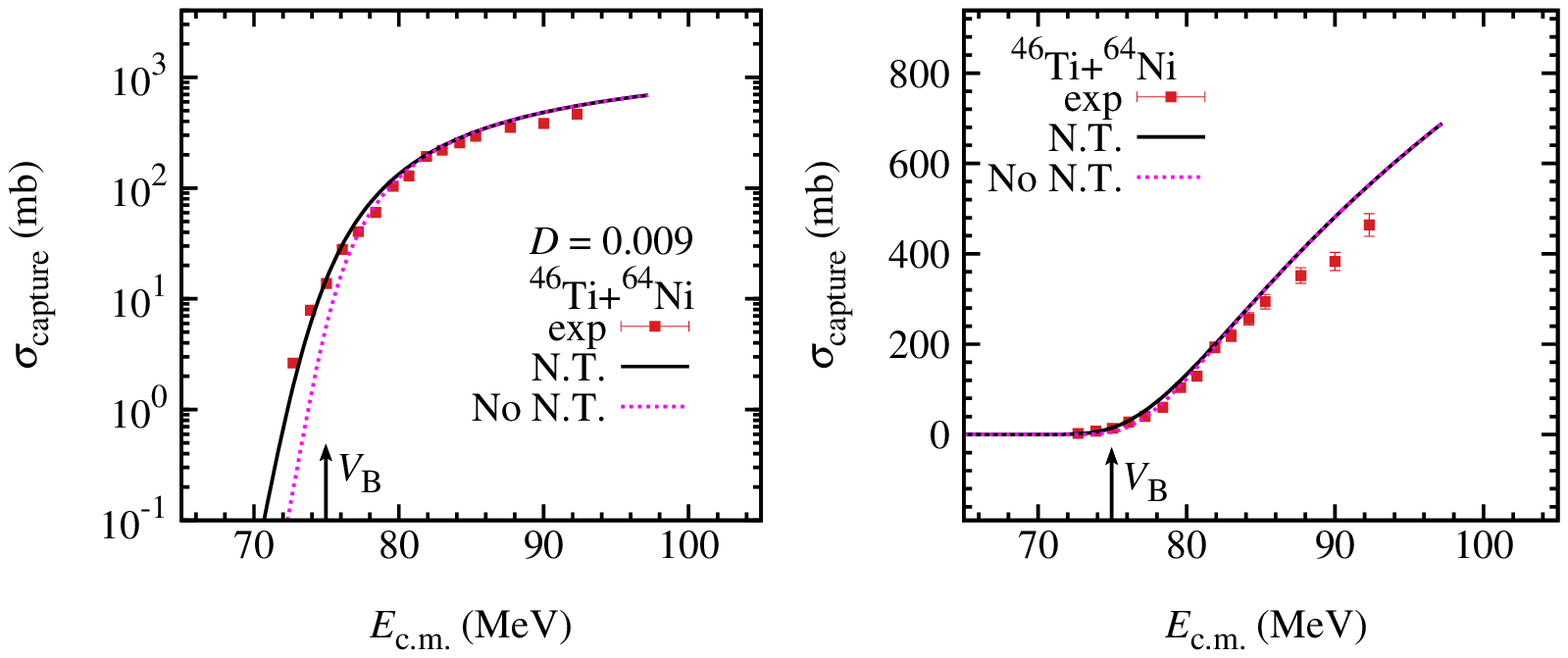}
\includegraphics[width=0.47\textwidth]{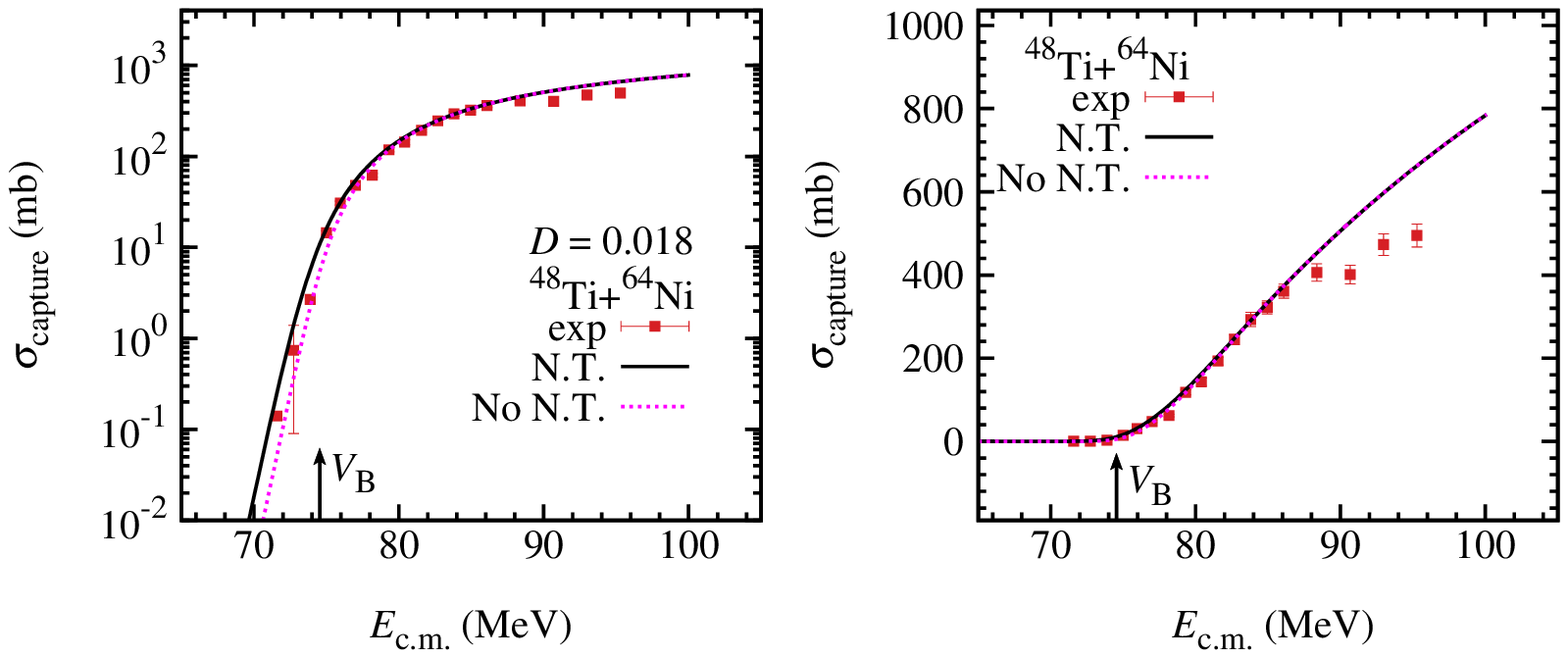}}
 \centerline{\includegraphics[width=0.47\textwidth]{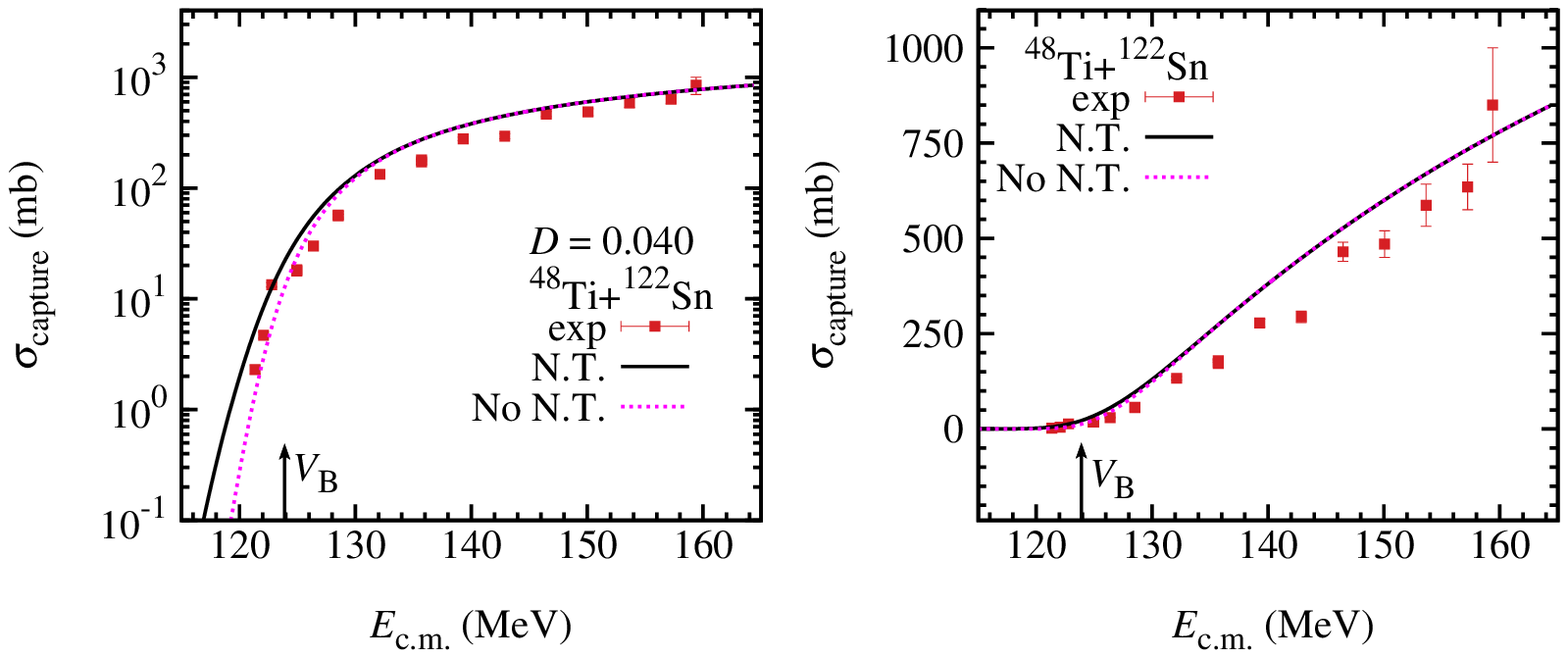}
\includegraphics[width=0.47\textwidth]{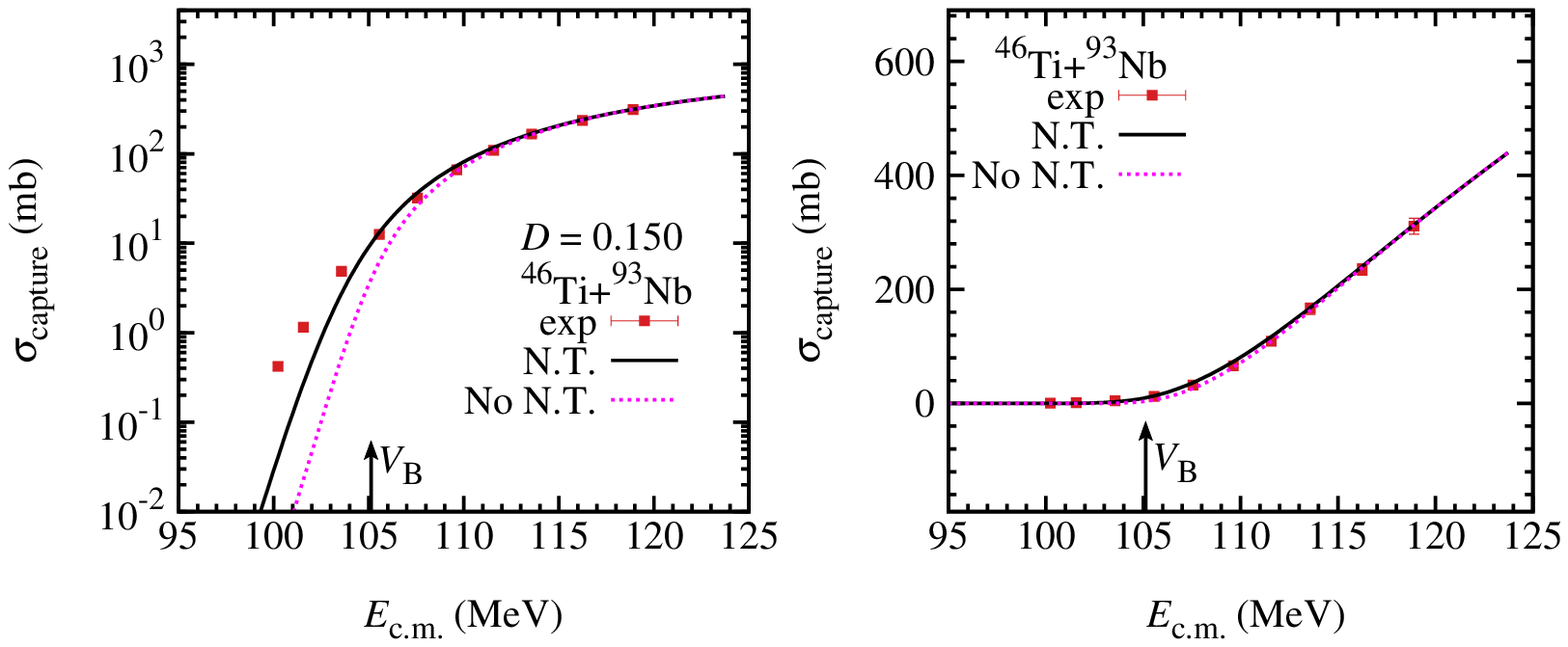}}
\centerline{\includegraphics[width=0.47\textwidth]{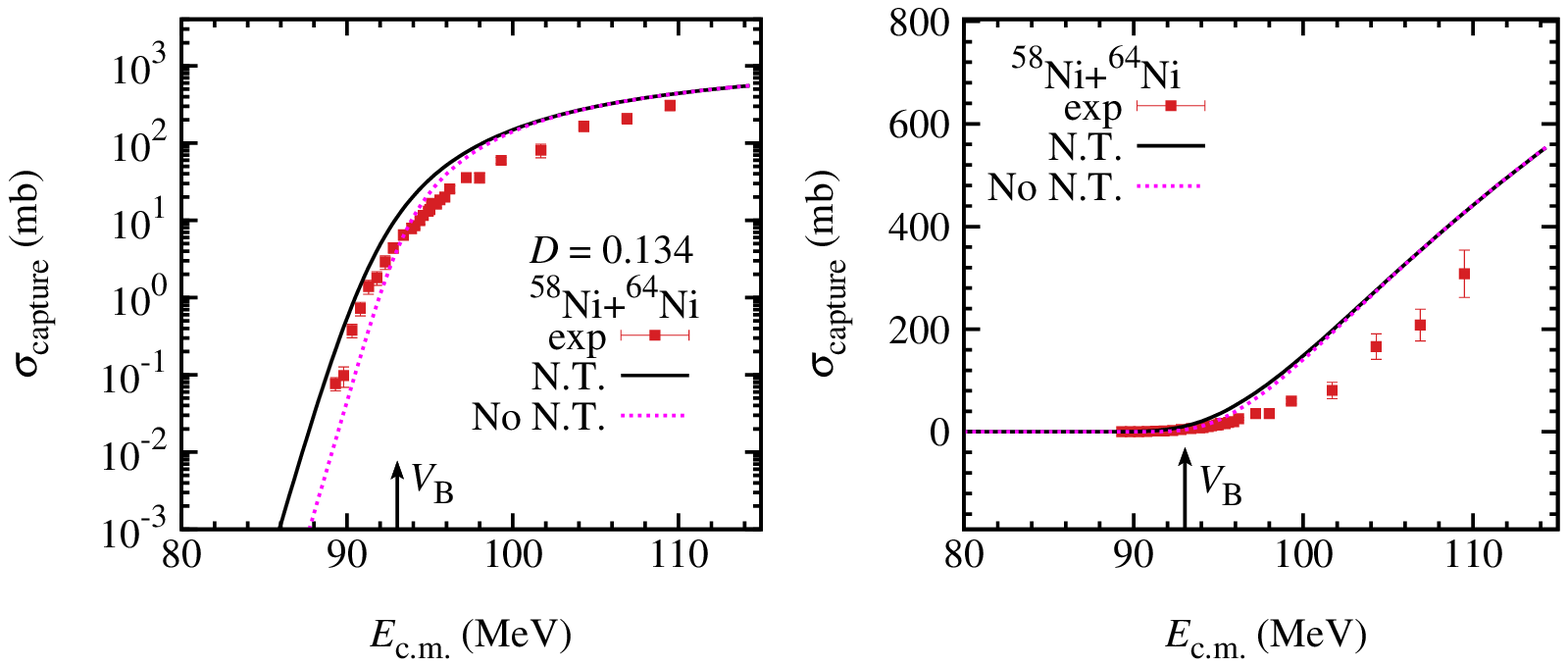}
\includegraphics[width=0.47\textwidth]{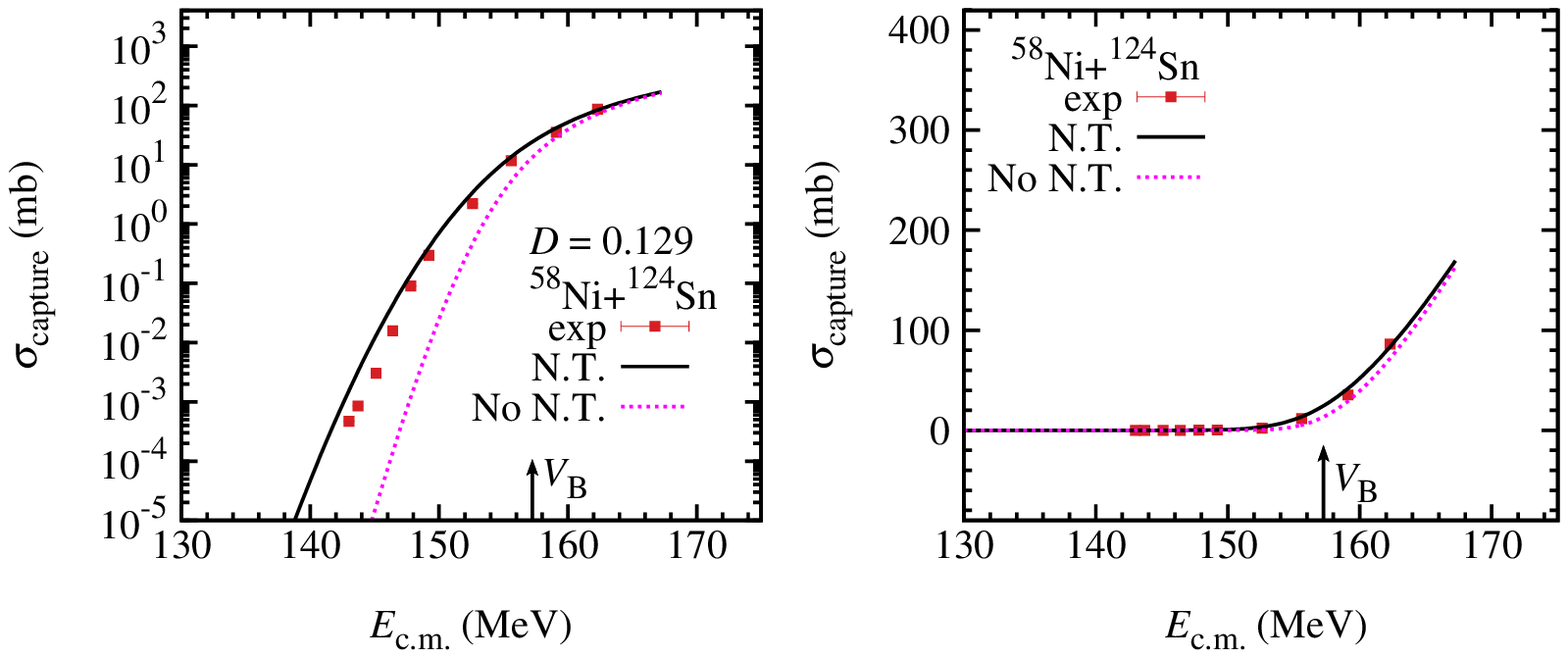}}
 \centerline{\includegraphics[width=0.47\textwidth]{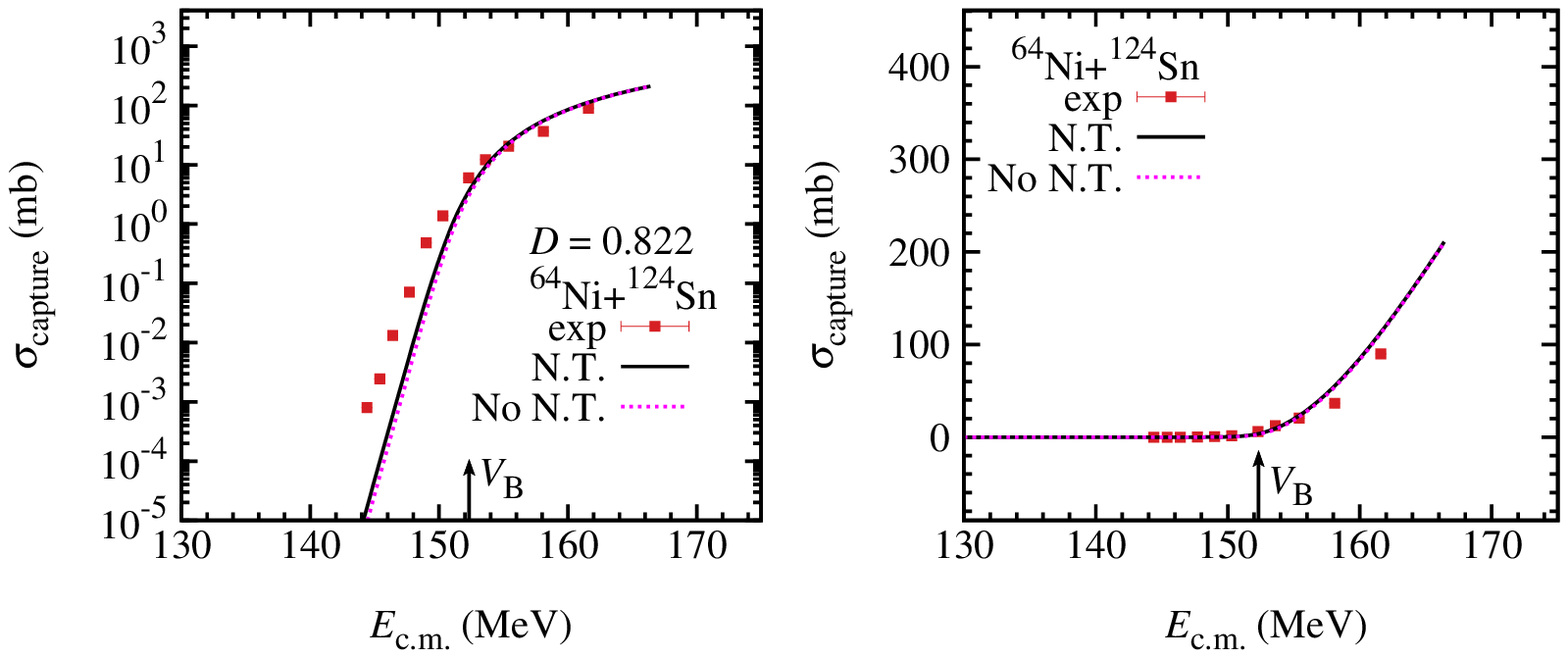}
\includegraphics[width=0.47\textwidth]{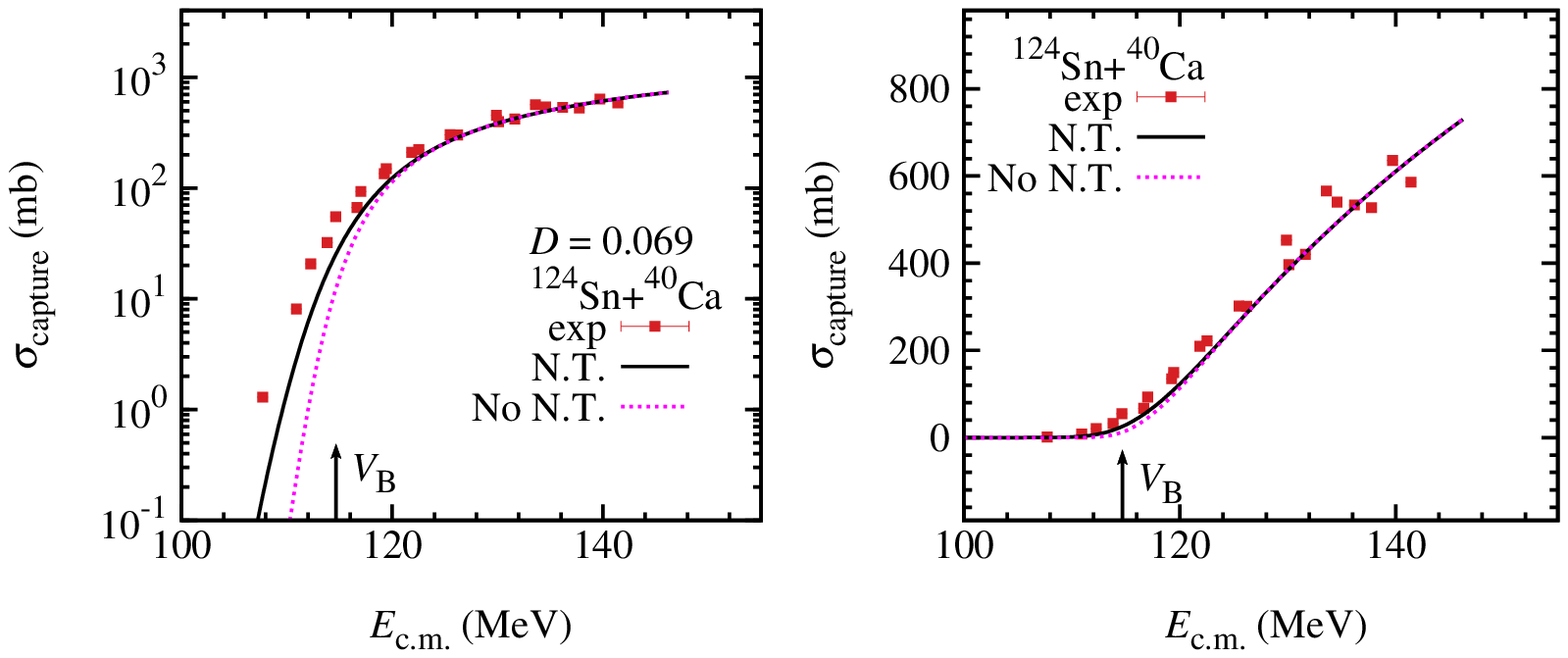}}
  \centerline{Graph 18}
 \end{Dfigures}
 \begin{Dfigures}[!ht]
  \centerline{\includegraphics[width=0.47\textwidth]{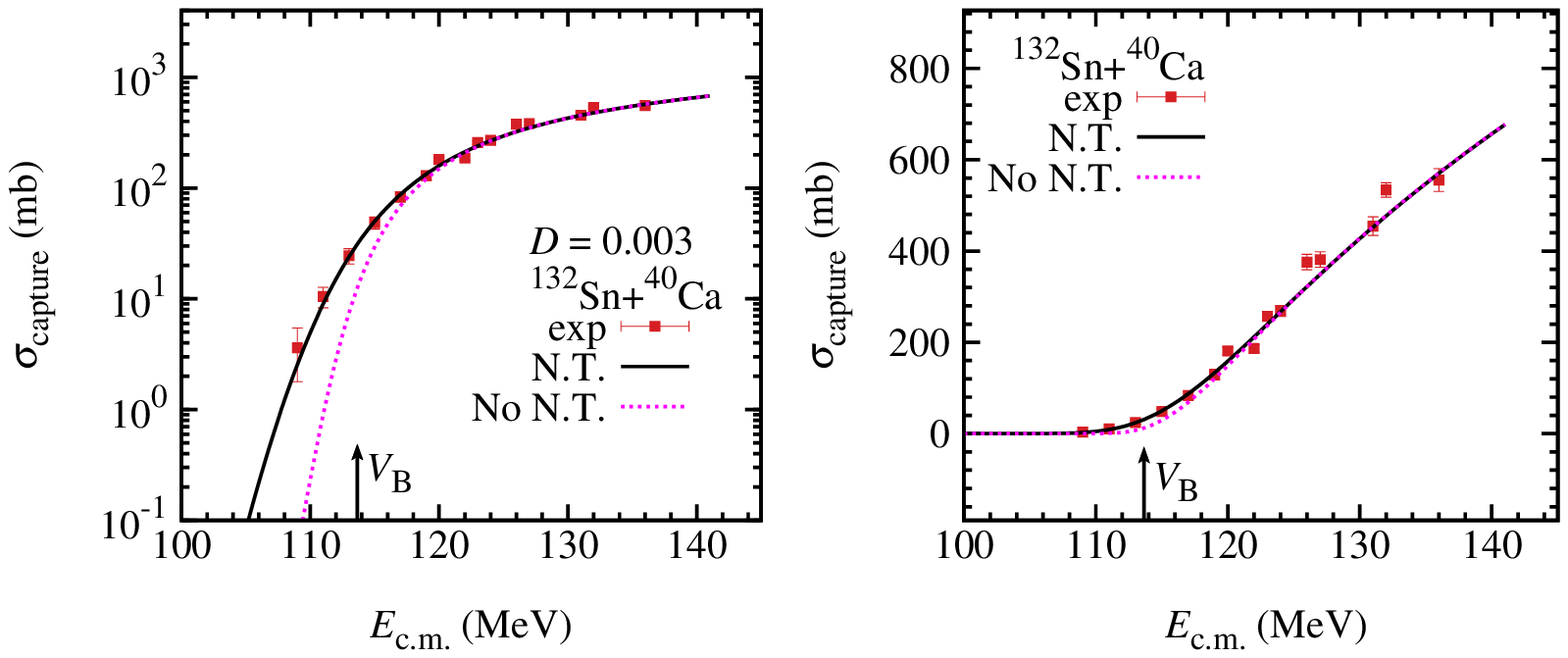}
  \includegraphics[width=0.47\textwidth]{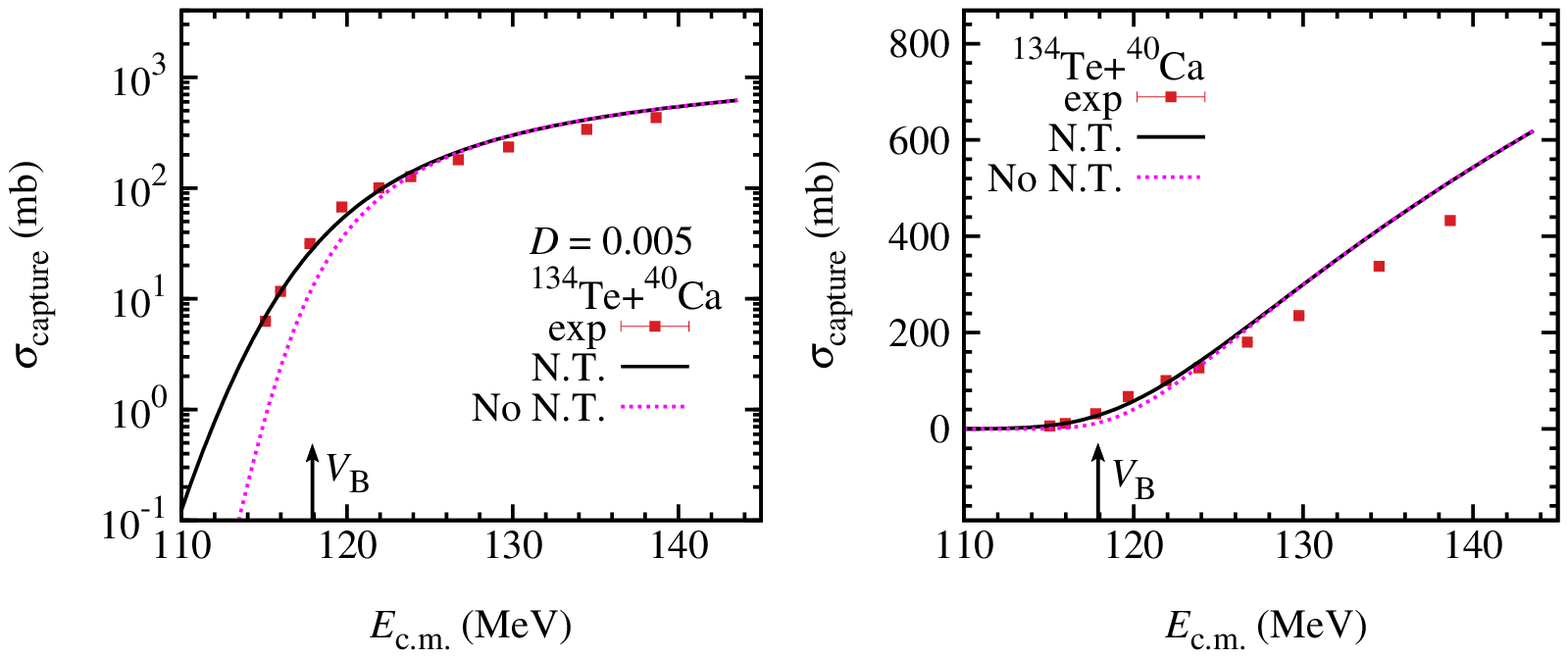}}
\centerline{\includegraphics[width=0.47\textwidth]{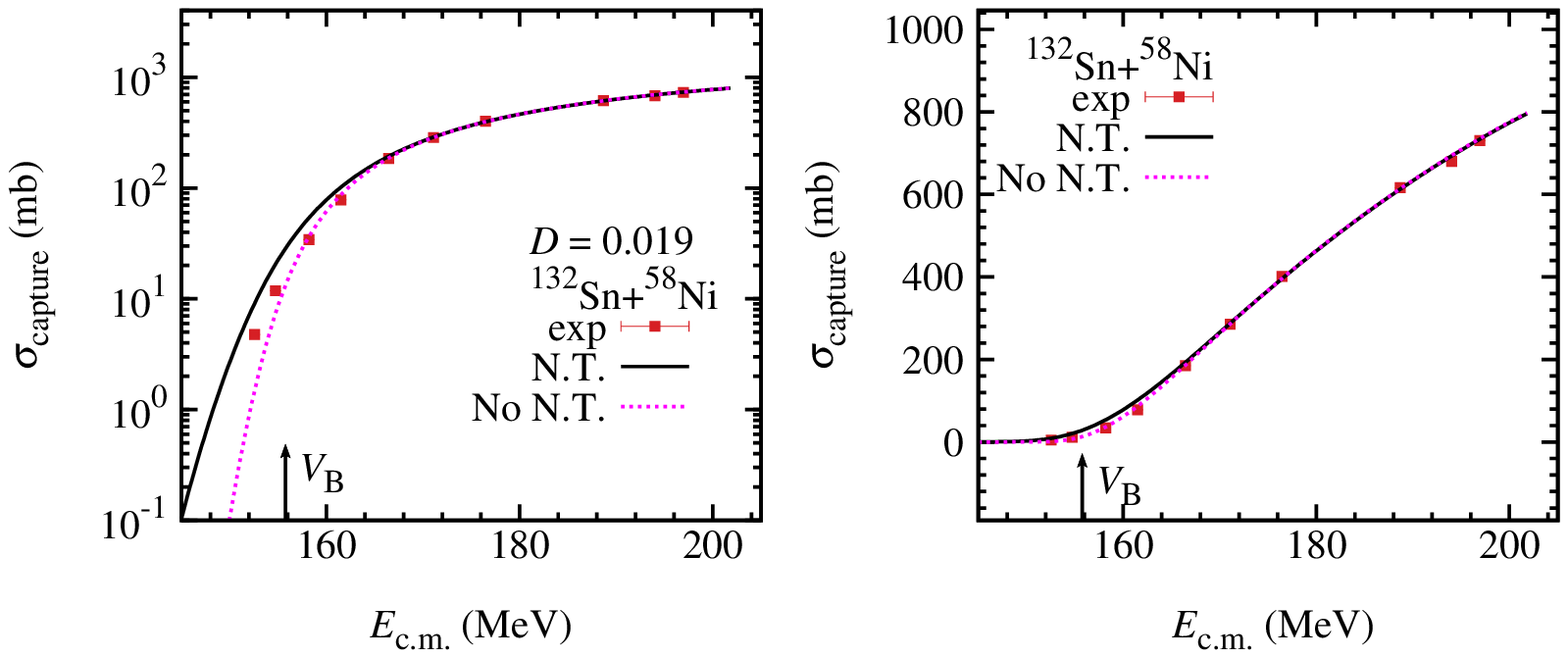}
\includegraphics[ width=0.47\textwidth]{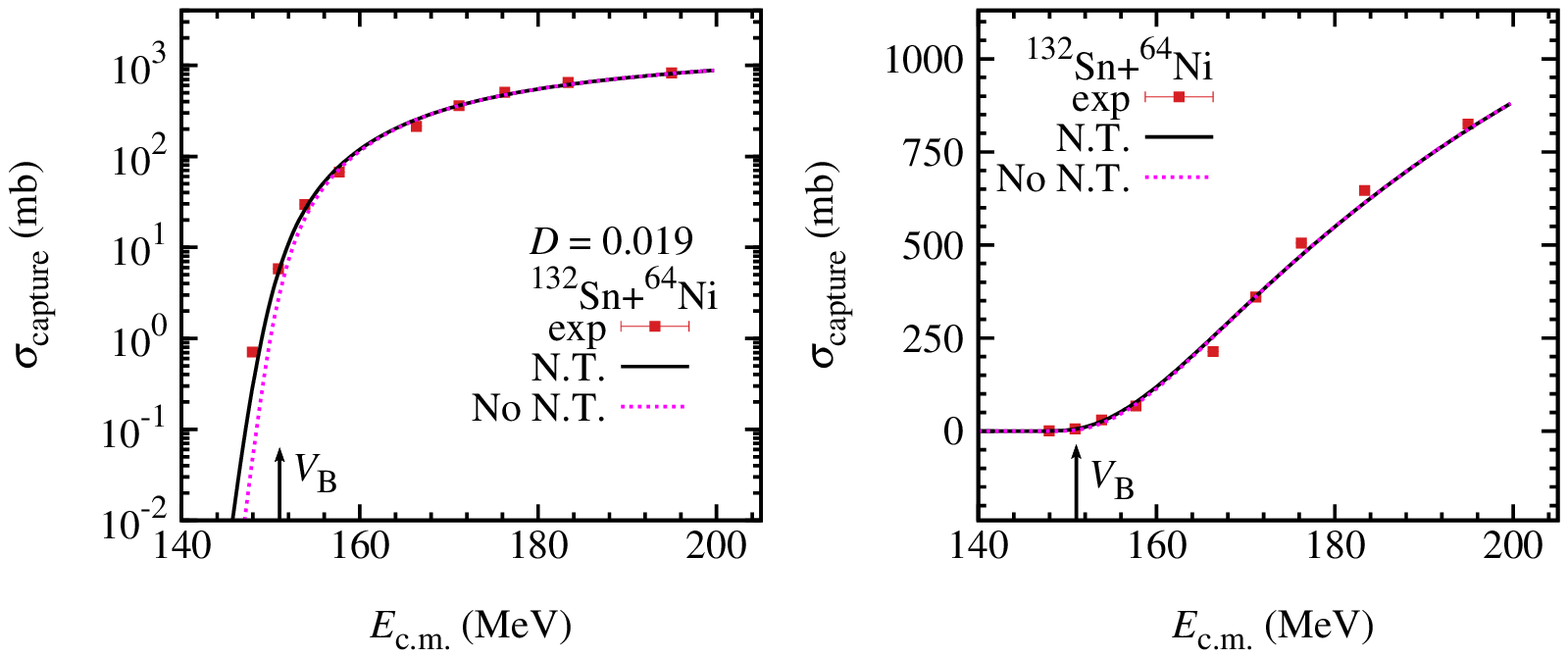}}
\centerline{\includegraphics[width=0.47\textwidth]{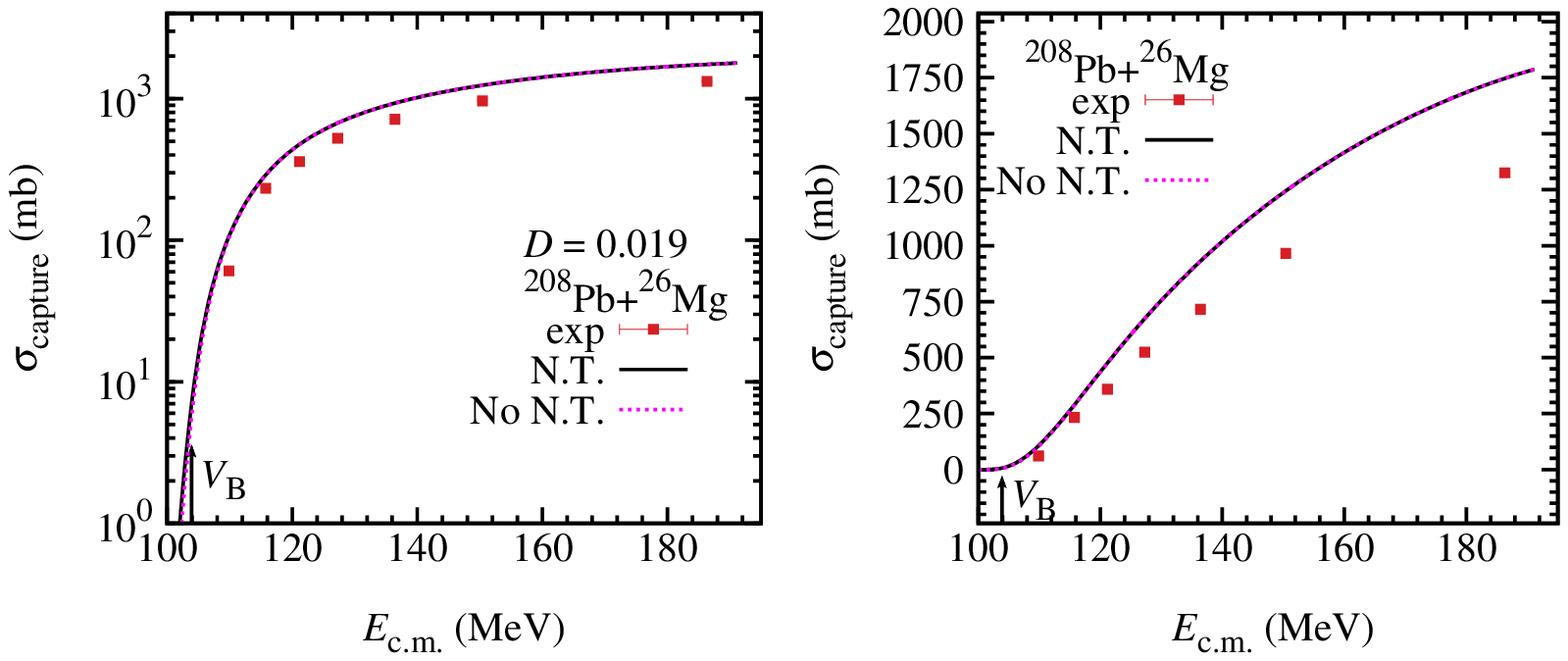}}
   \centerline{Graph 19}
 \end{Dfigures}

\clearpage
\TableExplanation
\label{TabExp}

\section*{Table 1. Fitted and calculated results for the parameters of
the barrier distribution and average deviations of fitted and calculated
cross sections from experimental values for a total of 220 reaction systems}

For these 220 reaction systems, we tabulate the ground state quadrupole
deformation parameters of
the projectile and the target used in calculating the potential, the position
and curvature of the potential approximated by a parabola,
the fitted and calculated results for the parameters of the barrier
distribution, the average deviations of fitted and calculated cross
sections from experimental values and the corresponding references, and the
$Q$ value for one neutron pair transfer.

\begin{center}
\begin{tabular}{ll}
 Reaction                  & Reaction system\\
 $\beta_{\rm P}$           & Ground state quadrupole deformation parameter of
the projectile  \\
 $\beta_{\rm T}$           & Ground state quadrupole deformation parameter of
the target\\
 $R_{\rm B}$               & Position of the barrier \\
 $\hbar\omega$             & Curvature of the parabolic barrier \\
 $B_{\rm m}^{\rm fit}$     & Fitted results for the central value of the
barrier distribution \\
 $\varDelta_1^{\rm f\/it}$ & Fitted results for the left width of the barrier
distribution  \\
 $\varDelta_2^{\rm f\/it}$ & Fitted results for the right width of the barrier
distribution  \\
 $\mathcal{D}^{\rm fit} $  & Average deviation of the fitted cross section from
data\\
 $B_{\rm m}^{\rm cal}$     & Calculated results for the central value of the
barrier distribution   \\
 $\varDelta_1^{\rm cal}$   & Calculated results for the left width of the
barrier
distribution \\
 $\varDelta_2^{\rm cal}$   & Calculated results for the right width of the
barrier
distribution\\
 $\mathcal{D}^{\rm cal}$   & Average deviation of the calculated cross section
from data \\
  $ Q(2n)$                 & $Q$ value obtained from experimental masses for the
one neutron pair transfer   \\
  Ref(s).                    &  References(s) where the data
are taken from \\
\end{tabular}
\end{center}

\section*{Table 2. The experimental and calculated excitation functions
for a total of 220 reaction systems}

For these 220 reaction systems, we tabulate the experimental excitation
function together with the corresponding references.
The detected particles in the experiment are the evaporation residual
(EvR), fission fragments (FF), and/or quasifission (QF).
There are two ways to extract the data:
``authors' graph'' denotes that the data are extracted from authors' graph
and
``authors' table'' denotes that the data are obtained from authors' table.
In the table,
we also indicate how the collision energy was given in the corresponding reference:
``$E_{\rm c.m.}$'' denotes that the collision energy was given in the center-of-mass frame
and then we calculate the energy in the laboratory frame by
$E_{\rm lab} = E_{\rm c.m.}(A_{\rm P}+A_{\rm T})/A_{\rm T}$;
``$E_{\rm lab}$'' denotes that the collision energy was given in the laboratory
frame and that in the center-of-mass frame is calculated by
$E_{\rm c.m.} = E_{\rm lab}A_{\rm T}/(A_{\rm P}+A_{\rm T})$;
``$E_{\rm lab}~\&~E_{\rm c.m.}$'' denotes that the collision energy was given
both in the laboratory frame and in the center-of-mass frame. The cross
section and its errors are in the unit of mb. The calculated cross section at
the corresponding energy is listed in the last column.

\begin{center}
\begin{tabular}{ll}
  Reaction            &   Reaction system \\
  Detected particles  &   Particles detected in the experiment\\
  Data obtained       &   Data taken from authors' table or extracted
 from authors' graph and the corresponding reference \\
 $E_{\rm lab}$         &   Collision energy in the laboratory frame \\
 $E_{\rm c.m.}$        &   Collision energy in the center-of-mass frame   \\
 $\sigma$              &   Cross section    \\
 $\delta\sigma$        &   Error of cross section     \\
$\sigma^{\rm cal}$     &  Calculated cross section  \\
\end{tabular}
\end{center}
\datatables
\LTright= 0pt
\LTleft = 0pt
\setlength{\tabcolsep}{0.5\tabcolsep}
\renewcommand{\arraystretch}{1.0}
\footnotesize
\begin{longtable}{@{\extracolsep\fill}lrrrrrrrcrrrrrc}
\caption{Fitted and calculated results for the parameters of the barrier
distribution and average deviations of fitted and calculated cross sections
from
experimental values for a total of 220 reaction systems. See
page~\pageref{TabExp} for Explanation of Tables.
\label{tab:total}}
 \multirow{2}{*}{Reaction} &
 \multirow{2}{*}{$\beta_{\rm P}$} &
 \multirow{2}{*}{$\beta_{\rm T}$} &
 $R_{\rm B}$ &
 $\hbar\omega$ &
 $B_{\rm m}^{\rm fit}$  &
 \multicolumn{1}{c}{$\varDelta_1^{\rm f\/it} $} &
 $\varDelta_2^{\rm f\/it} $  &
 \multirow{2}{*}{$\mathcal{D}^{\rm fit} $} &
 $B_{\rm m}^{\rm cal}$ &
 $\varDelta_1^{\rm cal}$  &
 $\varDelta_2^{\rm cal}$   &
 \multirow{2}{*}{$\mathcal{D}^{\rm cal}$} &
  $Q(2n)$ &
  \multirow{2}{*}{Ref(s).}\\
  &
  &
  &
  (fm)  &
  (MeV) &
  (MeV) &
 \multicolumn{1}{c}{(MeV)} &
  (MeV) &
  &
  (MeV) &
  (MeV) &
  (MeV) &
  &
 (MeV)  &  \\
\hline\noalign{\vskip3pt}
\endfirsthead
\multicolumn{15}{@{}c@{}}{Table \ref{tab:total} (continued)}\\[6pt]
 \hline\noalign{\vskip3pt}
 \multirow{2}{*}{Reaction} &
 \multirow{2}{*}{$\beta_{\rm P}$} &
 \multirow{2}{*}{$\beta_{\rm T}$} &
 $R_{\rm B}$ &
 $\hbar\omega$ &
 $B_{\rm m}^{\rm fit}$  &
 \multicolumn{1}{c}{$\varDelta_1^{\rm f\/it} $} &
 $\varDelta_2^{\rm f\/it} $  &
 \multirow{2}{*}{$\mathcal{D}^{\rm fit} $} &
 $B_{\rm m}^{\rm cal}$ &
 $\varDelta_1^{\rm cal}$  &
 $\varDelta_2^{\rm cal}$   &
 \multirow{2}{*}{$\mathcal{D}^{\rm cal}$} &
  $Q(2n)$ &
  \multirow{2}{*}{Ref(s).}\\
  &
  &
  &
  (fm)  &
  (MeV) &
  (MeV) &
 \multicolumn{1}{c}{(MeV)} &
  (MeV) &
  &
  (MeV) &
  (MeV) &
  (MeV) &
  &
 (MeV)  &  \\ [3pt]
 \hline\noalign{\vskip3pt}
\endhead
\hline
\noalign{\vskip3pt}
\multicolumn{15}{r}{\itshape Continued on next page}\\
\endfoot
\noalign{\vskip3pt}\hline
\endlastfoot
$^{ 12}{\rm C }$+$^{ 89}{\rm Y }$ & $-$0.320 &  0.035 & 10.53 &  4.18 &  31.46 &
 0.254 &  0.555 & 0.0362 &  30.64 &  0.182 &  1.700 &  0.101 &   $-$7.71 &
\cite{Palshetkar2010_PRC82-044608}  \\
$^{ 12}{\rm C }$+$^{ 92}{\rm Zr}$ & $-$0.320 &  0.053 & 10.65 &  4.19 &  31.24 &
 0.005 &  2.682 & 0.0073 &  31.19 &  0.202 &  1.882 &  0.021 &   $-$2.71 &
\cite{Newton2001_PRC64-064608}  \\
$^{ 12}{\rm C }$+$^{144}{\rm Sm}$ & $-$0.320 &  0.000 & 11.10 &  4.63 &  45.76 &
 0.028 &  1.856 & 0.0044 &  45.62 &  0.184 &  1.713 &  0.018 &   $-$6.00 &
\cite{Abriola1992_PRC46-244}  \\
$^{ 12}{\rm C }$+$^{152}{\rm Sm}$ & $-$0.320 &  0.243 & 12.25 &  4.39 &  47.15 &
 3.984 &  6.186 & 0.0001 &  43.79 &  0.691 &  6.453 &  0.011 &   $-$0.73 &
\cite{Broda1975_NPA248-356}  \\
$^{ 12}{\rm C }$+$^{154}{\rm Sm}$ & $-$0.320 &  0.270 & 12.40 &  4.36 &  42.97 &
 0.005 &  8.038 & 0.0044 &  43.62 &  0.788 &  7.358 &  0.005 &   $-$0.71 &
\cite{Gil1985_PRC31-1752}  \\
$^{ 12}{\rm C }$+$^{181}{\rm Ta}$ & $-$0.320 &  0.269 & 12.72 &  4.58 &  49.00 &
 0.007 & 13.890 & 0.0043 &  49.81 &  0.820 &  7.652 &  0.016 &   $-$1.10 &
\cite{Crippa1994_ZPA350-121}  \\
$^{ 12}{\rm C }$+$^{194}{\rm Pt}$ & $-$0.320 &  0.000 & 11.66 &  4.90 &  53.67 &
 0.106 &  4.015 & 0.0012 &  54.48 &  0.220 &  2.055 &  0.007 &   $-$1.49 &
\cite{Shrivastava2001_PRC63-054602}  \\
$^{ 12}{\rm C }$+$^{198}{\rm Pt}$ & $-$0.320 &  0.000 & 11.71 &  4.88 &  54.01 &
 0.044 &  3.264 & 0.0017 &  54.29 &  0.217 &  2.025 &  0.008 &   $-$0.28 &
\cite{Shrivastava2001_PRC63-054602}  \\
\end{longtable}
\clearpage
\LTright= 0pt
\LTleft = 0pt
\footnotesize
\begin{longtable}{@{\extracolsep\fill}lcc
D{.}{.}{-1} D{.}{.}{-1}D{.}{.}{-1}D{.}{.}{ -1}D{.}{.}{ -1}}
\caption{The experimental and calculated excitation functions for a total of
220
reaction systems. See page~\pageref{TabExp} for Explanation of Tables.}
\label{tab:data}
 Reaction  &
 Detected  particles &
 Data obtained &
 \multicolumn{1}{c}{$E_{\rm lab}$} &
 \multicolumn{1}{c}{$E_{\text{c.m.}}$} &
 \multicolumn{1}{c}{$\sigma$}  &
 \multicolumn{1}{c}{$\delta\sigma$} &
 \multicolumn{1}{c}{$\sigma^{\rm cal}$} \\
 &
 &
 &
  \multicolumn{1}{c}{(MeV)} &
  \multicolumn{1}{c}{(MeV)} &
  \multicolumn{1}{c}{(mb)} &
 \multicolumn{1}{c}{(mb)} &
 \multicolumn{1}{c}{(mb)}
   \\
\hline\noalign{\vskip3pt}
\endfirsthead
\multicolumn{8}{c}{Table \ref{tab:data} (continued)}\\[6pt]
 \hline\noalign{\vskip3pt}
Reaction  &
 Detected  particles &
 Data obtained &
 \multicolumn{1}{c}{$E_{\rm lab}$} &
 \multicolumn{1}{c}{$E_{\text{c.m.}}$} &
 \multicolumn{1}{c}{$\sigma$}  &
 \multicolumn{1}{c}{$\delta\sigma$} &
 \multicolumn{1}{c}{$\sigma^{\rm cal}$} \\
 &
 &
 &
  \multicolumn{1}{c}{(MeV)} &
  \multicolumn{1}{c}{(MeV)} &
  \multicolumn{1}{c}{(mb)} &
 \multicolumn{1}{c}{(mb)} &
 \multicolumn{1}{c}{(mb)}  \\ [3pt]
 \hline\noalign{\vskip3pt}
\endhead
\hline
\noalign{\vskip3pt}
\multicolumn{8}{r}{\itshape Continued on next page}\\
\endfoot
\noalign{\vskip3pt}\hline
\endlastfoot
$^{ 12}{\rm C }$+$^{ 89}{\rm Y }$  &  EvR      &  authors' graph  &       30.682
  &       27.037   &             0.033   &   \multicolumn{1}{c}{ \textemdash}
&         0.195 \\
   &     &  \cite{Palshetkar2010_PRC82-044608}  &       32.403   &       28.554
 &             0.668   &   \multicolumn{1}{c}{ \textemdash}   &         1.697 \\
   &     &  $E_{\rm lab}$  &       33.789   &       29.774   &             4.757
  &   \multicolumn{1}{c}{ \textemdash}   &         9.135 \\
   &     &     &       35.387   &       31.183   &            29.865   &
\multicolumn{1}{c}{ \textemdash}   &        47.223 \\
   &     &     &       36.912   &       32.526   &            96.919   &
\multicolumn{1}{c}{ \textemdash}   &       132.929 \\
   &     &     &       39.174   &       34.520   &           287.692   &
\multicolumn{1}{c}{ \textemdash}   &       302.095 \\
   &     &     &       41.019   &       36.145   &           369.296   &
\multicolumn{1}{c}{ \textemdash}   &       434.894 \\
   &     &     &       43.207   &       38.074   &           597.752   &
\multicolumn{1}{c}{ \textemdash}   &       575.375 \\
\end{longtable}

\end{document}